\documentclass[a4 paper,11pt]{article}
\usepackage{fancyhdr}
\usepackage{authblk}
\usepackage{latexsym}
\usepackage{mathrsfs}
\usepackage{cancel}
\usepackage{braket}
\usepackage{color}
\usepackage{multirow}
\usepackage{eso-pic}
\usepackage{amssymb}
\usepackage{graphicx}
\usepackage{setspace}
\usepackage{geometry}
\usepackage{enumitem}
\usepackage{amsthm}
\usepackage{hyperref}
\usepackage[intlimits]{amsmath}
\usepackage{hyperref}
\usepackage{amssymb,amsfonts}
\usepackage[all,arc]{xy}
\usepackage{mathrsfs}
\usepackage{bbm}
\usepackage{tikz-cd}
\usepackage{dsfont}
\usepackage{tikz}
\usetikzlibrary{decorations.pathmorphing}
\usetikzlibrary{shapes.geometric}
\usetikzlibrary{decorations.markings}
\usetikzlibrary{patterns,patterns.meta}
\usepgflibrary{patterns}
\usepackage{scalerel}
\usepackage{mathtools}
\usepackage{bm}

\graphicspath{{./figures/}}
\numberwithin{equation}{section}

\newcounter{sarrow}

\tikzstyle arrowstyle=[scale=1]
\tikzstyle directed=[postaction={decorate,decoration={markings,
    mark=at position .25 with {\arrowreversed[arrowstyle]{stealth}}}}]

\tikzdeclarepattern{
name=mylines,
parameters={
\pgfkeysvalueof{/pgf/pattern keys/size},
\pgfkeysvalueof{/pgf/pattern keys/angle},
\pgfkeysvalueof{/pgf/pattern keys/line width},
},
bounding box={
(0,-0.5*\pgfkeysvalueof{/pgf/pattern keys/line width})and
(\pgfkeysvalueof{/pgf/pattern keys/size},
0.5*\pgfkeysvalueof{/pgf/pattern keys/line width})},
tile size={(\pgfkeysvalueof{/pgf/pattern keys/size},
\pgfkeysvalueof{/pgf/pattern keys/size})},
tile transformation={rotate=\pgfkeysvalueof{/pgf/pattern keys/angle}},
defaults={
size/.initial=5pt,
angle/.initial=45,
line width/.initial=.4pt,
},
code={
\draw[line width=\pgfkeysvalueof{/pgf/pattern keys/line width}]
(0,0)-- (\pgfkeysvalueof{/pgf/pattern keys/size},0);
},
}

\usepackage{hyperref}

\newcommand{\ds}{\displaystyle}

\newcommand{\mb}{\mathbf}
\newcommand{\cf}{\emph{cf.}}
\newcommand{\ie}{\emph{i.e.}}
\newcommand{\eg}{\emph{e.g.}}
\newcommand{\ol}{\overline}
\newcommand{\wt}{\widetilde}

\newcommand{\Mod}{\text{-mod}}

\newcommand{\lp}{\left(}
\newcommand{\rp}{\right)}

\newcommand{\bmu}{\bm{\mu}}

\newcommand{\pd}{{\partial}}
\newcommand{\id}{{\mathds{1}}}

\newcommand{\Ba}{{\mathbf a}}
\newcommand{\Bb}{{\mathbf b}}
\newcommand{\Bc}{{\mathbf c}}
\newcommand{\BA}{{\mathbf A}}
\newcommand{\BB}{{\mathbf B}}

\newcommand{\BX}{{\mathbf X}}
\newcommand{\BY}{{\mathbf Y}}
\newcommand{\BZ}{{\mathbf Z}}

\newcommand{\BPsi}{{\mathbf \Psi}}
\newcommand{\BPhi}{{\mathbf \Phi}}

\newcommand{\C}{\mathbb C}
\newcommand{\R}{\mathbb R}
\newcommand{\Z}{\mathbb Z}
\newcommand{\Q}{\mathbb Q}

\newcommand{\N}{\mathbb N}

\newcommand{\V}{\mathbb{V}}
\newcommand{\W}{\mathbb{W}}
\newcommand{\cp}{\mathbb{CP}}

\newcommand{\fg}{\mathfrak{g}}
\newcommand{\fh}{\mathfrak{h}}

\newcommand{\CA}{{\mathcal A}}
\newcommand{\CB}{{\mathcal B}}
\newcommand{\CC}{{\mathcal C}}

\newcommand{\CF}{{\mathcal F}}

\newcommand{\CI}{{\mathcal I}}

\newcommand{\CK}{{\mathcal K}}
\newcommand{\CL}{{\mathcal L}}

\newcommand{\CN}{{\mathcal N}}
\newcommand{\CO}{{\mathcal O}}

\newcommand{\CT}{{\mathcal T}}

\newcommand{\CV}{{\mathcal V}}

\newcommand{\CY}{{\mathcal Y}}

\newcommand{\be}{\begin{equation}}
\newcommand{\ee}{\end{equation}}

\newcommand{\btik}{\begin{tikzcd}}
\newcommand{\etik}{\end{tikzcd}}

\newtheorem{Def}{Definition}[section]
\newtheorem{Thm}[Def]{Theorem}
\newtheorem{Conj}[Def]{Conjecture}
\newtheorem{Prop}[Def]{Proposition}

\title{Line Operators in 3d Holomorphic QFT: Meromorphic Tensor Categories and dg-Shifted Yangians}
\author[1]{Tudor Dimofte}
\author[2]{Wenjun Niu}
\author[1]{Victor Py}
\affil[1]{School of Mathematics and Maxwell Institute for Mathematical Sciences, University of Edinburgh, Edinburgh EH9 3FD, UK}
\affil[2]{Perimeter Institute for Theoretical Physics, Waterloo, ON N2L 2Y5, Canada}
\affil[ ]{\emph{tdimofte@gmail.com, wniu@perimeterinstitute.ca, victor.pycaimi@gmail.com}}
\date{\today}
\begin{document}

\maketitle
\begin{abstract}
We study line operators and their OPE's in perturbative 3d holomorphic-topological QFT's, including holomorphic-topological twists (quarter-BPS sectors) of 3d $\mathcal N=2$ theories. In particular, we develop the representation theory of the category $\mathcal C$ of perturbative line operators and its chiral tensor product, by generalizing techniques introduced by Costello and collaborators. 
We argue that lines are equivalent to modules for an $A_\infty$ algebra $\mathcal A^!$ that's Koszul-dual to bulk local operators. We further establish a non-renormalization theorem for the OPE's of lines in a large class of theories (dubbed quasi-linear), allowing an exact resummation of quantum corrections. Based on physics arguments, we propose axioms for the full algebraic~structure~on~$\CA^!$, calling it a ``dg-shifted Yangian,'' which controls the OPE of lines. A key part of the structure is a Maurer-Cartan element $r(z)\in \mathcal A^!\otimes \mathcal A^!(\!(z^{-1})\!)$ that satisfies an $A_\infty$ generalization of the Yang-Baxter equation. 
As examples, we consider 3d $\mathcal N=2$ gauge theories with arbitrary Chern-Simons levels, linear matter, and superpotential, and explicitly compute 1) perturbative bulk local operators (as $A_\infty$-chiral algebras); and 2) the Koszul-duals $\mathcal A^!$ (proving they are dg-shifted Yangians).
\end{abstract}

\newpage

\tableofcontents

\newpage

\section{Introduction} \label{sec:intro}

Holomorphic-topological (HT) QFT's are field theories defined on spacetimes locally of the form $\C^m\times \R^n$, whose correlation functions and underlying operator algebras have a hybrid local dependance on position: depending holomorphically on $\C^m$ coordinates, and being locally independent of (\ie\ depending topologically on) $\R^n$ coordinates. 
Globally, they may be defined on $(2m+n)$-manifolds with holomorphic foliations. Notable examples in 4d, on $\C\times \R^2$, include the HT twist (a particular BPS sector) of 4d $\CN=2$ theories \cite{Kapustin-holomorphic} and 4d Chern-Simons theory \cite{Costello-Yangian}. In 3d, every twist (or minimal BPS sector) of a 3d $\CN=2$ theory defines a HT QFT on $\C\times \R$ \cite{CDFKS,ACMV,CDGbdry}. Another class of  3d examples includes HT analogues of Poisson sigma-models \cite{KhanZeng}. The 3d HT twist can also be thought of as a lift of the ``half twist'' of 2d $\CN=(2,2)$ models \cite{Witten-mirror}.
3d HT QFT's also play a key role in celestial holography  \cite{CP-celestial,Zeng:2023qqp, Garner:2023izn}.

Recent years have seen many developments in the formal structure of HT theories, both physically and mathematically.
For example, it was shown that potential obstructions to perturbative quantization of a classical theory vanish as long as $n\geq 1$ \cite{GRW-quantization}. It was also shown that an analogue of the Kontsevich formality theorem holds and renders the perturbative algebra of local operators one-loop exact and free of nontrivial higher operations as long as $n\geq 2$ \cite{GRW-quantization, GaiottoKulpWu}.

Two important cases of the algebra $\CA$ of local operators are:
\begin{itemize}[leftmargin=*]
\item $m=1,n=2$ (locally $\C\times \R^2)$:  $\CA$ is an ($E_2$,chiral) algebra, which implies that its cohomology is a commutative vertex algebra, with a $\lambda$-bracket of cohomological degree (or ghost number) $-2$ \cite{OhYagi-Poisson}. In a twisted 4d $\CN=2$ superconformal theory, it was shown by \cite{OhYagi-Omega,Jeong-Omega} that $\CA$ ``quantizes'' to the 4d VOA's of \cite{BLLPRR}; its graded character computes the 4d Schur index \cite{KMMR,GRRW-Schur}.

\item $m=1,n=1$ (locally $\C\times \R$): $\CA$ is an ($E_1$,chiral) algebra, which implies that its cohomology is a commutative vertex algebra, with a $\lambda$-bracket of degree $-1$ \cite{OhYagi-Poisson}. Some of the first nontrivial examples were worked out in \cite{CDGbdry, Zeng-monopole}. 
In a twisted 3d $\CN=2$ theory, the graded character of $\CA$ computes its 3d index \cite{Kim-index,IY-index,KSV-index,KW-index}.
A general geometric perspective on such algebras, based on ``raviolo'' moduli spaces, was developed in \cite{GarnerWilliams}. A feature of the $n=1$ case that we will (by necessity) develop in this paper is that, in any interacting theory, $\CA$ is also endowed with higher $A_\infty$-like operations.%
\footnote{The higher operations should in principle be governed geometrically by a chain-level model of sheaves on the raviolo, as discussed in \cite{AlfonsiKimYoung}, though this has not yet been made fully concrete.}
\end{itemize}

Our main goal is to move beyond local operators, and initiate a study of \emph{line operators} in 3d HT theories, focusing on their local structure and associated algebraic properties. Since we are just interested in local structure, we'll fix spacetime to be $\C_z\times \R_t$. Lines operators extended along $\R_t$, placed at a point $z\in \C_z$, are expected to form an $A_\infty$ category $\CC$, whose morphisms are defined as spaces of local operators at junctions:
\be \raisebox{-.4in}{\includegraphics[width=4.2in]{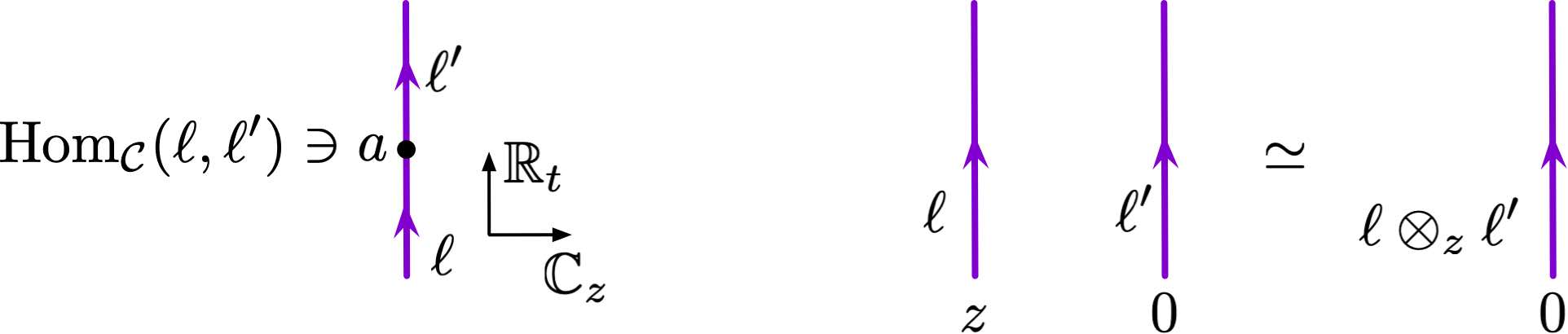}} \label{LineIntro} \ee
Moreover, the OPE of lines at nearby points should define a chiral tensor product in $\CC$, \ie\ a coherent family of tensor products $\otimes_z :\CC\boxtimes\CC\to\CC$ parameterized by $z\in \C^*$. Thus, roughly speaking, $\CC$ should be the categorical analogue of a VOA: the extra topological `time' direction $\R_t$ on which lines are extended increases the category level.

The main results of this paper, outlined in Sec. \ref{sec:perp-intro}, provide a concrete and computationally effective definition of the category $\CC$, including its OPE, in any perturbative 3d HT QFT satisfying some mild constraints --- including any HT-twisted 3d $\CN=2$ gauge theory, with Chern-Simons couplings, matter, and superpotential. Our approach generalizes the methods introduced by Costello and collaborators to study perturbative defects in QFT \cite{Costello-Yangian,CWY-I,CPkoszul}, reviewed in \cite{PaquetteWilliams}. A key feature of this approach is the emergence of an $A_\infty$ algebra $\CA^!$, Koszul-dual to local operators, such that
\be \CC\simeq \CA^!\text{-mod}\,. \label{A!-mod-intro} \ee
As we determine $\CA^!$, we'll find that in 3d HT QFT's it carries a new sort of algebraic structure that controls the OPE in $\CC$; we call this structure a \emph{dg-shifted Yangian}, by analogy with the (generalized) Yangians that appear in 4d HT QFT's \cite{Costello-Yangian, CWY-I, CWY-II, CautisWilliams, ANGrassmannian}.

Mathematically, categories like $\CC$ with a chiral tensor product have been formalized in several different ways. The formalism closest to our approach, and to the physical idea of OPE's, is Soibelman's notion of a \emph{meromorphic tensor category} \cite{SoibMero}, introduced in the 90's. This was directly motivated by representation theory of Yangians. A closely related notion is that of a \emph{chiral category}, also known as a \emph{factorization category}, which arose during the development of the geometric Langlands program \cite{Gaitsgory-Whittaker,Raskin-CPSI}, and later formalized in \cite{Gaitsgory-chiral, Raskin-chiral}. Chiral categories are not a single category, but rather a sheaf of categories defined over configuration spaces of points on an algebraic curve; the OPE $\otimes_z$ may (in principle) be reconstructed from this data. Chiral categories are to meromorphic tensor categories what Beilinson-Drinfeld's chiral algebras \cite{beilinson2004chiral} are to VOA's. We won't use chiral categories in this paper, though they should provide an alternative, conceptually powerful framework.%
\footnote{See \cite{HilburnRaskin} for a construction of line operators in topologically twisted 3d $\CN=4$ theories using the perspective of chiral categories.}

From a physics perspective, if a 3d HT theory is the twist of a 3d $\CN=2$ QFT, the objects of the category $\CC$ are just the half-BPS line operators in the $\CN=2$ theory. These have been widely studied in many contexts over the past two decades, albeit not (to our knowledge) as a full category with OPE's.
As codimension-two defects, half-BPS lines of 3d $\CN=2$ gauge theories are closely related to Gukov-Witten defects of 4d Yang-Mills \cite{GukovWitten}, and their ``wild'' generalizations \cite{Witten-wild}. BPS Wilson lines in 3d $\CN=2$ gauge theory were constructed in \cite{GaiottoYin}; and a basic set of vortex lines, induced from brane intersections, in \cite{DrukkerGomisYoung}. Some of the first computations of partition functions on $S^3$ and $S^2\times S^1$ with line operators inserted were performed in \cite{KWY,KWY-abelian,DrukkerOkudaPasserini}, later generalized to Seifert manifolds \cite{Closset:2012ru,ClossetKim}. In terms of our category $\CC$, the $S^2\times S^1$ partition functions compute characters of morphism spaces:%
\be \raisebox{-.7in}{\includegraphics[width=3.8in]{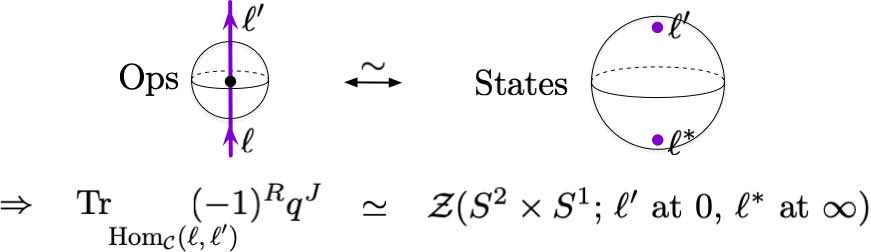}}  \label{S2-morph} \ee
This relation follows from a state-operator correspondence; \cf\ \cite{GangKohLee,CordovaGaiottoShao} for an analogous construction in 4d.

Another key observable related to BPS lines in a 3d $\CN=2$ theory is the algebra $\CK$ of loop operators wrapped on the fiber of a twisted $\C\times S^1$ compactification. 
The central role of $\CK$ in gauge theory was brought to light in the Bethe-gauge correspondence \cite{NS-3d}, where $\CK$ is computed as the Jacobian ring of an effective twisted superpotential, and identified with Bethe vacua.
In a sigma-model (\eg\ the IR limit of a gauge theory), $\CK$ becomes the quantum K-theory ring \cite{GiventalLee} of the target. The relation between quantum K-theory classes and loop operators was developed in \cite{JockersMayr,JMNT-algebras,ClossetKhlaif}, and is connected to the action of K-theory on elliptic stable envelopes \cite{AganagicOkounkov-ESE,AganagicOkounkov-quasimap}. We expect that $\CK$ is the cyclic Hochschild homology of our line-operator category $\CC$,
\be \CK \simeq HH_\bullet^{S^1}(\CC) \label{K-thy} \ee
with a product induced from the OPE in $\CC$. We hope to explore the connection between $\CK$ and $\CC$ in future work; establishing an isomorphism as in \eqref{K-thy} will likely require a fully nonperturbative description of $\CC$.

We finally note that an important motivation behind the initial work on 4d HT Chern-Simons and the Yangian \cite{Costello-Yangian, Costello:2013sla, CWY-I, CWY-II} was to establish new connections between HT QFT's and integrability. These connections have been expanded in vast directions in recent years, most involving string theory and/or higher-dimensional QFT's and defects. A reduction of these ideas to 3d HT theories appeared in \cite{Vicedo:2022mrm}, who showed that the state space of HT-twisted pure gauge theory with line operators reproduces the phase space of the finite Gaudin model. It would be very interesting to incorporate the categories and shifted Yang-Baxter equation of this paper in the Gaudin model and related integrable systems; we won't explore~this~here.

\medskip

We'll now describe in greater detail our methods and results.

\subsection{Structure and results}
\label{sec:perp-intro}

The main perspective on line operators used in this paper is to construct them by coupling bulk 3d HT QFT to 1d topological quantum mechanics.
This is a classic method for defining a large class of line operators. For example, Wilson lines are typically represented by coupling a bulk gauge connection to the state space of a 1d massive particle. Even disorder operators can often be represented this way, by using the coupling to 1d degrees of freedom to source a desired singularity in the bulk fields. In brane constructions of line operators, the 1d quantum mechanics is supplied directly by open-string states on brane intersections.
We will exclusively focus on a perturbative sector of 3d HT QFT's, where, by definition, \emph{every} line operator $\ell\in \CC$ admits a description as coupling to some quantum mechanics. 

In Section \ref{sec:setup}, we'll describe more precisely what we mean by a perturbative 3d HT QFT. We'll work in the BV-BRST formalism, which is well adapted to capture the field and operator content in an HT theory, naturally organizing fields into holomorphic-topological descendants. We'll review, following \cite{ACMV,CDGbdry}, how twisted 3d $\CN=2$ theories look in this formalism. We'll also review 1d topological quantum mechanics in a similar vein. Structurally, we will recall that in quantum mechanics the state space $V$ is a dg vector space and the operator algebra $\text{End}(V)$ is a dg algebra; and that interaction terms in the QM action are in 1-1 correspondence with Maurer-Cartan elements $\mu \in \text{End}(V)$, \ie\ $\mu$ of cohomological degree 1 satisfying $Q(\mu)+\mu^2 = 0$.

In Section \ref{sec:lines-QM}, we then couple 1d and 3d, following \cite{Costello-Yangian,CPkoszul,PaquetteWilliams}, to produce a uniform description of line operators. Given a 3d HT QFT $\CT$ with an algebra of bulk local operators $\CA$, we define objects $\ell\in \CC$ to be pairs $(V_\ell,\mu_\ell)$ where $V_\ell$ is a dg vector space (the state space of some 1d QM) and 
\be \label{bulk-MC} \mu_\ell \in \text{End}(V_\ell)\otimes \CA\,,\qquad \sum_{k=1}^\infty m_k(\underbrace{\mu_\ell,...,\mu_\ell}_k) = 0 \ee
is a Maurer-Cartan element in the product $\text{End}(V_\ell)\otimes \CA$, now satisfying the $A_\infty$ MC equation. We'll give some examples of how this translates to more familiar descriptions, including vortex lines for matter with and without superpotentials, and Wilson lines in gauge theories. 

The trickiest part of this description of line operators is understanding how and why bulk local operators $\CA$ form an $A_\infty$ algebra --- to feed into \eqref{bulk-MC}. The products we're talking about come from  bulk local operators placed along a fixed line in the $\R_t$ direction, colliding with each other. In $Q$-cohomology, the product should be associative, as well as commutative, due to the existence of the transverse $\C_z$ direction. However, associativity (and commutativity) only hold up to $Q$-exact terms, which are encoded in the aforementioned $A_\infty$ operations. Some of these operations were already observed in \cite{CDGbdry}, and a general scheme to compute them in perturbative HT theories was developed in \cite{GaiottoKulpWu}. 

We will ultimately compute $\CA$ explicitly as an $A_\infty$ algebra (in fact, an ($A_\infty$,chiral) algebra) in any perturbative HT theory in Section \ref{sec:Ainf}. We defer this until later in the paper because the analysis is a bit technical, and mostly not needed for discussing OPE's. We do the computation in Sec. \ref{sec:Ainf} by first reducing the 3d theory to a 2d B-model along a holomorphic cigar $\C_z$, and analyzing 2d Feynman diagrams. We introduce a mild constraint on bulk interaction terms, called \emph{quasi-linearity} (Sec. \ref{sec:quasi-linear}), which is satisfied by all HT-twisted 3d $\CN=2$ gauge theories with linear matter, superpotential, and CS terms. We'll argue that 

\medskip
\noindent \textbf{Proposition \ref{prop:Ainf}} \textit{In any perturbative quasi-linear 3d HT theory, both local operators $\CA$ and its Koszul dual $\CA^!$ (which we'll come to later) are determined exactly at tree level, with $A_\infty$ operations computed by derivatives of the 3d interaction terms.} 

\medskip
\noindent
For example, for 3d $\CN=2$ matter with superpotential $W$, $\CA$ turns out to be a chiral enhancement of the derived critical locus of $W$, \cf\ \eqref{dW-intro} below.

An important consequence of the structure of $A_\infty$ algebras in perturbative HT QFT's is that any line operator can be represented by \emph{linear} couplings to bulk local operators, \ie\ $\mu_\ell = \sum_i M^i \CO_i$ with $M^i\in \text{End}(V_\ell)$ and $\CO_i$ a linear functional of the bulk fields. (See Sec. \ref{sec:KD-MC}.) We'll assume linear couplings throughout earlier sections of the paper to simplify our analysis.

In Section \ref{sec:OPE}, we move on to OPE's. We consider a configuration of two operators $\ell(z), \ell'(0)$, placed at nearby points $z,0\in \C_z$, and define their OPE $\ell\otimes_z\ell'(0)$ by asking what line operator at $0\in\C_z$ would behave the same way in all correlation functions. We compute $\ell\otimes_z\ell'(0)$ explicitly by summing quantum corrections to the insertions of their couplings to 1d QM, and thus establish a key non-renormalization theorem:

\medskip
\noindent \textbf{Theorem \ref{thm:non-ren}} \textit{In a quasi-linear 3d HT QFT, if $\ell,\ell'$ are defined by (linear) MC elements $\mu_\ell,\mu_{\ell'}$, then the singular part of the OPE is given by a new MC element}
\be \label{OPE-intro}  \begin{array}{c} \ell(z) \;\leftrightarrow\; (V_\ell,\mu_\ell(z))\,,\qquad \ell'(0) \;\leftrightarrow\; (V_{\ell'},\mu_{\ell'}(0)) \\[.2cm]
\hspace{-1cm}\Rightarrow\qquad \ell\otimes_z\ell'(0) \;\leftrightarrow\; \big(V_\ell\otimes V_{\ell'},\, \mu_\ell(z)+\mu_{\ell'}(0)+r_{\ell,\ell'}(z)\big)\, \end{array} \ee
\textit{where $r_{\ell,\ell'}(z)$ is 1) given exactly by a sum of tree-level ``ladder'' diagrams, which are independent of bulk interactions; 2) independent of bulk fields; and 3) in axial gauge, an element of $\text{End}(V)\otimes \text{End}(V_{\ell'})[z^{-1}]$.
}

\medskip\noindent
In other words, the quantum correction to the OPE of lines is encoded in a correction $r_{\ell,\ell'}(z)$ to the naive MC element $\mu_\ell(z)+\mu_{\ell'}(0)$, which takes the same universal form independent of bulk interactions --- it can be computed in free field theory.

The remainder of the paper is devoted to working out the \emph{representation theory} underlying the category $\CC$ and its OPE, by applying and generalizing the Koszul-duality methods of \cite{Costello-Yangian,CPkoszul,PaquetteWilliams}. We frame the results from the perspective of Tannakian reconstruction in QFT, as reviewed in \cite{sparks}. (The Tannakian perspective also underlay the construction of the Yangian from Koszul duality in \cite{Costello-Yangian}.) 

In Section \ref{sec:Koszul}, for each perturbative 3d HT QFT, we introduce an $A_\infty$ algebra $\CA^!$ that's the mathematical Koszul dual of bulk local operators $\CA$. It's constructed physically by placing the 3d theory on a compact spatial geometry $\cp^1_z$, with a mild singularity `$\CB_\infty$' at $z=\infty$ that trivializes the state space, so $\text{States}(\cp^1,\CB_\infty)\simeq \C$. Then we define $\CA^!$ to be local operators supported at $z=\infty$: 
\be \CA^! :=\begin{array}{l} \text{$A_\infty$ algebra of local ops supported} \\ \text{on the `defect' $\CB_\infty$ at $z=\infty$} \end{array} \qquad \raisebox{-.5in}{\includegraphics[width=1.5in]{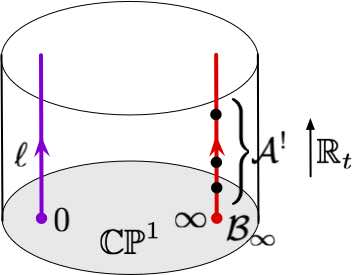}}  \ee
We argue that there is a faithful representation of the category of lines 
\be\label{eq:fiberF} \wt\CF:\CC \overset\sim\rightarrow \CA^!\text{-mod}\,,\qquad \wt\CF(\ell)=\text{States}(\cp^1,\ell(0),\CB_\infty) = V_\ell \ee
that sends each line $\ell$ to the vector space $V_\ell$ used to construct it by coupling to QM. We then seek to reconstruct algebraic structures on $\CA^!$ that encode the OPE of lines in $\CC$, ultimately leading (Section \ref{sec:KD-Yangian}) to the definition of a generalized \emph{dg-shifted Yangian}.

We say that $\CY$ is a dg-shifted Yangian if it's (roughly) 
\begin{itemize}
\item an $A_\infty$ algebra, triply graded by ghost number (cohomological degree), spin (loop rotation/conformal degree), and fermion number (parity), with
\item translation isomorphisms $\tau_z:\CY\to \CY[z]$ satisfying $\tau_z\circ\tau_w=\tau_{z+w}$,
\item a Maurer-Cartan element $r(z)\in \CY\otimes \CY(\!(z^{-1})\!)$, of degrees $(1,0,\text{odd})$
\item a coproduct $\Delta_z:\CY\to \CY\otimes_{r(z)} \CY(\!(z^{-1})\!)$ that's an $A_\infty$ algebra morphism, where the RHS denotes $\CY\otimes \CY(\!(z^{-1})\!)$ deformed by the MC element $r(z)$
\end{itemize}
We conjecture that

\medskip \noindent
\textbf{Conjecture \ref{conj:HT-Yang}} \textit{The Koszul-dual algebra $\CA^!$ representing line operators in any perturbative 3d HT QFT is a dg-shifted Yangian.}
\medskip

\noindent The coproduct and MC element $r(z)$ here simply repackage the OPE \eqref{OPE-intro} in a universal form. For example, evaluating $r(z)$ in any representation $V_\ell\otimes V_{\ell'}$ corresponding to line operators, it becomes $r_{\ell,\ell'}(z)$.

Remarkably, the Maurer-Cartan element $r(z)$ ends up satisfying many relations reminiscent of a classical r-matrix. In particular, associativity of the OPE requires $r_{23}(z)+(\text{id}\otimes\Delta_z)(r(z+w)) = r_{12}(w)+(\Delta_w\otimes\text{id})(r(z))  =  r_{12}(w)+ r_{13}(z+w) +r_{23}(z)$. This in turn implies an $A_\infty$ generalization of the \emph{Yang-Baxter} equation: letting $\text{MC}(x):=\sum_{k=1}^\infty m_k(x,...,x)$, we'll show that
\be \begin{array}{l} 0 = \text{MC}(r_{12}(w)+ r_{13}(z+w) +r_{23}(z)) - \text{MC}(r_{12}(w)) - \text{MC}(r_{13}(z+w))-\text{MC}(r_{23}(z)) \\[.1cm]
\phantom{0} = [r_{12}(w),r_{13}(z+w)]+[r_{12}(w),r_{23}(z)]+[r_{13}(z+w),r_{23}(z)]+ \text{higher $m_{k\geq 3}$ ops} \end{array}
\ee
We emphasize, though, that there is no actual braiding in the category $\CC$\,!

One way to interpret the relation between $r(z)$ and an R-matrix is suggested by Koszul duality itself. In a 3d  perturbative topological QFT, local operators $\CA$ have a $(-2)$-shifted Poisson bracket; their Koszul dual $\CA^!$ naturally carries an R-matrix that's dual to the Poisson bracket in $\CA$, and controls braiding for line operators \cite{Lurie-DAGVI}, \cf\ \cite[App. C]{sparks}. Similarly, in 4d HT QFT, such as holomorphic Chern-Simons the spectral R-matrix of $\CA^!$ is dual to the $(-2)$-shifted $\lambda$-bracket in $\CA$. In 3d HT QFT, we'd expect to find \emph{some} element of degree 1 in $\CA^!\otimes \CA^!$, behaving like a spectral R-matrix, that's dual to the $(-1)$-shifted $\lambda$-bracket in $\CA$ --- and it's our $r(z)$. Explicitly:
\be r(z) = \sum_{\alpha,\beta} \Phi^\alpha \otimes \Phi^\beta \int_0^\infty d\lambda \, e^{-\lambda z}\, \{ \Phi_\alpha {}_{\,\lambda\,} \Phi_\beta \} \label{r-lambda-intro} \ee
where $\{\Phi_\alpha\}$ are linear generators for $\CA$ and $\{\Phi_\alpha\}$ are dual generators for $\CA^!$ (App. \ref{app:r-lambda}).

Some further analogies between $r(z)$ and R-matrices are discussed in Section \ref{sec:Koszul}. More so, in the companion paper \cite{NP-Yangian}, the second two authors explain how dg-shifted Yangian structures arise naturally from quantization of \emph{1-shifted} Lie bialgebras.

Armed with the explicit computations of $\CA$ and $\CA^!$ from Section \ref{sec:Ainf}, we then prove in Section \ref{sec:proofs} that

\medskip\noindent
\textbf{Theorem \ref{thm:N2}} \textit{The $\CA^!$ algebras in HT-twisted 3d $\CN=2$ gauge theories, and matter with superpotential, do have the full structure of dg-shifted Yangians.}

\medskip\noindent
This is a purely algebraic statement and derivation. It's established by combining methods from $A_\infty$ algebras with contour-integral manipulations more familiar from VOA's. It would be interesting to prove an analogous result for the recent HT Poisson sigma models of \cite{KhanZeng} as well (we don't consider them here).

\subsection{The simplest example}
\label{sec:intro-example}

To make the abstract discussion above slightly more concrete, we consider the simplest possible example: the HT twist of a free 3d $\CN=2$ chiral multiplet. It leads to a category $\CC$ that's the anologue of a free-field VOA in 2d holomorphic QFT. However, just like free-field VOA's still have non-trivial (albeit simple) OPE's, line operators in $\CC$ also have nontrivial OPE's.

The twisted field theory in this case is determined by a bosonic field $X(z)$ and a fermionic field $\psi(z)$, whose equations of motion are $d'X=d'\psi=0$, with $d'=\pd_t dt+\pd_{\bar z}d\bar z$. If we expand in modes
\be X(z) = \sum_{n\in\Z} X_n z^{-n-1}\,,\qquad \psi(z) = \sum_{n\in \Z} \psi_n z^{-n-1} \ee
then local operators are constructed from the regular $n<0$ modes. Conversely, the Koszul dual $\CA^!$ is the free graded-commutative algebra generated by singular modes
\be \qquad \CA^!= \C[X_n,\psi_n]_{n\geq 0} \qquad \text{degrees:}\quad \begin{array}{c|c|c|c} & \text{ghost} & \text{spin} & \text{fermion} \\\hline
X_n & R & R/2-n-1 & \text{even} \\
\psi_n & 1-R & R/2-n & \text{odd} \end{array} \ee
(Here `$R$' is a free parameter, physically the $U(1)_R$ charge of the chiral multiplet.)

The simplest line operators are vortex lines $\V_N$, for $N\in \Z$, characterized by a singularity $X(z) \sim z^N\times\text{(regular)}$, $\psi(z)\sim z^{-N}\times\text{(regular)}$ near (say) $z=0$. They are the ``global vortex loops'' of \cite{KWY-abelian}.
Each $\V_N$ corresponds to the $\CA^!$ module
\be \V_N \;\simeq\; \begin{cases} \CA^!/(X_{m\geq 0},\psi_{n\geq 0}) \simeq \C & N= 0 \\
\CA^!/(X_{m\geq 0},\psi_{n\geq N}) \simeq \C[\psi_0,...,\psi_{N-1}] & N > 0 \\
\CA^!/(X_{m\geq |N|},\psi_{n\geq 0}) \simeq \C[X_0,...,X_{|N|-1}] & N < 0
\end{cases} \ee
The dg-shifted-Yangian structure on $\CA^!$, controlling the OPE, consists of:
\begin{itemize}
\item Shifts:\quad $\tau_z(X_n) = \sum_{m=0}^n \big(\begin{smallmatrix}n\\m\end{smallmatrix}\big) z^m X_{n-m}$\,,\quad $\tau_z(\psi_n) = \sum_{m=0}^n \big(\begin{smallmatrix}n\\m\end{smallmatrix}\big) z^m \psi_{n-m}$
\item MC element\quad $\ds r(z) = \sum_{n,m\geq 0} (-1)^n \bigg(\begin{array}{c}n+m\\m\end{array}\bigg) \frac{\psi_n\otimes X_m-X_n\otimes\psi_m}{z^{n+m+1}}$
\item Coproduct \quad $\Delta_z(X_n)=\tau_z(X_n)\otimes 1+1\otimes X_n$\,,\quad $\Delta_z(\psi_n)=\tau_z(\psi_n)\otimes 1+1\otimes \psi_n$
\end{itemize}
The coproduct and the MC element together determine the OPE of $\CA^!$-modules. Namely, to build an OPE $M\otimes_z M'$, we first deform the tensor product $M\otimes M'$ (of vector spaces) by adding a \emph{differential} $r(z)$, and then use the coproduct to define the action of $\CA^!$. For example, $\V_1\otimes_z\V_{-1}$ is the module $\C[\psi_0,X_0]$, with a differential $r_{1,-1}(z) = \frac{1}{z} \psi_0 X_0$\,.

The result of Theorem \ref{thm:non-ren} implies that the dg-shifted-Yangian structure of any perturbative HT QFT looks essentially the same, with coproduct and shifted r-matrix given by the same free-field formulas. The main difference in interacting theories is that the $A_\infty$ algebra $\CA^!$ becomes much richer. For example, adding a superpotential $W(X)= \tfrac{1}{d!}X^d$ to the free chiral induces $A_\infty$ operations $m_k$ for $1\leq k\leq d-1$ in $\CA^!$, each given by modes of $\pd_X^k W$:
\be
\begin{array}{r@{\;}c@{\;}l} Q(\psi_n) &=& \ds -\frac{1}{(d-1)!}\sum_{\substack{0\leq m_1+...+m_{d-1}\\=n+2-d}} X_{m_1}X_{m_2}\cdots X_{m_{d-3}}X_{m_{d-2}}X_{m_{d-1}} \\ {}
[\psi_{n_1},\psi_{n_2}] &=& \ds -\frac{1}{(d-2)!}\sum_{\substack{0\leq m_1+...+m_{d-2}\\=n_1+n_2+3-d}} X_{m_1}X_{m_2}\cdots X_{m_{d-3}}X_{m_{d-2}} \\
m_k(\psi_{n_1},...,\psi_{n_k}) &=& \ds -\frac{1}{k!(d-k)!} \sum_{\substack{0\leq m_1+...+m_{d-k} \\=n_1+...+n_k+k+1-d}}X_{m_1}X_{m_2}\cdots X_{m_{d-k}} \quad (k\geq 3)
\end{array} \label{dW-intro}
\ee
(For fermions $a,b$ our convention is that $[a,b]:=ab+ba$.)
\eqref{dW-intro} identifies $\CA^!$ with the (dual) chiral derived critical locus of $W$. Alternatively, in gauge theory with Lie algebra $\fg$, we'll find that $\CA^!$ is a deformation of the shifted cotangent loop algebra $(T^*[-1]\mathfrak g)[\lambda]$. 

\subsection{Boundary vertex algebras}
\label{sec:VA}

In this last part of the introduction, we comment/speculate briefly on an interesting connection between the structures in this paper and (boundary) vertex algebras and their conformal blocks. We won't discuss this further in the body of the paper, leaving a detailed study to future work.

Another natural way to study line operators in 3d HT theories involves holomorphic boundary conditions. More precisely, if one introduces a holomorphic boundary condition $\mathbf b$, supporting a boundary vertex algebra $\CV_{\mathbf b}$, then there is a functor from line operators of the bulk theory to $\CV_{\mathbf b}$ modules,
\be \CF_{\mathbf b}: \CC \to \CV_{\mathbf b}\text{-mod}\,,\qquad \CF_{\mathbf b}(\ell) = \text{Ops}(\mathbf b,\ell)\,,
\quad \raisebox{-.3in}{\includegraphics[width=1.45in]{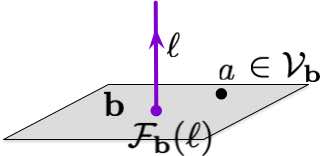}}
 \label{VA-functor} \ee
which sends each line to the vector space of local operators at its junction with the boundary. Moreover, if the boundary condition is rich enough, the functor \eqref{VA-functor} should be an equivalence. This idea was first used to relate line operators in Chern-Simons theories to modules of the boundary WZW VOA \cite{Witten-Jones,MS-taming,EMSS}; it has more recently found many applications in topological twists of 3d $\CN=4$ gauge theories, \eg\ \cite{Gaiotto:2016wcv,Costello:2018fnz,Costello:2018swh, CDGG, Ferrari:2023fez, BCDN}, as well as in the 3d-3d correspondence, using VOA's associated to 3-manifolds, \eg\ \cite{GGP-fivebranes, Cheng-3dMod, Cheng-3dChar}.

The theories we consider in this paper, such as HT twists of 3d $\CN=2$ gauge theories with superpotentials, all admit holomorphic boundary conditions \cite{GGP-fivebranes, Yoshida:2014ssa, DGPbdry}; their vertex algebras $\CV_{\mb b}$ were explicitly constructed in \cite{CDGbdry}.
An important and unusual feature of these $\CV_{\mb b}$ is that they don't admit stress tensors (\emph{a.k.a.} conformal vectors) unless the bulk 3d theory is fully topological.
It's then natural to pose the following questions: \emph{how is our formulation of $\CC$ related to boundary vertex algebras $\CV_{\mathbf b}$, and how is the OPE structure manifested from the perspective of $\CV_{\mathbf b}\Mod$}\;?

A useful example of a boundary condition $\mb b$ for HT-twisted $\CN=2$ theory with complexified gauge group $G$, Chern-Simons level $k$, matter representation $V$, and superpotential $W:V\to \C$ is ``Dirichlet'' for the gauge fields and ``Neumann'' for the matter fields \cite{CDGbdry}. Perturbatively, its boundary VOA $\CV_{\mb b}$ is a semidirect product of a Kac-Moody algebra $\hat \fg_{k'}$ at level $k'=k-h^\vee(G)-\tfrac12I_2(V)$ and the commutative loop algebra $V^*(\!(z)\!)$ (where $h^\vee$ is the dual Coxeter number and $I_2(V)$ is the quadratic index). There's also a BRST anomaly equal to $\oint W$, the residue of $W$, lifted to a function on the loop space $LV$.
Altogether, the category of modules should be described geometrically as $\hat \fg_{k'}$-equivariant matrix factorizations on the loop space $LV$, with potential $\oint W$,
\be \textstyle  \CF_{\mb b}: \CC \overset{\sim}{\to} \CV_{\mb b}\text{-mod} \simeq \text{MF}^{\hat \fg_{k'}}\big(LV,\oint W\big) \qquad \text{(perturbative).} \hspace{-.8in} \label{MF-intro} \ee
The dg-shifted Yangian of gauge theory from Section \ref{sec:proofs} provides an indirect construction of OPE's in this matrix-factorization category.

Schematically, one would expect the OPE in $\CV_{\mb b}\text{-mod}$ itself to be captured by some $\CV_{\mb b}$ fusion product.
A more precise relation to the analysis in the current paper lies in equation \eqref{eq:fiberF}. Here, the functor $\wt \CF$ sends a line operator $\ell$ to the state space on $\mathbb P^1$ with insertion of $\ell$ at $0$ and singularity $\CB_\infty$ at infinity. In terms of the vertex algebra $\CV_{\mathbf b}$, this state space is simply the \emph{derived} conformal blocks (CB) on $\mathbb P^1$, with corresponding module $\CF_{\mb b}(\ell)$ at $0$ and singularity $\CB_\infty$ at infinity,
\be \text{States}(\cp^1,\ell(0),\CB_\infty) \simeq \text{CB}(\cp^1,\CF_{\mb b}(\ell),\CF_{\mb b}(\CB_\infty)) \ee
The singularity ensures that the space of conformal blocks associated to the vacuum module is one-dimensional.  
The algebra $\CA^!$, moreover, acts naturally on any such space of conformal blocks (as it's the endomorphisms of $\CB_\infty$), so we get a lifted functor $\text{CB}(\cp^1,-,\CF_{\mb b}(\CB_\infty)):  \CV_{\mb b}\text{-mod} \to \CA^!\text{-mod} $. Altogether, this should fit in a commutative diagram
\be \raisebox{-.3in}{\includegraphics[width=2in]{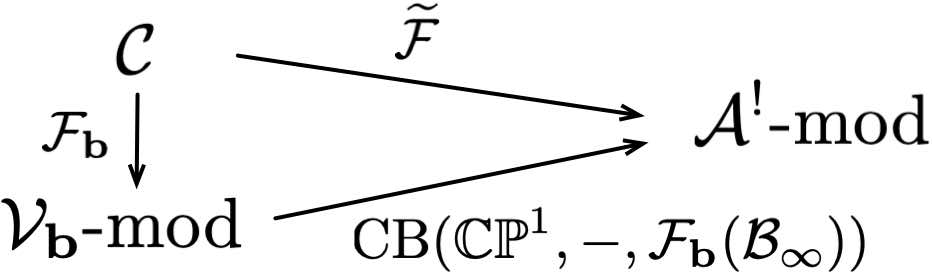}}
\ee
In the paper, we argue that the top arrow is an equivalence. On the bottom, one must be careful to consider CB as a derived functor of $A_\infty$ categories, since both $\CA^!$ and $\CV_{\mathbf b}$ have $A_\infty$ structures. We conjecture that, in this derived setting, the bottom arrow also becomes an equivalence.

In extremely nice situations, the category $\CV_{\mathbf b}\Mod$ admits a so-called vertex tensor category structure \cite{huang2014logarithmic1}. This vertex tensor structure (which we denote by $\times_z$) should be the vertex-algebra-theoretic counterpart of the OPE, as it satisfies
\be
\text{CB} (\mathbb{P}^1, \ell \times_z \ell', \CB_\infty )= \text{CB} (\mathbb{P}^1, \ell(z), \ell'(0), \CB_\infty ) \,.
\ee
It is therefore natural to conjecture that the functor CB intertwines the vertex tensor structure of \cite{huang2014logarithmic1} with the meromorphic tensor structure that we study. In the special case of gauge theory with no matter $(V=0)$, it would be the vertex tensor structure for Kac-Moody at generic level ($\hat{\mathfrak g}_{k'}\text{-mod}$) and our dg-shifted Yangian from Secs. \ref{sec:Ainf-gauge}, \ref{sec:Yangian-gauge}.

If this conjecture is true, then it should provide a powerful computational tool for the study of fusion products for vertex algebras. For example, the definition of vertex tensor product $\times_z$ requires many additional properties on the modules considered (\textit{e.g.} grading, finiteness, \textit{etc.}), whereas our definition of meromorphic tensor products requires only smoothness. Moreover, we should find (due to \eqref{OPE-intro}) that the functor CB maps the vertex tensor product $\ell\times_z\ell'$ to the tensor product of individual conformal blocks, deformed by a Maurer-Cartan element
\be\label{eq:Cnotmero}
\text{CB}(\ell\times_z\ell') \simeq \big[ \text{CB}(\ell)\otimes \text{CB}(\ell'),r_{\ell,\ell'}(z)\big]\,.
\ee
In particular, since characters are unchanged by the MC deformation, we recover an exact equality of characters
\be
\chi [\text{CB}(\ell\times_z\ell')]=\chi [\text{CB}(\ell)]\chi [\text{CB}(\ell')]\,. 
\ee
To obtain this good behavior, one must however consider derived fusion products and derived conformal blocks, rather than their classical counterparts.

In the case of 3d QFT's that are actually topological, such as HT-twisted 3d {$\CN=2$} Chern-Simons theory (or pure 3d Chern-Simons as in \cite{Witten-Jones}, or the topologically twisted 3d $\CN=4$ examples of \cite{Costello:2018fnz} (etc.)), there is additional structure around. Now the boundary vertex algebra $\CV_{\mb b}$ does have a stress tensor/conformal element $L$, \ie\ it is a \emph{vertex operator algebra}. The stress tensor $L$ trivializes the $z$-dependence in the vertex tensor product $\ell\times_z\ell'$, and gives rise to a flat connection on the space of conformal blocks. In the case of $\text{CB}(\mathbb P^1, -, \CF_{\mb b}(\CB_\infty))$, this is the famous Knizhnik-Zamolodchikov connection \cite{tsuchiya1987vertex}. This (schematically) makes $\CV_{\mb b}\text{-mod}$ an ordinary braided tensor category rather than a category with OPE's.

We expect that when an HT theory is actually topological, there will be an avatar of the stress tensor $L$ in the Koszul-dual algebra $\CA^!$. Namely, we would expect $\CA^!$ to admit a derivation $\CL$ of degree $-1$, such that $[Q, \CL]=T$, where $T$ is the infinitesimal translation operator, such that $\tau_z = \exp(z T)$. In particular, translations become trivial on the $Q$-cohomology of $\CA^!$. This would allow $z$-dependence in the OPE to be trivialized, and the resulting flat module would (ideally) be identical to the one defined by the KZ connection. We'll give an example of the derivation $\CL$ for perturbative Chern-Simons in Section \ref{sec:TQFT-gauge}. It would be interesting to explore this further in the future!

\subsection{Acknowledgements}
\label{sec:intro-ack}

We would like to thank Meer Ashwinkumar, Kevin Costello, Niklas Garner, Davide Gaiotto, Lukas M\"uller, Natalie Paquette, Surya Raghavendran, and Pavel Safronov for enlightening discussions and incisive questions on the topics of this work. VP would especially like to thank Lotte Hollands and Sasha Shapiro for discussions in the course of reviewing his Ph.D. thesis. TD’s research is supported by EPSRC Open Fellowship EP/W020939/1.
WN’s research is supported by the Perimeter Institute for Theoretical Physics, which in turn is supported in part by the Government of Canada through the Department of Innovation, Science and Economic Development Canada and by the Province of Ontario through the Ministry of Colleges and Universities.

\section{Setup}
\label{sec:setup}

In this section, we'll review some of the basic formalism and techniques that we use throughout the paper.

We begin (Section \ref{sec:3dHT}) with 3d HT QFT's in BV-BRST formalism, explaining what we mean by a perturbative theory, following \cite{GRW-quantization,GaiottoKulpWu}. We particularly emphasize the role of symmetries (ghost number/cohomological degree $R$, spin $J$, and fermion number $F$) and of holomorphic-topological descendants, which are helpful in organizing the field and operator content. We also recall (Section \ref{sec:3dN2}) how the HT twist of 3d $\CN=2$ gauge theories, with matter, superpotentials, and Chern-Simons interactions, fit into this formalism, following \cite{ACMV, CDGbdry}. We discuss where we expect to see additional nonperturbative effects in 3d $\CN=2$ theories that would go beyond the scope of this paper (Section \ref{sec:NP}).

In Section \ref{sec:prop}, we review the form of the propagator in perturbative 3d HT theories, in different gauges. We'll mostly use axial gauge for calculations of OPE's of line operators, but it's not always appropriate, \emph{e.g.} for insertions of local operators.

In Section \ref{sec:topQM}, we then consider 1d topological QFT's (\emph{a.k.a.} topological quantum mechanics), which will be the building blocks of (perturbative) line operators. We again introduce symmetries and descendants, and review how deformations of topological QM are classified by solutions to Maurer-Cartan equations. This is a well known subject, but central to the entire paper; other good recent review include \cite{PaquetteWilliams, GaiottoKulpWu}

Finally, in Section \ref{sec:line-setup}, we review some of the basic physical expectations for the category of line operators and its OPE $\otimes_z$ in a 3d HT QFT.

We call attention to two simple conventions introduced in this section that we use throughout the paper. Given any local operator $\CO$ in 3d HT QFT (resp. 1d topological QFT) we denote the sum of its descendants in boldface:
\be  \bm \CO := \sum_{k=0}^\infty \CO^{(k)} = \begin{cases} \CO+ \CO^{(1)}+\CO^{(2)} & \text{3d HT} \\ \CO + \CO^{(1)} & \text{1d top} \end{cases} \label{d-convention} \ee
Such an $\bm \CO$ transforms as a mixed-degree differential form on spacetime, with (potential) local components proportional to $1,d\bar z,dt, d\bar zdt$ (in 3d) or just $1,dt$ (in 1d).

For elements of an algebra, such as local operators in 1d QFT or the restriction of 3d local operators to a time-like line, sign conventions (including ``Koszul sign rules'' for $A_\infty$ algebras) are controlled by fermion number $F$ (\emph{a.k.a.} parity). This need not coincide with ghost number/cohomological degree. Commutators and anti-commutators are both written with square brackets; we define
\be [a,b] := ab - (-1)^{F(a)F(b)}ba\,. \label{b-convention} \ee

\subsection{3d HT QFT's}
\label{sec:3dHT}

A 3d HT theory may be defined on a 3-manifold with a transverse holomorphic foliation, such that the neighborhood of any point is isomorphic to $\C_z\times \R_t$. Since we're interested in local structure, we will just work on global $\C_z\times \R_t$.

A perturbative HT QFT --- one defined by quantizing a free theory, deformed by interactions --- has an action that in the BV-BRST formalism takes the general form
\be  S = \int_{\C\times \R} \big[ \mb p_i d' \mb x^i + \CI(\mb p,\mb x, \pd)  \big]\,. \label{HT-action} \ee
Here
\be d' := \partial_t dt + \partial_{\bar z} d\bar z \ee
is an exterior derivative acting on the complex $\Omega'{}^\bullet(\C\times \R):= \Omega^\bullet(\C\times \R)/(dz) \simeq C^\infty(\C\times \R)[dt,d\bar z]$, with $dt,d\bar z$ treated as fermionic (parity-odd) variables; while
\be \partial := \pd_z dz \ee
is the holomorphic derivative, with $dz$ treated as a bosonic (parity-even) variable. The kinetic term is formed using the differential operator $d'$; while arbitrary powers of $\pd$ may appear in the interactions $\CI$.

The fields of the theory are organized into $\Omega'$ multiforms
\be \label{xp-forms}  \mb x^i \in \Omega'{}^\bullet(\C\times \R)[R_i]dz^{J_i} \qquad \mb p_i \in \Pi \Omega'{}^\bullet(\C\times \R)[1-R_i]dz^{1-J_i}\,.\ee
There's some extra notation here that can be neglected on a first pass. The `$\Pi$' designates a fermion-number (parity) shift in $\mb p$; while $R$ and $J$ are shifts in cohomological degree and spin of the fields, which we'll discuss momentarily. Expanded in components, we have
\be \mb x^i = \mb x^i{}^{(0)} + \mb x^i{}^{(1)} + \mb x^i{}^{(2)} = (x^i{}^{(0)} + x^{i\,(1)}_{\bar z} d\bar z+ x^{i\,(1)}_{t} dt  + x_{\bar zt}^{i\,(2)} d\bar z dt)dz^{j_i}\,, \ee
and similarly for $\mb p_i$. The component fields $x^{(0)},x^{(2)}_{\bar z t}, p^{(1)}_{\bar z,t}$ are bosonic, while $p_{(0)},p_{(2),\bar z t}, x^{(1)}_{\bar z,t}$ are fermionic. Semi-classically, the BRST operator acts on fields as
\be Q = \big\{S,-\big\}_{\rm BV} = d' + \Big\{\int \CI,-\Big\}_{\rm BV}\,, \ee
where $\{-,-\}_{\rm BV}$ is the classical BV bracket, satisfying $\{\mb p_i(z,\bar z,t),\mb x^i(w,\bar w,s)\}_{\rm BV} \sim \delta^{(3)}(z-w,\bar z-\bar w,t-s)$. The interaction must satisfy $\big\{ \int \CI,\int\CI\big\}_{\rm BV}=0$ to avoid a semi-classical BRST anomaly, \ie\ for $Q$ to be nilpotent.

It is argued in \cite{GRW-quantization,WilliamsWang} that there are no further obstructions to perturbative quantization (at all orders) in a large family of 3d HT theories. This family includes all HT-twisted 3d $\CN=2$ theories.

\subsubsection{Symmetries}
\label{sec:sym}

In addition to fermion number $F\in \Z_2$, 3d HT theories have
\begin{itemize}
\item $Spin(2)_J \simeq U(1)_J$ symmetry from rotations of the $\C_z$ plane, with charge $J$, nominally valued in $\tfrac12 \Z$.
\item (potentially) a $U(1)_R$ ghost-number or cohomological symmetry, with charge $R$, nominally valued in $\Z$, which lifts fermion number in the sense that
\be F = 2J-R\quad (\text{mod 2})\,. \label{FJR} \ee
\end{itemize}
Conventions are such that $R$ counts $dt,d\bar z$ form degree while $J$ counts $dz$ degree,
\be \begin{array}{c|cccc}
 & d',dt,d\bar z & \pd & \text{Lagrangian} & Q \\\hline
F & \text{odd} & \text{even} & \text{even} & \text{odd} \\
J & 0 & 1 & 1 & 0 \\
R & 1 & 0 & 2 & 1
\end{array} \ee
and the Lagrangian must have $R=2,J=1$ in order to be integrated on $\C\times \R$. This implies that the linear operators $\mb x,\mb p$, and the interaction terms, satisfy
\be J(\mb x^i)+J(\mb p_i)=1\,,\quad R(\mb x^i)+R(\mb p_i) = 1\,;\quad J(\CI)=1\,,\quad  R(\CI)=2\,. \label{JR-const} \ee
We define $J_i=J(\mb x^i),\, R_i=R(\mb x^i)$. The square brackets $[...]$ appearing in \eqref{xp-forms} are cohomological shifts of the fields encoding their R-charges, while the $dz^\#$ encode shifts in spin. 

Interaction terms (and nonperturbative quantum effects) can break or modify symmetries. In this paper, we'll exclusively work with theories in which both spin and R-symmetry are preserved --- meaning there exists a choice of $\{J_i\}$ and $\{R_i\}$ for the $\mb x^i$'s such that $J(\CI)=1$ and $R(\CI)=2$ both hold. However, we'll allow the charges to take rational values
\be J_i,R_i \in \Q \qquad \text{(but $R_i-2J_i\in \Z$)}\,. \ee
This amounts to saying that certain covers of $U(1)_J,U(1)_R$ are actually acting.

 In the HT twist of a 3d $\CN=2$ theory, $U(1)_R$ is the standard R-symmetry of the underlying SUSY theory; while spin
 \be J = J_{phys}+\frac12 R \label{Jphys} \ee
 is a diagonal combination of physical spin and R-symmetry (it's the distinguished combination that leaves the twisting supercharge invariant). It is common for $R$ (and thus also $J$) to take rational, or even more general real values. For example, superconformal R-charges, satisfying Z-maximization \cite{Jafferis-Zmax}, are typically irrational. The choice of integer/rational/real affects the types of global 3-manifolds one can define an HT theory on, but are not important for the local features that we analyze.

\subsubsection{Descendants and universality}
\label{sec:descent}

The multiforms described above naturally organize fields of an HT theory into ``multiplets'' containing their descendants. We briefly review this idea, as it underlies the universality of the action \eqref{HT-action}.

In any HT QFT with a BRST differential $Q$, the theory does not need to be entirely independent of $t$ and $\bar z$; rather, $\pd_t$ and $\pd_{\bar z}$ need only act as zero on operators modulo $Q$-exact terms. For local operators, one way to make this precise is to introduce descendants: for each local operator $\CO$, transforming as a $dt,d\bar z$ form of some degree in spacetime, one requires the existence of a descendant $\CO^{(1)}$, such that
\be d' \CO = Q \CO^{(1)} + (Q\CO)^{(1)}\,. \label{descent-eq} \ee
In particular, for $Q$-closed operators, $d'\CO$ is $Q$-exact. (This generalizes descent in topological QFT's \cite{Witten-Donaldson,MooreWitten-u}.) One sets $\CO^{(2)}=(\CO^{(1)})^{(1)}$. There are no further descendants beyond $\CO^{(2)}$ since the $\Omega'$ complex tops out at degree two.

The structure \eqref{descent-eq}  is automatic in the HT twist of a 3d $\CN=2$ theory, where $Q$ is the twisting supercharge. There, $\CO^{(1)}$ is defined as $\wt Q \CO$, where $\wt Q$ is a particular combination of 3d $\CN=2$ supercharges satisfying $[Q,\wt Q] = d'$; see \cite[Sec. 3]{CDGbdry}. We adopt the convention from \eqref{d-convention}: for any local operator $\CO$, we will use boldface to denote the sum of its descendants, treated as a multiform on spacetime:
\be \bm\CO := \CO + \CO^{(1)} + \CO^{(2)}\,.  \ee

Descent is used to define the $\lambda$-bracket on local operators \cite{OhYagi-Poisson, GarnerWilliams}, generalizing $E_n$ brackets in topological theories \cite{Getzler-2d,descent}. Namely, for any local operators $\alpha,\beta\in\CA$, one sets
\begin{align} \{\alpha_{\,\lambda\,} \beta\}(w,s) &:= (-1)^{F(\alpha)} \oint_{S^2_{w,s}} e^{\lambda (z-w)} \alpha^{(1)}(z,t) \beta(w,s)  \label{def-lambda} \\
&=  (-1)^{F(\alpha)} \oint_{S^2_{w,s}} e^{\lambda (z-w)} \bm\alpha(z,t) \bm\beta(w,s)\,,  \notag \end{align}
where $S^2_{w,s}$ is a sphere centered at $w,s$. When $\alpha,\beta$ are $Q$-closed, the dependence on $S^2_{w,s}$ is topological, due to \eqref{descent-eq} and Stokes' Theorem.
 
Now, if a theory has a free limit, so that one can talk about fields and Lagrangians, then one expects fields of the free theory to obey descent equations as well. Suppose that in the free limit there are $Q$-closed fields $\phi_i$, which obey the free equation of motion $d'\phi_i=0$.
Let $\bm\Phi_i := \phi_i+\phi_i^{(1)}+\phi_i^{(2)}$. Then the descent equations imply $Q\bm\Phi_i = d' \bm\Phi_i$, \ie\ the BRST operator acts as $d'$ on this multiform. The free BV-BRST action must take the form $\int \mb p_i d' \mb x^i$, where, for various $i$, either $\mb\Phi_i=\mb x^i$ (with conjugate momentum $\mb p_i$) or  $\mb\Phi_i= \mb p_i$ (with conjugate momentum $\mb x^i$), depending on fermion number. This exercise explains (heuristically) how to construct an action in the form \eqref{HT-action} for any HT QFT with a free-field limit: just find the ``primary'' fields satisfying $Q\phi=0$ (or equivalently, the fields with EOM $d'\phi=0$) and group together their descendants.

\subsection{3d $\CN=2$ theories}
\label{sec:3dN2}

Our main examples will come from 3d $\CN=2$ QFT's in the HT twist. We review, following \cite{ACMV, CDGbdry} how these look once fields are reorganized in terms of descendants, and the action is recast in a canonical form \eqref{HT-action}.

A 3d $\CN=2$ free chiral multiplet reorganizes into a pair $(\mb p,\mb x) = (\BPsi,\BX)$ with action $S = \int_{\C\times \R} \BPsi d' \BX$. The physical complex ``chiral'' boson $X$ is the leading component of $\BX$; while a particular fermion $\psi$ in the anti-chiral multiplet, usually denoted $\ol\psi_-$, is the leading component of $\BPsi$. R-charge is undetermined, aside from the constraint $R(X)+R(\psi)=1$. Spin is determined by \eqref{Jphys} to be
\be \label{chiral-charges}  \begin{array}{c|cc}
 & \BX & \BPsi \\\hline
F & \text{even} & \text{odd} \\
J & \tfrac12 R(X) & 1-\tfrac12 R(X) \\
R & R(X) & 1-R(X) \end{array} \ee

Multiple chirals $(\BX^i,\BPsi_i)_{i=1}^n$ may interact via a 3d $\CN=2$ superpotential, a polynomial $W(X^1,...,X^n)$ with the constraint that $R(W)=2$. It enters the HT action directly:
\be  S = \int_{\C\times \R} \BPsi_i d' \BX^i + W(\BX)\,. \label{S-W} \ee
A natural generalization, perfectly compatible with all constructions in this paper, would be to allow $W$ to depend on $\BPsi$'s and holomorphic derivatives $\pd$ as well. Such terms make sense in the HT twist as long as $\{S,S\}_{\rm BV}=0$, though they are not compatible with the full 3d $\CN=2$ SUSY algebra.

A 3d $\CN=2$ Yang-Mills gauge theory with compact gauge group $G_c$ is reorganized in the HT twist as an analytically continued BF theory, for the complexified group $G=(G_c)_\C$. The fields $(\mb p,\mb x) = (\BA,\BB)$ are valued in \vspace{-.2in}
\be \begin{array}{l}\BA \in \Pi \mathfrak g \otimes \Omega'{}^\bullet[1] \\[.1cm]
  \;\;\; = c + A_t dt+A_{\bar z}\,d\bar z + ... \end{array}
,\quad  \raisebox{-.2cm}{$\begin{array}{l} \BB \in \mathfrak g^* \otimes \Omega'{}^\bullet dz \\[.1cm] 
 \;\;\; = \underbrace{B_zdz}_{B} + ... \end{array}$}
\qquad \text{with}\quad  \begin{array}{c|cc} & A & B \\\hline F & \text{odd} & \text{even} \\ J & 0 & 1 \\ R & 1 & 0 \end{array}\,. \ee
In particular, the 1-form $A = \mb A^{(1)} = A_t dt+A_{\bar z}d\bar z$ is a $G$ connection whose $dt$ component has been complexified by the 3d $\CN=2$ vectormultiplet scalar. The leading component $c = \BA^{(0)}$ is a ghost; while $B=\BB^{(0)}$ is a complex scalar cohomologous to a complex linear combination of the curvature $F_{zt}$ and the $\pd$ derivative of the physical vectormultiplet scalar. We will often choose a basis of the Lie algebra $\fg$ and a dual basis
\be \text{bases:} \quad \text{$\{t^a\}$ for $\fg$}\,,\quad \text{$\{t_a\}$ for $\fg^*$}\,,\qquad t^a\cdot t_b = \delta^a{}_b\,,\quad [t^a,t^b]=f^{ab}{}_c t^c\,, \ee
and expand
\be \BA = \BA_a t^a\,,\qquad \BB = \BB^a t_a\,. \ee
Note that the components $\BA_a$ transform as elements of $\fg^*$ and $\BB^a$ as elements of $\fg$.

The action of HT-twisted Yang-Mills theory takes the form
\be S = \int_{\C\times \R} \BB F'(\BA) = \int_{\C\times \R} \BA d' \BB - \tfrac12 \BB[\BA,\BA]  = \int_{\C\times \R} \BA_a d' \BB^a - \tfrac12 f^{ab}{}_c \BB^c \BA_a\BA_b\,.  \label{S-BF} \ee
Its EOM include $F'(A) = [D_{\bar z},D_t] =0$, which says that the holomorphic structure of the gauge bundle along $\C_z$ commutes with parallel transport in $\R_t$.

Gauge theory may be further deformed by a Chern-Simons interaction, which takes the form
\be S = \int_{\C\times \R} \BB F'(\BA) +\tfrac12 k(\BA,\pd \BA) = \int_{\C\times \R} \BB F'(\BA) +\tfrac 12k^{ab} \BA_a \pd \BA_b\,, \label{S-CS} \ee
where $k\in \text{Sym}^2(\fg^*)^\fg$ is an invariant symmetric bilinear form on $\fg$, with components $k^{ab}$. If $G$ is a simple reductive group, then $k$ is necessarily a multiple of the Killing form, which may be normalized to $\delta^{ab}$, so
\be k^{ab} = \frac{\wt k}{4\pi i}\, \delta^{ab}\,, \ee
where $\wt k\in \C$ is the more familiar numerical Chern-Simon level. Perturbatively, there is no constraint on $\wt k$. Non-perturbatively, $\wt k$ gets quantized to be an integer, or a half-integer if there are further matter fields around \cite{Witten-Jones,AHISS}.%

An important feature of 3d $\CN=2$ Yang-Mills-Chern-Simons theory at sufficiently large level $|\wt k|> h^\vee$ (where $h^\vee$ is the dual Coxeter number) is that it flows to pure Chern-Simons theory in the infrared, at shifted level $\wt k- h^\vee\text{sign}(\wt k)$. In the HT twist, RG flow is trivial, so Yang-Mills-Chern-Simons should be equivalent to pure Chern-Simons. Explicitly, this happens by identifying the missing $z$ component of the Chern-Simons gauge field as $A_z\sim \frac{1}{k} B$. 
Level shifts in the HT twist appear in perturbative computations of operator algebras, \eg\ boundary vertex algebras \cite{CDGbdry, Gwilliam:2019cbp} or bulk operator algebras (Section \ref{sec:Ainf-gauge}).

Finally, consider gauge theory with matter in representation $V$, and corresponding homomorphism $\varphi_V:\fg\to \text{End}(V)$, has matter fields
\be 
\BX\in V^*\otimes \Omega'{}^\bullet[R]dz^{R/2}\,,\qquad \BPsi \in \Pi V\otimes  \Omega'{}^\bullet[1-R]dz^{1-R/2}\,. \label{setup-X-fields} \ee
Given a basis and dual basis
\be \text{bases:}\quad \text{$\{v^i\}$ for $V$},\quad \text{$\{v_i\}$ for $V^*$},\qquad \varphi_V(t^a) v^i = (\varphi_V^a)^i{}_j v^j\,, \ee
we expand
\be \BX = \BX^i v_i\,,\qquad \BPsi = \BPsi_i v^i\,,\ee
with the components $\BX^i$ transforming as elements of $V$. This is usually what one means in physics by ``a chiral multiplet in representation $V$,'' despite that the field itself is an element of $V^*$. Different $R$-charges and spins may appear for each irreducible summand of $V$ in \eqref{setup-X-fields} as well.
The action then takes the form
\be S = \int_{\C\times \R} \BB F'(\BA) + \BPsi d_\BA' \BX = \int_{\C\times \R} \BA d' \BB + \BPsi d' \BX - \tfrac12 \BB[\BA,\BA]+\BPsi \varphi_V(\BA) \BX\,. \ee
Here $d'_\BA = d' + \varphi_V(\BA)$ is a covariant derivative.
This can be further deformed by Chern-Simons terms $\tfrac12\int k(\BA,\pd \BA)$ and/or a superpotential $\int W(\BX)$ such that $W:V\to \C$ is $G$-invariant and has $R=2$. 

For ease of reference, we can also spell out the BRST transformation in these theories. In the most general case of gauge theory with matter, CS terms, and superpotential, we have
\be \begin{array}{l} Q \BA = F'(\BA)\,, \\ Q \BB  = d_\BA' \BB + k\pd \BA- \varphi_V^*(\BPhi \BPsi)\,, \end{array} \qquad
\begin{array}{l} Q\BX = d'_\BA \BX\,, \\ Q \BPsi_i = d'_\BA\BPsi_i + \pd_{\BX^i} W(\BX)\,. \end{array} \label{Q-act} \ee
These semiclassical transformations refine to a rather nontrivial $A_\infty$ algebra structure on local operators in gauge theory, which we derive later in Section \ref{sec:Ainf}, summarized in \eqref{A-full}.

\subsubsection{Nonperturbative considerations}
\label{sec:NP}

Most twisted $\CN=2$ theories do have some nonperturbative aspects. We recall a bit of what they are, mainly as a warning --- as for the most part we will not be considering nonperturbative effects in this paper.

Given just matter with a superpotential, and trivial gauge group, we expect perturbation theory to give an exact analysis of local operators. This is because there are no instanton solutions to the equations of motion: in particular, the EOM for the scalar boson $\pd_t X =\pd_{\bar z} X=0$ don't admit solutions that are localized in time. We will describe the complete A-infinity algebra of local operators in Section \ref{sec:Ainf}. Similarly, we expect that line operators, as a category (objects and morphisms), are entirely perturbative. However, the OPE of lines might get nonperturbative corrections from solitons. They would be analogous to the instanton corrections in the half-twist of 2d $\CN=(2,2)$ theories \cite{DSWW,SilversteinWitten} (or the A-twist of 2d $\CN=(2,2)$ \cite{Witten-sigma}, \cf\ \cite{GMW} for local operators in theories with a superpotential), but extended along time $\R_t$.

Gauge theories have nonperturbative corrections \emph{both} to local operators (and thus to morphisms among lines) and to the OPE's of lines. Local operators in 3d HT-twisted gauge theories get enhanced by monopole operators --- roughly, by the chiral/BPS monopole operators that appear in the chiral ring \cite{AHISS} and by their holomorphic derivatives. Some explicit examples of monopole corrections in the HT twist appeared in \cite{Zeng-monopole, GW-analytic}, though they are not fully understood in general.
This must in turn change the nature of the category of line operators $\CC$, since (on one hand) lines are constructed by coupling QM to local operators; and (conversely) computing endomorphisms of the trivial line must reproduce local operators. In summary:
\be \quad  \begin{array}{c|ccc}
 & \text{free matter} & \text{matter + $W$} & \text{gauge theory} \\\hline
 \text{quantization/local operators} & \text{pert} & \text{pert} & \text{NP} \\
 \text{line operators (objects + morphisms)} & \text{pert} & \text{pert} & \text{NP} \\
 \text{line operators (OPE)} & \text{pert} & \text{NP?} & \text{NP}
 \end{array} \notag \ee

One basic nonperturbative effect in gauge theories that's easily incorporated in the twisted formalism above concerns the global nature of the gauge group. Compactness of $G_c$ (equivalently, reductiveness of $G$) implies that we should remove the constant mode of the ghosts $c$ from local operators, and implement invariance under constant $G$ gauge transformations by hand, rather than via BRST cohomology.

We also note that in pure $G_c=U(1)$ gauge theory, monopole operators have a perturbative definition as well, \cf\ \cite[Sec. 5.2]{GarnerWilliams}.
Physically, the chiral monopole operators are $M_n \sim e^{in\gamma}$ ($n\in \Z$), where $\gamma$ is the complexified dual photon. In our BV-BRST formalism, the field $B=\BB^{(0)}$ is cohomologous to a (complexified) component of the physical field strength, which in turn is equal to $\pd\gamma$. This we may define $\mb M_n$ at position $(w,s)\in \C\times \R$ by choosing a holomorphic contour $\eta_w$ in the $\C_z$ plane that ends at $w$ (with the other end off at infinity) and setting
\be \mb M_n(w,s) = \exp \int_{\eta_w\times\{s\}} \!\! n\, \BB\,,\qquad n\in \Z\,. \label{mon-def} \ee
In pure $U(1)$ theory, the leading component $M_n$ is $Q$-closed. (It generates a module for the Heisenberg raviolo vertex algebra introduced in \cite[Sec. 5.2]{GarnerWilliams}.)
In the presence of a Chern-Simons term or matter, $M_n$ is not $Q$-closed; \eg\ with Chern-Simons terms it obeys
\be Q M_n(w,s) = \bigg(\int_{\eta_w\times \{s\}} nk\,\pd \BA\bigg)M_n(w,s)  = nk\, c(w,s) M_n(w,s)\,, \ee
reproducing the well-known fact that the monopole acquires gauge charge.

In course of studying dualities, many interesting examples of 3d $\CN=2$ theories have been have been found with monopole operators in their \emph{superpotentials}. (Some of the first examples of this kind appeared in \cite{Karch-duality,Aharony-duality}.) This is yet another nonperturbative aspect of gauge theories that would be important to include in the future.

\subsection{The propagator}
\label{sec:prop}

The propagator $G(z,\bar z,t;w,\bar w,s)$ in a Lagrangian HT QFT is a form on the configuration space of two distinct points in spacetime, $\text{Conf}_2(\C\times \R)$, that satisfies%
\footnote{Including a factor of $2\pi i$ in \eqref{prop-eq} is mainly a matter of convention. The precise factor depends on the normalization of the kinetic terms in the action, and it's convenient to assume we've fix the normalization so that a $2\pi i$ appears in \eqref{prop-eq}.}
\be d'_{\bar z,t} G =- d'_{\bar w,s} G = 2\pi i \delta^{(3)}(z-w,\bar z-\bar w,t-s)\,. \label{prop-eq} \ee
Note that it must be a 1-form in the odd variables $d\bar z,dt,d\bar w,ds$, and a 1-form in the even $dz,dw$. It may be thought of as the 2-point function in free theory
\be G(z,\bar z,t;w,\bar w,s) = \big\langle \mb p(z,\bar z,t)\, \mb x(w,\bar w,s) \big\rangle\,,\ee
with form degrees in $G$ matching the multiforms $\mb p,\mb x$ on the RHS.

The differential operator $d'$ has a very large kernel, including all holomorphic functions. Correspondingly, there is a great deal of ``gauge freedom'' in choosing a propagator $G$. In order to be consistent with a computation of 2-point functions, each choice must be aligned with particular choices of boundary conditions on the fields. A common choice is the HT analogue of Lorentz gauge, in which fields (and the propagator) satisfy $\pd_t x_t+2\pd_z x_{\bar z} = 0$ and $\pd_t p_t+2\pd_z p_{\bar z} = 0$, giving (\cf\ \cite[Sec. 3]{Gwilliam:2019cbp}, \cite[Sec. 5]{CDGbdry})
\be G^{\rm Lor} = \frac{1}{2}\frac{(\bar{z}-\bar{w})d(t-s)-(t-s)d(\bar{z}-\bar{w})}{\left((t-s)^2+|z-w|^2\right)^{\frac{3}{2}}}\,dz^{J(p)}\,dw^{J(x)}\,. \label{G-Lor} \ee
This is compatible with balanced decay of the fields in both space and time.

Alternatively, we may consider an axial gauge, in which fields satisfy $x_{\bar z}=p_{\bar z}=0$ and the propagator takes the simpler form
\be G^{\rm ax} = \frac{\delta^{(1)}(t-s)}{z-w} dz^{J(p)}\,dw^{J(x)} 
  \label{G-hol} \ee
with $\delta^{(1)}(t-s) = \delta(t-s)(dt-ds)$.
This is usually not appropriate for computing correlation functions of local operators in the presence of (say) an asymptotic vaccum, due to the very slow decay of $\frac{1}{z-w}$ at spatial infinity. However, this propagator does capture the singular terms in the OPE of line operators, as two lines get very close to each other. We'll compare the two gauges further in Section \ref{sec:exact-gen}.

\subsection{Topological quantum mechanics}
\label{sec:topQM}

We will construct line operators by coupling a bulk 3d HT theory to 1d topological quantum mechanics. To set the stage, we collect a few results about topological quantum mechanics and its deformations.

Abstractly, the data of topological quantum mechanics is a vector space $V$ of states and an algebra $A$ of local operators. If $V$ is finite dimensional, then $A=\text{End}(V)$. If $V$ is infinite dimensional, then $A$ is typically a proper subalgebra of all linear endomorphisms of $V$; at the very least, $V$ should be treated as a topological vector space and endomorphisms in $A$ are required to be continuous.

In a cohomological setting, $V$ is further promoted to a dg vector space, \emph{i.e.} a chain complex with a BRST differential $Q$; and $A$ is promoted to a dg algebra, endowed with an induced differential $[Q,-]$. We'll assume we are working with theories that have a $\Z_2$ fermion number $F$, and a $\Z$ (or more generally $\Q$) valued cohomological grading with charge $R$, such that $Q$ is odd with $R(Q)=1$.

Just like in 3d HT theories, each local operator $\CO$ in topological QM comes with a descendant $\CO^{(1)}$, obeying
\be  Q \CO^{(1)} + (Q\CO)^{(1)} = d_t \CO\,, \ee
where $d_t = \pd_t dt$ is the 1d exterior derivative. Second and higher descendants of course vanish. We again use boldface to denote sums of descendants, $\bm \CO := \CO + \CO^{(1)}$.

\subsubsection{Deformations}
\label{sec:QM-def}

Deformations of topological QM will be the key to coupling a QM theory to a bulk QFT. Perturbative deformations can be analyzed from two different but ultimately equivalent perspectives. It is useful to be able to translate between them, so we review how this works. This is a classic result in topological field theory; see \emph{e.g.} \cite{PaquetteWilliams, GaiottoKulpWu, Gaiotto:2019wcc} for further discussion and references.

In a cohomological setting, the only nontrivial perturbative deformations are those that change the differential on the state space and operator algebra,
\be Q \to Q + M \label{QM-Qdef} \ee
for some local operator $M\in A$ with $F(M)=\text{odd}$ and $R(M)=1$. To avoid a BRST anomaly, \ie\ for the new differential to still be a differential, we must have $(Q+M)^2=0$, or
\be\text{Maurer-Cartan:}\qquad Q(M)+M^2 =  Q(M) + \tfrac12[M,M] = 0\,. \label{QM-MC} \ee
Deformations are classified by solutions to this Maurer-Cartan equation --- up to an equivalence relation that essentially says that $Q$-exact deformations act trivially.

Alternatively, deformations may be constructed by deforming the effective action of the QM theory. ``Deforming the action'' makes sense even if the original theory isn't Lagrangian, or the Lagangian is unknown; it is merely a prescription for modifying correlation functions. Suppose we change the action by
\be \delta S = \int_\R M' dt\,. \label{QM-S-def} \ee
for some local operator $M'$ with $F(M')=\text{even}$ and $R(M')=0$. Then, by definition, correlation functions of local operators $\langle a_1(t_1)a_2(t_2)...\rangle$ are modified by inserting time-ordered, integrated ``segments'' of the deformation inbetween each consectutive pair of operators:
\begin{align} \langle a_1(t_1)a_2(t_2)...\rangle_{\delta S} &:= 
   \Big\langle T\Big[e^{-\int_{\R\backslash\{t_1,t_2,...\}} M' dt}a_1(t_1)a_2(t_2)...\Big]\Big\rangle \label{QM-corr-def}  \\
   &=  \Big\langle T\Big[ e^{-\int_{t_1+\epsilon}^\infty M' dt}\Big] a_1(t_1) T\Big[ e^{-\int_{t_2+\epsilon}^{t_1-\epsilon} M' dt}\Big] a_2(t_2) T\Big[ e^{-\int_{...}^{t_2-\epsilon} M' dt}\Big]...\Big\rangle\,. \notag
\end{align}
Here `$T$' denotes time ordering, and in the second line we've assumed $t_1>t_2>...$.

This sort of expression should not be unfamiliar: the deformation of the action of quantum mechanics by a Hamiltonian (essentially $M'$ here) is well known to modify correlation functions in precisely this way! Each $T\Big[ e^{-\int_{t+\epsilon}^{t'-\epsilon} M' dt}\Big]$ is ``time evolution.''

The key result we now need is that a deformation \eqref{QM-S-def} is free of BRST anomalies precisely when $M'dt$ is the first descendant of a local operator $M$ (with $F(M)=\,$odd, $R(M)=1$) and $M$ satisfies the MC equation,
\be M'dt  = M^{(1)}\qquad\text{and}\qquad Q(M)+M^2=0\,. \label{QM-def-desc} \ee
Moreover, the effect of the deformation on the operator algebra in this case is simply to deform the differential $Q_{\delta S} = Q+[M,-]$ just like in \eqref{QM-Qdef}. Thus, anomaly-free deformations of the action and deformations of the differential are equivalent.

To understand the equivalence a bit better, let's assume $M'dt=M^{(1)}$ and consider a the BRST variation of a segment of deformation $T\Big[ e^{-\int_{0}^{1} M^{(1)}}\Big]$ between (say) $t=0$ and $t=1$. When de-exponentiating the deformation, one must be careful to regularize so that multiple operators $M^{(1)}$ are never integrated across coincident points. We have:
\begin{align} Q\, T\Big[ e^{-\int_{0}^{1} M^{(1)}}\Big] &:= Q\Big( -\int_{1>t>0} M^{(1)}(t) +\frac12 \int_{1>t_1\neq t_2>0} T\big[ M^{(1)}(t_1)M^{(1)}(t_2)\big] + \ldots \Big) \notag \\
& = - \int_{1>t>0} Q M^{(1)}(t) +  \int_{1>t_1> t_2>0}  Q\big(M^{(1)}(t_1)M^{(1)}(t_2)\big) + \ldots \notag \\
&= - \int_{1>t>0} dM + \int_{1>t>0} (QM)^{(1)} +  \!\int_{t_1>t_2} d\big(M(t_1)M^{(1)}(t_2)-M^{(1)}(t_1)M(t_2)\big) \notag  \\
&= -M(1)+M(0)+\int_{1>t>0} \Big( (QM)^{(1)} +[M^{(1)},M]\Big) +\ldots \label{QM-def-expand} 
\end{align}
where the descent equation $QM^{(1)}+(QM)^{(1)}=dM$ and Stokes' theorem leads to (1) boundary terms at $t=0,1$ and (2) bulk terms proportional to the integral of $(QM)^{(1)} +[M^{(1)},M]$. The latter is just the descendant of the MC equation itself: $\big( Q\,M+M^2\big)^{(1)} = (Q\,M)^{(1)}+M^{(1)} M- M M^{(1)} = (Q\,M)^{(1)} + [M^{(1)},M] $; so if \eqref{QM-def-desc} holds then the bulk terms vanish. The cancellations of bulk terms continue to all orders (see \cite{PaquetteWilliams, Gaiotto:2019wcc}), and the entire variation resums to
\be   Q\, T\Big[ e^{-\int_{0}^{1} M^{(1)}}\Big]  = -M(1) T\Big[ e^{-\int_{0}^{1} M^{(1)}}\Big] + T\Big[ e^{-\int_{0}^{1} M^{(1)}}\Big] M(0) \label{Q-def-segment} \ee
In turn, \eqref{Q-def-segment} guarantees that the BRST variation of a deformed correlator \eqref{QM-corr-def} is equivalent to deforming the $Q+[M,-]$ variation of the original correlator:
\begin{align} \big\langle Q_{\delta S}(a_1(t_1)a_2(t_2)...) \big\rangle_{\delta S} &:= \Big\langle Q\Big( T\Big[ e^{-\int_{t_1+\epsilon}^\infty M^{(1)}}\Big] a_1(t_1) T\Big[ e^{-\int_{t_2+\epsilon}^{t_1-\epsilon} M^{(1)}}\Big] a_2(t_2) T\Big[ e^{-\int_{...}^{t_2-\epsilon} M^{(1)}}\Big]...\Big)\Big\rangle \notag \\
& \hspace{-1.3in}= \Big\langle T\Big[ e^{-\int_{t_1+\epsilon}^\infty M^{(1)}}\Big] \Big(Q\,a_1(t_1)+M(t_1+\epsilon)a_1(t_1)-(-1)^{F(a_1)} a_1(t_1)M(t_1-\epsilon)\Big) T\Big[ e^{-\int_{t_2+\epsilon}^{t_1-\epsilon} M^{(1)}}\Big] a_2(t_2) ... \notag \\
&\hspace{-1.2in} + T\Big[ e^{-\int_{t_1+\epsilon}^\infty M^{(1)}}\Big] a_1(t_1) T\Big[ e^{-\int_{t_2+\epsilon}^{t_1-\epsilon} M^{(1)}}\Big] (-1)^{F(a_1)}\Big( Q\,a_2(t_2)+ M(t_2+\epsilon) a_2(t_2)- \ldots \Big)\cdots \Big\rangle \notag \\
&\hspace{-1.3in}=  \big\langle Q(a_1(t_1)a_2(t_2)...) + [M,a_1](t_1) a_2(t_2)... + (-1)^{F(a_1)} a_1(t_1)[M,a_2](t_2)...+\cdots\big\rangle_{\delta S} \notag \\
&\hspace{-1.3in}= \big\langle (Q+[M,-])\cdot (a_1(t_1)a_2(t_2)...)\big\rangle_{\delta S}
\end{align}
This ties the deformation of the action by $M^{(1)}$ to the algebraic deformation \eqref{QM-Qdef}.

\subsubsection{Free fields}
\label{sec:QM-free}

Many topological quantum mechanics (including all QM with finite-dimensional state spaces) can be engineered by quantizing free fields, deformed by interactions. We'll use free-field models in many of our examples of line operators.

Analogous to \eqref{HT-action}, 1d free fields with interactions have general action
\be S = \int_{\R_t} \Ba^i d_t\bar\Ba_i  + I(\Ba,\bar\Ba) \label{QM-action} \ee
where $\Ba,\bar\Ba$ are multiforms on $\R_t$, and $d_t = \partial_t dt$ is the 1d exterior derivative. Conventions are such that $dt$ is odd with $R(dt)=1$; also the 1d Lagrangian must be odd, with $R=1$, $J=0$, to be consistently integrated along $\R_t$. This means for each $i$ that either
\be \Ba^i \in \Omega^\bullet(\R_t)[R_i]dz^{J_i}\,,\quad \bar\Ba_i \in \Omega^\bullet(\R_t)[-R_i]dz^{-J_i} \label{QM-bos} \ee
or
\be \quad \Ba^i \in \Pi\Omega^\bullet(\R_t)[R_i]dz^{J_i}\,,\quad \bar\Ba_i \in \Pi\Omega^\bullet(\R_t)[-R_i]dz^{-J_i} \label{QM-ferm} \ee
(\ie\ the leading components of $\Ba^i$ and $\bar\Ba_i$ are both bosonic or both fermionic) and
\be F(I)=\text{odd}\,,\qquad R(I) = 1\,,\qquad J(I)= 0\,.\ee

When $I=0$, the BRST operator in \eqref{QM-action} acts as $Q_0=d_t$ on all fields. When $I\neq 0$, the BRST operator acts as
\be  Q\Ba^i = d_t\Ba^i +\pd_{\bar\Ba_i} I(\Ba,\bar\Ba)\,,\qquad Q \bar\Ba_i = d_t \bar\Ba_i + \pd_{\Ba^i} I(\Ba,\bar\Ba) \,. \label{QM-Qact} \ee
Having $Q^2=0$, \ie\ avoiding a BRST anomaly, is typically a nontrivial constraint on $I$.

The quantization of \eqref{QM-action} at $I=0$ can be given a state space $V$ consisting of polynomials in the leading components $a^i$ of the $\Ba^i$ (either bosons or fermions), and an operator algebra generated by the leading components $a^i$ and $\bar a_i$,
\be V = \C[a^i]\,,\qquad A = \C[a^i,\bar a_i]/\big([\bar a_i,a^j]=\delta_i{}^j\big)\,. \label{QM-VA} \ee
with canonical commutation relations in the operator algebra
\be [\bar a_i,a^j]=\delta_i{}^j\,. \ee
(This is the $Q_0$-cohomology of the state space and operator algebra.) Recall that $[a^i,\bar a_j]:=  a^i\bar a_j - (-1)^{F(a^i)F(\bar a_j)} \bar a_j a^i$ denotes either a commutator or anti-commutator, depending on fermion numbers. When acting on $V$, $a^i$ acts as left multiplication and $\bar a_i$ acts as a derivative $\pd_{a^i}$.

Adding the interaction term $I(\Ba,\bar\Ba)$ deforms $V$ to a complex with differential $Q = I(a,\bar a)$ and deforms $A$ to an algebra with differential $[Q,-]$. The BRST anomaly vanishes, and $Q^2=0$, if any only if the Maurer-Cartan equation holds,
\be \underbrace{Q_0 I(a,\bar a)}_{=0} +  I(a,\bar a)^2=   I(a,\bar a)^2  = 0 \quad \text{(as an element of $A$)}\,. \ee

An alternative description of theories of type \eqref{QM-action} is as B-type topological twists of 1d $\CN=2$ quantum mechanics with either chiral multiplets \eqref{QM-bos} or fermi multiplets \eqref{QM-ferm}, together with a fermionic superpotential $I(\Ba,\bar\Ba)$. The fermionic superpotential that appears here includes both standard E and J terms of 1d $\CN=2$ theories \cite{DineSeiberg-02,Witten-phases,KapustinLi}. For example, if $(\bar\Bb_j,\Bb^j)$ are fermi multiplets and the remaining $(\bar\Ba_i,\Ba^i)$ are chirals, standard E and J superpotential terms would be written
\be I(\Ba,\Bb,\bar\Ba,\bar\Bb) = \bar\Bb_j E^j(\Ba) + \Bb^j J_j(\Ba)\,. \label{QM-EJ} \ee
The MC equation becomes $I(a,b,\bar a,\bar b)^2 = 2 E^j(a)J_j(a) = 0$, a familiar constraint.

\subsection{The category of line operators}
\label{sec:line-setup}

Now, let's consider the expected structure of line operators in a 3d HT QFT on $\C_z\times \R_t$ in a bit more detail.

Lines inserted at each point $z\in \C$ and extended along $\R_t$ form a linear category $\CC_z$. As explained in the introduction, its objects are the local ``labels'' $\ell$ that define line operators, and its morphism spaces $\text{Hom}_{\CC_z}(\ell,\ell')$ are the vector spaces of local operators at a junction of lines:
\be \raisebox{-.4in}{\includegraphics[width=3.8in]{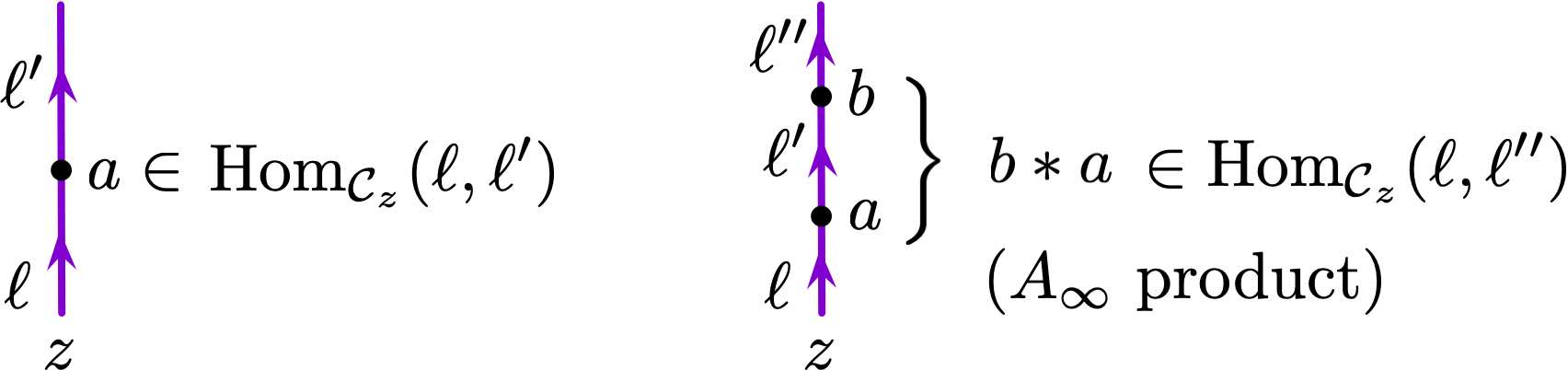}} \ee
Since we assume we have a cohomological theory, with a BRST operator $Q$, the spaces $\text{Hom}_{\CC_z}(\ell,\ell')$ become a dg vector space, with differential $Q$. They are endowed with all three gradings/symmetries discussed in Section \ref{sec:sym}: fermion number $F\in \Z_2$, cohomological $R$, and rotation $J$ in the $\C$ plane about the point $z$. Moreover, collision of junctions in the $\R_t$ direction defines a product on local operators there (\ie\ a composition of morphisms) that is associative up to $Q$-exact terms (up to homotopy). Altogether, this endows $\CC_z$ with the structure of an $A_\infty$ category. In particular, each endomorphism space $\text{End}_{\CC_z}(\ell):=\text{Hom}_{\CC_z}(\ell,\ell)$ is an $A_\infty$ algebra. If we take $Q$-cohomology, then composition of morphisms becomes associative and $H^\bullet(\text{End}_{\CC_z}(\ell),Q)$ becomes a standard associative algebra; but simply taking $Q$-cohomology is a forgetful operation that may neglect nontrivial higher operations.

A special case is the identity, or trivial line $\ell = \id$. Its endomorphisms are bulk local operators (inserted at $z$)
\be \text{End}_{\CC_z}(\id) \simeq \CA =\text{bulk local operators}. \label{End1A} \ee

In terms of an HT-twisted 3d $\CN=2$ theory, the objects $\ell\in \CC_z$ correspond to half-BPS line operators of the 3d $\CN=2$ theory. Namely, they preserve the supercharge $Q$ that defines the HT twist together with a second supercharge $\ol Q$ that satisfies $[Q,\ol Q]=\pd_t$. (In the conventions of \cite{DGPbdry}, $Q=\ol Q_-$ and $\ol Q = Q_+$.) Similarly, after passing to cohomology, local operators at junctions correspond to quarter-BPS local operators in the 3d $\CN=2$ theory that preserve $Q$.

Translation invariance of our QFT implies that for any two points $z,w\in \C$, the categories $\CC_z,\CC_w$ are isomorphic, via $A_\infty$ functors
\be T_{z-w}:\CC_w \overset\sim\longrightarrow \CC_z\,. \label{trans-C} \ee
such that in turn $T_z \circ T_{z'} \simeq T_{z+z'}$.
We will often take the isomorphisms \eqref{trans-C} for granted, and just write `$\CC$' for the category at any given point in $\C$.

In addition, lines in a 3d HT QFT should have an OPE, encoded in a family of functorial tensor products
\be \otimes_{z}: \CC_{w+z} \boxtimes \CC_w  \to  \CC_w \quad (z\in \C^*) \quad \text{or simply} \quad \otimes_z :\CC\boxtimes\CC\to \CC\,. \label{oz-intro} \ee
This collects several pieces of physical data, including:
\begin{enumerate}
\item The tensor product of objects $\ell\otimes_z\ell'$, defined as the effective line operator at $w\in \C$ that is equivalent to an insertion of $\ell$ at $w+z$ and $\ell'$ at $w$ in all correlation functions (with all other operators inserted further away): 
\be \raisebox{-.35in}{\includegraphics[width=2.2in]{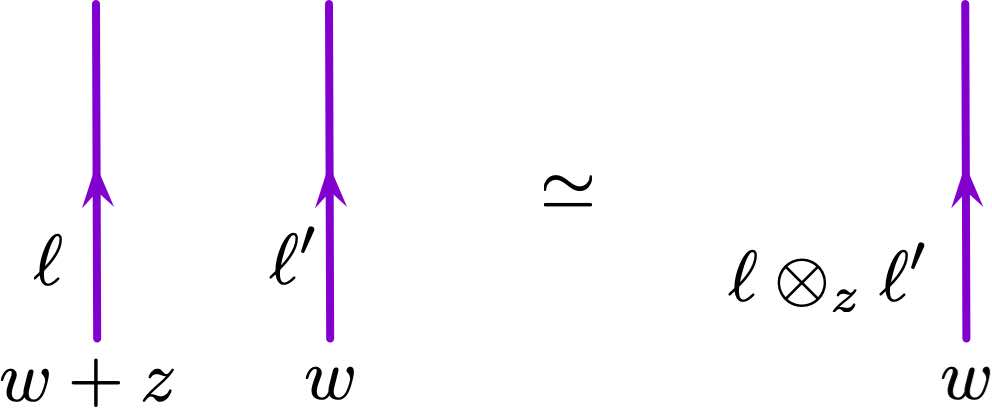}} \ee

\item The tensor product of morphisms (local operators), families of linear maps
\be \begin{array}{cccccc}
\otimes_z:&\text{Hom}(\ell_1,\ell_2) &\otimes & \text{Hom}(\ell_1',\ell_2') & \to & \text{Hom}(\ell_1\otimes_z\ell_1',\ell_2\otimes_z\ell_2') \\
& a && a' &\mapsto & a\otimes_z a' \end{array} \label{oz-hom-intro} \ee
for all $\ell_{1,2},\ell_{1,2}'$ that preserve the $A_\infty$ compositions of junctions.
The OPE $a\otimes_z a'$ is defined as the local operator at the junction of $\ell_1\otimes_z\ell_1'$ and $\ell_2\otimes_z\ell_2'$ that behaves the same way as inserting $a$ and $a'$ separately in all correlation functions:
\be \raisebox{-.35in}{\includegraphics[width=2.6in]{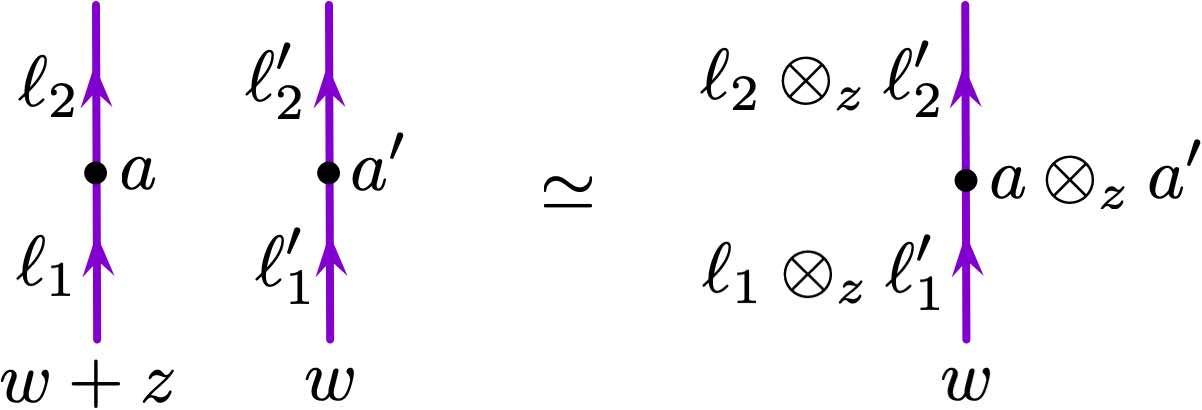}} \ee

This OPE of junctions most closely matches the OPE of local operators in a vertex algebra. In particular, if $\ell_1=\ell_2=\ell_1'=\ell_2'=\id$ are all the identity line, then \eqref{oz-hom-intro} produces a family of maps $\otimes_z:\CA\otimes \CA\to \CA$ on bulk local operators, encoding the (nonsingular) OPE in the vertex algebra of bulk local operators. 
In general, $a\otimes_z a'$ has local holomorphic dependence on $z$ and may be singular at $z=0$. 

\item Natural isomorphisms $(\ell\otimes_z \ell')\otimes_{z'} \ell'' \simeq \ell\otimes_{z+z'}(\ell'\otimes_{z'} \ell'')$ reflecting OPE associativity.
\end{enumerate}

\noindent As prefaced in the introduction, a main goal in the rest of the paper will be to understand the OPE on the category $\CC$. We will do this by combining several different physically meaningful representations of $\CC$, each of which make different features manifest.

\section{Line operators from quantum mechanics}
\label{sec:lines-QM}

We now combine the background of Section \ref{sec:setup} to give a systematic construction of line operators in perturbative 3d HT theories, by coupling a bulk 3d QFT to 1d topological quantum mechanics. Abstractly, one arrives at the statement (\cf\ \cite{Costello-Yangian, CPkoszul}) that 
\be \text{perturbative line operators $\ell$} \quad\leftrightarrow\quad  \text{pairs}\; (V_\ell,\mu_\ell)  \label{pertl-Vmu} \ee
where $V_\ell$ is the dg vector space of a 1d quantum mechanics, and $\mu_\ell$ is a Maurer-Cartan element that encodes the coupling of $V_\ell$ to the bulk. It satisfies an $A_\infty$ MC equation in $\text{End}(V_\ell)\otimes\CA$, where $\CA$ is the bulk operator algebra.

After reviewing this in Section \ref{sec:lines-QM-idea}, we devote the rest of the section to explaining how some familiar (and some less familiar) examples of line operators in HT-twisted 3d $\CN=2$ theories arise in this formalism.  We consider free matter theory in Section~\ref{sec:line-chiral} rediscover flavor vortex lines (dual to Wilson lines under particle-vortex duality, or Witten's $SL(2,\Z)$ action on 3d theories \cite{Witten:2003y,KWY-abelian}). Then in Section \ref{sec:line-W} we consider the effect of adding a superpotential $W$. This restricts the set of non-anomalous vortex lines in a way related to tropical geometry, as we find a new constraint that the residue $\oint W$ (computed on an infinitesimal circle linking the line) must vanish. We also encounter the first truly nontrivial solutions to the MC equation for couplings $\mu$, which we illustrate in the XYZ model.

Finally, in Section \ref{sec:line-gauge} we make some brief observations about perturbative line operators in gauge theory. We recall how Wilson lines are defined in the formalism \eqref{pertl-Vmu}. Various vortex lines can also be constructed, but keeping them BRST-anomaly-free (making sure $\mu_\ell$ satisfies the MC equation) depends strongly on the Chern-Simons levels and matter content. We give some examples of hybrid Wilson/vortex lines in abelian theories with matter. We will find a more systematic construction of Wilson/vortex lines in Section \ref{sec:gauge-loop}, using representation theory of the Koszul dual $\CA^!$.

\subsection{General idea}
\label{sec:lines-QM-idea}

Let $\CT$ be a 3d holomorphic-topological QFT, with fermion number $F$, ghost number (homological degrees) $R$, and spin $J$. We'd like to construct a line operator along $\R_t\times \{0\} \subset \R_t\times \C_z$, \ie\ at position $z=0$ in the holomorphic plane.

Let $\CA$ denote the algebra of local operators of $\CT$, restricted to the real line $z=0$.
We recall from Section \ref{sec:descent} that (the cohomology of) $\CA$ is in general a (-1)-shifted Poisson vertex algebra \cite{OhYagi-Poisson, GarnerWilliams,AlfonsiKimYoung}, graded by $F,R,J$. The restriction to the line $z=0$ ignores the vertex-algebra structure (holomorphic OPE's) and leaves behind the structure of an $A_\infty$ algebra, with differential $Q$ and gradings $F,R,J$. 

It's useful to compare the structure of the algebra here to that in topological quantum mechanics (Section \ref{sec:topQM}). \emph{Both} are special cases of $A_\infty$ algebras. In purely 1d QM, local operators form a dg algebra, which we think of as an $A_\infty$ algebra with trivial higher operations. Its cohomology is an associative algebra. It's also maximally non-commutative, in the sense that it should have a trivial center. 
In a 3d HT theory, local operators at $z=0$ form an $A_\infty$ algebra that can have nontrivial higher operations due to interactions with the ``bulk'' away from $z=0$. Its cohomology is associative and necessarily commutative, due to the extra bulk directions.

We can create a perturbative line operator $\ell$ in $\CT$ that preserves $F,R,J$ by
\begin{itemize}
    \item[1)] Choosing a 1d topological quantum mechanics, labeled by a state space $V_\ell$ and a (dg) operator algebra $A_\ell \subseteq V_\ell$. We additionally assume this QM has an abelian global symmetry with charge $J$.
    \item[2)] Choosing an element $\mu_\ell \in A_\ell\otimes \CA$ with $F(\mu_\ell)=\text{odd}$, $R(\mu_\ell)=1$, $J(\mu_\ell)=0$ that we use to deform the action
\be \delta S = \int_{\R_t\times \{z=0\}} \mu_\ell^{(1)} =  \int_{\R_t\times \{z=0\}} \bm \mu_\ell \,. \label{line-def} \ee
 \end{itemize}

Here $\mu_\ell^{(1)}$ denotes a topological descendant (in the $A_\ell$ component) and an HT descendant (in the $\CA$ component). The integral just picks off the $dt$ component. For example, if $\mu_\ell = M \otimes \CO$ with $M\in \text{End}(V_\ell)$, $\CO\in \CA$, then
    \be \mu_\ell^{(1)} = M^{(1)} \otimes \CO + M\otimes \CO^{(1)} = dt[ M^{(1)}_t \otimes \CO - M\otimes \CO^{(1)}_t] + \ldots \ee
 We also have $\mu_\ell^{(1)}\big|_{dt} = \bm \mu_\ell \big|_{dt}$ with $\bm \mu_{\ell} = \mb M \otimes \bm \CO$, where as usual the multiform descendants are $\mb M = M+M^{(1)}$, $\bm \CO = \CO+\CO^{(1)}+\CO^{(2)}$.

To ensure that the deformation \eqref{line-def} is (perturbatively) free of BRST anomalies, we must require that $\mu_\ell$ satisfies a Maurer-Cartan equation. If $\CA$ is a dg algebra (no higher $A_\infty$ operations) then this is simply the Maurer-Cartan equation \eqref{QM-MC}, namely $Q(\mu_\ell)+(\mu_\ell)^2=0$. More generally, when $\CA$ (and thus $A_\ell\otimes \CA$) is a nontrivial $A_\infty$ algebra, the correct generalization is the $A_\infty$ Maurer-Cartan equation
\be \sum_{k=1}^\infty m_k(\mu_\ell,...\mu_\ell) = 0 \qquad\text{in $A_\ell\otimes\CA$}\,,  \label{line-MC-gen} \ee
where $m_1(a)=Q(a)$ is the differential, $m_2(a,b) = ab$ is the multiplication, and $m_{\geq 3}$ are the higher products.
See \cite{PaquetteWilliams, Gaiotto:2019wcc} and \cite[App. A]{GMW} for reviews of how the deformation analysis from Section \ref{sec:QM-def} is generalized in the presence of higher operations. Roughly, higher operations contribute to higher-order terms in \eqref{QM-def-expand}.

That is essentially it. One finds, schematically, that \emph{perturbative line operators in $\CT$ are in 1-1 correspondence with pairs $(V_\ell,\mu_\ell)$, where $V_\ell$ is a dg vector space and $\mu_\ell\in \text{End}(V_\ell)\otimes \CA$ is a solution of the MC equation}, up to a suitable notion of isomorphism on both sides.

In practice, one may want to be more precise about the form of $\mu_\ell$ --- \ie\ what sorts of couplings are allowed. For example, $\CA$ is typically infinite dimensional, but has spin $J$ bounded from below. One may want to allow couplings of the form
\be \mu_\ell = \sum_{n=0}^\infty M_n \otimes \CO_n\,,\qquad M_n\in A_\ell\,,\; \CO_n \in \CA \label{inf-mu} \ee
with infinitely many independent $\CO_n$'s, but for each $j\in \R$ only finitely many satisfying $J(\CO_n)\leq j$. The MC equation still makes sense in this case. Formally, one would say that $\mu_\ell$ belongs to a completed tensor product $A_\ell \widehat\otimes \CA$. Physically,  such infinite couplings tend to correspond to allowing very ``heavy'' lines that represent boundary conditions wrapped on an infinitesimal circle. Including them is a choice.

\subsubsection{Linear coupings}
\label{sec:linear}

We want to further simplify the form of the couplings.

In Section \ref{sec:Ainf} we'll see explicitly how algebras $\CA$ of local operators in perturbative 3d HT theories are generated  \emph{linearly} by the elementary fields and their $\pd_z$ derivatives.
Explicitly, for a Lagrangian theory of the form \eqref{HT-action} with fields $\{\mb p_i,\mb x^i\}$ and some interaction $\CI(\mb p,\mb x,\pd)$, we find a model for $\CA$ that is generated (as an algebra) by the subspace of linear operators
\be \CA_1 := \C\langle \pd^np_i, \pd^nx^i\rangle_{n\geq 0} \subset \CA\,, \label{def-A1} \ee
containing the leading components of $\mb p_i,\mb x^i$ and their derivatives. $\CA$ itself is generated, as a vector space, by polynomials in the linear operators from \eqref{def-A1}; it has various relations in its products and higher products determined by the interaction $\CI$.

In such a situation, we might hope that every perturbative line operator is equivalent to one that's defined by purely linear couplings, \ie\ one with some $V_\ell$ and $\mu_\ell$ of the form
\be \mu_\ell = \sum_{i,n} \alpha_{i,n} \pd^n x^i + \beta^i_n \pd^n p_i\,,\qquad \alpha_{i,n},\,\beta^i_n\in \text{End}(V_\ell)\,. \label{mu-linear} \ee
This hope turns out to be correct, but it's rather nontrivial to prove in general. The cleanest argument (conceptually) involves  showing that perturbative lines are equivalent to modules for the Koszul-dual algebra $\CA^!$, and then showing that $\CA^!$ is linearly generated by the shifted linear dual of $\CA_1$. We'll develop this chain of reasoning in Section \ref{sec:KD-MC} to prove that linear couplings \eqref{mu-linear} are always sufficient.

For some \emph{intuition} of why any coupling is equivalent to a linear one, consider the following example. Suppose we have a line operator constructed from quantum mechanics with space $V$ and
\be \mu = \mu_0 + M\CO\CO'\,,\qquad M\in \text{End}(V)\,,\quad \CO,\CO'\in \CA\,, \ee
where $\mu_0$ and $M\CO\CO'$ separately satisfy the MC equation, and $\CO,\CO'$ are $Q$-closed bulk local operators that commute with each other. This represents a non-linear coupling, which we want to simplify. We define a second line operator by adding an extra 1d free boson $(\mb b,\bar{\mb b})$ and fermion $(\mb a,\bar{\mb a})$ (as in Section \ref{sec:QM-free}), with a 1d potential $I = \mb a\mb b$, so that the QM vector space becomes $\wt V= V\otimes\C[a,b]$, $Q=Q_V+ab$. (Note that $\wt V$ and $V$ are quasi-isomorphic.) We also introduce a new coupling
\be \wt \mu = \mu_0 + M\,\CO\,a -b\,\CO'\,,\qquad Ma,b\in\text{End}(\wt V)\,, \quad \CO,\CO'\in \CA\,, \ee
which has split up the bulk operators. (This assumes $\CO'$ is fermionic; otherwise swap $a$ and $b$.)
To see that the modified $(\wt V,\wt\mu)$ line is equivalent to the original $(V,\mu)$ line, it's helpful to consider the total deformation to the 1d action:
\be \delta S_{\rm new} = \int_\R \mb a d\mb{\bar a} + \mb b d\mb{\bar b} + \wt{\bm\mu}  = \int_\R \mb a d\mb{\bar a} + \mb b d\mb{\bar b} + \bm\mu_0  + \mb M  \CO\hspace{-.3cm}\CO \mb a + \mb b\mb a - \mb b\CO\hspace{-.3cm}\CO' \ee
The quadratic term $\mb b\mb a$ lets us integrate out both new 1d multiplets, and the equation of motion for $\mb b$ (say) sets $\mb a = \CO\hspace{-.3cm}\CO'$, recovering 
\be \delta S_{\rm old} = \int_\R \bm \mu_0 + \bmu  \CO\hspace{-.3cm}\CO \CO\hspace{-.3cm}\CO' = \int_\R \bm \mu\,. \ee

By iterating this procedure, multilinear couplings --- at least those whose linear factors all commute --- can be split up into linear couplings. Generalizing the argument above to the case where $\CO$ and $\CO'$ don't commute can be done by adding extra terms on both sides to satisfy the MC equations;  this seems tricky to carry out in full generality.

\subsection{Example: free chiral and vortex lines}
\label{sec:line-chiral}

We'll now give examples of a few simple families of line operators in HT-twisted 3d $\CN=2$ theories, relating the construction via coupling to quantum mechanics to more familiar physical descriptions.

Consider an HT-twisted free chiral multiplet, on $\C_z\times \R_t$. As introduced in Section \ref{sec:3dN2}, it has an action
\be S = \int_{\C\times \R} \BPsi d' \BX\,, \ee 
where $\BX,\BPsi$ are multiform-valued superfields with leading components $X$ (a complex boson) and $\psi$ (a complex scalar fermion), with R-charges \eqref{chiral-charges}.

The free chiral also has a $U(1)_f$ global ``flavor'' symmetry, under which $\BX,\BPsi$ have weights $+1,-1$. In the HT twist, the symmetry gets enhanced to the infinite-dimensional group of holomorphic $\C^*$-valued functions on $\C_z$. Given such a function $g(z)$, the fields transform as
\be (\BX,\BPsi) \mapsto (g(z)\BX,g(z)^{-1}\BPsi)\,. \label{free-flavor} \ee
One can create nontrivial line operators, sometimes called \emph{flavor vortex lines}, by acting with a singular flavor transformation --- one that's not $\C^*$-valued. For $N\in \Z$, let $\V_N$ denote the vortex line created at (say) $z=0$ by the flavor transformation $g(z) = z^N$.%
\footnote{We are only interested in behavior near $z=0$. Globally, any meromorphic function $g(z)$ that behaves like $z^N$ near $z=0$ and is $\C^*$-valued away from $z=0$ will do.}
Put differently, $\V_N$ is a disorder operators that requires field to have a singular profile
\be \V_N : \quad \left\{ \begin{array}{l}\BX(z) \sim z^N\times(\text{regular)} \\[.1cm] \BPsi(z) \sim z^{-N}\times(\text{regular}) \end{array}\right. \quad\text{near $z= 0$} \qquad (N\in \Z)\,. \label{free-VN-vortex} \ee
In the underlying 3d $\CN=2$ theory, these are half-BPS line operators, called ``global vortex loops'' in \cite{KWY-abelian}.

We can construct such vortex lines by coupling to quantum mechanics in a standard way.%
\footnote{This is similar to the analysis of vortex lines in A-twisted 3d $\CN=4$ theories in~\cite[Sec.~4]{lineops}.} %
It's useful to recall (\eg\ from \cite{CDGbdry}) that the bulk operator algebra $\CA$ is quasi-isomorphic to a commutative vertex algebra with zero differential, generated by $X(z),\psi(z)$ and their holomorphic $\pd$ derivatives. Restricting to $z=0$, we get a (super)commutative polynomial algebra
\be \CA = \C[\pd^nX(0),\pd^n\psi(0)]_{n=0}^\infty \ee
Then:
\begin{itemize}
    \item If $N=0$, $\V_0=\id$ is the trivial/identity line, so there is nothing to do.
    \item If $N>0$, introduce $N$ 1d fermi multiplets $(\Ba_n,\bar\Ba_n)_{n=0}^{N-1}$, and a Maurer-Cartan element $\mu_N = \sum_{n=0}^{N-1} \tfrac{1}{n!}a_n \pd^n X(0)$. Thus, we insert in the bulk path integral
    \begin{subequations} \label{free-N}
    \be \hspace{-.2in} \V_N(0) = \int \prod_{n=0}^{N-1} D\Ba_n D\bar\Ba_n \;\exp \!\!  \int_{\{0\}\times \R_t} \sum_{n=0}^{N-1} \Big[\Ba_n d\bar \Ba_n  + \underbrace{\frac{1}{n!}\Ba_n \pd^n \BX}_{\bm \mu_N}\Big]\,,\quad (N>0)\,. \label{free-posN}\ee
    \item If $N<0$, introduce $|N|$ 1d bosonic multiplets $(\Bb_n,\bar\Bb_n)_{n=0}^{|N|-1}$, and a Maurer-Cartan element $\mu_N = \sum_{n=0}^{|N|-1} \tfrac{1}{n!}b_n  \pd^n \psi(0)$. Thus, we insert in the bulk path integral
    \be \hspace{-.2in} \V_N(0) = \int \prod_{n=0}^{|N|-1} D\Bb_n D\bar\Bb_n \;\exp \!\!  \int_{\{0\}\times \R_t} \sum_{n=0}^{|N|-1} \Big[\Bb_n d\bar \Bb_n  + \underbrace{\frac{1}{n!}\Bb_n \pd^n \BPsi}_{\bm \mu_N}\Big]\,, \quad (N<0)\,. \label{free-negN} \ee
    \end{subequations}
\end{itemize}

To see why these couplings actually match the desired flavor vortices, consider the total action with a $\V_N$ for (say) $N>0$ inserted:
\be S_{3d}+S_{1d} = \int_{\C_z\times \R_t}\BPsi d' \BX + \delta^{(2)}(z,\bar z)\sum_{n=0}^{N-1} \bigg( \Ba_n d\bar{\Ba}_n + \frac{1}{n!} \Ba_n\,\pd^n\BX\bigg)\,. \ee
The 1d equations of motion for the $\Ba$'s (or in BV formalism, the $Q$ variation of $\bar\Ba$'s) effectively sets  $\pd^n\BX|_{z=0}=0$, $n=0,...,N-1$, enforcing the constraint that $\BX(z)\sim z^N\times(\text{regular})$. Dually, the bulk EOM for $\BPsi$ becomes $Q(\BPsi)=\{S+S_{1d},\BPsi\}_{BV} = d'\BPsi-\sum_{n=0}^{N-1}(-1)^n\pd^n\delta^{(2)}(z,\bar z) \Ba_n = 0$, which now has an extra source term that sets
\be \BPsi(z) =\sum_{n=0}^{N-1} \frac{\Ba_n}{z^{n+1}} + (\text{regular})\,. \ee
Thus the 1d fermions we introduced become identified with the additional polar modes of $\BPsi$ at the vortex.

Matching \eqref{free-negN} with vortex lines $\V_N$ for $N<0$ works the same way: the EOM for $\bar\Bb$'s effectively set regular modes of $\BPsi$ to zero, while the $\Bb$'s become coefficients for new polar modes of $\BX$.

We note that the charges of the 1d fields above are completely determined by the requirement that $F(\mu_N)=\text{odd}$, $R(\mu_N)=1$, $J(\mu_N)=0$; namely,
\be \begin{array}{c|cccc} & \Ba_n & \bar\Ba_n & \Bb_n & \bar\Bb_n \\\hline
 F & \text{odd} & \text{odd} & \text{even} & \text{even} \\
 R & 1-r & r-1 & r & -r \\
 J & -\tfrac{r}{2}-n & \tfrac{r}{2}+n & \tfrac{r}{2}-1-n & 1-\tfrac{r}{2}+n \end{array} \qquad r:=R(X)  \hspace{-.5in} \label{chiral-QM-charges} \ee
Moreover, the MC equation $Q(\mu_N)+\mu_N^2=0$ is trivially satisfied for each $N$. The QM operators $a_n,b_n$ involved in the couplings are $Q$-closed and they all super-commute, since no $\bar a_n$ or $\bar b_n$ appear; and the bulk algebra $\CA$ has $Q=0$ and is super-commutative; so
\be Q(\mu_N) = 0\,,\qquad \mu_N^2 = 0\,. \ee

One can imagine constructing many other line operators in a similar way. As long as the QM operators that appear in a coupling $\mu_\ell$ are $Q$-closed and (super)commutative, it is guaranteed that the MC equation will be satisfied. Simple generalizations include vortex lines that ``skip'' modes;  for example, we could consider
\be \V_{[0,2]}:= \int D\Ba_0D(...)\exp\!\! \int_{\C\times\{0\}} \Big[\Ba_0 d\bar \Ba_0+\Ba_2 d\bar \Ba_2 + \Ba_0 \BX + \tfrac12\Ba_2 \pd^2 \BX\Big]\,, \label{free-modeskip} \ee
which effectively sets $X(z)\sim c\,z + z^3\times\text{(regular)}$ and $\psi(z) \sim \frac{a_2}{z^3}+\frac{a_0}{z} + \text{(regular)}$ near $z=0$. %
It might be noted that mode-skipping lines such as \eqref{free-modeskip} partially break holomorphic flavor transformations \eqref{free-flavor}, whereas the standard flavor vortices \eqref{free-N} preserve the full holomorphic flavor symmetry. However, if one only require that line operators preserve $F,R,J$, then \eqref{free-modeskip} is a perfectly reasonable example.

A feature that emerges is that it is not possible to give aribtrary poles and zeros to both $\BX$ and $\BPsi$. By coupling to 1d operators, any pole created for one of the bulk fields gets ``balanced'' by a zero for the other. There is a more precise characterization of this phenomenon in terms of a shifted symplectic structure on the space of modes of $X,\psi$ that will appear later --- in the kinetic term of the $S^1$ reduction of a 3d HT theory \eqref{S2d-gen}. Line operators will need to be supported on Lagrangian submanifolds with respect to the symplectic structure (because they define boundary conditions for the $S^1$ reduction).

\subsection{Example: matter with a superpotential}
\label{sec:line-W}

Next, consider the HT twist of a theory with $d$ chiral matter multiplets $(\BX^i,\BPsi_i)_{i=1}^d$ and a polynomial superpotential $W(X)$, whose action is
\be S_{3d} = \int_{\C\times \R} \BPsi_i d' \BX^i + W(\BX)\,. \ee

The presence of the superpotential restricts the set of vortex lines in the theory. A qualitative way to see this is the following. Suppose we attempt to define a disorder line operator, by specifying a singularity near $z=0$ for the HT equations of motion. The BRST transformation of $\BX$'s and associated EOM is unchanged $(Q\BX = d'\BX)$, so we still need a holomorphic profile in the $\C_z$ plane, with some singularity at $z=0$. However, if we remove a neighborhood of $z=0$, the BRST transformation of the action may acquire a new boundary term due to the superpotential,
\be Q\,S_{3d}\big|_{\C^*\times \R} = \int_{\C^*\times \R} d' W(\BX) = \int_\R\oint_{S^1_\epsilon} W(\BX)\,, \ee
where $S^1_\epsilon$ is an infinitesimal circle around $z=0$.
We find that disorder operators have a  BRST anomaly unless
\be \boxed{\oint_{S^1_\epsilon} W(\BX) \;\;\text{vanishes for the specified singularity in $\BX^i(z)$}\,.} \ee

This is an analogue of the ``Warner problem'' \cite{Warner} for 2d $\CN=(2,2)$ Landau-Ginzburg models in the presence of a boundary, and is the reason that the simplest B-type boundary conditions for 2d LG models are ones on which the superpotential vanishes, \cf\ \cite{HoriIqbalVafa}. More general boundary conditions in 2d are matrix factorizations of the superpotential \cite{KapustinLi}.

In our case, we expect the full category of operators for a 3d theory with matter $V$ and superpotential $W$ to be equivalent to matrix factorizations of $\oint W$ on the loop space $LV$, as in \eqref{MF-intro}. This isn't a perspective that we'll use in the current paper, but it does provide some additional intuition. The vortex lines with singular profiles
\be \V_{\vec N}:\quad  \left\{ \begin{array}{l}\BX^i(z) \sim z^{N_i}\times(\text{regular)} \\[.1cm] \BPsi_i(z) \sim z^{-N_i}\times(\text{regular}) \end{array}\right. \quad\text{near $z= 0$} \label{W-VN-vortex} \ee
for $\vec N\in \Z^d$ such that $\oint W(X) =0$ correspond to elementary objects in the matrix-factorization category: they are the structure sheaves of the loci $z^{\vec N} \C[\![z]\!] \subset LV$, with trivial matrix factorizations, since the superpotential $\oint W$ vanishes on their support.

\subsubsection{Tropical polytopes and flavor symmetry}
\label{sec:tropical}

To get an idea of what the constraint $\oint W =0$ means, consider the ``XYZ model,'' three chirals $(\BX,\BPsi_X),\,(\BY,\BPsi_Y),\,(\BZ,\BPsi_Z)$, with $W = \BX\BY\BZ$\,. There are putative vortex lines $V_{N_1,N_2,N_3}$ for all $(N_1,N_2,N_3)\in \Z^3$. However, for $W$ to have vanishing residue, we need
\be N_1+N_2+N_3 \geq 0\,. \label{XYZ-W-0} \ee

The sublattice $\Z^2\subset\Z^3$ defined by $N_1+N_2+N_3=0$, which bounds the region \eqref{XYZ-W-0}, has line operators of a special type. They are all flavor vortices, generated, just as in \eqref{free-flavor}, by singular flavor transformations for the  $U(1)^2\subset U(1)^3$ flavor symmetry that is preserved by the superpotential. The interior points of the region \eqref{XYZ-W-0} are not flavor vortices, but are still perfectly legitimate line operators.

We can also consider a general polynomial superpotential, a sum of monomials
\be W = \sum_\alpha c_\alpha  (X^1)^{p_{\alpha,1}}(X^2)^{p_{\alpha_2}}\cdots (X^d)^{p_{\alpha,d}}\,. \ee
The constraint $\oint W =0$ requires the singularities in vortex lines $\V_{\vec N}$ to satisfy
\be \text{min}_\alpha \big\{ p_{\alpha,1}N_1 + \ldots + p_{\alpha,d}N_d\big\} \geq 0\,. \ee
This describes a \emph{tropical lattice polytope}; the equation on the LHS is so-called the tropicalization of $W$.

\subsubsection{Solving the MC equation for the XYZ model}
\label{sec:QM-XYZ}

In order to construct lines by coupling to quantum mechanics in a 3d QFT with superpotential, one need to know the full $A_\infty$ structure on the algebra $\CA$ of 3d bulk local operators. It is derived in Section \ref{sec:Ainf-matter}, where it is described geometrically as a chiralization of the ``derived critical locus'' of $W$. In general, when $W$ is a polynomial of degree $d$, the algebra $\CA$ has nontrivial higher products $m_k$ for all $k\leq d-1$.

Here we'll illustrate how things work for a special example: the XYZ model defined in Section \ref{sec:tropical}. We'd like to express any of the basic vortex lines $\V_{(N_1,N_2,N_3)}$ with $N_1+N_2+N_3\geq 0$ (the anomaly-free domain, as in  \eqref{XYZ-W-0}) via quantum mechanics.

The bulk algebra $\CA$ for the XYZ model is generated by $X,Y,Z,\psi_X,\psi_Y,\psi_Z$ and their $\pd_z$ derivatives at (say) $z=0$. All $m_{k\geq 3}$ operations vanish because the superpotential is cubic; but there is a differential (controlled by the first derivatives of $W$) as well as a nontrivial commutator (controlled by the second derivatives of $W$):
\be \begin{array}{c} Q\psi_X=YZ\,,\quad Q\psi_Y=XZ\,,\quad Q\psi_Z = XY\,, \\[.1cm]  [\psi_X,\psi_Y]=\pd Z\,,\quad [\psi_Y,\psi_Z]=\pd X\,,\quad [\psi_X,\psi_Z]=\pd Y\,. \end{array} \label{XYZ-alg} \ee
The bosons $X,Y,Z$ are all $Q$-closed and central.  The action of $Q$ here could have been deduced from the semi-classical action of $Q$ on fields \eqref{Q-act}, but the commutators are a purely quantum effect. They were already observed in \cite[Eqn. 5.18]{CDGbdry}, and are a special case of the general $A_\infty$ structure in Sec. \ref{sec:Ainf-matter}.

When all of $N_1,N_2,N_3\geq 0$ (individually), $\V_{N_1,N_2,N_3}$  is constructed the same way same as for free matter. We introduce $N_1+N_2+N_3$ 1d fermi multiplets $\{\Ba_n,\bar\Ba_n\}_{n=0}^{N_1-1}$, $\{\Bb_m,\bar\Bb_m\}_{m=0}^{N_2-1}$, $\{\Bc_p,\bar\Bc_p\}_{p=0}^{N_3-1}$, coupled to the bulk via an MC element
\be  \mu = \sum_{n=0}^{N_1-1} \frac{a_n}{n!} \pd^n X(0) +  \sum_{m=0}^{N_2-1} \frac{b_m}{m!} \pd^m Y(0)  +  \sum_{p=0}^{N_3-1} \frac{c_p}{p!} \pd^p Z(0)\,. \ee 
This trivially satisfies $Q\mu = \mu^2 = 0$, and thus satisfies the MC equation.

When one of the $N$'s is negative, say (WLOG) $N_1<0$ and $N_2,N_3\geq 0$, we start by introducing $|N_1|$ 1d bosonic multiplets $\{\Ba_n,\bar\Ba_n\}_{n=0}^{|N_1|-1}$ and $N_2+N_3$ fermi multiplets $\{\Bb_m,\bar\Bb_m\}_{m=0}^{N_2-1}$, $\{\Bc_p,\bar\Bc_p\}_{p=0}^{N_3-1}$, coupled by an MC element
\be  \mu_0 = \sum_{n=0}^{|N_1|-1} \frac{a_n}{n!} \pd^n \psi_X(0) +  \sum_{m=0}^{N_2-1} \frac{b_m}{m!} \pd^m Y(0)  +  \sum_{p=0}^{N_3-1} \frac{c_p}{p!} \pd^p Z(0)\,. \ee 
However, this now satisfies $Q(\mu_0) = \sum_{n=0}^{|N_1|-1} \tfrac{1}{n!}a_n \pd^n (YZ)(0) \neq 0$ due to \eqref{XYZ-alg}, as well as $(\mu_0)^2=0$. To fix this, we look for a second term $\mu_1$ that obeys $Q(\mu_1)=0$ and $[\mu_0,\mu_1]+(\mu_1)^2 = -Q(\mu_0)$, so that the sum $\mu=\mu_0+\mu_1$ will obey $Q(\mu)+\mu^2=0$. Consider
\be \mu_1 = -\sum_{m=0}^{N_2-1} \bar b_m E_m - \sum_{p=0}^{N_3-1} \bar c_p F_p\,. \ee
If we assume that the $E$'s and $F$'s are functions only of $a,Y,Z$ and their derivatives, then we've got
\be  -[\mu_0,\mu_1]-(\mu_1)^2 = \sum_{m=0}^{N_2-1} \frac{E_m}{m!} \pd^m Y + \sum_{p=0}^{N_3-1} \frac{F_p}{p!}\pd^p Z = Q(\mu_0) = \sum_{m=0}^{|N_1|-1} a_n \pd^n(YZ)\,. \label{XYZ-mu01system} \ee
The inequality $|N_1|\leq N_2+N_3$ guarantees that there is always a solution with $E,F$ linear in the bulk fields. It's unique when $|N_1|= N_2+N_3$, and otherwise underconstrained --- but all solutions appear to be related by redefinition of the 1d fields, and are thus ultimately equivalent. One solution that sets to zero as many $F$'s as possible is
\be E_m= \sum_{i=0}^{|N_1|-1-m} \frac{a_{m+i}}{i!} \pd^iZ\quad (0\leq m< N_2)\,,\qquad F_p = \sum_{j=N_2}^{|N_1|-1-p} \frac{a_{p+i}}{i!} \pd^i Y \quad (0\leq p < N_2-|N_1|)\,. \ee

For example, the vortex line $\V_{(-1,1,0)}$ is uniquely represented by the linear coupling
\be \mu = a \psi_X + b Y  -  \bar b (a Z) \label{V-110} \ee
involving a bosonic multiplet $(\Ba,\bar\Ba)$ and a fermi multiplet $(\Bb,\bar\Bb)$. On the other hand, the vortex line $\V_{(-1,1,1)}$ has an entire space of representations with linear couplings to a bosonic multiplet $(\Ba,\bar\Ba)$ and two fermi multiplets $(\Bb,\bar\Bb)$, $(\Bc,\bar\Bc)$. Two possibilities are
\be \mu = a\psi_X + b Y + c Z  -  \bar b (a Z)\,,\qquad \text{or}\qquad \mu' =  a\psi_X + b Y + c Z  -  \bar c (a Y)  \ee
To go from $\mu$ to $\mu'$ we perform the 1d field redefinition
\be (\Ba,\bar\Ba,\Bb,\bar\Bb,\Bc,\bar\Bc) \mapsto  (\Ba,\bar\Ba+\bar\Bb\bar\Bc,\Bb-\bar\Bc\Ba,\bar\Bb,\Bc+\bar\Bb\Ba,\bar\Bc)\,,  \ee
noting that this preserves the kinetic term $S_{1d} = \int_\R \bar\Ba d \Ba + \bar \Bb d\Bb + \bar \Bc d\Bc\,.$

Finally, when two $N$'s are negative, say $N_1<0,N_2<0$ and $N_3>0$, the commutation relations between bulk fermions play an important role in the MC equation. We introduce $|N_1|+|N_2|$ 1d bosonic multiplets $\{\Ba_n,\bar\Ba_n\}_{n=0}^{N_1-1}$, $\{\Bb_m,\bar\Bb_m\}_{m=0}^{N_2-1}$ and $N_3$ fermi multiplets $\{\Bc_p,\bar\Bc_p\}_{p=0}^{N_3-1}$, with the naive coupling
\be  \mu_0 = \sum_{n=0}^{|N_1|-1} \frac{a_n}{n!} \pd^n \psi_X(0) +  \sum_{m=0}^{|N_2|-1} \frac{b_m}{m!} \pd^m \psi_Y(0)  +  \sum_{p=0}^{N_3-1} \frac{c_p}{p!} \pd^p Z(0)\,, \ee 
which fails to satisfy the MC equation because
\be Q\mu_0 =  \sum_{n=0}^{|N_1|-1} \frac{a_n}{n!} \pd^n (YZ) +  \sum_{m=0}^{|N_2|-1} \frac{b_m}{m!} \pd^m(XZ)\,,\qquad (\mu_0)^2 =  \sum_{n=0}^{|N_1|-1}  \sum_{m=0}^{|N_2|-1} \frac{a_n b_m}{(n+m+1)!} \pd^{n+m+1}Z\,. \ee
We satisfy the MC equation by correcting the MC element to $\mu = \mu_0+\mu_1$, with
\be \mu_1 = - \hspace{-.1cm} \sum_{p=0}^{|N_1|-1} \bar c_p \hspace{-.2cm}  \sum_{i=0}^{|N_1|-1-p} \frac{a_{p+i}}{i!}\pd^iY - 
  \sum_{p=0}^{|N_2|-1} \bar c_p\hspace{-.2cm}  \sum_{i=0}^{|N_2|-1-p} \frac{b_{p+i}}{i!}\pd^iX 
  - \sum_{n=0}^{|N_1|-1}  \sum_{m=0}^{|N_2|-1} a_n b_m\bar c_{n+m+1}\,.  \ee
Technically speaking, the third term, cubic in 1d fields and containing no bulk fields at all, is a deformation of the differential in the 1d quantum mechanics itself; while the first two terms are part of the coupling to the bulk.

For example, the vortex line $\V_{(-1,-1,2)}$ is created by introducing two bosonic multiplets $(\Ba,\bar\Ba)$, $(\Bb,\bar\Bb)$ and two fermi multiplets $(\Bc_0,\bar\Bc_0)$, $(\Bc_1,\bar\Bc_1)$, with
\be \mu = a \psi_X + b\psi_Y + c_0 Z + c_1\pd Z - \bar c_0 a Y - \bar c_0 b X - ab\bar c_1\,. \label{V-1-12} \ee
The final term deforms the differential on the QM itself; \eg\ if the 1d state space is the Fock space $V= \C[a,b,c_1,c_2]$ the differential becomes $Q =-ab\pd_{c_2}$.

\subsection{Example: gauge theories}
\label{sec:line-gauge}

As a final example, we'll consider line operators in HT-twisted 3d $\CN=2$ gauge theories, and their quantum-mechanics realization. There is a great richness of line operators in gauge theory.  We expect to find Wilson lines (namely, half-BPS Wilson lines \cite{GaiottoYin}, which are compatible with the HT twist), as well as disorder operators at which the gauge connection or dual-photon fields --- or more generally, operators parameterizing the Coulomb branch --- have a specified singularity \cite{KWY-abelian, DrukkerOkudaPasserini}. In the presence of matter, we also expect to find vortex-like defects for the matter fields, as long as their profiles are gauge invariant. On the other hand, in the presence of Chern-Simons terms, a pure gauge theory should be equivalent to a topological 3d Chern-Simons theory, so all line operators should become equivalent to standard Wilson lines. We'd like to indicate how this comes about. 

\subsubsection{Bulk local operators}

We'll only consider coupling 1d QM to perturbative bulk local operators (no monopoles). Nevertheless, the algebra  of perturbative bulk operators in gauge theory is remarkably subtle, and we need to know it to solve the MC equation. There is a full derivation in Section \eqref{sec:Ainf-gauge}. We'll summarize here the result without matter.

Recall from Section \ref{sec:3dN2} that the fields of gauge theory with group $G$ are $\{\BA_a,\BB^a\}_{a=1}^{\text{dim}\,\fg}$. The perturbative algebra $\CA$ of local operators at $z=0$ is a dg algebra, generated by the zero-form components $\{c_a,B^a\}_{a=1}^{\text{dim}\,\fg}$ and their $\pd$ derivatives. Since the potential terms in the action \eqref{S-CS} are cubic, higher $A_\infty$ operations vanish, but there is a differential and nontrivial commutation relations when $G$ is nonabelian:
\be \begin{array}{c} 
  \Big[B^a, c_b\Big] = f^{ac}{}_b \pd c_c\,,\qquad \Big[B^a,B^b\Big] = -f^{ab}{}_c  \pd B^c\,,  \\[.2cm]
  Q c_a = -\tfrac12 f^{bc}{}_a c_b c_c\,, \\[.2cm]
   Q B^a = (k+h^\vee)\pd c_a - f^{ab}{}_c c_b B^c = (k- h^\vee)\pd c_a- f^{ab}{}_c B^c c_b\,. \end{array} \label{gauge-A-0}
\ee
Here we've written the Chern-Simons form as $k\delta^{ab}$, with $\delta^{ab}$ the Killing form in a suitable basis; 
$h^\vee$ is the dual Coxeter number, and shifts by it appear due to the nontrivial commutator between $B$ and $c$. 
The differential in \eqref{gauge-A-0} can be deduced from the action of $Q$ on fields \eqref{Q-act}, but the commutators are a a purely quantum effect.

\subsubsection{Wilson lines and more}
\label{sec:Wilsonlines}

To construct a Wilson line we choose a 1d state space $V$ with a collection of topological operators $J^a\in \text{End}(V)$, $a=1,...,\text{dim}\,\fg$, and introduce a coupling
\begin{subequations} \label{Wilsonline}
\be  \mu = J^a c_a  \quad \leftrightarrow\quad \delta S = \int_\R J^a\BA_a\,.  \ee
As explained already in \cite{Costello-Yangian, CPkoszul, PaquetteWilliams}, the MC equation in this case requires
\be 0 = Q(\mu)+\mu^2 = -\tfrac12 J^a f^{bc}{}_a c_b c_c + \tfrac12c_a c_b[J^a,  J^b] \quad\Rightarrow\quad [J^a,J^b] = f^{ab}{}_c J^c\,. \ee
\end{subequations}
In other words, the $J^a$'s must furnish a representation of $\fg$. 

To avoid confusion, it may be useful to keep in mind that this description of Wilson lines is, strictly speaking, perturbative, in the following sense. The local operator $c$ (the zero-mode of the ghost) only truly exists in the theory if we take the entire gauge ``group'' to consist of infinitesimal $\fg$-valued gauge transformations, ignoring the global form of the group $G$. In this context, it makes sense to use the descent procedure to relate couplings to MC elements --- and we find that Wilson lines are representations of $\fg$, with no knowledge of $G$.

Nonperturbatively, if $G$ is a reductive group (the complexification of a compact physical gauge group), the zero-mode $c$ should be dropped from the algebra of local operators, and invariants with respect to global gauge transformations should be taken by hand. Then the deformation to the action is $\delta S = \int_\R J^a A_t dt$  still makes sense, but it does not come from an MC element. Moreover, invariance under large gauge transformation requires the $J^a$ to integrate to a representation of the actual group $G$, rather than just the Lie algebra $\fg$.

We will continue using a perturbative description (involving $c_a$'s) and supplement by hand the MC equation to require integrability of operators coupling to them.

The Wilson lines \eqref{Wilsonline} are anomaly-free for any gauge theory, even with matter. 
More generally, we could introduce couplings to $\BB$'s as well as to higher derivatives of $\BA$'s and $\BB$'s. For example, for an \emph{abelian} $G=\C^*$ theory at level $k=0$ we can choose $r\in \Z$ and $N \in \N$, introduce bosonic 1d multiplets $(\Bb_n,\bar\Bb_n)_{n=1}^{N}$,  and define
\be \W_{r,N}: \quad \mu = r c + \sum_{n=1}^{N} \tfrac{1}{n!} b_n \pd^n c \,. \label{WrN}  \ee
At $N=0$, this would just be a Wilson line of charge $r$. For $N> 0$, this line operator effectively requires the first $N$ derivatives of the connection to vanish at $z=0$ while simultaneously allowing $\BB$ to be singular, $\BB \sim \frac{\Bb_N}{z^{N+1}} + \frac{\Bb_{N-1}}{z^{N}}  + \cdots + \frac{\Bb_1}{z^2} + \frac{r}{z} + (\text{regular})$. Now, the chiral monopole operators in abelian gauge theory are $v_\pm \sim e^{\pm \gamma}$ \cite{AHISS}, where $\gamma$ is the complexified dual photon, satisfying $\pd \gamma = B$. Thus, in the presence of the line operator \eqref{WrN}, the monopole operators get an essential singularity
\be v_\pm \sim z^{\pm r} \exp\big[ z^{-N}\times(\text{regular})\big]\,. \ee

Dually, in an abelian theory at $k=0$, we could introduce 1d fermi multiplets $(\Ba_n,\bar\Ba_n)_{n=0}^{N-1}$ and a line operator
\be \W_{0,-N} :\quad \mu = \sum_{n=0}^{N-1} \frac{1}{n!}  a_n \pd^n B \label{W0-N} \ee
This effectively sets to zero $\BB$ and its first $N-1$ derivatives (\ie\ the first $N$ derivatives of the complexified dual photon) while allowing the connection to have a singular profile
\be \BA \sim \frac{\Ba_{N-1}}{z^N} + \cdots + \frac{\Ba_0}{z} + \text{(regular)}\,. \ee
One could generalize this to non-abelian theories as well, as long as \eqref{W0-N} is modified to be gauge-invariant. See for example \cite[Sec. 2.3.2]{Vicedo:2022mrm}.

\subsubsection{Chern-Simons terms}

Adding a Chern-Simons level in abelian theory will remove all the line operators \eqref{WrN}, \eqref{W0-N} when $N\neq 0$. Namely, the $\W_{0,-N}$ lines no longer satisfy the MC equation because $Q(\pd^n B) = k \pd^{n+1} c \neq 0$. Dually, in the $\W_{r,N}$ lines all the terms $b_n\pd^n c$ and associated deformations $\int_\R \Bb_n \pd^n \BA$ become Q-exact; thus any $\W_{r,N}$ line becomes equivalent to the standard Wilson line $\W_{r,0}$. This is the behavior we were looking for, since the theory at $k\neq 0$ is equivalent to pure abelian Chern-Simons. 

In abelian Chern-Simons theory we also expect an equivalence of Wilson lines $\W_{r,0}\simeq \W_{r+k,0}$ with charges shifted by the level. The equivalence comes from acting with a singular gauge transformation $g(z)=z$ in the neighborhood of $z=0$ to transform $\W_{r,0}$ to $\W_{r+k,0}$ (similar to \eqref{free-flavor}, except now the symmetry is gauged). Alternatively, the equivalence could be seen coming from a monopole operator that defines an invertible junction between $\W_{r,0}$ and $\W_{r+k,0}$, producing an isomorphism in the category of lines. This feature is difficult to see from the current perturbative perspective, but can be added by hand.

For non-abelian $G$, we similarly expect that when the Chern-Simons level satisfies $|k|> h^\vee$, the HT-twisted theory is equivalent to pure topological Chern-Simons (at level $k-h^\vee\text{sign}(k)$), so only Wilson lines should remain. A perturbative proof of this statement would involve showing that the algebra \eqref{gauge-A-0} is quasi-isomorphic to a finite-dimensional algebra $\C[c_a]_{a=1}^{\text{dim}\,\fg}$ generated just by the zero-modes of the ghosts. This is easily conceivable: even in the non-abelian case, no combinations of $B$'s will be $Q$-closed due to the Chern-Simons term, and all derivatives $\{\pd^{n}c_a\}_{n\geq 1}$ seem to become $Q$-exact for the same reason. (In Section \ref{sec:TQFT-gauge}, we'll give an explicit form of this argument for the Koszul-dual algebra $\CA^!$, proving that derivatives $\pd_z$ become $Q$-exact.)

\subsubsection{Adding matter}

With additional matter, line operators in a gauge theory become extremely rich. All the vortex lines from Sections \ref{sec:line-chiral}--\ref{sec:line-W} are in play, as long as they can be made gauge-invariant --- by coupling the 1d QM fields to the bulk connection in the appropriate representation. They can further be superposed with Wilson lines, or gauge singularities of the form \eqref{WrN}, \eqref{W0-N}.

The complete perturbative bulk algebra for gauge theory with matter is in Sec.~\ref{sec:Ainf-full}. As an illustrative example, let's consider $G_c=U(1)$ gauge theory, with a chiral of charge $q$, so the fields $(\BX,\BPsi)$ have charges $(q,-q)$. The bulk algebra at $z=0$ is generated by bosons $B,X$ and fermions $c,\psi$ (and their derivatives), with
\be \begin{array}{c}  [B,X]=-q\pd X\,,\quad [B,\psi]=q\pd \psi\,,\quad [X,\psi]=-q\pd c \\[.1cm]
[B,B]=[B,c]=[c,c] = [c,X]=[c,\psi]=0 \\[.1cm]
     Q\,c = 0\,,\quad Q\,B = \big(k+\tfrac12 q\big)\pd c - q\psi X\,,\qquad Q\,X=q\,cX\,,\quad Q\,\psi = -q\,c\psi\,. \end{array}
 \ee

In this theory, one can try to create a matter vortex $\V_N$ of charge $N>0$ (say) just as in \eqref{free-posN}, by introducing $N$ 1d fermi multiplets $(\Ba_n,\bar\Ba_n)_{n=0}^{N-1}$ with a coupling
\be \mu_0 = \sum_{n=0}^{N-1} \frac1{n!}a_n \pd^n X\,. \label{gaugeqvortex} \ee
This doesn't satisfy the MC equation, since $Q(\mu_0) = \sum_{n=0}^{N-1} \frac{q}{n!}a_n\pd^n(cX)$; but with a correction
\be  \mu_1 = -q  \sum_{n=0}^{N-1}  \bar a_n \sum_{i=0}^{N-1-n} \frac{\pd^i c}{i!} a_{n+i}   \ee
the total MC element $\mu=\mu_0+\mu_1$ does satisfy the MC equation. 
The correction simply couples the QM multiplets to the bulk gauge field, deforming the 1d action by $ \int_\R \bm\mu_1=-q\int_\R \bar\Ba_0 \BA \Ba_0 -q \int_\R \bar\Ba_1\BA\Ba_1 - q\int_\R \bar\Ba_0 \pd\BA \Ba_1-...,$ which ensures that the original coupling \eqref{gaugeqvortex} is invariant under axial gauge transformations near $z=0$. We have effectively created the vortex
\be \V_N:\quad \BX \sim z^N\times(\text{regular})\,,\qquad \BPsi = \frac{\Ba_{N-1}}{z^{N}} + \ldots + \frac{\Ba_{0}}{z}+(\text{regular}) \sim z^{-N}\times(\text{regular})\,. \ee

We can further superpose a matter vortex with a Wilson line of charge $r\in \Z$ by adding an `$rc$' term to the coupling, \eg\
\be \V_N^{(r)}: \quad \mu = rc+ \sum_{n=0}^{N-1} \frac1{n!}a_n \pd^n X - q  \sum_{n=0}^{N-1}  \bar a_n \sum_{i=0}^{N-1-n} \frac{\pd^i c}{i!} a_{n+i} \quad (N> 0)\,. \ee
Nonperturbatively, one expects to be able to ``screen'' matter vortices: acting with a singular gauge transformation $g(z)=z^p$ in a neighborhood of $z=0$ should induce an equivalence between $\V_N^{(r)}$ and $\V_{N+qp}^{(r+kp)}$, where $k$ is the bulk Chern-Simons level. The equivalence seems difficult to see from the current perspective.

\section{OPE's and non-renormalization}
\label{sec:OPE}

In this section, we take the general construction of line operators via quantum mechanics from Section \ref{sec:lines-QM}, and use it to analyze their perturbative OPE's.

The basic problem we'd like to address is the following. 
We consider a general perturbative 3d HT QFT $\CT$, with action $S = \int \mb p_i d' \mb x^i + \CI(\mb p,\mb x,\pd)$, and a pair of line operators  $\ell,\ell'$, constructed from 1d state spaces $V,V'$ and couplings $\mu_\ell,\mu_{\ell'}$. Let's insert $\ell$ at a point $z\in \C$ and $\ell'$ at $w\in \C$ (both extended along $\R_t$), by taking the respective couplings
\be \mu_\ell = \mu_\ell(z) \in \text{End}(V)\otimes \CA\big|_z\,,\qquad \mu_{\ell'} = \mu_{\ell'}(w) \in \text{End}(V')\otimes \CA\big|_w \ee
to depend on separate copies of the bulk algebra $\CA$, restricted to lines $\{z\}\times \R_t$ and $\{w\}\times \R_t$ in $\C\times \R_t$.
In terms of path integrals, we're tensoring with $V\otimes V'$ and inserting
\be \ell(z)\otimes \ell'(w) = \exp \int_{\R_t} \bm \mu_\ell(z) \; \exp \int_{\R_t} \bm \mu_{\ell'}(w) \qquad \raisebox{-.4in}{\includegraphics[width=.8in]{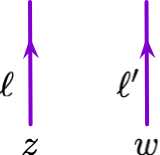}} \label{LineOPEzw} \ee

When $z$ is close to $w$, we expect this insertion to be equivalent --- in all correlation functions --- to a single insertion at $w$:
\begin{align}  \label{OPE-general}
(\ell\otimes_{z-w} \ell')(w)  &\simeq \exp \int_{\R_t} \Big[ \bm \mu_\ell(z) + \bm \mu_{\ell'}(w) + \mb r_{\ell,\ell'}(w,z-w) \Big] \\
 &\simeq  \exp \int_{\R_t} \bigg[  \sum_{n=0}^\infty \frac{1}{n!}(z-w)^n \pd^n \bm\mu_\ell(w)+ \bm \mu_{\ell'}(w) + \mb r_{\ell,\ell'}(w,z-w) \bigg] \notag
   \end{align}
Here $r_{\ell,\ell'} \in \text{End}(V)\otimes \text{End}(V')\otimes \CA\big|_w \otimes C^\infty(D^*_{z-w})$ captures potential quantum corrections due to interactions between the two lines. In general, its coefficients may be any smooth functions in an infinitesimal punctured neighborhood of $z=w$, denoted $D^*_{z-w}$, with an arbitrary singularity as $z\to w$.

The main result of this section is that the quantum corrections $r_{\ell,\ell'}$ are highly structured. For the strongest form of the result, we'll introduce the notion of a \emph{quasi-linear} theory (Section \ref{sec:quasi-linear}): roughly, one whose fields can be split into multiple subsets, such that each interaction terms is at-most-linear in the bosons \emph{or} the fermions from one of the subsets. All twisted 3d $\CN=2$ gauge theories, with linear matter, CS terms, and arbitrary superpotentials, are quasi-linear. We then prove the following in Section~\ref{sec:non-renorm}:

\begin{Thm} \label{thm:non-ren}
Suppose we have either a quasi-linear 3d HT QFT, or one with at most cubic interactions; and line operators~$\ell,\ell'$ that preserve gradings $F,R,J$ and
are defined by coupling \emph{{linearly}} to elements of the perturbative bulk algebra $\CA$.  Then $ (\ell\otimes_{z-w} \ell')(w)   \simeq  \exp \ds \int_{\R_t}  \bigg[ \bm\mu_\ell(z) +\bm \mu_{\ell'}(w) +  \mb r_{\ell,\ell'}(z,z-w) \bigg]\,,$
where perturbative quantum corrections are give by the sum of single contractions
\be \int_\R   \mb r_{\ell,\ell'}(z,z-w) = \sum\;  \raisebox{-.3in}{\includegraphics[width=1.3in]{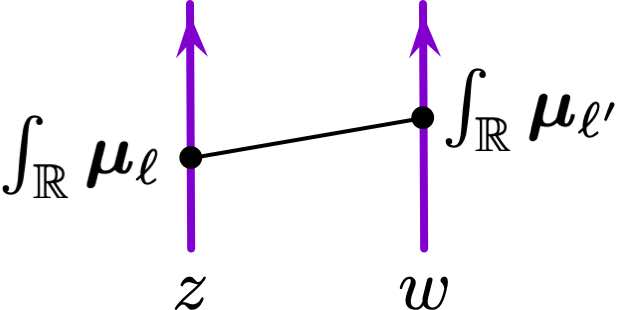}} \label{OPE-thm} \ee
\end{Thm}

In particular, bulk interactions terms $\CI(\mb x,\mb p,\pd)$  play no direct role in the quantum correction \eqref{OPE-thm} --- there are no Feynman diagrams with bulk vertex insertions.
Bulk interactions are of course important for solving the MC equations that define the line operators themselves. Indeed, we will see later in Sections \ref{sec:Koszul}--\ref{sec:proofs} that there exist highly nontrivial constraints on the form of $r(z-w)$ that involve bulk interactions. These constraints ensure, for example, that the OPE of lines is associative.

In Section \ref{sec:exact-gen} we explicitly compute the correction to the OPE. It follows directly from Theorem \ref{thm:non-ren} that, in axial gauge, the correction is meromorphic and takes the exact form
\begin{align}  r_{\ell,\ell'}(z-w) &= \sum_i \sum_{n,m=0}^\infty \bigg( \frac{\pd \mu_\ell}{\pd (\pd^n x^i)}  \frac{\pd \mu_{\ell'}}{\pd (\pd^m p_i)} -\frac{\pd \mu_{\ell'}}{\pd (\pd^m x^i)}   \frac{\pd \mu_\ell}{\pd (\pd^n p_i)}  \bigg) \frac{(-1)^n (n+m)!}{(z-w)^{n+m+1}}  \label{r-thm} \\
&\in \text{End}(V)\otimes \text{End}(V') \otimes \C[(z-w)^{-1}]\,.  \notag \end{align}
(We've used a linear basis  $\{\pd^n x^i,\pd^n p_i\}$ for bulk local operators as in \eqref{def-A1}.) We comment on the general structure of $r_{\ell,\ell'}$ in other gauges in Sections \ref{sec:othergauge}, \ref{sec:Lorentz}, arguing that axial gauge captures leading singularities, and that other gauges may deviate from axial gauge by $Q$-exact terms.

\subsection{Warm-up: free chiral}
\label{sec:OPE-free}

To get a feeling for how OPE's work, let's consider a simple example. Take the bulk theory $\CT$ to be a free chiral multiplet, with the two simplest flavor vortex lines of charge $\pm$ from Section \ref{sec:line-chiral},
\be \V_1: \quad V_1 = \C[a]\,,\quad \mu_1 = a X \qquad\text{and}\qquad \V_{-1}:\quad V_{-1}=\C[b]\,,\quad \mu_{-1} = b \psi\,, \ee
engineered by coupling (respectively) to a 1d fermi multiplet $(\Ba,\bar\Ba)$ and bosonic multiplet $(\Bb,\bar\Bb)$. Recall that, effectively, $\V_1$ gives $X$ a first-order zero and $\psi$ a first-order pole and $\V_{-1}$ does the opposite. 

We'd like to find the OPE $(\V_1\otimes_{z-w} \V_{-1})(w)$, so we consider the insertion
\be \V_1(z)\otimes \V_{-1}(w) = \int D\Ba D\bar\Ba D\Bb D\bar\Bb \;e^{\int_\R \Ba d\bar\Ba + \Ba \BX(z)}\;e^{\int_\R \Bb d\bar\Bb + \Bb \BPsi(w)}\,.  \ee
inside the bulk path integral, or a bulk correlation function. We care about the interaction $e^{\int_\R  \Ba \BX(z)}\; e^{\int_\R \Bb \BPsi(w)}$ and its behavior in bulk correlation functions, possibly in the presence of \emph{arbitrary collections of other operators}. We'll still denote the correlation function simply as $\big\langle e^{\int_\R  \Ba \BX(z)}\; e^{\int_\R \Bb \BPsi(w)}\big\rangle$. De-exponentiating this, we get
\be 
\Big\langle\sum_{n,m=0}^\infty \frac{1}{n!m!} \int_{t_1,...,t_n}\int_{s_1,...,s_m} \Ba(t_1)\BX(t_1,z)\cdots \Ba(t_n)\BX(t_n,z)\;\Bb(s_1)\BPsi(s_1,w)\cdots \Bb(s_n)\BPsi(s_m,w)\Big\rangle \label{OPE-chiral-2} \ee
We do not need to worry about time-ordering or regularizing singularities when multiple operators coincide in this simple example, because as long as $z\neq w$ all the operators here have non-singular collisions.

We now contract each pair of bulk fields, replacing
\be \BX(t_i,z) \BPsi(s_j,w) \;\leadsto\; :\!\BX(t_i,z) \BPsi(s_j,w)\!: + \;G(z,\bar z,t_i; w,\bar w,s_j)\,, \ee
where $G$ is the free-field propagator from Section \ref{sec:prop}. Recall that, in this superfield formalism, $G$ is a 1-form on the configuration space of two points in $\C\times \R$. It is symmetric in the two sets of variables, $G(z,\bar z,t_i; w,\bar w,s_j)=G(w,\bar w,s_j;z,\bar z,t_i)$. The normal-ordered combination $:\!\BX(t_i,z) \BPsi(s_j,w)\!:$ can simply be defined as $\BX(t_i,z) \BPsi(s_j,w)-G(z,\bar z,t_i; w,\bar w,s_j)$; it has the usual property that it's non-singular as insertion points of $\BX$ and $\BPsi$ collide, and comes with a prescription that it should not have any internal contractions when inserted in correlation functions with other operators. Summing over all possible contractions in \eqref{OPE-chiral-2}, we find
\begin{align} \eqref{OPE-chiral-2} &= \Big\langle\sum_{n,m=0}^\infty \frac{1}{n!m!} \int_{t_1,...,t_n}\int_{s_1,...,s_m}  \sum_{k=0}^{\text{min}(n,m)} 
 : \prod_{i=1}^{n-k} \Ba(t_i)\BX(t_i,z) \prod_{j=1}^{m-k} \Bb(s_j) \BPsi(s_j,w): \notag \\ 
 & \hspace{.4in}  \times k! \bigg(\!\!\begin{array}{c} n\\k\end{array}\!\!\bigg)\bigg(\!\!\begin{array}{c} m\\k\end{array}\!\!\bigg) \prod_{i=0}^{k-1} \Ba(t_{n-i})\Bb(s_{m-i}) G(z,\bar z,t_{n-i};w,\bar w,s_{m-i})\Big\rangle\,. \label{OPE-chiral-3}  \end{align}
Here `$k$' counts the number of pairs of fields we replace with a propagator. For fixed $k$, there are $\small  \bigg(\!\!\begin{array}{c} n\\k\end{array}\!\!\bigg)$ choices of $\BX$'s and $\small \bigg(\!\!\begin{array}{c} m\\k\end{array}\!\!\bigg)$ choices of $\BPsi$'s, as well as $k!$ ways to match $\BX$'s to $\BPsi$'s, providing the prefactor. By symmetry, we then relabel/reorder the time variables so that the contractions only involve the final $k$ $t$'s and $s$'s.  

To keep things simple for the moment, let's use the axial-gauge propagator from \eqref{G-hol}, $G^{\rm ax}(z,\bar z,t_{n-i};w,\bar w,s_{m-i}) = \frac{\delta^{(1)}(t_{n-i}-s_{m-i})}{z-w} dz^{\frac r2}dw^{1-\frac r2}$. We can strip off the powers of $dz$ and $dw$, since they simply cancel against $dz^{-\frac r2}$ from each $\Ba$ and $dw^{\frac r2-1}$ from each $\Bb$; the cancellation is guaranteed by the fact that the couplings $\Ba\BX$ and $\Bb\BPsi$ have spin zero. Then each contraction allows us to do an integration, using the delta-function to set $t_{n-i}=s_{m_i}$, and we're left with
\begin{align} \eqref{OPE-chiral-3} &= \Big\langle\sum_{n,m=0}^\infty \sum_{k=0}^{\text{min}(n,m)} 
 :  \prod_{i=1}^{n-k} \int_{t_i} \Ba(t_i)\BX(t_i,z)  \prod_{j=1}^{m-k} \int_{s_j} \Bb(s_j) \BPsi(s_j,w): \notag \\ 
 & \hspace{.7in} \frac{1}{k!(n-k)!(m-k)!} \prod_{i=0}^{k-1}\; \int_{t_{n-i}} \Ba(t_{n-i})\Bb(t_{n-i}) \frac{1}{z-w}\Big\rangle \hspace{.7in} \end{align}
\vspace{-.3in}
\begin{align} \hspace{.5in} & = \Big\langle \sum_{k=0}^\infty \sum_{n,m=k}^\infty : \frac{1}{(n-k)!}\bigg[ \int_\R \Ba \BX(z)\bigg]^{n-k} \frac{1}{(m-k)!} \bigg[ \int_\R \Bb \BPsi(w)\bigg]^{m-k} : \frac{1}{k!}\bigg[ \int_\R \frac{1}{z-w} \Ba\Bb \bigg]^k \Big\rangle \notag  \\
 & = \Big\langle \sum_{k=0}^\infty \sum_{n,m=0}^\infty : \frac{1}{n!}\bigg[ \int_\R \Ba \BX(z)\bigg]^{n} \frac{1}{m!} \bigg[ \int_\R \Bb \BPsi(w)\bigg]^{m} : \frac{1}{k!}\bigg[ \int_\R \frac{1}{z-w} \Ba\Bb \bigg]^k \Big\rangle \notag  \\
 &= \Big\langle \exp \int_\R \Big[ \Ba\BX(z) +\Bb\BPsi(w) +\frac1{z-w}\Ba \Bb \Big] \Big\rangle\,.
 \label{OPE-chiral-4}  \end{align}
We further expand the coupling $\Ba\BX(z)$ around $w$ to get the final form
\be \hspace{-.4in} (\V_1\otimes_{z-w}\V_{-1})(w) = \int D\Ba D\bar\Ba D\Bb D\bar \Bb\, \exp\int_\R \Big[\Ba d\bar\Ba +\Bb d\bar\Bb + \sum_{n=0}^\infty \frac1{n!} (z-w)^n \Ba \pd^n \BX(w) + \Bb \BPsi(w) + \frac1{z-w}\Ba\Bb \Big] \label{OPE-chiral-5} \ee

Comparing to \eqref{OPE-general} and Theorem \ref{thm:non-ren}, we identify the new interaction term
\be r_{1,-1} = \frac{1}{z-w} ab \ee
This is induced from the Feynman diagram
\be \raisebox{-.3in}{\includegraphics[width=2.1in]{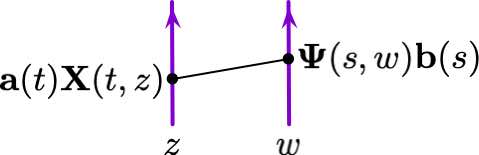}} \ee
which is the only \emph{connected} Feynman diagram involving the interaction vertices $\Ba\BX$ and $\Bb\BPsi$.
The chain of manipulations above \eqref{OPE-chiral-4} simply rederive an elementary result from field theory, that the sum of all Feynman diagrams involving arbitrary numbers of $\Ba\BX,\,\Bb\BPsi$ vertices is the exponential of the sum of fully connected Feynman diagrams.

\subsubsection{Fusion to the identity line}
\label{sec:chiral-fusion}

We comment briefly on why the quantum correction in \eqref{OPE-chiral-4} is not just reasonable but in fact \emph{necessary} for analyzing the behavior of line operators $\V_1$ and $\V_{-1}$ as they get close to each other.

One might intuitively expect that as $z\to w$ the vortex lines $\V_1(z)$ and $\V_{-1}(w)$ annihilate each other, leaving behind the identity $\V_0 = \id$.  Such limits turn out to be very subtle. If we include local operators on the lines, and evaluate their correlation functions,  the $z\to w$ limit is typically singular. Nevertheless, we could assume that there are \emph{no} local operators inserted anywhere else, and attempt to take the limit.

Consider \eqref{OPE-chiral-5}. As long as $z\neq w$ (and there are no local operators inserted), we can freely redefine the quantum-mechanics fields as
\be  (\Ba,\bar\Ba,\Bb,\bar\Bb)\;\mapsto\; ( (z-w)\Ba, (z-w)^{-1}\bar\Ba,\Bb,\bar\Bb)\,, \ee
while preserving the QM kinetic terms. This defines an isomorphic line operator, which has a regular limit as $z\to w$,
\begin{align}  & \int D\Ba D\bar\Ba D\Bb D\bar \Bb\, \exp \int_\R \Big[\Ba d\bar\Ba +\Bb d\bar\Bb + \sum_{n=0}^\infty \frac1{n!} (z-w)^{n+1} \Ba \pd^n \BX(w) + \Bb \BPsi(w) + \Ba\Bb \Big] \notag \\
 & \hspace{.8in} \underset{z\to w}\longrightarrow   \int D\Ba D\bar\Ba D\Bb D\bar \Bb\, \exp \int_\R \Big[\Ba d\bar\Ba +\Bb d\bar\Bb +  \Bb \BPsi(w) +\Ba\Bb \Big]\,.  \end{align}
The coupling $\Ba\Bb$ makes all the QM fields massive, and we can now integrate them out, using the EOM for $\Bb$ to set $\BPsi(w) = -\Ba$. (This is not a constraint on $\BPsi$; it just says how $\BPsi$ is related to the QM field that's been removed.) After integrating out the QM fields, we are left with a trivial line operator, as expected.

Alternatively, we could have rescaled $(\Bb,\bar\Bb)\mapsto ((z-w)\Bb,(z-w)^{-1}\bar\Bb)$ in \eqref{OPE-chiral-5}, then taken  the $z\to w$ limit, to recover a trivial line operator. Either way, it's the quantum correction $\Ba\Bb$ that ultimately allows us to fully integrate out 1d fields.

In contrast, had the quantum correction proportional to $\Ba\Bb$ been \emph{absent}, there would be no way to integrate out the QM fields. A naive collision limit of $\V_1$ and $\V_{-1}$ (without the correction) would have led to a nontrivial superposition of the two vortices, rather than the identity line. Different rescaling of QM fields during the limit could remove couplings to the bulk, but there would never be a way to fully integrate out the QM degrees of freedom themselves.

As might already be evident here, collision limits of line operators in HT theories are very subtle. Unlike in the case of 3d topological QFT's, such collisions are not ``functorial'': they are typically not unique (\eg\ many rescalings of QM fields are possible), and they do not preserve spaces of local operators and their correlation functions. We refer to such collision limits as \emph{fusions}, denoted $\odot$. For example, we argued above that there exists a fusion
\be \V_1\odot \V_{-1} \simeq \id\,. \ee
More examples of fusion, including in the XYZ model and SQED, appear in the PhD thesis \cite{VictorThesis}; we hope to revisit fusion more systematically in the future.

\subsection{Quasi-linear field theories}
\label{sec:quasi-linear}

We'd like to generalize the OPE analysis of \ref{sec:OPE-free} to any perturbative HT theory (as in Section \ref{sec:3dHT}), with fields $(\mb p_i,\mb x^i)$ and action $\int_{\C\times \R}\big[ \mb p_i d' \mb x^i + \CI(\mb p,\mb x,\pd)\big]$, where the interaction is some polynomial in the fields and their holomorphic derivatives.
To obtain a strong non-renormalization theorem, we find that we must impose a mild constraint on the interactions, which we call \emph{quasi-linearity}. We define the constraint here and argue that all HT-twisted 3d $\CN=2$ gauge/matter theories satisfy it. We'll explain how to use the constraint in Section \ref{sec:non-renorm}.

We say that a 3d HT theory is \emph{quasi-linear} if 1) we can write the interaction as a sum of terms
\be \CI(\mb x,\mb p,\pd) = \sum_{\alpha\in \Lambda} \CI^{(\alpha)}(\mb x,\mb p,\pd)\,; \label{I-sum}  \ee
and 2) we can find a corresponding partition of the set $S$ that indexes the pairs of superfields (in some chosen basis)
\be (\mb x^i,\mb p_i)_{i\in S}\,,\qquad  S = \bigsqcup_{\alpha\in \Lambda} S_\alpha\,,  \label{S-split} \ee
such that, for each $\alpha\in \Lambda$ and for each $i\in S_\alpha$, either
\begin{itemize}[leftmargin=*]
\item $\CI^{(\alpha)}(\mb x,\mb p,\pd)$ is at most linear in $\mb x^i$, and all other $\CI^{(\beta\neq\alpha)}(\mb x,\mb p,\pd)$  are independent~of~$\mb x^i$;~or
\item $\CI^{(\alpha)}(\mb x,\mb p,\pd)$ is at most linear in $\mb p_i$, and all other $\CI^{(\beta\neq\alpha)}(\mb x,\mb p,\pd)$ are independent of~$\mb p_i$\,.
\end{itemize}
(By ``at most linear'' we mean a polynomial of degree one.)

Consider, for example, HT-twisted gauge theory with arbitrary CS term, linear matter representation, and superpotential. The action from Section \ref{sec:3dN2} is
\be \int_{\C\times \R} \mb A d'\mb B + \BPsi d' \mb X - \tfrac12 \mb B[\mb A,\mb A]+\tfrac12 k(\mb A,\pd \mb A)+\BPsi\varphi_V(\mb A)\mb X +W(\mb X)\,. \label{gauge-int-nr} \ee
We take
\be \begin{array}{l@{\qquad}l@{\quad}l} \CI = \CI^{(1)}+\CI^{(2)}\,,&  \CI^{(1)} =- \tfrac12\mb B[\mb A,\mb A]+\tfrac12 k(\mb A,\pd \mb A)\,, & \CI^{(2)} = \BPsi\varphi_V(\mb A)\mb X +W(\mb X)\,; \\[.2cm]
 S=S_1\sqcup S_2\,, &  (\mb B,\mb A)\in S_1\,, & (\mb X,\BPsi)\in S_2\,.\end{array} \label{gauge-split-nr}  \ee
This is quasi-linear because
\begin{itemize}
\item for $\alpha=1$: $\CI^{(1)}$ is at most linear in $\mb B$ and $\CI^{(2)}$ is independent of $\mb B$\,;
\item for $\alpha=2$: $\CI^{(2)}$ is at most linear in $\BPsi$ and $\CI^{(1)}$ is independent of $\BPsi$\,.
\end{itemize}

\subsection{Perturbative non-renormalization}
\label{sec:non-renorm}

Let's now return to the computation of OPE's, in a perturbative HT theory with action $\int_{\C\times \R}\big[ \mb p_i d' \mb x^i + \CI(\mb p,\mb x,\pd)\big]$. We consider two general perturbative line operators $\ell,\ell'$ defined by solving the Maurer-Cartan equation to couple quantum mechanics with state spaces $V,V'$ to bulk local operators (as in Section \ref{sec:lines-QM}). We assume that \emph{the couplings are represented linearly} in terms of bulk local operators. Thus they take the form
\be \mu_\ell  = \sum_{i,n\geq 0} \tfrac{1}{n!}\big(\alpha_{i,n} \pd^n x^i + \beta^i_n \pd^n p_i\big)\,,\qquad  \mu_{\ell'}  = \sum_{j,m\geq 0}\tfrac1{m!} \big(\alpha_{j,m}' \pd^m x^j + \beta'{}^j_m \pd^m p_j\big)\,, \label{non-ren-couplings} \ee
for collections of QM operators $\alpha_{i,n},\beta^i_n\in \text{End}(V)$ and $\alpha_{j,m}',\beta'{}^j_m\in \text{End}(V')$.
We'd like to establish Theorem \ref{thm:non-ren}

\subsubsection{General Feynman diagrams}
\label{sec:nr-Feynman}

We analyze Feynman diagrams that can give perturbative quantum corrections to the insertion
\begin{align}  \ell(z)\otimes \ell'(w) 
&= \exp\int_{t\in \R} \bm \mu_\ell(z,t) \; \exp\int_{s\in \R}\bm \mu_{\ell'}(w,s) \label{non-ren-ll}  \\
 &= \sum_{a,b=0}^\infty \frac{1}{a!b!} \int_{t_1,...,t_a} \text{T} \Big[ \bm\mu_\ell(z,t_1)...\bm \mu_\ell(z,t_a)\Big] \int_{s_1,...,s_b}  \text{T}\Big[ \bm\mu_{\ell'}(w,s_1)...\bm\mu_{\ell'}(w,s_b)\Big]\,, \notag
\end{align}
where T denotes time-ordering. It is sufficient to consider connected Feynman diagrams; all others are products of connected diagrams, and the final re-exponentiated answer should simply contain the sum of connected diagrams. A connected Feynman diagram could have
\begin{itemize}
\item some number $v_\ell$ of vertices along $\ell$ or $\ell'$, where propagators will contract with a single copy of $\pd^n \mb x(z),\pd^n\mb p(z),\pd^m\mb x(w),$ or $\pd^m\mb p(w)$
\item some number $v$ of bulk vertices (insertions of the interaction $\CI(\mb p,\mb x,\pd)$)
\item some numbers $e$ of bulk-bulk propagators (connecting $v$ vertices); $e_\ell$ of bulk-line propagators (connecting $v$ vertices to $v_\ell$ vertices); and $e_{\ell\ell}$ of line-line propagators (connecting $v_\ell$ vertices on different lines)
\end{itemize}
Moreover, there is a fundamental constraint from preserving cohomological degree, \emph{a.k.a.} ghost number symmetry. For a connected diagram to contribute to the effective, re-exponentiated action, it must be a one-form and thus have degree $R=1$ (just like any interaction in QM). Line vertices also have $R=1$, while bulk vertices have $R=2$ (for their $dt$, $d\bar z$ indices). Propagators contracting pairs of bulk fields, whether in the bulk or on lines, effectively have $R=-1$. Thus, we require
\be 2v + v_\ell - (e+e_\ell+e_{\ell\ell}) = 1\,.  \label{OPE-R} \ee

The simplest connected diagram satisfying \eqref{OPE-R}  has $v_\ell=2,e_{\ell\ell}=1$, and contracts a $\pd^n \mb x(z)$ on $\ell$ with a $\pd^m\mb p(w)$ on $\ell'$ (or vice versa), with a single line-line propagator:
\be \raisebox{-.3in}{\includegraphics[width=4in]{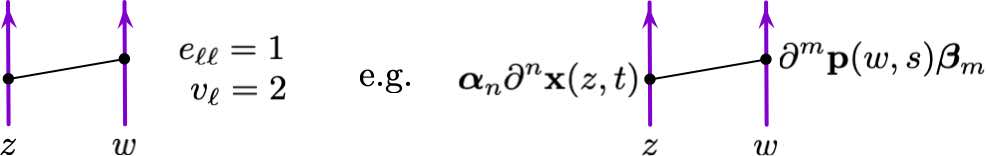}} \label{tree-diag} \ee
This gives a ``tree-level'' quantum correction, which we already found in the free-chiral example. We'd like to show that there are no other admissible diagrams.

Note that, since the couplings \eqref{non-ren-couplings} are linear in bulk fields, the vertices on the lines are necessarily univalent. Thus the diagram \eqref{tree-diag} is the unique connected diagram with $e_{\ell\ell}> 0$. To search for other connected diagrams, let us assume that $e_{\ell\ell}=0$. Then each univalent vertex on a line is connected to a unique bulk-line propagator, so $v_\ell = e_\ell$, and the degree constraint \eqref{OPE-R} simplifies to
\be \boxed{2v = e +1}\,. \label{OPE-R-bulk} \ee
Here's an example of a diagram that satisfies \eqref{OPE-R-bulk}, with $v = 5$ and $e=9$:
\be \raisebox{-.3in}{\includegraphics[width=2.3in]{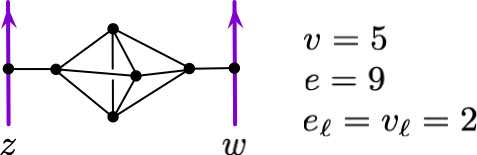}} \label{bipyramid} \ee
To rule out \eqref{bipyramid} (and all other possibilities), we need to further constrain potential interactions.

\subsubsection{Constraint from quasi-linear theories}
\label{sec:nr-quasi-linear}

Suppose we have a quasi-linear theory, as defined in Section \ref{sec:quasi-linear}.
Let's split the bulk vertices and propagators as
\be v = \sum_{\alpha\in \Lambda} v_\alpha\,,\qquad e = \sum_{\alpha\in \Lambda} e_\alpha\,,\ee
where $v_\alpha$ is the number of insertions of $\CI^{(\alpha)}$ (or a term therein) and $e_\alpha$ is the number of propagators connecting any $(\mb x^i,\mb p_i)$ pair within the set $S_\alpha$. The quasi-linearity condition then forces
\be e_\alpha \leq v_\alpha\quad\forall\,\alpha\in \Lambda\,. \label{EV-linear-nr} \ee
To see this, let's assume for simplicity that the quasi-linearity condition below \eqref{S-split} is satisfied entirely with $\mb x^i$ fields. (If instead it's satisfied with $\mb p_i$'s, just relabel $\mb x^i\leftrightarrow \mb p_i$; the bose/fermi distinction between these two fields is not important here.) Thus $\CI^{(\alpha)}$ is at most linear in $\mb x^i$ and all other $\CI^{(\beta\neq \alpha)}$ are independent of $\mb x^i$, for all $i$ in $S_\alpha$.
 Then each `$\mb x$' end of a propagator counted by $e_\alpha$ must connect to a bulk vertex of type $\CI^{(\alpha)}$, and no two propagators can connect to the same $\CI^{(\alpha)}$ vertex this way because $\CI^{(\alpha)}$ contains $\mb x$ at most linearly. This implies \eqref{EV-linear-nr}.

To illustrate this in gauge theory \eqref{gauge-split-nr}, let $v_1$ be the number of $\mb B\mb A\mb A$ and $k\mb A\pd\mb A$ vertices; let $v_2$ be the number of $\BPsi\mb A\mb X$ and $W(\mb X)$ vertices; let $e_1$ be the number of $\mb B_{a}\!-\!\mb A^a$ propagators; and let $e_2$ be the number of $\mb X^i\!-\!\BPsi_{i}$ propagators. Since $\BB$ only appears linearly in $\mb B\mb A\mb A$ vertices (and not at all in the others), we have $e_1\leq v_1$. Since $\BPsi$ only appears linearly in $ \BPsi\mb A\mb X$ vertices (and not at all in the others), we also have $e_2 \leq v_2$.

Summing \eqref{EV-linear-nr} over $\alpha$ we obtain $e\leq v$.
Combined with \eqref{OPE-R-bulk}, we get $2v=1+e\leq 1+v$, whence 
\be  \boxed{v\leq 1}\,. \ee

The remaining possibilities can quickly be ruled out. For $v=0$, the only available diagram is our desired \eqref{tree-diag}. For $v=1$, \eqref{OPE-R-bulk} requires $e=1$, and the only diagram is
\be \raisebox{-.3in}{\includegraphics[width=2.2in]{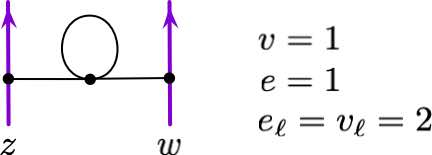}} \label{v=1} \ee
This does not contribute because the bulk interaction vertices $\CI$ are (by definition) already normal-ordered, so there are no contractions of fields within the same vertex.

\subsubsection{Constraint for cubic field theories}
\label{sec:cubic}

An alternative to considering quasi-linear theories is to simply require all interactions terms to be at most cubic in the fields $\mb x,\mb p$. In this case, bulk vertices will be at most 3-valent, leading to the constraint
\be 2e \leq 3 v\,. \ee 
This in turn implies $2v \leq \frac32 v+1$, or $\boxed{v\leq 2}$.

We've already ruled out $v=1$ in \eqref{v=1}. At $v=2$, the degree constraint \eqref{OPE-R-bulk} requires $e=3$. Diagrams with a propagator connecting a vertex to itself do not contribute, for the same reason as in \eqref{v=1}. The only remaining diagram with $v=2$ is
\be  \raisebox{-.3in}{\includegraphics[width=2.7in]{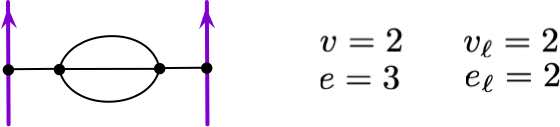}} \label{melon} \ee
Here, however, we observe that the propagator $\langle \mb p_i(z,\bar z,t)\mb x^j(w,\bar w,s)\rangle= \delta_i{}^jG_i(z,\bar z,t;w,\bar w,s)$ for free fields is a one-form on configuration space $\text{Conf}^{\,2}(\C\times \R) = (\C\times \R)^2\backslash\Delta$ (it's a one-form in $dt,d\bar z,ds,d\bar w$ indices). Moreover, given any gauge condition that treats all $\mb x$'s and $\mb p$'s, for all $i$, \emph{symmetrically} --- such as Lorentz gauge, or axial gauge --- the propagator will be independent of $i$ and will be symmetric, $G(z,\bar z,t;w,\bar w,s) = G(w,\bar w,s;z,\bar z,t)$. In any such gauge, the diagram \eqref{melon} necessarily vanishes, because three identical one-forms get wedged with each other. Therefore, the only remaining option is $v=0$, establishing Theorem \ref{thm:non-ren}.

\subsubsection{Non-linear couplings}
\label{sec:non-linear}

As a caveat, we note that our assumption above that line operators are represented by linear couplings to bulk operators is \emph{essential} for the non-renormalization theorem. Without linear couplings, the vertices on the lines can have any valency. Then there are infinitely many admissible connected diagrams aside from \eqref{tree-diag}, such as
\be \raisebox{-.5in}{\includegraphics[width=3.8in]{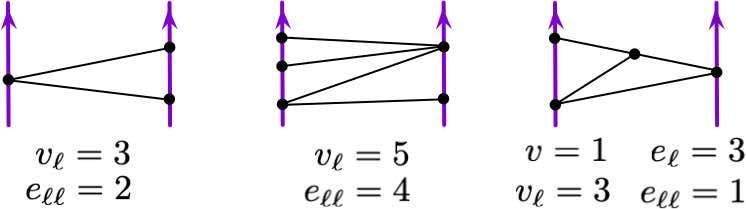}} \label{nonlinear-diag} \ee

It's instructive to analyze this more general situation. The degree constraint generalizing \eqref{OPE-R-bulk} becomes
\be 2v+v_\ell = e+e_\ell + e_{\ell\ell} + 1 \ee
and the inequality due to quasi-linearity becomes
\be   e+ e_\ell+e_{\ell\ell} \leq v + v_\ell \ee
which together imply $2v+v_\ell \leq v+v_\ell + 1$, whence $\boxed{v\leq 1}$. Thus, there is still at most one bulk vertex. However, there can be arbitrarily many vertices on the lines, as in the diagrams of \eqref{nonlinear-diag}.

\subsection{The exact tree-level correction}
\label{sec:exact-gen}

Having established that the only diagrams contributing to the OPE (in quasi-linear theories or cubic theories) are the direct line-line contractions \eqref{tree-diag}, with no bulk vertices inserted, we can go ahead and compute the exact perturbative quantum correction to the RHS of \eqref{non-ren-ll}. Since bulk interactions play no role, the answer looks essentially the same as in the free theory of Section \ref{sec:OPE-free}.

\subsubsection{Axial gauge}
\label{sec:exact-r}

Let's first work in axial gauge, with propagator $G^{\rm ax}(z,\bar z,t_{n-i};w,\bar w,s_{m-i}) = \frac{\delta^{(1)}(t_{n-i}-s_{m-i})}{z-w} dz^{\frac r2}dw^{1-\frac r2}$.

We note that, beyond tree level, axial gauge has been known to give incorrect answers for calculations in holomorphic field theories.%
\footnote{We thank Kevin Costello for pointing this out, in the context of NHMV amplitudes computed using twistor space. The phenomenon unfortunately does not seem to be well documented in the literature.} %
In the present setting, the combinatorial non-renormalization theorem from Section \ref{sec:non-renorm} guarantees that there are no higher-loop corrections --- in fact, no corrections with bulk insertions at all --- so axial gauge should be a valid choice.

More so, we expect axial gauge to capture the leading singularities in the OPE. If we place $\ell$ at $z_1$ and $\ell'$ at $0$, then bringing $z_1$ close to $0$ can be effected by a coordinate rescaling $z\to \epsilon z$ in the $\C_{z,\bar z}$ plane. For very small $\epsilon$, the `$d\bar z$' components of all fields ($x_{\bar x},p_{\bar z}$, etc.) will become vanishingly small relative to the `$dt$' components; thus all fields will effectively be cast into axial gauge. (We'll return to this argument in Section \ref{sec:othergauge}.)

A contraction between a single $\bm \mu_\ell(z,t) =  \sum_{i,n\geq 0}\frac{1}{n!} \big[\bm \alpha_{i,n}(t) \pd^n \mb x^i(z,t) + \bm \beta^i_n(t) \pd^n \mb p_i(z,t)\big]$ and $\bm\mu_{\ell'}(w,s)= \sum_{j,m\geq 0} \frac{1}{m!}\big[\bm \alpha_{j,m}'(s) \pd^m \mb x^j(w,s) + \bm\beta'{}^j_m(s) \pd^m \mb p_j(w,s)\big]$ in \eqref{non-ren-ll} 
evaluates~to
\begin{align} \hspace{-.3in} \bm \mu_\ell\overbracket{(z,t)\bm\mu_{\ell'}(w},s) &= \sum_{i,j}\sum_{m,n\geq 0} \frac{1}{n!m!} \Big[ \bm \alpha_{i,n}(t)\bm\beta'{}^j_m(s)  \pd^n \overbracket{\mb x^i(z,t)\pd^m\mb p_j}(w,s)  - \bm \alpha_{j,m}'(s)\bm \beta^i_n(t) \pd^m\overbracket{ \mb x^j(w,s)\pd^n \mb p_i}(z,t)\Big] \notag \\
&=  \sum_{i,j}\sum_{m,n\geq 0}\frac{1}{n!m!}\Big[ \bm \alpha_{i,n}(t)\bm\beta'{}^j_m(s) \delta_{ij} \pd_z^n\pd_w^m \frac{\delta^{(1)}(t-s)}{z-w} - \bm \alpha_{j,m}'(s)\bm \beta^i_n(t)\delta_{ij} \pd^m_w\pd^n_z  \frac{\delta^{(1)}(s-t)}{w-z} \Big] \notag \\
&=   \sum_{m,n\geq 0} \frac{ (-1)^n(n+m)!}{n!m!}  \frac{ \sum_i \big[\bm \alpha_{i,n}(t)\bm\beta'{}^i_m(t) -  \bm \alpha_{i,m}'(t)\bm \beta^i_n(t)\big]}{(z-w)^{n+m+1}} \delta^{(1)}(t-s)  \label{mumu}
 \end{align}
Note that the minus sign in the second term in the first line comes from commuting the fermionic operators $\bm \alpha$ and $\mb p$ past each other. We've also stripped off $dz$'s and $dw$'s, as they just cancel against corresponding terms from $\bm\alpha$ and $\bm\beta$, keeping track of spin.
We denote the QM operator from the last line as
\begin{align} \mb r(t, z-w) &:= \sum_{m,n\geq 0} (-1)^n \bigg(\!\!\begin{array}{c} n+m\\n\end{array}\!\!\bigg)  \frac{  \sum_i \big[\bm \alpha_{i,n}(t)\bm\beta'{}^i_m(t) -  \bm \alpha_{i,m}'(t)\bm \beta^i_n(t)\big]}{(z-w)^{n+m+1}}   \label{non-ren-r} \\
 & \in \text{End}(V)\otimes\text{End}(V')\otimes \C[[ (z-w)^{-1}]]\,.   \notag \end{align}

We can now sum over all contractions in \eqref{non-ren-ll} exactly the same way we did for a free chiral. Indeed, the derivation in \eqref{OPE-chiral-3}--\eqref{OPE-chiral-5} carries over nearly identically, with the replacements $\Ba\BX\leadsto \bm\mu_\ell$, $\Bb\BPsi\leadsto \bm\mu_{\ell'}$, $\frac{\Ba\Bb}{z-w}\leadsto \mb r$, and the inclusion of extra time-ordering along each line. We find that, in correlation functions (potentially with insertions of other operators),
\begin{align} (\ell\otimes_{z-w} \ell')(w) &=  \sum_{a,b=0}^\infty \frac{1}{a!b!} \int_{t_1,...,t_a} \text{T} \Big[ \bm\mu_\ell(z,t_1)...\bm \mu_\ell(z,t_a)\Big] \int_{s_1,...,s_b}  \text{T}\Big[ \bm\mu_{\ell'}(w,s_1)...\bm\mu_{\ell'}(w,s_b)\Big] \notag \\
 &= \sum_{a,b=0}^\infty \sum_{k=0}^{\text{min}(a,b)} \frac{k!}{a!b!} \bigg(\!\!\begin{array}{c} a\\k\end{array}\!\!\bigg)\bigg(\!\!\begin{array}{c} b\\k\end{array}\!\!\bigg) \text{T} \Big[ \int_{t_1,...,t_{a-k}} :\prod_{i=1}^{a-k} \bm \mu_\ell(z,t_i) \int_{s_1,...,s_{b-k}} \prod_{i=1}^{b-k} \bm \mu_{\ell'}(w,s_j): \notag \\
 & \hspace{2in} \times \int_{t_{a-k+1},...,t_a} \prod_{i=1}^k \mb r(t_{a-i},z-w) \;\Big]  \notag \end{align}
 \vspace{-.3in}
 \begin{align}\hspace{0in}&= \exp \int_{t\in\R} \bm \mu_\ell(z,t) + \bm \mu_{\ell'}(w,t) +\mb r(t,z-w)  \label{non-ren-OPE} \\
 &= \exp \int_{t\in\R}  \sum_{n=0}^\infty \frac{(z-w)^n}{n!} \pd^n \bm \mu_\ell(w,t)  + \bm \mu_{\ell'}(w,t) +\mb r(t,z-w)\,.  \notag
\end{align}

\subsubsection{Other gauges: universal behavior}
\label{sec:othergauge}

Working in axial gauge is not always convenient, or appropriate. For example, in a global spacetime, the choice of gauge we use must be compatible with global boundary conditions. Here we comment a bit further on the structure of quantum corrections to the OPE of line operators in \emph{any} gauge.

Since our HT theory must ultimately be independent of gauge at a local level, we might expect that the total sum of all corrections to the OPE induced by computing it in a different gauge is cohomologically trivial --- \ie\ $Q$-exact after de-exponentiating all time-ordered exponentials and keeping careful track of boundary terms. 
This seems difficult to check/prove in full generality, though we'll indicate later how it comes about in Lorentz gauge.

We can also use symmetry to partially constrain the behavior of any propagator. Let's denote our general propagator as $G(z,\bar z,t;w,\bar w,s)$. It is a (distributional) 1-form on configuration space $\text{Conf}^{\,2}(\C\times \R)$, though we only care about its $dt$ and $ds$ components, as all others will vanish upon insertion in the diagram \eqref{tree-diag} and integration of the two vertices along the support of the lines. If we assume that we use a gauge that's invariant under translations, then these components of the propagator will take the form
\be G(z,\bar z,t;w,\bar w,s)_{s,t} = g(z-w,\bar z-\bar w,t-s)d(t-s) \ee
for some function $g$. The OPE will get contributions from contractions of derivatives $\pd^n\mb x(z,t)$ and $\pd^m\mb p(w,s)$, as in \eqref{tree-diag}, which evaluates to
\be (-1)^m \pd_z^{m+n} g(z-w,\bar z-\bar w,t-s)d(t-s)\,. \ee
It is convenient to use translation symmetry to set $w=\bar w=s=0$, and thus analyze the distribution $(-1)^m\pd_z^{m+n} g(z,\bar z,t)dt$. Its behavior is constrained as follows:

\begin{itemize}
\item Due to the scaling argument discussed at the beginning of Section \ref{sec:exact-r}, every gauge must approach axial gauge in the limit $z\to 0$. Thus, the leading singularity is given by the holomorphic-gauge propagator
\be (-1)^m\pd_z^{m+n} g(z,\bar z,t)dt \sim \frac{(-1)^n(n+m)!}{z^{n+m+1}} \delta(t) dt +\text{less singular in $z$} \label{dng-expand} \ee
\item If $t\neq 0$ then the $z\to 0$ limit must be regular (because the local operators $\pd^n\mb x$ and $\pd^m\mb p$ can have a singular 2-point function only when their supports coincide). Thus, the subleading singular terms are distributions supported at $t=0$, and should be proportional to $\delta(t)dt$ and various derivatives $\pd_t^p\delta(t)dt$.
\item The entire expansion \eqref{dng-expand} must have spin $J=n+m$. This constrains it to be an expansion in $|z|$ (which has $J=0$), so that subleading terms are of the form $C \frac{1}{z^{n+m}} |z|^a\times \pd_t^p\delta(t)dt$ for various $a$ and $p$ and constants $C$.\end{itemize}
Going back through the computation of Section \ref{sec:exact-r}, we conclude that in \emph{any} gauge the OPE of line operators takes the general form 
\begin{align} & (\ell\otimes_{z-w}\ell')(w) \sim \label{Lor-OPE}  \\
 & \hspace{.4in} \exp\int_{\R}\big[ \bm \mu_\ell(w)+\bm\mu_{\ell'}(w) + \mb r(z-w) +\text{(subleading, non-meromorphic, singular terms)}\big]\,, \notag \end{align}
modulo terms regular as $z\to w$, which may be nonlocal along $\R$.

\subsubsection{Lorentz gauge}
\label{sec:Lorentz}

We can be more precise if we specialize to Lorentz gauge \eqref{G-Lor}. The time components of its propagator give
\be   (-1)^m\pd_z^{m+n} g(z,\bar z,t) = \frac{(-1)^n}{2} \pd_z^{n+m} \frac{\bar z\, dt}{(t^2+|z|^2)^{\frac32}}\,. \ee
We claim that there exists an expansion near $z\to 0$
\be \frac12 \pd^n \frac{\bar z\, dt}{(t^2+|z|^2)^{\frac32}} = \pd^n \frac{1}{z}\delta(t)dt + C_{n,1} \frac{\bar z}{z^n} \delta''(t)dt + \ldots + C_{n,n} \frac{\bar z^n}{z} \pd_t^{2n}\delta(t)dt + o(|z|^n)\,, \label{Ltoh-dist}  \ee
for some numerical coefficients $C_{n,k}$. To see this, we integrate against a test function~$f(t)$,
\begin{align} \frac12 \int_\R f(t)\, \pd_z^n \frac{\bar z\, dt}{(t^2+|z|^2)^{\frac32}} &= \frac{(-1)^n (2n+1)!!}{2^{n+1}} \int_\R f(t)\, \frac{\bar z^{n+1}\, dt}{ (t^2+|z|^2)^{\frac32+n}} \notag \\
&= \frac{(-1)^n (2n+1)!!}{2^{n+1} z^{n+1}} \int_\R \frac{ f(|z|t)\, dt}{ (t^2+1)^{\frac32+n}} \label{Lor-integral}
\end{align}
From here, we use Taylor's theorem with remainder to approximate $f(|z|t) = f(0) + f'(0)|z|t + ... + \frac{1}{(2n)!} f^{(2n)}(0) |z|^{2n}t^{2n} + \frac{1}{(2n+1)!}f^{(2n+1)}(0) |z|^{2n+1}t^{2n+1} + h(|z|t) |z|^{2n+1}t^{2n+1}$, for some function $h$ that tends to zero as $|z|\to 0$. Moreover, it's reasonable to assume that $h$ grows slowly enough as $t\to \infty$ so that the integral $\int \frac{h(|z|t) t^{2n+1}}{(t^2+1)^{3/2+n}}$ converges --- since the original integral \eqref{Lor-integral} is assumed to converge. Then, noting that only even powers of $t$ give a nonzero integral, we get
\be \eqref{Lor-integral} =  (-1)^n n!\frac{1}{z^{n+1}} f(0) + \frac{(-1)^n (n-1)!}{4} \frac{\bar z}{z^{n}} f''(0) +\ldots+ \frac{(-1)^n}{2^n(2n)!!} \frac{\bar z^n}{z}  f^{(2n)}(0)  + o(|z|^{n})\,, \ee
which justifies \eqref{Ltoh-dist}.

Deriving further regular terms in the series is tricky, because the Taylor expansion under the integrand cannot be continued beyond order $2n+1$ (since the integral won't converge). Instead, the complete $|z|\to 0$ asymptotics of \eqref{Lor-integral} can be derived be Fourier-transforming $f(|z|t)$, then doing the $dt$ integral to get modified Bessel functions, and taking the $|z|\to 0$ asymptotics of the Bessel functions. The asymptotics of Bessel functions at small argument involve non-analytic, logarithmic terms, which ultimately result in corrections to \eqref{Ltoh-dist} at order $|z|^{n+1}$ and higher that are not local --- they are not proportional to delta functions and their derivatives at $t=0$. This produces the regular corrections to \eqref{Lor-OPE}.

We note that, in Lorentz gauge, the nonlocal corrections to \eqref{Lor-OPE} are not just regular but \emph{vanishing} as $z\to w$.

There is one final observation to make. In Lorentz gauge, all subleading singular corrections \eqref{Ltoh-dist} are proportional to derivatives of delta functions. Therefore, the correction to the OPE of lines with couplings $\bm \mu_\ell(z,t) =  \sum_{i,n\geq 0}\frac{1}{n!} \big[\bm \alpha_{i,n}(t) \pd^n \mb x^i(z,t) + \bm \beta^i_n(t) \pd^n \mb p_i(z,t)\big]$ and $\bm\mu_{\ell'}(w,s)= \sum_{j,m\geq 0} \frac{1}{m!}\big[\bm \alpha_{j,m}'(s) \pd^m \mb x^j(w,s) + \bm\beta'{}^j_m(s) \pd^m \mb p_j(w,s)\big]$ involves singular terms of the form
\be \exp\int_\R\bigg[\bm\mu_\ell+\bm\mu_{\ell'}+ \mb r + \sum_{n,m}\sum_{k=1}^{n+m}  C_{n+m,k} \frac{\bar z^k}{z^{n+m+1-k}}  \big[(\pd_t^{2k}\bm\alpha_{i,n}) \bm\beta'{}^i_m- (\pd_t^{2k}\bm\alpha'_{i,n}) \bm\beta^i_m\big] \bigg] \ee
(modulo signs). In topological quantum mechanics, we may rewrite $\pd_t = [Q,\wt Q]$ in terms of the BRST and descent operators.  If we moreover impose the simplifying assumption that all the QM operators $\bm \alpha,\bm\beta,\bm\alpha',\bm\beta'$ are $Q$-closed, then the corrections immediately become $Q$-exact:
\be (\pd_t^{2k}\bm\alpha_{i,n}) \bm\beta'{}^i_m- (\pd_t^{2k}\bm\alpha'_{i,n}) \bm\beta^i_m = Q\big[(\pd_t^{2k-1}\wt Q\bm\alpha_{i,n}) \bm\beta'{}^i_m- (\pd_t^{2k-1}\wt Q\bm\alpha'_{i,n}) \bm\beta^i_m\big]\,. \ee
This is in line with the expectation that the deviation from axial gauge should be $Q$-exact.

\section{Koszul duality and dg-shifted Yangians}
\label{sec:Koszul}

In this section, we introduce Koszul-dual algebras $\CA^!$ as a way to represent perturbative line operators, adapting \cite{Costello-Yangian} and the general procedure outlined in \cite{CPkoszul, PaquetteWilliams} to the context of 3d HT QFT's.%
\footnote{For mathematical background on Koszul duality, we refer the reader to \cite{loday2012algebraic}.} %
This perspective has several virtues. First, it lets us describe the entire category of perturbative line operators $\CC$ (including its morphisms) as a category of $A_\infty$-modules for the Koszul-dual,
\be \CC \simeq \CA^!\text{-mod}\,. \ee
Second, it lets us reinterpret (using Tannakian principles) the data of OPE's of lines, and their associativity relations, in terms of universal structure in $\CA^!$: the structure of a ``dg-shifted Yangian.'' 

We'll begin in Section \ref{sec:KD-perspective} with two concrete physical constructions of the Koszul-dual algebra $\CA^!$ in a perturbative 3d HT QFT $\CT$. The first construction (Section \ref{sec:cp1}) involves compactifying space to a $\cp^1_z$, with a particular singularity $\CB_\infty$ at $z=\infty$ that makes the state space one-dimensional; then we define $\CA^!$ to be the algebra of local operators supported on the timelike line at $z=\infty$. Equivalently (Section \ref{sec:KD-2d}) we may first reduce $\CT$ on a spatial circle, to obtain a 2d B-model $\CT[S^1]$ with infinitely many fields --- the modes of the original 3d fields. In $\CT[S^1]$, the identity line $\id\in \CC$ and the singularity $\CB_\infty$ both define natural boundary conditions, the former where only regular modes survive, and the latter where only polar modes survive. We again define $\CA^! = \text{End}(\CB_\infty)$ to be local operators on $\CB_\infty$.

Explicitly, if we expand 3d fields in modes on $\C^*_z$ as $ \mb x^i = \sum_{n\in \Z} \mb x^i_n z^{-n-1}\,,\; \mb p_i = \sum_{n\in \Z} \mb p_{i,n} z^{-n-1}$, we find that $\CA$ and $\CA^!$ are generated by regular and polar modes:
\be \CA = \C[x^i_{-n-1} ,p_{i,-n-1}]_{n\geq 0} 
\,,\qquad  \CA^! = \C[p_{i,n},x^i_n]_{n\geq 0}\,. \ee
Bulk interactions $\CI$ induce further $A_\infty$ operations on both $\CA$ and $\CA^!$, studied in Section~\ref{sec:Ainf}.

The key (physical) results of Section \ref{sec:KD-perspective} are that $\cp^1$ compactification must define $A_\infty$ functors on configurations of lines
\be \CF: \begin{array}{ccc} \CC_{z_1}\boxtimes \CC_{z_2}\boxtimes \cdots \boxtimes \CC_{z_n} &\to& \CA^!\text{-mod} \\
  \ell_1(z_1)\otimes\ell_2(z_2) \otimes \cdots \ell_n(z_n) &\mapsto& \text{States}(\cp^1,\ell_1(z_1)\otimes\cdots \ell_n(z_n),\CB_\infty) \label{F-multi-intro} \end{array}
\ee
and that the restriction of the functor to a \emph{single} line 
\be \CF: \CC \overset\sim\to \CA^!\text{-mod} \label{equiv-sec-intro} \ee
is an equivalence of categories (Section \ref{sec:KD-MC}).
Underlying the equivalence is the existence of a ``universal'' Maurer-Cartan element $\mu\in\CA^!\otimes \CA$, given explicitly by
\be \mu = \sum_{n\geq 0,i}\big[- p_{i,n}\otimes x^i_{-n-1} + x^i_n \otimes p_{i,-n-1}\big]\,. \label{univ-MC-intro} \ee
The MC element $\mu$ enters physically in the transversality and completeness relations of $\id$ and $\CB_\infty$, viewed as boundaries of the 2d theory $\CT[S^1]$.
It sets up a 1-1 correspondence, up to isomorphisms
\be \begin{array}{c@{\;}c@{\;}c}
 \left\{ \begin{array}{c} \text{perturbative lines $\ell\in \CC$, \ie} \\ \text{spaces $V_\ell$ with $\mu_\ell\in \text{End}(V_\ell)\otimes \CA$} \end{array} \right\}
 & \leftrightarrow &
  \left\{ \begin{array}{c}\text{$\CA^!$-modules, \ie\ spaces $V_\ell$ } \\  \text{with an action $\rho_\ell:\CA^!\to\text{End}(V_\ell)$}\end{array}\right\} \\
V_\ell,\,  \mu_\ell=(\rho_\ell\otimes 1)(\mu) & \reflectbox{$\mapsto$} & V_\ell,\, \rho_\ell \end{array}  \label{corresp-MC}  \ee
such that $\mu_\ell = (\rho_\ell\otimes 1)(\mu)$. In other words, a line operator with linear couplings $\mu_\ell = \sum_{i,n} \frac{1}{n!}( \alpha_{i,n}\pd^n x^i+\beta^i_n\pd^np_i)$ corresponds to the representation $\rho_\ell(p_{i,n},x^i_n) = (-\alpha_{i,n}, \beta^i_n)$.

The correspondence \eqref{corresp-MC}, together with the fact that $\CA^!$ is linearly generated,  further guarantees that all perturbative line operators can be represented by linear couplings to $\CA$. We spell this out in Section \ref{sec:KD-MC}. We've assumed this important property in earlier sections of the paper.

To make \eqref{corresp-MC} seem a bit less abstract, we give some examples of line operators as $\CA^!$-modules, for the free chiral and the XYZ model, in Section \ref{sec:KD-eg}.
A few gauge-theory examples are discussed later in the paper (Sec. \ref{sec:Ainf-gauge}).

In Sections \ref{sec:KD-Yangian} and \ref{sec:Yangian-summary}, we then return to the matter of OPE's. We consider
\begin{enumerate}
\item  translations $T_{z}:\CC_w\overset\sim\to \CC_{z+w}$ of perturbative lines;
\item the OPE with quantum corrections $\mb r_{\ell,\ell'}$;
\item associativity of the OPE for arbitrary collections of lines
\end{enumerate}
and reconstruct this data as additional structure (and properties thereof) in $\CA^!$. We arrive this way at the definition of a dg-shifted Yangian (Def. \ref{def:Yang}), and \textbf{Conjecture~\ref{conj:HT-Yang}}: the Koszul-dual algebra $\CA^!$ in any perturbative 3d HT QFT is a dg-shifted Yangian.

Here's a short summary of the results. First, translations simply induce dual translations $\tau_z:\CA^!\to \CA^!$. Next, a key observation is that in the OPE of lines from Section~\ref{sec:OPE},
\be  (\ell\otimes_{z} \ell')(0) \sim \exp\int_\R \big[ \bm\mu_\ell(z) + \bm\mu_{\ell'}(0) + \mb r_{\ell,\ell'}(z)\big]\,, \label{OPE-sum4K} \ee
has a universal form, when using axial gauge. Namely, the quantum correction $r_{\ell,\ell'}(z) \in \text{End}(V\otimes V') [ z^{-1} ]$ can be expressed as $r_{\ell,\ell'}(z) = (\rho_\ell\otimes \rho_{\ell'})(r(z))$ via \eqref{corresp-MC}, where
\be r(z) = \sum_{m,n\geq 0} \bigg(\begin{array}{@{}c@{}} n+m \\[-.1cm] n\end{array}\bigg)  \frac{ p_{i,n}  \otimes x{}^i_m- x^i_n \otimes  p_{i,m}   }{(-1)^n z^{n+m+1}} \in \CA^!\otimes \CA^! [\![z^{-1}]\!]\,. \label{r-intro} \ee
More so, assuming that $\ell,\ell'$ were BRST-anomaly-free line operators, the quantum-corrected coupling on the RHS of the OPE \eqref{OPE-sum4K} must also define a BRST-anomaly-free line operator. This means that for every $\ell,\ell'$, $\mu_\ell(z)+\mu_{\ell}(0)+r_{\ell,\ell'}(z)$ is an MC element in $\text{End}(V_\ell\otimes V_{\ell'})\otimes \CA$, an extremely nontrivial constraint.  (We give some examples in Section \ref{sec:OPE-MC}.) For Koszul-dual algebras, it translates to the facts that 1) $r(z)$ is an MC element in $\CA^!\otimes \CA^!$ and 2) the coproduct, defined on linear elements by
\be  \Delta_z(x^i_n) := \tau_z(x^i_n) \otimes 1 + 1\otimes x^i_n\,,\qquad \Delta_z(p_{i,n}) = \tau_z(p_{i,n})\otimes 1+1\otimes p_{i,n}\,, \ee
is a map of $A_\infty$ algebras $\Delta_z:\CA^!\to \CA^!\otimes_{r(z)} \CA^!$, where the RHS is deformed by $r(z)$.

Finally, associativity of the OPE (Sec. \ref{sec:KD-assoc}) is essentially the statement that 
\be  \ell_1(z_1)\otimes \ell_2(z_2) \otimes \cdots \otimes \ell_n(z_n) \sim \exp \int_\R \bigg[ \sum_{i=1}^n \bm \mu_{\ell_i}(z_i) +\sum_{1\leq i<j\leq n} \mb r_{\ell_i,\ell_j}(z_i-z_j) \bigg] \ee
in any correlation function, no matter in which order (and around which $z_i$'s) we choose to expand the LHS. In terms of $\CA^!$, associativity is fully captured by the property
\be
 r_{23}(z)+(\text{id}\otimes\Delta_z)(r(z+w)) = r_{12}(w)+(\Delta_w\otimes\text{id})(r(z))=  r_{12}(w)+ r_{13}(z+w) +r_{23}(z) \ee
 in $\CA^!{}^{\otimes 3}[\![z^{-1},w^{-1},(z-w)^{-1}]\!]$, which is reminiscent of quasi-triangularity. Associativity also implies that $r(z)$ satisfies an $A_\infty$ analogue of the Yang-Baxter equation (YBE).
 
We collect all these structures/properties of $\CA^!$ in a definition of dg-shifted Yangians. Physically, the analogy between our $\CA^!$ and (generalized) Yangians comes from comparing lines in 3d HT theories to lines in 4d HT theories, as developed in \cite{Costello-Yangian, CWY-I, CWY-II} (see also \cite{Kapustin-holomorphic, CautisWilliams, ANGrassmannian, ANyangian}). Both are controlled by algebras with a chiral coproduct~$\Delta_z$. In 4d, quantum corrections similar to \eqref{r-intro} give rise to a spectral R-matrix of cohomological degree zero, which quantizes a symmetric monoidal structure to a braided monoidal structure --- and thus satisfies the quantum YBE.  In 3d, quantum corrections give rise to our dg-shifted ``spectral r-matrix'' $r(z)$ of degree $1$, controlling singular terms in the OPE. It thus ``quantizes'' a symmetric monoidal structure to a meromorphic monoidal structure --- and satisfies our $A_\infty$ YBE.%
\footnote{One might also compare with 5d HT theories on $\C\times \C\times \R$, whose line operators are represented as modules for (generalized) affine Yangians. In this case, quantum corrections similar to \eqref{r-intro} enter the coproduct $\Delta_z$ itself, \cf\ \cite[Eqn. 5.4]{Gaiotto:2023ynn}, \cite[Eqn. 3.5]{Ashwinkumar:2024vys}.}

The analogy with standard Yangians is tightest for HT-twisted gauge theory with group $G$, matter $V$, and vanishing superpotential. Then $\CA^!$ is a quantization of a dg-shifted Lie bialgebra structure on the positive loop algebra of the cotangent Lie algebra $T^*[-1](\mathfrak g\ltimes \Pi V)$. The bialgebra structure arises precisely from a shifted version of Yang's r-matrix. This perspective is developed in \cite{NP-Yangian}.

Another analogy with Yangians and quantum groups is explored in Appendix \ref{app:r-lambda}, where we show that the shifted r-matrix $r(z)$ in $\CA^!$ inverts the shifted $\lambda$-bracket in $\CA$.

\medskip

\noindent \textbf{Caveat:} 
This section and the remainder of the paper invoke many tensor products of $A_\infty$ algebras. In general, products of $A_\infty$ algebras are only defined up to quasi-isomorphism, \cf\ \cite{Loday-diagonal}.
Nevertheless, in perturbative HT QFT, tensor products of $\CA$ and $\CA^!$ are canonical: both $\CA$ and $\CA^!$ are constructed as deformations of free commutative algebras by bulk interactions (Section \ref{sec:Ainf}); so $\CA^{\otimes n}\otimes \CA^!{}^{\otimes m}$ can be defined similarly, as a deformation of a free commutative algebra by a sum of $(n+m)$ independent bulk interaction terms.

\subsection{Koszul duality in 3d HT theories}
\label{sec:KD-perspective}

We outline here how a canonical Koszul dual algebra $\CA^!$ arises in perturbative 3d HT QFT's, and why its modules represent perturbative line operators. For further details, we refer the reader to \cite{Costello-Yangian, CPkoszul, PaquetteWilliams,Zeng:2023qqp}, mathematical background in \cite{loday2012algebraic}, as well as the review of Tannakian principles in physics in \cite{sparks}. Our analysis in particular generalizes Sections 6-7 of \cite{PaquetteWilliams}, which consider 3d HT free matter and pure gauge theories.

We'll begin here by defining $\CA^!$ as an algebra of operators acting at $z=\infty$. Then, in Section \ref{sec:KD-2d}, we'll use an equivalent perspective via 2d reduction to relate this to couplings $\mu_\ell$ that we've used to define perturbative lines so far. 

\subsubsection{Boundary condition on $\cp^1$}
\label{sec:cp1}

A starting point for Koszul duality is to find a boundary condition near spatial infinity --- an asymptotic vacuum --- that trivializes the state space, making it one-dimensional. Such a boundary condition always exists for perturbative 3d HT QFT's, and is in fact canonical. To make it more concrete, it's useful to think of ``compactifying'' the spatial $z,\bar z$ plane to a $\cp^1$, and imposing a boundary condition at $z=\infty$.

Suppose that a perturbative theory has fields $(\mb x^i,\mb p_i)$. Recall that $\mb x^i$ and $\mb p_i$ transform as sections of the $\bar\pd$-complex on $\cp^1$, valued in the $J(x^i)$ and $J(p_i)$ powers of the canonical bundle, respectively. (Here $J(x^i),J(p_i)$ are the spins of the fields.)
Letting $z'=1/z$ be a local coordinate near infinity, we'll impose
\be \CB_0:\begin{cases} \mb x^i(z)\to \text{finite} \\
\mb p_i(z)\to \text{finite}\end{cases}  \text{as $z\to 0$}\qquad
     \CB_\infty: \begin{cases} (z')^{2J(x^i)-1} \mb x^i(z') \to \text{finite}   \\  (z')^{2J(p_i)-1} \mb p_i(z') \to \text{finite} \end{cases}
      \text{as $z'\to 0$}  \label{bc-P1} \ee
The `$\CB_0$' conditions just impose regularity of the fields at $z=0$. The `$\CB_\infty$' conditions allow a singularity at $z=\infty$ when $J$ is positive, and require some order of vanishing when $J$ is negative. Fields with this behavior are equivalent to regular sections of the $\bar\pd$ complex valued in $\CO(-1)$. (The equivalence just comes from redefining fields $\wt {\mb x}^i(z') :=(z')^{2J(x^i)-1} \mb x^i(z')$ in a patch near infinity, and similarly for $\mb p_i$.)

Now, in a free field theory, the equations of motion along $\cp^1$ require $\mb x^i,\mb p_i$ to be elements of $\bar\pd$-cohomology. But the bundle $\CO(-1)$ has vanishing cohomology: once $\mb x^i,\mb p_i$ are locally holomorphic, the respective boundary conditions \eqref{bc-P1} force $\mb x^i,\mb p_i\equiv 0$ on $\cp^1$. The state space is a quantization of the unique solution $\mb x^i,\mb p_i\equiv 0$ (geometrically, a point), and is thus one-dimensional, as desired.

 In an interacting theory, the EOM are deformed. However, if the interactions are treated perturbatively, the interacting state space can be constructed by starting with the free-field state space and deforming it --- \emph{e.g.} by adding appropriate differentials corresponding to the interaction terms. When the free-field state space is already one-dimensional, there is no room for interactions to alter it, so we expect it to remain one-dimensional in general:
\be \text{States}(\cp^1,\CB_0,\CB_\infty)\;=\;\;\raisebox{-.55in}{\includegraphics[width=2in]{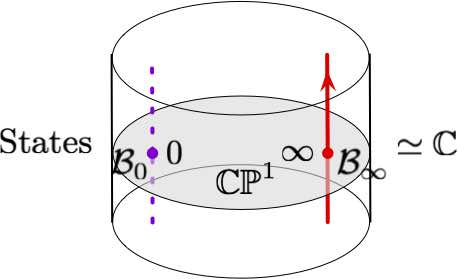}}  \ee
(It's important here that that we're working perturbatively, with infinitesimally small interactions; otherwise new global solutions on $\cp^1$ could be introduced and the argument would break down.)

The reason that trivializing the state space is important is that it allows us to ``measure'' the QM state spaces that create line operators. Suppose that $\ell$ is a line operator constructed by coupling bulk local operators to QM with state space $V$, using a coupling~$\mu$. If we insert $\ell$ at $z=0$ (or in fact any other finite point along $\cp^1$), the $\CB_\infty$ boundary condition kills all the bulk fields, \emph{decoupling} them from the QM, and recovering
\be \text{States}(\cp^1,\ell,\CB_\infty) \;=\;\;\raisebox{-.55in}{\includegraphics[width=2in]{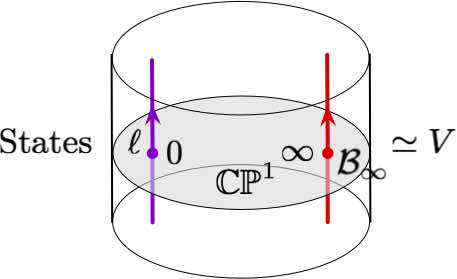}} \,.\ee
Moreover, if we define
\be \CA^! :=\begin{array}{l} \text{$A_\infty$ algebra of local ops supported} \\ \text{on the `defect' $\CB_\infty$ at $z=\infty$} \end{array} \qquad \raisebox{-.5in}{\includegraphics[width=1.3in]{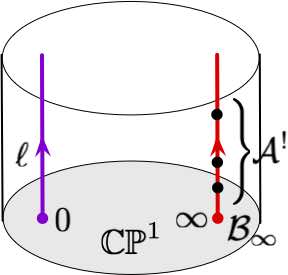}} \label{CP1-algebra} \ee
then
\begin{itemize}
\item $\CA^!$ naturally acts on every state space $\text{States}(\cp^1,\ell,\CB_\infty)$, for all $\ell$ in the category of line operators $\CC$
\item Given two lines $\ell,\ell'\in \CC$ and a local operator $\CO \in \text{Hom}_\CC(\ell,\ell')$ at their junction, we get a linear map of state spaces $\CO_* :V\to V$. The action of $\CA^!$ commutes with $\CO_*$, since they are supported at different points on $\cp^1$:
\be \raisebox{-.5in}{\includegraphics[width=4in]{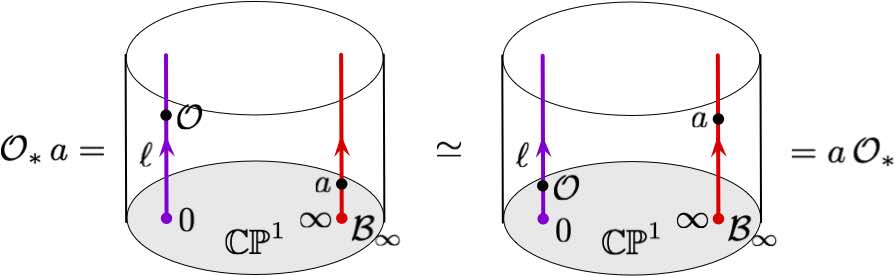}} \ee
\end{itemize}
Indeed, the action of $\CA^!$ will also commute with the differential on local operators at junctions of lines, as well as with higher $A_\infty$ operations of local operators at junctions, because these are all defined by physical correlation functions supported away from $z=\infty$, whereas $\CA^!$ is supported at $z=\infty$. All this commutativity amounts to the mathematical statement that there is an $A_\infty$ functor%
\be \CF: \begin{array}{ccc} \CC & \to & \CA^!\text{-mod} \\
 \ell & \mapsto & \text{States}(\cp^1,\ell,\CB_\infty)\,. \end{array} \label{A!-functor} \ee
 
\subsubsection{Tannakian perspective}
\label{sec:KD-Tannakian}

To put the current construction in a representation-theoretic context, it may be helpful to interpret the role/appearance of $\CA^!$ from a Tannakian perspective.
We start with a superficial, lightning review of Tannakian ideas; see \cite{sparks}, especially the introduction and many further references therein, for a broader discussion of Tannakian constructions in QFT.
In Tannakian formalism, one represents a category $C$ by producing a functor $F$ from $C$ to vector spaces --- a so-called fiber functor --- and defining $A:=\text{End}(F)$ to be the algebra of natural transformations of the functor. It is then automatic that each vector space gets an action by $A$, so there's a representation
\be F : C \to A\text{-mod}\,. \ee
Moreover, if $F$ preserves additional structures on $C$ (\emph{e.g.} a tensor product), $A$ gets endowed with corresponding structures (\emph{e.g.} coproducts).

Now, the procedure of measuring a state space on $\cp^1$, with $\CB_\infty$ at infinity, defines an analogue of a so-called fiber functor on our category of line operators $\CC$. The measurement actually makes sense for multiple line operators as well: Given lines $\ell_1,...\ell_n\in \CC$ placed at $z_1,...,z_n\in \C= \cp^1\backslash\{\infty\}$ we get a state space
\be   \text{States}\big(\cp^1, \ell_1(z_1)\otimes \ell_2(z_2)\otimes\cdots\otimes \ell_n(z_n),\CB_\infty\big) \in \text{Vect}\,, \ee
where `Vect' is the category of dg vector spaces.
The state spaces define functors of categories, for every collection of points on $\C$,
\be  \CF_{(z_1,...,z_n)}:\CC_{z_1}\boxtimes \CC_{z_2}\boxtimes \cdots \boxtimes \CC_{z_n} \to \text{Vect}\,. \ee

The functors for different collections of points on $\C$ cannot be completely independent. We know that translation isomorphisms can be used to shift the points around. Moreover, the OPE of lines contains (in principle) precisely the information that one needs to relate $\CC_{z_1}\otimes \CC_{z_2}$ to $\CC_{z_2}$ (etc.), and thus to relate $\CF_{z_1,z_2}$ to $\CF_{z_2}$ (etc.). Altogether, what one has is an entire system of functors $\CF_\bullet$ for all possible collections of points, and relations among them induced by translations and the OPE.

The endomorphism algebra of the entire system of functors should be precisely $\CA^!$,
\be \text{expect:} \qquad \boxed{\text{End}(\CF_\bullet) = \CA^!} \ee
The physical reason to expect this is that $\text{End}(\CF_\bullet)$ represents operators on $\cp^1$ that 1) are localized in time (so they act on $\cp^1$ state spaces); and 2) commute with all local operators at junctions of any collection of lines at any finite points $(z_1,...,z_n)$ on $\cp^1$. Property (2) is what makes some $a\in \text{End}(\CF_\bullet)$ a natural transformation. In order for (2) to be satisfied, the operators should be localized near infinity, and we recover the definition of $\CA^!$ from \eqref{CP1-algebra}.

We warn the reader that $\CF_\bullet$ does \emph{not} preserve tensor products of lines: $\CF_{z_1,z_2}\big(\ell_1(z_1)\otimes \ell_2(z_2)\big)\;/\hspace{-2.1ex}\simeq \CF_{z_1}(\ell_1(z_1))\otimes  \CF_{z_2}(\ell_1(z_2))$ (etc.). Thus it is not a monoidal fiber functor. Its failure to preserve products, and the structure induced in order to correct this, will be the subject of Section \ref{sec:KD-Yangian}.

\subsection{Koszul duality via 2d reduction}
\label{sec:KD-2d}

An important question about the functor \eqref{A!-functor} for a \emph{single} line operator remains:
\begin{itemize}
\item Is the functor $\CC\to\CA^!\text{-mod}$ an \emph{equivalence} of categories?
\end{itemize}
We'll argue that it is, for \emph{perturbative} 3d HT QFT's.
Moreover, we'll show how to explicitly ``invert'' the functor, going from the structure of an $\CA^!$-module on $V_\ell$ to the MC element $\mu_\ell$ used to couple $V_\ell$ to bulk local operators, in order to define a perturbative line $\ell$ as in Section \ref{sec:lines-QM}. In other words, we'll give a concrete physical realization of the correspondence \eqref{corresp-MC}.

To proceed, it's very helpful to work with a 2d rewriting of a 3d HT theory --- a reduction to 2d that keeps all modes, and is thus equivalent to the original theory. In the 2d reduction, the singularity $\CB_\infty$ on $\cp^1$ will become an ordinary boundary condition; we will be able to place it at finite distance, and in particular to ``invert'' it (by gluing to a second, dual boundary condition) in order to invert the functor $\CC\to \CA^!\text{-mod}$.

\subsubsection{Reduction to 2d}
\label{sec:2d-reduction}

Consider a Lagrangian theory with fields $(\mb x^i,\mb p_i)$ and interaction $\CI(\mb x,\mb p,\pd)$ on $\C_z^*\times \R_t$. We'll remove the origin for the moment, because we may want to place nontrivial line operators there. We let
\be z = re^{i\theta}\,,\qquad \bar z = r e^{-i\theta} \ee
and consider the circle fibration (\cf\ \cite[Sec. 6.5]{PaquetteWilliams}, \cite[Sec. 2]{Zeng:2023qqp})
\be \begin{array}{ccc} S^1& \to & \C_z^* \times \R_t \\ &&\downarrow \\ &&  \;\;\R_r\times \R_t\,. \end{array} \label{S1fib} \ee

Schematically, the 3d action is
\be S = \int_{\C^*_z\times \R_t} \mb L \,=  \int_{\C^*_z\times \R_t} \mb p_i d' \mb x^i + \CI(\mb x,\mb p,\pd)\,, \label{S3d-reduce} \ee
with multiforms $\mb x^i\in \Omega'{}^\bullet (\C\times \R)[R_i]dz^{J_i}$, $\mb p_i\in \Omega'{}^\bullet (\C\times \R)[1-R_i]dz^{1-J_i}$, where $\Omega'{}^\bullet(\C\times \R) = \Omega^\bullet(\C\times \R)/(dz)$ is the de Rham complex of $\C\times \R$, modulo $dz$.
If we change coordinates from $(z,\bar z,t)$ to $(z,r,t)$ then each multiform field transforms as, \eg,
\be \begin{array}{c} \mb x = x + x_t dt+x_{\bar z} d\bar z+x_{\bar z t} d\bar z dt = x + x_t dt+ x_r dr+x_{rt} dr dt\quad(\text{mod}\, dz) \\[.2cm]
\displaystyle \Rightarrow\qquad  x_{\bar z} = \frac{z}{2r} x_r\,,\qquad x_{\bar zt} =  \frac{z}{2r} x_{rt}\,. \end{array}
 \ee
The exterior derivative simply becomes $d' = \pd_r dr+\pd_t dt$. The Lagrangian itself (the part that contributes to the action) has $R=2$ and $J=1$, and thus must transform as $\mb L = L_{\bar z t} d\bar z dt  dz = L_{rt} dr dt dz$ with $ L_{\bar z t} = \frac{z}{2r} L_{rt}$. The action then becomes
\be S = \int_{\C_z^*\times \R_t} L_{\bar z t} d\bar z dt dz = \int_{\C_z^*\times \R_t} dr dt dz\,  L_{rt}(z,r,t) = \int_{\{r> 0\}\times \R_t} dr dt \oint_{S^1} dz\,  L_{rt}(z,r,t)\,.  \ee
Let's now expand the multiform fields into modes along $S^1$:
\be \label{xp-expand} \begin{array}{l@{\,}l@{\,}ll}
 \mb x^i(z) &\displaystyle = \sum_{n\in \Z} \mb x^i_n(r,t)\, z^{-n-1} & \displaystyle =  \sum_{n\in \Z} \mb x^i_n(r,t) r^{-n-1}\,e^{-i(n+1)\theta} &\quad  \in \Omega^\bullet(\R^2)(\!(z)\!)\,[R_i]dz^{J_i}\,, \\[.2cm]
  \mb p_i(z) &\displaystyle = \sum_{n\in \Z} \mb p_{i,n}(r,t)\, z^{-n-1} & \displaystyle =  \sum_{n\in \Z} \mb p_{i,n}(r,t) r^{-n-1}\,e^{-i(n+1)\theta} &\quad  \in \Omega^\bullet(\R^2)(\!(z)\!)\,[1-R_i]dz^{1-J_i}\,. \end{array}
\ee
We use integer-valued moding in order to be compatible with inserting the trivial line $\id$ at $z=0$, which requires all fields to be regular.%
\footnote{Other choices of moding may be possible, and would detect other line operators that have no morphisms to those in the category $\CC$ described here. We don't consider them in this paper.} %
Assuming that the interaction $\CI(\mb x,\mb p,\pd)$ is a polynomial in the fields and $\pd$, let
\be \CI = \sum_{n\in \Z} \CI_n(r,t) z^{-n-1} \ee
be its corresponding mode expansion. Each $\CI_n$ is a (typically infinite) sum of monomials in the $\mb x^i_k$ and $\mb p_{i,k}$. Then we find (ignoring factors of $2\pi i$)
\be \label{S2d-gen} S 
=  \int_{\{r > 0\}\times \R_t} \bigg[ \sum_{n+m=-1} \mb p_{i,n} d \mb x^i_m + \CI_0(\mb x^i_k,\mb p_{i,k})\Bigg] \;=: S_{2d} \ee
Here $d = \pd_r dr+\pd_t dt$ is the 2d exterior derivative.

Altogether, the 3d theory on $\C^*_z\times \R_t$ becomes equivalent to a 2d theory with infinitely many pairs of fields $(\mb x^i_n,\mb p_{i,n})$, and interaction $\CI_0 = \oint_{S^1} \CI$. The kinetic term $\sum_{n} \mb p_{i,-n-1} d \mb x^i_n$ identifies the 2d theory as a generalized B-model, whose ``target space'' is the infinite-dimensional space with bosonic coordinates $x^i_n$.

\subsubsection{Boundaries and algebras for the 2d theory}
\label{sec:KD-bdy}

Now let's come back to line operators. The category of line operators $\CC$ in the 3d theory, placed at $z=0$ and viewed as an $A_\infty$ category (forgetting translations and the OPE), is equivalent to the category of \emph{boundary conditions} of the effective 2d theory at $r=0$. This equivalence can be viewed as a state-operator correspondence for line operators:
\be \raisebox{-.6in}{\includegraphics[width=4.3in]{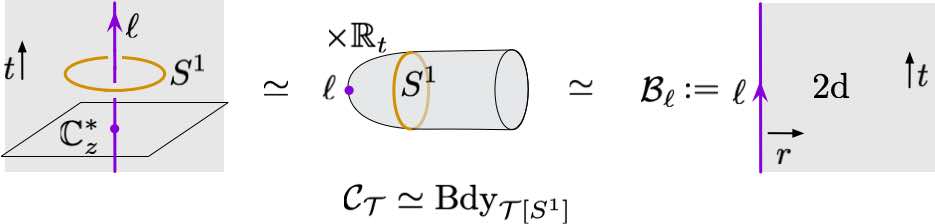}} \label{stateop} \ee
In particular, the identity line $\id\in \CC$ corresponds to the boundary condition $\CB_0$ (the same one as in \eqref{bc-P1}) that makes all fields regular at $z=0$, namely
\be \id\; \leftrightarrow\; \CB_0 : \quad \mb x^i_n|_{r=0}=\mb p_{i,n}|_{r=0} = 0 \quad\text{for $n\geq 0$}\,; \label{def-B0} \ee
and the bulk algebra of local operators restricted to $z=0$ becomes identified with local operators on the boundary $\CB_0$ of the 2d reduction:
\be \begin{array}{c@{}c@{}c@{}c@{}c}  \CA & \simeq &  \text{End}_\CC(\id) & \simeq &  \text{End}_{\text{Bdy}}(\CB_0) \\
 \text{(bulk local ops, at $z=0$)} && \text{(local ops on the line $\id$)} && \text{(local ops on the boundary $\CB_0$)} \end{array} \ee

In a free theory, the algebra $\CA$ is a free, graded-commutative polynomial algebra
\be \CA \simeq \C[x^i_{-n-1},p_{i,-n-1}]_{n\geq 0}\,, \label{A-xp} \ee
generated by the zero-form components of the modes that aren't set to zero by the boundary condition \eqref{def-B0}. (Only the zero-form parts survive in the cohomology of $Q=d$.) Of course we knew this: from a 3d perspective, $x^i_{-n-1}=\frac{1}{n!}\pd^n x(0)$ and $p^i_{-n-1}=\frac{1}{n!}\pd^n p(0)$ the operators we've been using all along to couple to QM, as in Section \ref{sec:line-chiral}. If we introduce interactions $\CI(\mb x,\mb p,\pd)$, then, \emph{perturbatively}, the algebra $\CA$ is still generated by the modes in \eqref{A-xp}, but has corrections to its BRST operator, commutation relations, and higher $A_\infty$ operations. We'll compute the corrections in Section \ref{sec:Ainf}.

Now, the boundary condition at $z=\infty$ on $\cp^1$ described in \eqref{bc-P1} \emph{also} corresponds to a boundary condition of the 2d theory. After circle reduction, the radial $r$ direction is topological, so we can bring $r=\infty$ in to a finite point, say `$r_1$'. (This is a key simplifying feature of the reduction.) Written in terms of modes \eqref{xp-expand}, the spin dependence drops out, and the boundary condition ``at infinity'' just becomes
\be \CB_\infty:\quad \mb x^i_n|_{r=r_1}=\mb p_{i,n}|_{r=r_1} = 0 \quad\text{for $n< 0$}\,, \label{def-Binf} \ee
and
\be \begin{array}{ccc}  \CA^! & \simeq &   \text{End}_{\text{Bdy}}(\CB_\infty) \\
 \text{(local ops near $z=\infty$)}  && \text{(local ops on the boundary $\CB_\infty$)} \end{array} \ee
We immediately deduce that in a free theory $\CA^!$ will be a free, graded-commutative polynomial algebra
\be \CA^! = \C[x^i_n,p_{i,n}]_{n\geq 0}\,. \label{A!-xp} \ee
Once again, if we introduce \emph{perturbative} interactions $\CI$, then $\CA^!$ is still generated by $x^i_n,p_{i,n}$ as in \eqref{A!-xp}, but with deformed BRST differential, commutators, and $A_\infty$ operations.
\footnote{
There is a small caveat here. Boundary conditions for the reduced 2d theory $\CT[S^1]$ on a half-space $H$ have a potential BRST anomaly when the interaction term $\CI_0$ fails to vanish at the boundary, since $Q\int_H \CI_0 = \int_H d\CI_0 = \int_{\pd H} \CI_0$. On the $\CB_0$ boundary condition, we always have $\CI_0|_{\CB_0}=0$, consistent with the fact that the identity line $\id$ in 3d is anomaly-free. On the $\CB_\infty$ boundary condition, $\CI_0|_{\CB_\infty} = 0$ \emph{unless} the original 3d interaction $\CI(\mb x,\mb p,\pd)$ has linear terms in $\mb x$, $\mb p$ (independent of $\pd$). In this latter case, the anomaly can be incorporated into representation theory: a version of Koszul duality still holds, but $\CA^!$ must be treated as a curved $A_\infty$ algebra, with $m_0 = \CI_0|_{\CB_\infty}$. We won't consider anomalous examples.\label{foot:anomaly}
}

Since the $r$ direction is topological in 2d, it is very easy to see that the state space on a strip bounded by $\CB_0$ and $\CB_\infty$ --- which is the same as the state space on $\cp^1$ in 3d --- must be one-dimensional:
\be \raisebox{-.3in}{\includegraphics[width=5.2in]{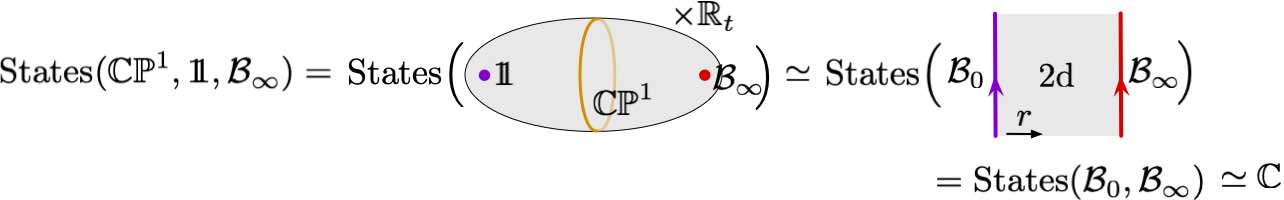}} \label{CP1-2d}  \ee
Namely, using topological invariance, we can squeeze the strip to infinitesimal width while keeping the state space unchanged (up to isomorphism). In the infinitesimal strip, the boundary conditions force $\mb x^i_n = \mb p_{i,n}=0$ for all $n$, independent of bulk interactions. There is a single massive vacuum, and its quantization gives state space $\C$.

\subsubsection{Completeness and MC elements}
\label{sec:KD-MC}

Next, consider placing any other line operator $\ell\in \CC$ at $z=0$. We make no assumption about whether $\ell$ is constructed by coupling to QM or not. Upon circle compactification, $\ell$ becomes equivalent to a boundary condition $\CB_\ell$ for the 2d theory at $r=0$; and measuring the $(\CB_\ell,\CB_\infty)$ strip state space defines our functor \eqref{A!-functor}
\be \CF: \ell\mapsto \raisebox{-.27cm}{$\begin{array}{l} \text{States}(\CB_\ell,\CB_\infty) \\   \simeq \text{States}(\cp^1,\ell,\CB_\infty) \end{array}$} \in \CA^!\text{-mod} \ee

We'd like to argue that the representation $\CF:\CC\to\CA^!\text{-mod}$ is faithful. To do so, we'll place the $(\CB_\ell,\CB_\infty)$ strip on the interval $[0,r_1]$ next to a second copy of the 2d theory on a half-space $[r_1,\infty)$, with boundary condition $\CB_0$ at $r=r_1$, as on the top-left:
\be \raisebox{-.8in}{\includegraphics[width=4.6in]{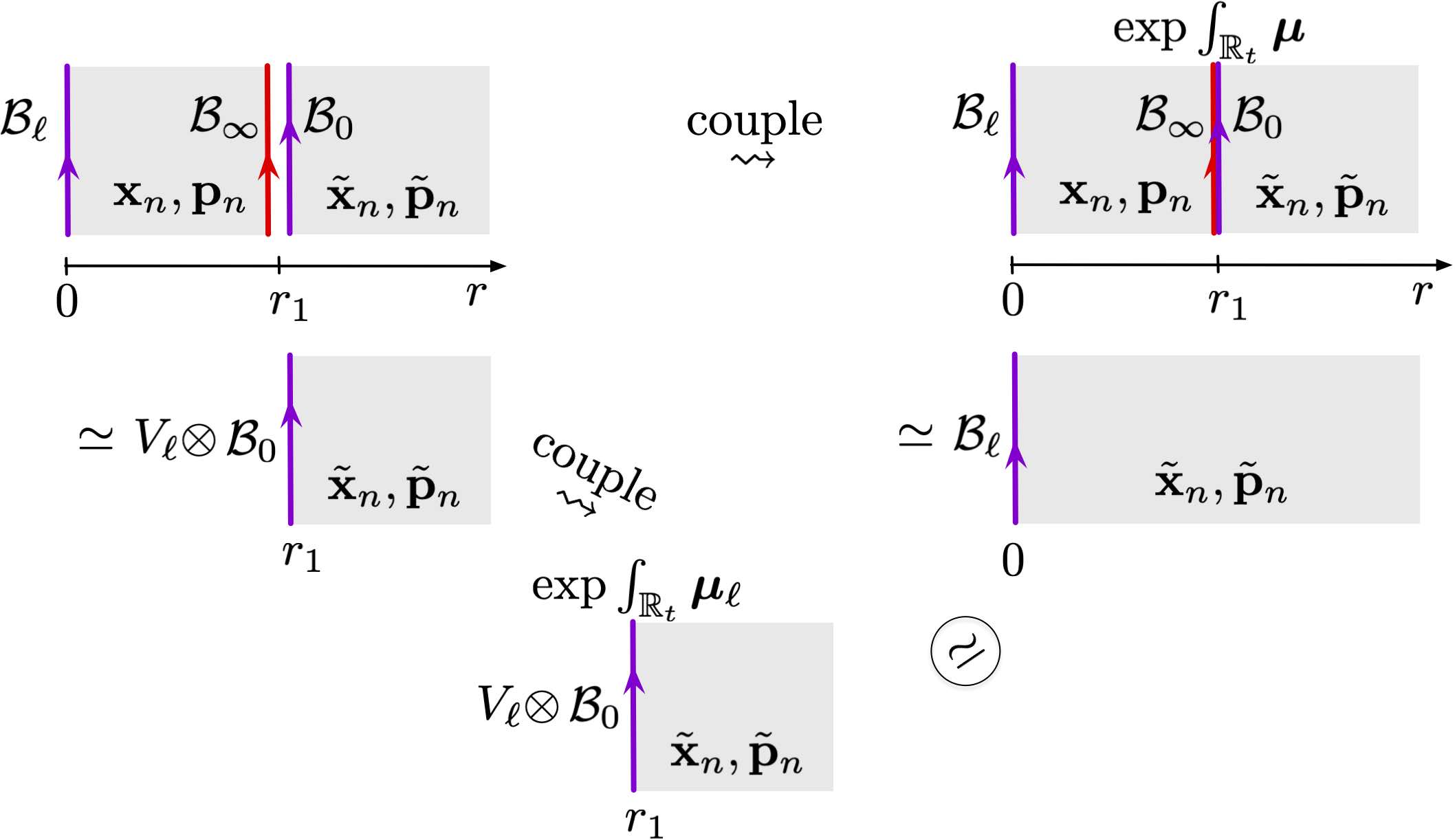}} \label{B-completeness} \ee
We label the fields in the half-space $\tilde{\mb x}^i,\tilde{\mb p}_i$, to distinguish them from the strip.
Then we add a ``universal'' coupling $\exp \int_{\R_t} \bmu$ at $r=r_1$ (shown on the top-right), where
\be \bmu = \sum_i \sum_{n\geq 0}\Big[ (-1)^{F(x^i)}\mb x^i_n \tilde{\mb p}_{i,-n-1} + (-1)^{F(p_i)} \mb p_{i,n} \tilde {\mb x}^i_{-n-1} \Big]\,. \label{def-M} \ee

The full bulk+interface theory near $r=r_1$ has a combined (classical) action
\begin{align}& S_{\text{int}}  =  \int_{\R_r\times \R_t} \sum_{i,n\geq 0} \Big[ \big(\mb x_{-n-1}d\mb p_n+\mb p_{-n-1}d\mb x_n +\CI_0\big)\chi_{\leq r_1}(r) \\
& - \big(\mb x_n\delta^{(1)}(r-r_1)\tilde{\mb p}_{-n-1} + \mb p_n\delta^{(1)}(r-r_1)\tilde{\mb x}_{-n-1}\big) 
+ \big(\tilde{\mb x}_nd\tilde{\mb p}_{-n-1}+\tilde{\mb p}_nd\tilde{\mb x}_{-n-1} +\tilde \CI_0 \big)\chi_{\geq r_1}(r) \Big]\,,  \notag \end{align}
where $\chi_{\leq r_1}(r)=1$ if $r\leq r_1$ and $0$ otherwise, and similar for $\chi_{\geq r_1}$.
The EOM near $r=r_1$ have interface terms that set $\mb x_n=\tilde{\mb x}_n$ and $\mb p_n=\tilde{\mb p}_n$ at $r=r_1$ for all $n\in \Z$, effectively gluing the theory back together. For example, for $n\geq 0$, $\{S_{\rm int},\tilde{\mb x}_n(r)\}_{\rm BV} =  \mb x_n \delta^{(1)}(r-r_1) + d\tilde{\mb x}_n-\tilde{\mb x}_n\delta^{(1)}(r-r_1)$ sets $\mb x_n=\tilde{\mb x}_n$ at $r=r_1$.  Therefore,  the universal coupling $\exp\int \bmu$ deforms the decoupled interface on the LHS of \eqref{B-completeness} to a transparent, or identity interface.%
\footnote{This property was referred to as ``completeness'' of the $(\CB_\infty,\CB_0)$ pair, in the Tannakian context of \cite{sparks}. It's important to note that completeness only holds \emph{perturbatively} here. For example, in a gauge theory, additional sums over bundles and insertions of boundary monopole operators are required to glue a theory back together from two halves --- the universal coupling given by \eqref{def-M} is not sufficient.} %

We can interpret the deformed theory in two ways:
\begin{itemize}
\item[1)] If we first use the deformation $\exp\int \bmu$ to get a transparent interface, we recover the 2d theory on a full half-space $[0,\infty)$ with the original boundary condition $\CB_\ell$ at $r=0$ (bottom-right of \eqref{B-completeness}). Thus we've inverted the functor $\CF:\CC\to \CA^!\text{-mod}$, showing that it must be faithful.
\item[2)] If we first squeeze the $[0,r_1]$ strip to infinitesimal width, we get the vector space $V_\ell:= \text{States}(\CB_\ell,\CB_\infty)$ there, with its $\CA^!$ action (bottom-left of \eqref{B-completeness}). Suppose that the generators of $\CA^!$ act by some endomorphisms
\be \begin{array}{lcl} x^i_n & \mapsto & (-1)^{F(x^i)} \beta^i_n \ \\  p_{i,n} & \mapsto & (-1)^{F(p_i)}\alpha_{i,n} \end{array},\qquad \beta^i_n,\alpha_{i,n} \in \text{End}(V_\ell) \qquad (n\geq 0) \label{rep-univ} \ee
Then turning on the deformation $\exp\int \bmu$  produces a theory on the half-space $[r_1,\infty)$, where at $r=r_1$ the boundary condition $\CB_0$ has been tensored with $V_\ell$, and coupled with the MC element
\be \mu \big|_{\text{evaluated on \eqref{rep-univ}}} = \sum_{i,n\geq 0} \Big[ \alpha_{i,n}\tilde x^i_{-n-1} + \beta^i_n\tilde p_{i,-n-1} \Big] =: \mu_\ell  \label{mu-univ} \ee
\end{itemize}
Comparing (1) and (2), we find that --- if we work perturbatively --- \emph{every} $\ell\in \CC$ can be represented by coupling to QM with linear couplings! The QM vector space required is $V_\ell = \text{States}(\CB_\ell,\CB_\infty)$, and the couplings are computed by  \eqref{rep-univ}--\eqref{mu-univ}.

We have argued that $\CF:\CC\to \CA^!\text{-mod}$ is faithful. To see that it's in fact an equivalence, we define $\CF^\vee:\CA^!\text{-mod}\to\CC$ for any $\CA^!$-module $\rho_M:\CA^!\to \text{End}(M)$ by $\CF^\vee(M,\rho_M) := (M,\mu_M)$, with
\be \mu_M=(\rho_M\otimes 1)(\mu) = \sum_{i,n} \Big[ (-1)^{F(x^i)} \rho_M(x^i_n) p_{i,-n-1}+ (-1)^{F(p_i)} \rho_M(p_{i,n}) x^i_{-n-1}\Big]\,. \label{muM} \ee
The construction in \eqref{B-completeness} shows $\CF^\vee\circ\CF \simeq \text{id}_{\CC}$. Dually, we have that $\CF\circ \CF^\vee(M,\rho_M) = \CF(M,\mu_M)$ is represented by states on a $(\CB_0,\CB_\infty)$ strip, tensored with $M$ and deformed by $\mu_M$. On one hand, transversality \eqref{CP1-2d} says that $\CF(M,\mu_M)\simeq M\otimes \C=M$ with trivial deformation (since the right boundary sets $x^i_{-n-1}=p_{i,-n-1}=0$ and thus $\mu_M=0$). On the other hand, at the left boundary, the EOM in the bulk-boundary action $S=\int_{\{0\}\times \R_t} \bm \mu_M + \int_{\R_{r\geq 0}\times \R_t}\sum_{n\geq 0}\big(\mb x^i_{-n-1}d\mb p_{i,n}+\mb p_{i,-n-1}d\mb x^i_n +\CI_0\big) $ set $x^i_n\big|_0=\rho_M(x^i_M)$ and $p_{i,n}\big|_0=\rho_M(p_{i,n})$ for $n\geq 0$ (also these fields are constant through the bulk and equal to their values on the right boundary), which determines the action of $\CA^!$ on the strip state space to be given by $\rho_M$. Thus $\CF\circ\CF^\vee\simeq \text{id}_{\CA^!\text{-mod}}$.

There's one final picture that's useful to consider. Consider two strips of the $\CT[S^1]$ theory, one bounded by $\CB_0$ and its dual (180$^\circ$ rotation) on both sides, and one bounded by $\CB_\infty$ and its dual on both sides. The \emph{state spaces} on the respective strips can be identified with the underlying vector spaces of $\CA$ and $\CA^!$, respectively, by state-operator correspondence. More so, local operators on the boundaries give $\CA$ the structure of an $\CA\otimes \CA^{\rm op}$ module, and give $\CA^!$ the structure of an $\CA^!{}^{\rm op}\otimes \CA^!$ module, as on the LHS:
\be \raisebox{-.3in}{\includegraphics[width=5.7in]{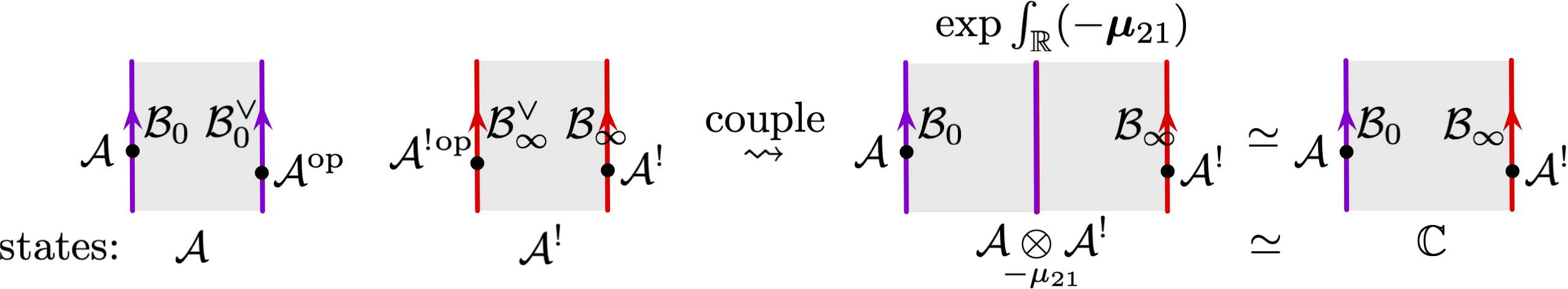}} \label{B-MC} \ee
Let's glue the two strips to each other, using a deformation $\exp\int_\R (-\bm \mu_{21})$ with the orientation-reversal of the universal MC element \eqref{def-M}. Then transversality \eqref{CP1-2d} implies that we must recover the trivial vector space, as on the RHS.

Algebraically, we learn that the vector space $\CA\otimes \CA^!$, treated as a left $\CA\otimes \CA^!$ module, is deformed by the MC element $-\mu_{21} \in \CA^{\rm op}\otimes \CA^!{}^{\rm op}$ (acting on the right) to a trivial module:
\be  (\CA\otimes \CA^!,(-\mu_{21})\cdot)\simeq \C \qquad \text{as an $\CA\otimes \CA^!$ module}\,. \ee
This is in fact easy to see from the explicit form of $\mu$ in \eqref{def-M}

\subsubsection{Summary}
\label{sec:KD-summary}

The setup here has provided all the ingredients required for a mathematical manifestation of Koszul duality. Let's summarize. We've found two $A_\infty$ algebras $\CA,\CA^!$, both linearly generated by respective subspaces $\CA_1 = \C\langle x^i_{-n-1},p_{i,-n-1}\rangle_{n\geq 0}$, $\CA_1^!\langle p_{i,n},x^i_n\rangle_{n\geq 0}$, satisfying
\be \CA_1^! = (\Pi \CA_1[1])^*\,, \ee
where $\Pi$ denotes parity shift, $[1]$ denotes cohomological shift, and $()^*$ is linear dual. Namely, $p_{i,n}$ is the dual of $x^i_{-n-1}$ and $x^i_n$ is the dual of $p_{i,-n-1}$.

The universal coupling \eqref{def-M} provides our universal MC element $\mu\in \CA^!\otimes \CA$ as in \eqref{univ-MC-intro}.
In particular, by using both gluing/completeness and transversality, we find that $(\CA\otimes \CA^!,(-\mu_{21})\cdot)\simeq \C$ as an $\CA\otimes \CA^!$ module. This ensures mathematically that the algebras $\CA,\CA^!$ are indeed Koszul duals, and that there's an equivalence of categories $\CC\simeq \CA^!\text{-mod}$. The the objects of $\CC$ can be thought of as pairs $(V_\ell,\mu_\ell)$ (deformations of $V_\ell\otimes \CA$), and they correspond to modules $\rho_\ell:\CA^!\to \text{End}(V_\ell)$ via $\mu_\ell = (\rho_\ell\otimes 1)(\mu)$ as in \eqref{corresp-MC}. The gluing in \eqref{rep-univ}--\eqref{mu-univ} is the physical realization of this correspondence.

In order to be precise about the equivalence $\CC\simeq \CA^!\text{-mod}$, one should further impose (and carefully match) finiteness conditions on both sides. There are multiple consistent choices. Physically, we are often interested in line operators that can be constructed with a finite number of linear couplings, which corresponds to \emph{smooth} modules for $\CA^!$ --- meaning all $x^i_n,p_{i,n}$ act as zero for sufficiently large $n$. We also typically want to couple to quantum mechanics that's not necessarily finite dimensional, but can be built from a finite number of free fields, which corresponds to \emph{finite-rank}, a.k.a. \emph{coherent} modules for $\CA^!$. All the examples in this paper are both smooth and coherent.

\medskip

\noindent\textbf{Convention:} It is convenient to remove the $(-1)^F$'s from \eqref{def-M}, \eqref{rep-univ}, etc. by twisting (redefining) the algebra $\CA^!$ by the automorphism $(-1)^F$, which flips the sign of any odd element. We will implement this in the remainder of the paper. Note that twisting also changes the overall sign of the MC element $r(z)$ introduced in Section \eqref{sec:KD-Yangian} below.

\subsection{Interlude: some examples}
\label{sec:KD-eg}

We include a few concrete examples to illustrate how some of the basic line operators considered earlier in the paper translate to modules for a Koszul-dual algebra. Some further examples for gauge theory are included later in Section \ref{sec:gauge-loop}. Readers mainly interested in the theoretical developments may skip this.

With a view to our next goal of combining Koszul duality and OPE's, we also include in Section \ref{sec:OPE-MC} an example of the nontrivial constraints imposed on quantum corrections, essentially saying that $\mu_\ell(z)+\mu_{\ell'}(w)+r_{\ell,\ell'}(z-w)$ is itself an MC element for any pair of lines. We illustrate this for two interesting vortex lines in the XYZ model.

\subsubsection{Line operators as modules: free chiral}

For a free chiral, bulk local operators $\CA=\C[X_{-n-1},\psi_{-n-1}]_{n\geq 0}$ are a graded-commutative algebra freely generated by the regular modes $X_{-n-1}=\tfrac1{n!}\pd^n X,\;\psi_{-n-1}=\tfrac1{n!}\pd^n\psi$ at (say) $z=0$. The Koszul-dual $\CA^!=\C[X_n,\psi_n]_{n\geq 0}$ is a graded-commutative algebra freely generated by the `polar' modes at $z=0$. After a $(-1)^F$ twist, the universal MC element is $\mu = \sum_{n\geq 0} \psi_n X_{-n-1} + X_n\psi_{-n-1}$, and trivially satisfies the MC equation because $Q=0$ and the algebras are commutative.

Consider the simple flavor vortex lines $\V_N$ from Section \eqref{sec:line-chiral}. They correspond to the following representations of $\CA^!$. The identity line is (as always) the trivial module
\be \V_0 =\id:\qquad V = \C\,,\qquad \rho(X_n)=\rho(\psi_n) = 0\quad \forall n\geq 0\,, \ee
while
\be \V_{N> 0}:\quad V = \C[a_0,...,a_{N-1}]\,,  \quad  \begin{array}{l} \rho(\psi_k) = a_k \;\;\text{(if $k<N$)},\\ \rho(X_n)=\rho(\psi_n)=0\;\;\text{(otherwise)} \end{array} \ee
\be \V_{N< 0}:\quad V = \C[b_0,...,b_{|N|-1}]\,,  \quad  \begin{array}{l} \rho(X_k) = b_k \;\;\text{(if $k<|N|$)},\\ \rho(X_n)=\rho(\psi_n)=0\;\;\text{(otherwise)} \end{array} \ee
are represented on finitely-generated fermionic and bosonic Fock spaces, respectively.

\subsubsection{Line operators as modules: XYZ model}
 
The XYZ model is more interesting. The bulk algebra $\CA$ from \eqref{XYZ-alg}  is a nontrivial dg algebra. It is generated by regular modes $(X_n,Y_n,Z_n,\psi_{X,n},\psi_{Y,n},\psi_{Z,n})_{n<0}$, with
 \be \begin{array}{c} \ds Q \, \psi_{X,n} 
   = \hspace{-.15in}  \sum_{m+p=n-1} \hspace{-.15in} Y_m Z_p\,, \qquad
   Q \, \psi_{Y,n}= \hspace{-.15in}  \sum_{m+p=n-1} \hspace{-.15in} X_m Z_p\,,\qquad Q \, \psi_{Z,n}=  \hspace{-.15in}  \sum_{m+p=n-1} \hspace{-.15in} X_m Y_p\,, \\[.1cm]
  [\psi_{X,n},\psi_{Y,m}] = Z_{n+m}\,,\qquad  [\psi_{Y,n},\psi_{Z,m}] = X_{n+m}\,,\qquad  [\psi_{X,n},\psi_{Z,m}] = Y_{n+m}\,.
  \end{array}  \label{A-XYZ} \ee
We'll see in Section \ref{sec:Ainf-matter} that the Koszul-dual algebra in a theory with superpotential looks nearly identical, with a few signs flipped. Here, after a $(-1)^F$ automorphism twist, $\CA^! = \C[X_n,Y_n,Z_n,\psi_{X,n},\psi_{Y,n},\psi_{Z,n}]_{n\geq 0}$ is generated by `polar' modes, with
 \be \hspace{-.2in} \begin{array}{c}  \ds Q \, \psi_{X,n} 
   = - \hspace{-.15in}  \sum_{m+p=n-1} \hspace{-.15in} Y_m Z_p\,, \qquad
   Q \, \psi_{Y,n}= - \hspace{-.15in}  \sum_{m+p=n-1} \hspace{-.15in} X_m Z_p\,,\qquad Q \, \psi_{Z,n}= -  \hspace{-.15in}  \sum_{m+p=n-1} \hspace{-.15in} X_m Y_p\,, \\[.1cm]
  [\psi_{X,n},\psi_{Y,m}] = -Z_{n+m}\,,\qquad  [\psi_{Y,n},\psi_{Z,m}] = -X_{n+m}\,,\qquad  [\psi_{X,n},\psi_{Z,m}] = -Y_{n+m}\,.
  \end{array}  \label{A!-XYZ} \ee
All higher $A_\infty$ operations vanish.

As always, the identity line $\id$ corresponds to the trivial module $\C$, where $\rho$ maps all generators to zero.

More interestingly, consider the two elementary vortex lines $\ell \simeq \V_{(-1,1,0)}$ and $\ell' \simeq \V_{(2,-1,-1)}$, which were  constructed by coupling to QM in \eqref{V-110} and \eqref{V-1-12}, respectively. We've got:
\be \begin{array}{ll} \ell :  &V=\C[a,b]\,,\quad \mu_\ell = a \psi_X + b Y - \bar b a Z \\[.1cm]
 \ell' : &V' = \C[c,d,e_0,e_1]\,,\quad\underbrace{c\psi_Y + d\psi_Z + e_0 X + e_1 \pd X - \bar e_0(cZ+dY)}_{\mu_{\ell'}} \underbrace{- \, \bar e_1cd}_{\text{deform $V'$}}\,, \end{array} \label{MC-OPE-ells} \ee
with bosonic QM operators $a,\bar a = \pd_a, c,\bar c = \pd_c,d,\bar d=\pd_d$ and fermionic QM operators $b,\bar b=\pd_b,e_0,\bar e_0=\pd_{e_0},e_1,\bar e_1=\pd_{e_1}$.

Our convention is that the ``couplings'' $\mu_\ell$, $\mu_{\ell'}$ are the parts of the MC elements in \eqref{MC-OPE-ells} that are strictly linear in the bulk fields. For $\ell$, that's the entire MC element. For $\ell'$, it excludes the final term $-\bar e_1cd$, which is independent of bulk fields. We instead think of $-\bar e_1cd$ as deforming the QM space $V'$ (and thus the algebra $\text{End}(V')$), by adding a differential $Q_{V'} = -cd\pd_{e_1}$.

Writing the coupling for $\ell$ in modes (generators of $\CA$) as
\be \mu_\ell =  a\psi_{X,-1}+bY_{-1} - \bar b a Z_{-1}\,, \ee
we can translate the line operator to a representation $\rho_\ell:\CA^!\to \C[a,b,\bar a,\bar b] \subseteq \text{End}(V)$ that sends
\be X_0 \mapsto a\,,\qquad \psi_{Y,0}\mapsto b\,,\qquad \psi_{Z,0}\mapsto -\bar b a\,, \ee
with all other generators vanishing. Let's check that this is a map of dg algebras. From \eqref{A!-XYZ} we have $Q(X_0) = Q(\psi_{Y,0}) = Q(\psi_{Z,0})=0$, consistent with the fact that $Q$ vanishes on $\text{End}(V)$. However, there's a nontrivial relation $[\psi_{Y,0},\psi_{Z,0}]=-X_0$, which maps correctly under $\rho_\ell$ to $[b, -\bar ba] = a$.

Similarly writing the coupling of $\ell'$ in modes as
\be \mu_{\ell'} = c\psi_{Y,-1} + d\psi_{Z,-1} + e_0 X_{-1} + e_1 X_{-2} - \bar e_0(cZ_{-1}+dY_{-1})\,, \ee
we find that this encodes a representation $\rho_{\ell'}:\CA^!\to \text{End}(V')$ with
\be Y_0\mapsto c\,,\quad Z_0\mapsto d\,,\quad \psi_{X,0}\mapsto e_0\,,\quad \psi_{X,1}\mapsto e_1\,,\quad \psi_{Z,0}\mapsto -\bar e_0 c\,,\quad \psi_{Y,0}\mapsto -\bar e_0 d \ee
and other generators vanishing. To see this is a map of dg algebras, we compare
\be \begin{array}{c} Q(\rho_{\ell'}(\psi_{X,1})) = Q(e_1) = [-\bar e_1cd,e_1] = -cd\quad\text{vs.}\quad \rho_{\ell'}(Q(\psi_{X,1})) = \rho_{\ell'}(-Y_0Z_0)=-cd \\[.1cm]
[\rho_{\ell'}(\psi_{X,0}),\rho_{\ell'}(\psi_{Z,0})] = [e_0,-\bar e_0 c]=-c  \qquad\text{vs.}\qquad  \rho_{\ell'}[\psi_{X,0},\psi_{Z,0}] = \rho_{\ell'}(-Y_0) = -c \\[.1cm]
[\rho_{\ell'}(\psi_{X,0}),\rho_{\ell'}(\psi_{Y,0})] = [e_0,-\bar e_0 d]=-d  \qquad\text{vs.}\qquad  \rho_{\ell'}[\psi_{X,0},\psi_{Y,0}] = \rho_{\ell'}(-Z_0) = -d \end{array}
 \ee
Note that the nontrivial differential $Q = [-\bar e_1cd,-]$ in $\text{End}{V'}$ is essential in the first line.

\subsubsection{OPE's and the MC equation}
\label{sec:OPE-MC}

Finally, we'd like to illustrate in a concrete example how the MC equation interacts with OPE's. This is simply motivation for the subsequent sections, in which we systematize all this structure using dg-shifted Yangians.

The key point we want to illustrate is that in any OPE of line operators $\ell,\ell'$, represented by spaces and couplings $(V,\mu_\ell)$ and $(V',\mu_{\ell'})$,
\be  (\ell\otimes_{z-w} \ell')(z) \sim \exp\int_\R \big[ \bm\mu_\ell(z) + \bm\mu_{\ell'}(w) + \mb r_{\ell,\ell'}(z-w)\big]\,, \label{OPE-MC-eg} \ee
the fact that the RHS is also effectively a line operator requires the combination $\mu_\ell(z) + \mu_{\ell'}(w) +  r_{\ell,\ell'}(z-w)$ to be an MC element, in $\text{End}(V\otimes V')\otimes \CA[(z-w)^{-1}]$. Moreover, the MC equation has multiple, noninteracting contributions. Let $\text{MC}(x):=\sum_{k=0}^\infty m_k(x^{\otimes k})$. We already know that $\text{MC}(\mu_\ell(z))=\text{MC}(\mu_{\ell'}(w))=0$ (for the individual lines $\ell,\ell'$).
Since the quantum correction $r_{\ell,\ell'}$ is independent of bulk fields, it must also satisfy an MC equation by itself,
\be \text{MC}(r_{\ell,\ell'})=Q(r_{\ell,\ell'}) + ( r_{\ell,\ell'})^2 = 0 \qquad\text{in}\; \text{End}(V\otimes V')[(z-w)^{-1}]\,. \label{OPE-MC-r} \ee
The remaining terms in the MC equation for $\mu_\ell(z) + \mu_{\ell'}(w) +  r_{\ell,\ell'}(z-w)$ look like
\be [\mu_\ell(z),\mu_{\ell'}(w)]+[\mu_{\ell}(z)+\mu_{\ell'}(w),r_{\ell,\ell'}(z-w)]+\text{(higher $A_\infty$ ops)} = 0\,. \label{OPE-MC-mm} \ee

We'd like to illustrate how these hold using the vortex lines $\ell = \V_{(-1,1,0)}$ and $\ell' = \V_{(2,-1,-1)}$ in the XYZ model above, defined by couplings \eqref{MC-OPE-ells}. Using the general form of quantum corrections $r_{\ell,\ell'}$ in \eqref{non-ren-OPE} we get
\be (\ell\otimes_{z-w}\ell')(w) \sim \exp\int_\R \big[ \mu_\ell(z) + \mu_{\ell'}(w)  \underbrace{- \frac{ae_0}{z-w}-\frac{ae_1}{(z-w)^2}+\frac{bc}{z-w} -\frac{\bar ba d}{z-w}}_{r(z-w)}\bigg]\,. \label{OPE-MC-example} \ee
The quantum correction come (as usual) from contractions $\psi_X(z)X(w)\sim \frac{1}{z-w}$, etc. 

To compute the MC equation for the RHS of \eqref{OPE-MC-example}, where bulk operators are placed at different points $z,w$, we need a slight generalization of the bulk operator algebra $\CA$. It can be derived from the mode expansions \eqref{A-XYZ}. The algebra including commutators at generic points looks like (see Sec. \ref{sec:Ainf-matter} for details)
\be \begin{array}{c} Q\psi_X(z)=Y(z)Z(z)\,,\quad Q\psi_Y(z)=X(z)Z(z)\,,\quad Q\psi_Z(z) = X(z)Y(z)\,, \\[.2cm] 
 \ds [\psi_X(z),\psi_Y(w)]= \frac{Z(z)-Z(w)}{z-w}\,,\qquad\quad   [\psi_Y(z),\psi_Z(w)]= \frac{X(z)-X(w)}{z-w}\,,\\[.2cm]  \ds [\psi_X(z),\psi_Z(w)]= \frac{Y(z)-Y(w)}{z-w}\,. \end{array} \label{XYZ-alg-zw}  \ee
 
Now let's check that \eqref{OPE-MC-r}--\eqref{OPE-MC-mm} hold, as they must due to self-consistency of  perturbative QFT! The MC equation for $r$ is
\be Q(r) = \frac{acd}{(z-w)^2}\,,\qquad r^2= -\frac{[b,\bar b]cad}{(z-w)^2} = \frac{cad}{(z-w)^2} \quad \Rightarrow \quad Q(r)+r^2=0\,. \ee
(Note that the first term comes from the differential $Q_{V'}$ in $\text{End}(V')$.) We also have
\be \begin{array}{c} \ds [\mu_\ell(z),\mu_{\ell'}(w)] \overset{\eqref{XYZ-alg-zw}}{=} ac\frac{Z(z)-Z(w)}{z-w} +ad\frac{Y(z)-Y(w)}{z-w}\,, \\[.4cm]
\ds [\mu_\ell(z),r] = -\frac{adY(z)}{z-w}-\frac{acZ(z)}{z-w}\,,\qquad [\mu_{\ell'}(w),r] = \frac{a(cZ(w)+dY(w))}{z-w}\,. \end{array} \ee
ensuring that $[\mu_\ell(z),\mu_{\ell'}(w)]+[\mu_\ell(z)+\mu_{\ell'}(w),r] = 0$. There are no higher terms in \eqref{OPE-MC-mm}, since the bulk algebra $\CA$ in the XYZ model has no higher $A_\infty$ operations.

The cancellations that lead to \eqref{OPE-MC-r}--\eqref{OPE-MC-mm}  here were quite nontrivial. We want to systematize and reinterpret them via Koszul duality.

\subsection{Translations, OPE, and associativity}
\label{sec:KD-Yangian}

We'd next like to explain what additional structures the dual algebra $\CA^!$ in an HT-twisted theory should contain, by virtue of the fact that the category of lines $\CC$ that $\CA^!$ represents has translation symmetries and an associative OPE. We'll derive these one at a time, starting with translations (which induce translation automorphisms $\tau_z$ in $\CA^!$), then the OPE (which induces a chiral coproduct $\Delta_z$, deformed by a shifted r-matrix $r(z)$); and finally explaining how chiral-commutativity of the OPE implies co-commutativity of $\Delta_z,r(z)$ and how associativity of the OPE implies identities for $\Delta_z,r(z)$ that resemble a quasi-triangular structure. 

Along the way, we'll begin expressing generators for $\CA$ and $\CA^!$ in terms of generating functions \eqref{xp-gen}, analogous to ``fields'' of vertex algebras. This introduces a convenient, compact formalism that we'll adopt for the remainder of the paper.

\subsubsection{Translation isomorphisms}
\label{sec:trans}

Recall that the category $\CC$ is actually a collection of categories $\CC_w$, labelled by the points $w\in \C$ where line operators are inserted. Since translations in the complex plane are a symmetry, there must be corresponding isomorphisms $T_z:\CC_w\overset\sim\to \CC_{w+z}$ for each $z\in \C$, with $T_z\circ T_{z'}\simeq T_{z+z'}$.

Dually, suppose we're representing each $\CC_w$ as $\CA^!\text{-mod}$ in a perturbative HT QFT, using the functor
\be \CF_w:\CC_w\to \CA^!\text{-mod}\,,\qquad \CF_w(\ell) = \text{States}(\cp^1,\ell(w),\CB_\infty) \ee
that inserts a line at $w$.
In computing the state space on the RHS, it cannot matter where on $\cp^1$ the line $\ell$ is actually inserted, since shifting coordinates $w\mapsto w+z$ on $\cp^1$ is a symmetry that preserves the point at infinity. Formally, this means that there should exist natural isomorphisms $T_z:\CF_w\overset\sim\to \CF_{w+z}$, which relate every pair of state spaces $T_{z,\ell}:\CF_w(\ell)\overset\sim\to \CF_{w+z}(T_z(\ell))$. In turn, these induce automorphisms $\tau_z:\CA^!\to\CA^!$  (for each $z\in \C$) of the algebra $\CA^!$ such that, for all $a\in \CA^!$ and $\ell\in \CC_w$, acting with $\tau_z(a)$ on $\CF_w(\ell)$ becomes equivalent to acting with $a$ on $\CF_{w+z}(T_z\ell)$.
\be  T_z\circ \tau_z(a) \big|  \raisebox{-.1cm}{$\CF_w(\ell)$} =   a\big| \raisebox{-.1cm}{$ \CF_{w+z}(T_z \ell) $}\,.  \ee
If we think of $a$ itself as a natural transformation $a:\CF_w\to \CF_w$, then we simply have:
\be \tau_z(a) := T_{-z}\circ a \circ T_z\,. \ee

In a perturbative HT QFT, we can find an explicit form for $\tau_z$ by looking investigating MC elements. Suppose we have fields $(\mb x^i,\mb p_i)$, with modes $(x^i_n,p_{i,n})_{n<0}$ generating bulk local operators $\CA$ and modes $(x^i_n,p_{i,n})_{n\geq 0}$ generating the Koszul-dual $\CA^!$. Let's group these in formal generating functions
\be \begin{array}{ll} \CA:\quad  &x^i(u)_+ := \sum_{n< 0} x^i_n u^{-n-1}\,,\qquad p_i(u)_+ := \sum_{n< 0} p_{i,n} u^{-n-1} \\[.2cm]
 \CA^!:\quad & x^i(u)_- := \sum_{n\geq 0} x^i_n u^{-n-1}\,,\qquad p_i(u)_- := \sum_{n\geq 0} p_{i,n} u^{-n-1} \end{array}  \label{xp-gen} \ee
Then the MC element associated to a representation $\rho_\ell:\CA^!\to \text{End}(V)$, representing a line operator inserted at $0\in \C$, has the universal form $\mu_\ell(0) = \rho_\ell(\mu(0))$, with
\begin{align} \mu(0) &= \sum_{n\geq 0} \Big[ p_{i,n}\tfrac{1}{n!} \pd^n x^i(0)+ x^i_n \tfrac{1}{n!} \pd^np_i(0)\Big] =  \sum_{n\geq 0}\Big[ p_{i,n}x^i_{-n-1} + x^i_n p_{i,-n-1}\Big]  \notag \\
&=\oint_u \Big[p_i(u)_-x^i(u)_+ + x^i(u)_- p_i(u)_+ \Big] \label{KD-univ-mu0} \end{align}
The contour integral should be interpreted algebraically, as picking out the coefficient of $u^{-1}$. We observe that the integrand only has singularities at $u=0$ (due to $p(u)_-,x(u)_-$) and at $u=\infty$ (due to $x(u)_+,p(u)_+$) --- the formal series expansion around any other finite point $u=u_0$ is regular --- so the contour can be taken to run along any counterclockwise circle encircling the origin in the $u$-plane.

The same line $\ell$ inserted at $w\in \C$ is represented by $\mu_\ell(w) = \rho_\ell(\mu(w))$, with
\be
\mu(w) = \sum_{i,n\geq 0} \Big[ p_{i,n} \tfrac{1}{n!}\pd^n x^i(w) + x^i_{n} \tfrac{1}{n!}\pd^n p_i(w) \Big]  =  \oint_u   \Big[p_i(u)_-x^i(u+w)_+ + x^i(u)_- p_i(u+w)_+ \Big]\,, \label{shift-1}
\ee
noting that the integrand still has singularities only at $u=0$ and $u=\infty$ (since $x_+,p_+$ are regular functions), and the contour encircles $u=0$.
Now for any $w,z\in\C$, we have
\begin{align} \mu(w+z) &= \oint_u   \Big[p_i(u)_-x^i(u+w+z)_+ + x^i(u)_- p_i(u+w+z)_+ \Big] \notag \\
 & \qquad = \oint_{u>z}   \Big[p_i(u-z)_-x^i(u+w)_+ + x^i(u-z)_- p_i(u+w)_+ \Big]\,. \label{shift-2}
\end{align}
After shifting integration variables $u\mapsto u-z$, the new integrand has singularities at $u=z$ and $u=\infty$, and the new contour must encircle (take the algebraic residue of) the singularity at $u=z$. This can be accomplished by placing the contour in a region where $|u|>|z|$, which we indicate by writing $\oint_{u>z}$. This means algebraically that the terms $\frac{1}{(u-z)^n}$ in $p_-$ and $x_-$ can be expanded as series in $z/u$ --- but \emph{not} in $u/z$.

For any line $\ell$, the translation $T_z :\ell(w)\mapsto \ell(w+z)$ relates its insertions at $w$ and $w+z$; and thus the automorphism $\tau_z$ in $\CA^!$ must satisfy $\rho_\ell(\tau_z\, \mu(w)) = \rho_\ell(\mu(w+z))$. Since $\ell$ is arbitrary, this implies
\be \tau_z\, \mu(w) = \mu(w+z)\,.  \label{shift-MC} \ee
Then by comparing the RHS of  \eqref{shift-1} and \eqref{shift-2} we read off
\be \tau_z: \;x^i(u)_-,p_i(u)_- \mapsto x^i(u-z)_-,p_i(u-z)_-\qquad \text{(expanded in $z/u$)\,.}  \label{def-shift-fields} \ee
Expanding $\sum_{n\geq 0} x_n (u-z)^{-n-1} = \sum_{n,p\geq 0} x_n u^{-n-1} \big(\begin{smallmatrix} n+m \\ m \end{smallmatrix} \big) (z/u)^m$ and taking modes, we get
\be \tau_z: \;\;x^i_n\mapsto  \sum_{m=0}^n \begin{pmatrix} n \\ m \end{pmatrix} z^m x^i_{n-m}\,,\quad p_{i,n}\mapsto  \sum_{m=0}^n \begin{pmatrix} n \\ m \end{pmatrix} z^m p_{i,n-m}\,. \label{def-shift-modes} \ee
Note that if we think of $z$ as a formal parameter (rather than a fixed number) the translation $\tau_z$ in \eqref{def-shift-modes} sends any element of $\CA^!$ to a finite polynomial in $z$, \ie\
\be \tau_z:\CA^!\to \CA^![z]\,. \label{trans-A!-poly} \ee
Infinitesimally, $\tau_z$ is generated by a derivation $T:\CA^!\to \CA^!$ representing $-\pd_z$:
\be \tau_z = \exp(z T)\,,\quad \bigg\{\!\!\!\begin{array}{r@{\,}c@{\,}l} T(x^i_n) &=& n\,x^i_{n-1} \\ T(p_{i,n}) &=&n\,p_{i,n-1} \end{array}\,,\qquad R(T)=0\,,\;F(T)=\text{even},\;J(T)=1\,. \label{inf-T} \ee

\subsubsection{OPE, coproduct, and shifted r-matrix}
\label{sec:OPE-r}

Next, consider any two lines $\ell,\ell'$ placed%
\footnote{We could have placed them at $z,w$; but up to a translation, we may assume that $w=0$.} %
at points $z,0\in\C$. The OPE provides a family of functors $\CC_z\boxtimes \CC_0 \to \CC_0$,
\be \ell(z)\otimes \ell'(0) \mapsto \underbrace{(\ell \otimes_z \ell')}_{\text{OPE}}(0)\qquad (z\neq 0)\,. \ee
Consider, then, the state space on $\cp^1$ in the presence of the two lines, which we expect to be preserved by taking the OPE,
\be \begin{array}{ccc}
  \CF_0 ( \ell\otimes_z\ell')   &=& \text{States}\big(\cp^1,  (\ell\otimes_z\ell')(0), \CB_\infty) \\[-.3cm]
   \rotatebox{-90}{$\simeq$} && \rotatebox{-90}{$\simeq$} \\[.2cm]
    \CF_{(z,0)}(\ell(z)\otimes \ell'(0)) &=& \text{States}(\cp^1,\ell(z)\otimes \ell'(0),\CB_\infty)\,. \end{array}
\ee

Suppose that the functor $\CF$ were monoidal with respect to the OPE. This means that there are natural isomorphisms of vector spaces
\be \qquad  J_{z;\ell,\ell'}: \;\CF_0(\ell\otimes_z \ell') \overset\sim\to \CF_z(\ell)\otimes \CF_0(\ell') \label{Jz}\qquad (?) \ee
Physically, being monoidal means that putting a theory on $\cp^1$ with $\CB_\infty$ at infinity trivializes all the bulk dynamics, to the extent that there are no longer any interactions between $\ell(z)$ and $\ell'(0)$ --- so that evaluating the state space just reduces to the product of the individual quantum mechanics associated to $\ell$ and to $\ell'$.
Then, by a standard result in representation theory (\cf\ \cite{EGNO}), there would be an induced family of coproducts on $\CA^!$,
\be   \Delta_z:\CA^!\to \CA^!\otimes \CA^! \qquad (z\neq 0)\,, \label{D-naive} \ee
such that $a$ acts on $\CF_0(\ell\otimes_z\ell')$ the same way that $\Delta_z(a)$ acts on $\CF_z(\ell)\otimes \CF_0(\ell')$, \ie\
\be  J_{z;\ell,\ell'}\circ\, a\big|\raisebox{-.1cm}{$\CF_0(\ell\otimes_z\ell')$} =  \Delta_z(a)\big| \raisebox{-.1cm}{$J_{z;\ell,\ell'}\big(\CF_z(\ell)\otimes \CF_0(\ell')\big)$}\,. \ee
Formally, one would define $\Delta_z(a) := J_{z;\ell,\ell'} \circ a\circ J_{z;\ell,\ell'}^{-1}$.

In 4d holomorphic-topological theories, such as the holomorphic Chern-Simons theory studied in \cite{Costello-Yangian, CWY-I}, this is exactly what happens: a vacuum at infinity trivializes the bulk dynamics sufficiently so that evaluating state spaces becomes monoidal with respect to OPE. Then the Koszul-dual algebra gains a family of standard coproducts $\Delta_z$. (In \cite{Costello-Yangian, CWY-I}, the Koszul dual is Drinfeld's Yangian.) The 4d fiber functor \emph{does} fail to preserve quantum corrections to the braiding of lines as they move past each other in the topological plane; this failure is  compensated by introducing a meromorphic R-matrix in $\CA^!$.

In 3d HT theories, the perturbative analysis of the OPE from Section \ref{sec:OPE} shows that it's the basic product that isn't preserved. Let's work in axial gauge, so the OPE is given exactly by Thm. \ref{thm:non-ren} and \eqref{non-ren-r}--\eqref{non-ren-OPE}. 
Suppose we have line operators $\ell,\ell'$ in a perturbative HT QFT that are constructed by coupling to dg vector spaces $V,V'$ (respectively), with linear MC elements $\mu_\ell,\mu_{\ell'}$ --- and corresponding representation $\rho_\ell,\rho_{\ell'}$ of $\CA^!$. Then the OPE is
\be (\ell\otimes_z \ell')(0) \simeq \exp\int_\R \big[ \bm\mu_\ell(z) + \bm\mu_{\ell'}(0) +  \mb r_{\ell,\ell'}(z)\big]\,, \label{KD-OPE-gen} \ee
where $\mu_\ell,\mu_{\ell'}$, and $r_{\ell,\ell'}(z)$ \emph{all} take a universal form.  Namely, $\mu_\ell(z) = \rho_\ell(\mu(z))$, and $ \mu_{\ell'}(0) = \rho_{\ell'}(\mu(0))$ are given by \eqref{KD-univ-mu0} and \eqref{shift-1}, while  $ r_{\ell,\ell'}(z) = (\rho_\ell\otimes \rho_{\ell'})(r(z))$ with
\begin{align} 
 r(z) &= \sum_{m,n\geq 0} \bigg( \begin{array}{@{}c@{}} n+m \\ n \end{array} \bigg) \frac{ x^i_n p'_{i,m} -p_{i,n} x'{}^i_m   }{(-1)^n z^{n+m+1}} \quad \in \CA^!\otimes \CA^!  \label{KD-univ-r}
 \end{align}
(There's an extra minus sign in $r$ compared to Sec. \ref{sec:OPE} due to twisting $\CA^!$ by $(-1)^F$.) 

The r-matrix has a contour-integral formulation, similar to the universal MC element $\mu$ in \eqref{shift-1}. However, since the singularity structure of the `fields' $x(s)_-,p(s)_-$ in $\CA^!$ is different from the `fields'  $x(s)_+,p(s)_+$ in $\CA$, its interpretation is a bit different. We have: \vspace{-.5cm}
\be r(z) = \oint_{s<z} \Big[x^i(s)_- p'_i(s+z)_- - p_i(s)_- x'{}^i(s+z)_-\Big] 
 =  - \oint_{s<z} \Big[x^i(s-z)_- p'_i(s)_- - p_i(s-z)_- x'{}^i(s)_-\Big] \label{r-conv} \ee
On the LHS, the integrand has singularities at $s=0$ and $s=-z$, and we're only taking the residue at $s=0$. Thus the contour lies in a region $|s|<|z|$. In particular, terms $\frac{1}{(s+z)^n}$ in $p',x'$ are expanded as series in $s/z$ (not $z/s$). On the RHS, we can ``shift'' integration variables $s\mapsto s-z$, but this actually involves flipping the contour so that it picks up the residue from $p',x'$ rather than $x,p$; the contour is flipped at the cost of a minus sign, as indicated. We also have
\be  r(z) = \oint_{s,s'<z} \frac{x^i(s)_-p_i'(s')_- - p_i(s)_-x'{}^i(s')_-}{z+s-s'}\,. \label{r-Drin-s} \ee
Here we expand $(z+s-s')^{-1} = \sum_{k\geq 0} (s'-s)^k z^{-k-1}$ and take the coefficient of $s^{-1}$ and $s'{}^{-1}$. The relation between \eqref{r-Drin-s} and \eqref{r-conv} comes from $(z+s-s')^{-1} = \delta(s'-s-z)$.

Now consider the state space $\CF_0(\ell\otimes_z\ell') = \text{States}(\cp^1,\ell\otimes_z\ell'(0),\CB_\infty)$, The boundary conditions on bulk fields $x_+,p_+$ set $\mu_\ell=\mu_{\ell'}=0$ in \eqref{KD-OPE-gen}; but $r_{\ell,\ell'}(z)$, which is independent of bulk fields, survives.  We find
\begin{align} \CF_0(\ell\otimes_z\ell')
 &=V\otimes V' \quad\text{deformed by the MC element $r_{\ell,\ell'}(z)\in \text{End}(V)\otimes\text{End}(V')$} \notag \\
 &=:  V\otimes_{r_{\ell,\ell'}(z)} V'  \end{align}
in contrast to $\CF_z(\ell)\otimes \CF_0(\ell') = V\otimes V'$ (undeformed). Thus the functor $\CF$ is not quite monoidal, and its failure to be monoidal can be compensated by $r(z)$.

Each coupling $\mu_\ell(z)+\mu_{\ell'}(0)$ may be viewed as an MC element in $\text{End}(V)\otimes_{r_{\ell,\ell'}(z)} \text{End}(V')\otimes \CA$.  It determines a representation of $\CA^!$ in $\text{End}(V)\otimes_{r_{\ell,\ell'}(z)} \text{End}(V')$ in the usual way, by sending generators $(x_n,p_n)$ to the coefficients of $(p_{-n-1},x_{-n-1})$ in the coupling $\mu_\ell(z)+\mu_{\ell'}(0)$. 
Writing out
\be \mu_\ell(z)+\mu_{\ell'}(0) = (\rho_\ell\otimes\rho_{\ell'}) \oint_{u>z} \Big[ \big( p_i(u-z)_-+p'_i(u)_-\big) x^i(u)_+ + \big(x^i(u-z)_- +x'{}^i(u)_-\big)p_i(u)_+\Big] \ee
for all $\ell,\ell'$, we see that this representation of $\CA^!$  is induced by a family of coproducts $\Delta_z$, given explicitly on generators by
\be 
 \begin{array}{ll} \Delta_z x^i(u)_- = x^i(u-z)_-\otimes 1 + 1\otimes x^i(u)_- &\quad  = \tau_z\, x^i(u)_-\otimes 1 + 1\otimes x^i(u)_- \\[.1cm]  \Delta_z p_i(u)_- = p_i(u-z)_-\otimes 1 + 1\otimes p_i(u)_-  &\quad = \tau_z\, p_i(u)_-\otimes 1+1\otimes p_i(u)_- \end{array}
 \label{D-formula-gen} \ee
(Borrowing Hopf-algebra language, one might say that the linear generators $x^i,p_i$ are \emph{primitive elements} for this chiral coproduct.)
However, since $\text{End}(V)\otimes_{r_{\ell,\ell'}(z)} \text{End}(V')=(\rho_\ell\otimes\rho_{\ell'})(\CA^!\otimes_{r(z)}\CA^!)$, the coproduct is now a map of $A_\infty$ algebras
\be  \Delta_z :\CA^!\to \CA^!\otimes_{r(z)} \CA^! \qquad (z\ne 0)\,, \label{D-rz} \ee
where the RHS denotes $\CA^!\otimes\CA^!$ deformed by the MC element $r(z)$.

If we view $z$ as a formal parameter, then
\be r(z) \in \CA^!\otimes \CA^![\![z^{-1}]\!]\quad\text{and}\quad \Delta_z : \CA^! \to \CA^!\otimes_{r(z)} \CA^![\![z^{-1},z]  \label{rD-param} \ee
where on the RHS we adjoin formal power series in $z^{-1}$ (as required for the the deformation by $r(z)$ to make sense) and finite polynomials in $z$ (as required for the shifts $\tau_z$ in \eqref{D-formula-gen} to make sense). With $z$ a formal parameter of degrees $J(z)=-1$, $R(z)=0$, $F(z)=\text{even}$, the MC element $r(z)$ becomes homogeneous, with
\be  R(r(z)) = 1\,,\qquad J(r(z))=0\,,\qquad F(r(z))=\text{odd}\,. \label{r-degrees} \ee

We'll refer to $r(z)$ as a \emph{shifted r-matrix}, by analogy with (classical) $r$-matrices in the theory of Hopf algebras, in particular in Yangians and their generalizations. Both $\Delta_z$ and $r(z)$ satisfy a surprising number of properties that have standard analogues for Yangians, despite the fact that $r(z)$ is not manifestly related to braiding, and has homological degree 1 (rather than 0). We describe these properties next.

\subsubsection{Co-commutativity}
\label{sec:comm}

Consider the fact that we can take an OPE of $\ell(z)$ and $\ell'(0)$ in two different ways. We have been ``bringing'' $\ell(z)$ to $0\in \C$, but we could equally well have ``brought'' $\ell'(0)$ to the point $z\in \C$. We expect these to be equivalent, up to a translation:
\be T_z \cdot (\ell \otimes_z \ell')(0)  \simeq (\ell'\otimes_{-z}\ell)(z) \label{comm-OPE} \ee
Comparing MC elements as in \eqref{KD-OPE-gen} for the two OPE's, we must have
\be  \mu_\ell(z) + \mu_{\ell'}(0) + r_{\ell,\ell'}(z)  \;\overset{?}{=} \;    \mu_{\ell'}(0)+ \mu_\ell(z) + r_{\ell',\ell}(-z)  \ee
This is true for all $\ell,\ell'$ if and only if
\be r_{21}(-z) = r(z)\,. \label{r-op} \ee
(We've also implicitly used translation invariance $(\tau_w\otimes\tau_w)r(z) = r(z)$ for any $w$.) 
Property \eqref{r-op} does hold in any perturbative theory, as seen easily from \eqref{r-Drin-s}.

The identity \eqref{comm-OPE} for the OPE now translates to a symmetry of $\Delta_z$:
\be (\tau_{-z}\otimes \tau_{-z})\circ \Delta_{z} = \Delta_{z}\circ \tau_{-z} = \Delta_{-z}^{\rm op}\,. \label{weak-comm} \ee
This is easy to check from the explicit formulas \eqref{D-formula-gen}.  The equality makes sense because \eqref{r-op} allows us to identify  $\CA^!\otimes_{r(z)}\CA^!$ (where the LHS lands) with $P(\CA^!\otimes_{r(-z)}\CA^!)$ (where the RHS lands), where $P$ permutes the factors. This property is sometimes called \emph{weak co-commutativity} of $\Delta_z$.

\subsubsection{Associativity and $A_\infty$ Yang-Baxter}
\label{sec:KD-assoc}

Finally, we describe the consequences of associativity, starting with associativity of three lines and generalizing to arbitrary configurations of $n$ lines. We'd like to explain that associativity of any number of lines is controlled by a simple set of identities \eqref{Dr-1}--\eqref{Dr-2} for the coproduct in $\CA^!$, and that it in turn implies an $A_\infty$ analogue of the Yang-Baxter equation for $r(z)$.

For three lines $\ell(z+w)\otimes\ell'(z)\otimes \ell''(0)$, associativity of the OPE means that
\be \ell \otimes_{z+w} (\ell'\otimes_z \ell'')(0) \simeq  (\ell\otimes_w \ell') \otimes_z \ell''(0)\,. \label{triple-lines}  \ee
Let's examine this in terms of MC elements, starting with the LHS.

There are two ways to construct the LHS. On one hand, we observe that the OPE $(\ell'\otimes_z \ell'')(0)$ has an MC element $\mu_{\ell'}(z)+\mu_{\ell''}(0) + r_{\ell',\ell''}(z)$. Further taking its OPE with $\ell(z+w)$ must add the MC element $\mu_\ell(z+w)$ of $\ell''$, together with the interaction $r_{\ell,\ell'}(w)$ between $\ell$, $\ell'$ (the result of contracting bulk fields in $\mu_\ell$ with bulk fields in $\mu_{\ell'}$) and the interaction $r_{\ell,\ell'}(z+w)$ between $\ell$, $\ell''$ (from contractions between $\mu_\ell$ and $\mu_{\ell''}$). Thus:
\be \ell \otimes_{z+w} (\ell'\otimes_z \ell'')(0):\quad \exp\int_\R\Big[ \bm\mu_\ell(z+w)+ \bm \mu_{\ell'}(z)+\bm\mu_{\ell''}(0)+ \mb r_{\ell,\ell'}(w)+\mb r_{\ell,\ell''}(z+w)+\mb r_{\ell',\ell''}(z)\Big] \label{triple-MC} \ee
On the other hand, we must get this same interaction term, in a universal way, by repeatedly applying the coproduct in $\CA^!$ and working ``from the outside in.'' We first apply $\Delta_{z+w}$ to ``split off'' $\ell$ from $\ell'\otimes_z\ell''$, and add a correction $r(z+w)$ to the MC element in $V\otimes(V'\otimes V'')$. Then we apply $1\otimes \Delta_z$ to ``split off'' $\ell'$ from $\ell''$, and add a correction $r_{\ell',\ell''}(z)$ to the MC element in $V'\otimes V''$. For the two calculations to agree for all triples $\ell,\ell',\ell''$, we must have
\be r_{12}(w)+r_{13}(z+w)+r_{23}(z) = (\text{id} \otimes \Delta_z) r(z+w) + r_{23}(z)\,. \label{Dr-1} \ee
Note that $r_{23}(z)$ cancels out, to give a familiar formula  from Hopf algebras $(\text{id} \otimes \Delta_z) r(z+w)=r_{12}(w)+r_{13}(z+w)$ (albeit now playing a different role).
It's easy to check this is true, using formula \eqref{D-formula-gen} for the coproduct and \eqref{r-conv} for the r-matrix:
\begin{align}
(\text{id} \otimes \Delta_z) r(z+w) &= (\text{id} \otimes \Delta_z) \oint_{s<z+w} \Big[ x^i(s)_- p'_i(s+z+w)_- - p_i(s)_-x'{}^i(s+z+w)_-\Big] \notag \\
&= \oint_{s<z+w} \Big[ x^i(s)_- \Big(p'_i(s+w)_-\otimes 1+1\otimes p''_i(s+z+w)\Big) \notag \\
&  \hspace{.8in}  - p_i(s)_-\Big(x'{}^i(s+w)_-\otimes 1+1\otimes x'{}^i(s+z+w)_-\Big)\Big] \notag \\
& =  r_{12}(w)+r_{13}(z+w)\,.
\end{align}

Similarly, there are two ways to construct the RHS of \eqref{triple-lines}. Successively taking OPE's and computing corrections to the MC elements again produces the symmetric expression \eqref{triple-MC}. (Thus, associativity holds for a triple of lines.)
On the other hand, in terms of coproducts, we're led to the second identity 
\be r_{12}(w)+r_{13}(z+w)+r_{23}(z) = (\Delta_w\otimes \text{id}) r(z) + r_{12}(w)\,. \label{Dr-2} \ee
This is easily verified from formulas \eqref{D-formula-gen} and \eqref{r-conv}.

Now let's generalize to an arbitrary collection of $n$ lines. What we'd like to show is that for any $\ell_1(z_1)\otimes\cdots \otimes \ell_n(z_n)$, with OPE's taken in any order and expanded around any point, the total MC element takes the manifestly symmetric form
\be \ell_1(z_1)\otimes\cdots \otimes \ell_n(z_n):\quad  \exp \int_\R \rho_{\ell_1}\otimes...\otimes\rho_{\ell_n} \Big[ \sum_{1\leq i\leq n} \bm \mu_i(z_i)  + \mb r^{(n)} \Big]\,, \ee
with total (universal) quantum correction
\be r^{(n)} := \sum_{\substack{1 \leq i < j \leq n}}   r_{ij}(z_i-z_j) \;\; \in\; \CA^!{}^{\otimes n} [\![ (z_i-z_j)^{-1} ]\!]_{i<j}\,.  \label{rn} \ee
Here $r_{ij}$ denotes a copy of the r-matrix \eqref{r-conv} using the $i^{th}$ and $j^{th}$ factors of $\CA^!{}^{\otimes n}$.
We claim that, at a purely algebraic level, the two identities \eqref{Dr-1}--\eqref{Dr-2} alone are sufficient to guarantee this.

 Suppose the statement is true for any $n-1$ lines, indexed as $\ell_1(z_1)\otimes\cdots \otimes \ell_{k-1}(z_{k-1}) \otimes \ell_{k+1}(z_{k+1})\otimes\cdots\otimes \ell_n(z_n)$, with symmetric MC element
\be r^{(n-1)} = \sum_{\substack{1 \leq i < j \leq n \\ i,j\neq k}}   r_{ij}(z_i-z_j) \;\; \in\; \CA^!{}^{\otimes n-1} [\![ (z_i-z_j)^{-1} ]\!]_{i<j;i,j\neq k}\,.  \ee
To produce the MC element for $n$ lines, we act with the coproduct $\Delta_{z_k-z_{k+1}}$ on the $k$-th factor and add $r_{k,k+1}(z_k-z_{k+1})$ (for new quantum corrections). We need to show
\be r_{k,k+1}(z_k-z_{k+1})+  (\text{id}^{\otimes k-1}\otimes \Delta_{z_k-z_{k+1}} \otimes \text{id}^{\otimes n-k-1})\, r^{(n-1)}  = r^{(n)}  \ee
This follows immediately  from \eqref{Dr-1} and \eqref{Dr-2}. (The cases $k=1$ and $k=n$ must be considered separately, but still follow directly  from  \eqref{Dr-1}--\eqref{Dr-2}.)

A very concise, schematic, way to summarize associativity is
\begin{align} r^{(n)} & = (r+\Delta)^{n-1}(0) \\ & = r+\Delta(r + \Delta(r+\ldots   +\Delta(r)...))\,. \notag  \end{align}
That is, if we act $n-1$ times with the coproduct and add $r$, acting on any tensor factors in any order, we will always obtain the same symmetric combination $r^{(n)}$. Its parameters $\{z_i\}$ will depend on the parameters $r(w)+\Delta_w$ at each step.

An interesting consequence of associativity is a shifted, $A_\infty$ analogue of the Yang-Baxter equation. Since the twisted coproduct $\Delta_z:\CA^!\to \CA^!\otimes_{r(z)}\CA^![\![z^{-1},z]$ is a morphism of $A_\infty$ algebras, each operator
\be r_{k,k+1}(z) +  \text{id}^{\otimes k-1}\otimes (\Delta_z)\otimes \text{id}^{\otimes n-k-1} : \CA^!{}^{\otimes n-1} \to  \CA^!{}^{\otimes n}[\![z^{-1},z] \ee
must preserve MC elements. Therefore, each $r^{(n)}$ in \eqref{rn} must satisfy the MC equation in $\CA^!{}^{\otimes n}[\![(z_i-z_j)^{-1}]\!]$.%
\footnote{An alternative way to argue this is that $r^{(n)}$ is the universal quantum correction to an OPE of $n$ lines, and is independent of bulk fields, and so must satisfy an MC equation on its own.} %
Let's write the MC equation as
\be \text{MC}(x) := \sum_{d=1}^\infty m_d(x^{\otimes d}) = Q(x) + \tfrac12[x,x] + \sum_{d=3}^\infty m_d(x,...,x) \ee
Since $r^{(2)}=r$ and $r^{(3)}$ satisfy $\text{MC}(r^{(2)}) = \text{MC}(r^{(3)})=0$, we have
\begin{align}
0&= \text{MC}(r^{(3)}(z+w,z,0)) - \text{MC}(r_{12}(w)) - \text{MC}(r_{13}(z+w)) - \text{MC}(r_{23}(z))  \notag \\
&  = \text{MC}(r_{12}(w)+ r_{13}(z+w) +r_{23}(z)) - \text{MC}(r_{12}(w)) - \text{MC}(r_{13}(z+w)) - \text{MC}(r_{23}(z)) \notag \\[.2cm]
&  = [r_{12}(w),r_{13}(z+w)]+[r_{12}(w),r_{23}(z)]+[r_{13}(z+w),r_{23}(z)] \label{YB} \\[.2cm]
 & \hspace{.1in} + \sum_{d\geq 3} m_d\big[(r_{12}(w)+ r_{13}(z+w) +r_{23}(z))^{\otimes d} -r_{12}(w)^{\otimes d}-r_{13}(z+w)^{\otimes d}-r_{23}(z)^{\otimes d}\big] \notag
\end{align}
in $\CA^!{}^{\otimes 3}[\![z^{-1},w^{-1},(z+w)^{-1}]\!]$. When higher $m_{d\geq 3}$ operations vanish, we are left with
\be [r_{12}(w),r_{13}(z+w)]+[r_{12}(w),r_{23}(z)]+[r_{13}(z+w),r_{23}(z)] =0 \ee
which is the classical \emph{Yang-Baxter} equation, for our shifted r-matrix. In general, we find an $A_\infty$ generalization of Yang-Baxter equation given by \eqref{YB}.

The remaining equations $\text{MC}(r^{(n)}) =0$ for $n> 4$ all follow from $\text{MC}(r^{(2)})=\text{MC}(r^{(3)}) =0$ when  higher $m_{\geq 3}$ operations vanish. When higher $A_\infty$ operations are present, they seem to provide addition nontrivial constraints.

\subsection{dg-shifted Yangians}
\label{sec:Yangian-summary}

The structures in Section \ref{sec:KD-Yangian} motivate the following abstract definition. Let $R,F,J$ as usual denote cohomological, parity (fermion number), and spin gradings; and let $z$ be a formal parameter with $R(z)=0$, $F(z)=\text{even}$,  $J(z)=-1$.

\begin{Def} \label{def:Yang}  A (generalized) \emph{dg-shifted Yangian} is an $A_\infty$ algebra $\CY$, with gradings $R,F,J$, and further endowed with
\begin{subequations}
\be \text{\emph{translation `automorphisms':}} \quad \tau_z:\CY\to \CY[z]\,,\qquad \tau_z=\exp(z\,T) \ee
generated by an $A_\infty$ derivation $T:\CY\to \CY$ of degrees $(0,\text{odd},1)$ (so $\tau_z$ are graded $A_\infty$ morphisms satisfying $\tau_z\tau_w=\tau_{z+w}$);
\be \begin{array}{ll} \text{\emph{a Maurer-Cartan element:}} \quad  & r(z) \in \CY\otimes \CY[\![z^{-1},z] \quad \text{of degrees}\; (1,\text{odd},0) \\[.2cm]
 \text{\emph{and a twisted coproduct:}}\quad & \Delta_z:\CY\to \CY\otimes_{r(z)}\CY[\![z^{-1},z] \end{array}  \ee
that's a graded $A_\infty$ algebra morphism; and a unit `$1$' and
\be \text{\emph{counit:}}\quad  \varepsilon:\CA^!\to \C \ee
that's also an $A_\infty$ morphism, where $\C$ denotes the trivial algebra.
These satisfy
\be \text{\emph{counit axioms:}} \quad \varepsilon(1)=1\,,\quad (\varepsilon\otimes\text{id})\circ\Delta_z = \text{id}\,,\quad (\text{id}\otimes\varepsilon)\circ\Delta_z \simeq \tau_z \label{counit-ax} \ee
\vspace{-.4in}
\be \begin{array}{l}\text{\emph{translation invariance}} \\ \text{\emph{ \& weak co-commutativity:}} \end{array}
\quad \begin{array}{c} \varepsilon(T(a))=0\,,\quad \varepsilon(\tau_z(a))=\varepsilon(a)\quad \forall\, a\in \CY \\[.1cm]
  (\tau_w\otimes\tau_w)r(z)=r(z)\,,\quad r(-z)=r_{21}(z) \\[.1cm] (\tau_z\otimes \tau_z)\circ \Delta_{-z} = \Delta_{-z}\circ \tau_z = \Delta_z^{\rm op} \end{array}  \label{comm-def} \ee 
and
\be \begin{array}{l} \text{\emph{co-associativity:}}\quad  r_{23}(z)+(\text{id}\otimes\Delta_z)(r(z+w)) = r_{12}(w)+(\Delta_w\otimes\text{id})(r(z))  \\[.1cm]
 \hspace{2in} =  r_{12}(w)+ r_{13}(z+w) +r_{23}(z)\,. \end{array}  \label{ass-def} \ee
\end{subequations}
\end{Def}

\noindent We also recall from the end of Section \ref{sec:KD-assoc} that $\Delta_z$ being a (co)associative algebra morphism implies that $r(z)$ will satisfy the $A_\infty$ Yang-Baxter equation \eqref{YB}.

We argued in Section \ref{sec:KD-Yangian}, by consistency of OPE's in 3d HT QFT's, that

\begin{Conj} \label{conj:HT-Yang} (Physics theorem) In a perturbative 3d HT QFT, the $A_\infty$ algebra $\CA^!$ that's Koszul-dual to local operators has the structure of a dg-shifted Yangian.
\end{Conj}

Let's quickly summarize. In perturbative HT QFT with local operators $\CA$, the Koszul dual $\CA^!$ is such that the universal coupling
\be \mu = \oint_{u}\Big[p_i(u)_- x^i(u)_+ + x^i(u)_-p_i(u)_+\Big] \ee
is an MC element in $\CA^!\otimes \CA$. We represent line operators as $\CA^!$-modules. In particular, a line $\ell(z) = T_z\ell(0)$ has MC element $\rho_\ell(\mu(z)) = \rho_\ell (\tau_z \mu)$, with $\tau_z$ acting on $\CA^!$ via \eqref{def-shift-fields}.
We did not mention the counit previously; it is simply defined by the action of $\CA^!$ on the identity line $\id$, whose state space \eqref{CP1-2d} is trivial, so $\varepsilon:=\rho_{\id}:\CA^!\to \C$. Identities \eqref{counit-ax}--\eqref{comm-def} for the counit are induced by $T_z(\id)\simeq \id$ and $\id\otimes_z\ell\simeq \ell$, $\ell\otimes_z\id \simeq T_z\ell$ in the category of lines.
The coproduct is given by
\be \Delta(a)=\tau_z(a)\otimes 1+1\otimes a\quad\forall\;\text{linear generators $a\in\CA^!$} \ee
and the universal r-matrix by
\be r(z) = \oint_{s<z} \Big[ x^i(s)_- p_i'(s+z)_- - p_i(s)_- x'{}^i(s+z)_- \Big] \ee
which manifestly satisfy \eqref{comm-def}, \eqref{ass-def}.

The universal coupling $\mu(z)+\mu'(0)+r(z)$ associated to an OPE of two lines must itself satisfy an MC equation. The part independent of bulk fields is an MC equation for $r(z)$ alone, which says that $r(z)$ is an MC element that can deform the algebra $\CA^!\otimes \CA^!$. The MC equation for the entire sum  $\mu(z)+\mu'(0)+r(z)$ then becomes equivalent to the statement that $\Delta_z:\CA^!\to \CA^!\otimes_{r(z)} \CA^![\![z^{-1},z]$ is a morphism of $A_\infty$ algebras.

These MC equations are highly nontrivial statements in a given QFT. They depend on the precise $A_\infty$ structures of $\CA$ and $\CA^!$. We prove that they do indeed hold in examples of HT-twisted (perturbative) 3d $\CN=2$ theories in Section \ref{sec:proofs}.

\subsection{The special case of TQFT}
\label{sec:TQFT}

We include a short remark about topological theories, and additional structure that $\CA^!$ acquires in this case.

Some QFT's that are defined as holomorphic-topological theories may in fact be fully topological. Examples include topological twists of 3d $\CN=4$ theories, expressed as deformations of HT twists (as described in \cite{Garner:2022vds,HilburnRaskin}); or HT-twisted 3d $\CN=2$ pure Chern-Simons theory at non-critical level. Perturbatively, the HT twist of gauge theory with nonzero level $k$ is equivalent to perturbative Chern-Simons at level $k-h^\vee$. In these cases, translation functors $T_z:
\CC_w\to \CC_{w+z}$ on the category of lines should be ``trivial," in the sense that they are all isomorphic to the identity functor.

From the prespective of boundary vertex algebras, the triviality of translations is encoded in the existence of a stress-energy tensor (\emph{a.k.a.} conformal element) $L(z)$. In particular, the action of $\exp (z L_{-1})$ identifies $T_z$ with the identity functor $T_0$. For non-critical level Chern-Simons theory, this stress-energy tensor is given by the famous Segal-Sugawara construction.

One can ask for an analogous trivialization of  $T_z$ when representing  $\CC\simeq \CA^!\Mod$. The minimal structure we expect to be present is the following: we would like the derivation $T$ that generates translations $\tau_z = \exp(z\,T)$ in $\CA^!$ to be exact. In other words, for a TQFT, we expect that there exists a second derivation
\begin{subequations} \label{TQL}
\be \CL:\CA^!\to\CA^! \qquad \text{of degrees $R(\CL)=-1$, $F(\CL)=\text{odd}$, $J(\CL)=1$} \ee
such that
\be T = [Q,\CL] = Q\CL + \CL Q\,. \ee
\end{subequations}

The existence of this $\CL$ implies that $T=0$ on the cohomology of $Q$, since if an element $a\in \CA^!$ is $Q$-closed, then $T(a)$ is $Q$-exact: $T(a)=[Q, \CL]a=Q(\CL\,a)$. 
A sophisticated reformulation of this statement is that the derivation $T$, which defines a class in the Hochschild cohomology of $\CA^!$, is homotopic to the zero derivation. In principle, this means that the functor $T_z$ generated by $T$ should be equivalent to the identity functor (after all, the Hochschild cohomology of $\CA^!$ is the space of endomorphisms of the identity functor on $\CA^!\Mod$). We leave it to future work to develop this point more precisely.

\section{$A_\infty$ structure on local operators and their dual}
\label{sec:Ainf}

Our last major goal in this paper is to verify explicitly that the Koszul-dual algebras $\CA^!$ in perturbative HT QFT's really do have the structure of dg-shifted Yangians, as introduced in Section \ref{sec:KD-Yangian}. In order to do this, we must be able to compute and utilize the full $A_\infty$ structure on $\CA^!$. In this section, we'll explain how to do so.

In fact, computing $\CA^!$ as an $A_\infty$ algebra in a perturbative HT QFT turns out to be just as easy (or hard) as computing bulk local operators $\CA$ as an $A_\infty$ algebra. We'll tackle both at the same time.

The approach we'll take is to compactify a 3d HT QFT to two dimensions along $\C^*$, as discussed in Section \ref{sec:2d-reduction}. Then the 3d HT action
\be S = \int_{\C^*_z\times\R_t} \mb p_i d' \mb x^i + \CI(\mb x,\mb p,\pd)\,, \label{S3d-reduce-2} \ee
reduces to a 2d action for a topological B-model involving infinitely many fields
\be \label{S2d-gen-2} S_{\rm 2d}
=  \int_{\R_r\times \R_t} \bigg[ \sum_{n+m=-1} \mb p_{i,n} d \mb x^i_m + \CI_0(\{\mb x^i_k\},\{\mb p_{i,k}\})\Bigg]\,,  \ee
where $\mb x^i = \sum_{n\in \Z} \mb x^i_n z^{-n-1}$, $\mb p_i = \sum_{n\in \Z} \mb p_{i,n} z^{-n-1}$, and $\CI_0 = \oint_z \CI$ is the coefficient of $z^{-1}$ in the expansion of the original interaction $\CI$. The algebras $\CA$ and $\CA^!$ are now on the same footing: they are endomorphism algebras of two boundary conditions for the 2d topological theory
\be \CA = \text{End}(\CB_0)\,,\qquad \CA^! = \text{End}(\CB_\infty) \qquad\quad  \raisebox{-.2in}{\includegraphics[width=1.2in]{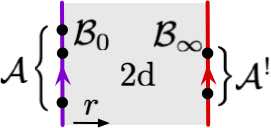}} \label{AABB} \ee
where%
\footnote{Technically speaking, $\CB_0$ and $\CB_\infty$ are objects of different categories, since in the compactification of \eqref{CP1-2d} $\CB_0$ appears as a left boundary and $\CB_\infty$ appears as a right boundary. They are related by $180$-degree rotation, or a dualization operation. In 2d conventions, we should be careful to do computations with all boundary conditions placed on the same side, in the same orientation. Then we actually have $\CA^!=\text{End}(\CB_\infty^\vee)^{\rm op}$, where $\CB_\infty^\vee$ is $\CB_\infty$ viewed as a left boundary.\label{foot:dual}}
\be \begin{array}{ll} \CB_0: & \mb x^i_n\big|_{\rm bdy} =  \mb p_{i,n}\big|_{\rm bdy}  = 0\quad\text{for $n\geq 0$} \\[.1cm]
  \CB_\infty: & \mb x^i_n\big|_{\rm bdy} =  \mb p_{i,n}\big|_{\rm bdy} = 0 \quad\text{for $n< 0$}\,. \end{array} \label{B-modes-def}
 \ee

Our problem then boils down to computing the boundary algebras $\CA,\CA^!$ in the 2d B-model \eqref{S2d-gen-2}. It's well known that the $A_\infty$ operations on boundary algebras in a B-model are controlled by derivatives of the interaction $\CI_0$; mathematically, $\CI_0$ is called a ``cyclic function'' that generates a deformation of an $A_\infty$ algebra, \cf\ \cite[Rmk. 10.1.5]{KS-notes}. An explicit QFT-based computational framework was described in the lecture notes \cite{PIMS}. We'll generalize the approach of \cite{PIMS} to the present setting,~as~follows.

We begin by proving a non-renormalization result (generalizing the non-renormali\-zation theorem of \cite{Gwilliam:2019cbp} for Chern-Simons theory):

\begin{Prop} \label{prop:Ainf}
The perturbative algebras $\CA$ and $\CA^!$ are deformations of the free graded-commutative algebras $\C[x^i_n,p^i_n]_{n<0}$ and $\C[x^i_n,p^i_n]_{n\geq0}$ (respectively), with $A_\infty$ operations computed exactly by tree-level and one-loop Feynman diagrams involving the interaction $\CI_0$. For quasi-linear theories, the one-loop diagrams vanish and $A_\infty$ operations are exact at tree level, given by the diagrams in \eqref{bdy-tree} below.
\end{Prop}

\noindent We establish the general result in Section \ref{sec:bdy-nonren} by degree counting, and specialize to quasi-linear theories in Section \ref{sec:bdy-loop}. Recall that quasi-linear theories, introduced in Section \ref{sec:quasi-linear}, include all HT twists of 3d $\CN=2$ gauge/matter/Chern-Simons theories with superpotentials. Our argument for vanishing of the one-loop correction to $\CA^!$ is slightly non-rigorous, but becomes rigorous after proving Koszul duality of $\CA$ and $\CA^!$ for twisted 3d $\CN=2$ theories in Section \ref{sec:proofs}.

In the remainder of the section, we give explicit formulas for the $A_\infty$ operations in $\CA$ and $\CA^!$, both in terms of modes $x^i_n,p_{i,n}$ and generating functions or ``fields" $x^i(z),p_i(z)$. We compute the operations in a general perturbative QFT in Section \ref{sec:Ainf-gen}. In Section \ref{sec:Ainf-matter} we specialize to matter with a superpotential, showing how $\CA$ becomes a chiralization of the derived critical locus of the superpotential. In Section \ref{sec:Ainf-gauge}, we then consider gauge theory with Chern-Simons terms and matter, but vanishing superpotential. Such gauge theories can compactly be rewritten as gauge theories for ``Takiff algebras'' $\mathfrak h=\mathfrak g\ltimes V$, and $\CA$ and $\CA^!$ take the form of loop algebras, deformed by Chern-Simons terms:
\be \CA \approx \lambda^{-1}(T^*[-1]\mathfrak h)[\lambda^{-1}]\,,\qquad  \CA^! \approx (T^*[-1]\mathfrak h)[\lambda]\,. \ee
We'll also mention (Sec. \ref{sec:gauge-loop}) how Wilson lines, and some simple generalizations in the presence of matter, appear as modules for $\CA^!$. We reincorporate superpotentials, for completely general HT-twisted $\CN=2$ Lagrangian theories, in Section \ref{sec:Ainf-full}.

Our results generalize the Koszul duals in \cite[Secs. 6-7]{PaquetteWilliams} for 3d HT free matter and pure gauge theory at critical CS level, obtained with slightly different techniques.

Finally, in Section \ref{sec:TQFT-gauge}, we consider the special case of perturbative pure gauge theory (no matter), with a non-degenerate Chern-Simons level. This is a topological theory, whose line operators should to be equivalent to modules for a boundary Kac-Moody algebra,
\be \CC,\otimes_z \;\simeq\; (\CA^!,\Delta_z)\text{-mod} \;\simeq\; (\hat{\mathfrak g}_k \text{-mod},\otimes_{\hat{\mathfrak g}_k})\,. \ee
In particular, the OPE $\otimes_z$, which maps to the fusion product for $\hat{\mathfrak g}_k$, should be locally constant.
We'll write down the explicit derivation $\CL:\CA^!\to \CA^!$ that makes the product in $\CA^!$-mod locally constant.

We remark that, once $\CA$ and $\CA^!$ are known as $A_\infty$ algebras, the MC element
\be \mu_{13}(z)+\mu_{23}(0)+r_{12}(z) \in \CA^!\otimes \CA^!\otimes \CA\, [\![z^{-1},z]\!] \label{mmr-control} \ee
associated to an OPE of two lines (as in \eqref{OPE-MC-eg}, \eqref{KD-OPE-gen}) controls most of their remaining algebraic structure. (The subscripts above refer to the tensor factors that the $\mu$'s and $r$ are elements of.) In Section \ref{sec:proofs} we'll prove that \eqref{mmr-control} is indeed an MC element in twisted $\CN=2$ theories, and conclude that $\CA^!$ has the structure of a dg-shifted Yangian, by dualizing \eqref{mmr-control} to an $A_\infty$ algebra map $\Delta_z:\CA^!\to \CA^!\otimes_{r(z)}\CA^![\![z^{-1},z]$. Alternatively, we can dualize the two $\CA^!$ factors to obtain an $A_\infty$ algebra map
\be \CA \otimes \CA \to \CA(\!(z)\!)\,. \ee
This is the \emph{chiral} product on bulk local operators, turning $\CA$ into an ($A_\infty$,chiral) algebra.

\subsection{Non-renormalization for boundary algebras}
\label{sec:bdy-nonren}

In a free theory $(\CI=0)$, the algebras of local operators on both $\CB_0$ and $\CB_\infty$ are graded-commutative algebras, freely generated by the zero-form components of the fields that survive at the boundary,
\be \CA = \C[x^i_n,p_{i,n}]_{n<0}\,,\qquad \CA^! = \C[x^i_n,p_{i,n}]_{n\geq 0}\,. \label{AA!-free} \ee
When bulk interactions are turned on, the differential, product, and higher operations in these algebras are deformed perturbatively. Specifically, each $m_d$ operation (where $m_1=Q$ is the differential) is corrected by connected Feynman diagrams with
\begin{itemize}
\item $d$ boundary insertions of $\mb x^i_n$ or $\mb p_{i,n}$ (with $n<0$ for $\CB_0$ or $n\geq 0$ for $\CB_\infty$)
\item $v$ bulk insertions of the interaction $\CI_0 = \oint \CI(\mb x^i,\mb p_i,\pd)$
\item  $e_\pd=d$ propagators connecting boundary to bulk vertices; and $e$ propagators connecting bulk to bulk vertices.
\end{itemize}
Schematically, for example:
\be  m_4(x^{i_1}_{n_1},p_{i_2,n_2},x^{i_3}_{n_3},x^{i_4}_{n_4})=  \raisebox{-.5in}{\includegraphics[width=1.2in]{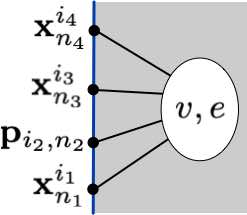}}  \ee
In particular, there cannot be any boundary-boundary propagators. This is because the 2d propagator corresponding to  \eqref{S2d-gen-2} contracts $\mb p_{i,n}$ with $\mb x^i_{-n-1}$, so one end or the other necessarily vanishes on both $\CB_0$ and $\CB_\infty$.

Furthermore, in these diagrams, the bulk vertices are integrated over a 2d half-space. Of the boundary vertices, which are ``time ordered'' along the boundary, the first and last are fixed, but all intermediate vertices are integrated along the boundary while preserving the ordering. This picks out 1-form descendants of the boundary $\mb x^i_n,\mb p_{i,n}$ insertions, and integrates their product along a $(d-2)$-simplex. The boundary integration is required in order for the $m_d$ operations defined this way to satisfy $A_\infty$ relations, \cf\ \cite[Sec.~4.5]{PIMS}. 

Now, the $m_d$ operation should have cohomological degree $R(m_d) = 2-d$ (lowering the degree of the boundary insertions --- whatever they are --- by $d-2$). By the same calculus as in Section \ref{sec:nr-Feynman}, each bulk vertex contributes $+2$ to the degree of a diagram and each propagator contributes $-1$, so this means that
\be 2 v -e_\pd - e = 2-d\,, \ee
which (using $e_\pd = d$) implies
\be 2v-e=2 \label{2d-diag-R} \ee

Now let's assume that our 3d HT theory is quasi-linear, as defined in Section \ref{sec:quasi-linear}. Then the reduced 2d theory will be quasi-linear as well (by simply placing all $n\in \Z$ modes of a given field within the same partition $S_\alpha$). We obtain from \eqref{EV-linear-nr}, after summing over partitions, a second relation
\be e\leq v\,. \label{EV-linear} \ee
Combining this with \eqref{2d-diag-R}, we get
\be 2 = 2v-e \geq 2v-v = v \quad \Rightarrow\quad v\leq 2 \ee

This leaves just two possibilities. One is $v=1$ and (due to \eqref{2d-diag-R}) $e=0$, which leads to the tree-level diagrams
\be \raisebox{-.5in}{\includegraphics[width=1.0in]{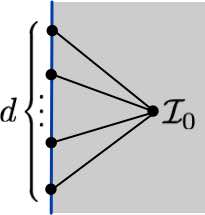}} \label{bdy-tree} \ee
The other is $v=2$ and (due to \eqref{2d-diag-R}) $e=2$. There are no self-contractions of bulk vertices, because they are (by definition) normal-ordered; so the $e=2$ edges must connect the $v=2$ bulk vertices to each other. This leads to 1-loop diagrams of the form
\be \raisebox{-.5in}{\includegraphics[width=1.03in]{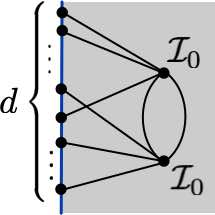}} \label{bdy-loop} \ee
with some $d_1$ boundary vertices connected to one bulk vertex, and $d_2$ connected to the other, $d=d_1+d_2$.

\subsection{Vanishing of the one-loop correction}
\label{sec:bdy-loop}

We'd next like to argue that all one-loop diagrams of the form \eqref{bdy-loop} actually vanish, in the presence of either $\CB_0$ or $\CB_\infty$ boundary conditions.

The choice of boundary conditions is important. The boundary condition $\CB_0$ is defined by circle reduction of the bulk 3d theory with no line operators inserted. Indeed, when we are computing the algebra $\CA=\text{End}(\CB_0)$, we're just computing the algebra of local operators in the smooth 3d bulk, arranged along a particular (arbitrary) time-like line. Thus, the contribution of 1-loop diagrams \eqref{bdy-loop} could also have been computed directly in 3d, using the 3d Lorentz-gauge propagator \eqref{G-Lor}. The 3d computation would insert two copies of the interaction $\CI$, contracted by two propagators:
\be  \raisebox{-.4in}{\includegraphics[width=.9in]{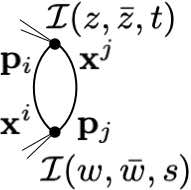}}   \begin{array}{ll}\sim &  \langle \mb p_i(z,\bar z,t) \mb x^i(w,\bar w,s)\rangle \langle \mb p_j(w,\bar w,s) \mb x_i(z,\bar z,t)\rangle  \\ 
&=\; G^{\rm Lor}(z,\bar z,t;w,\bar w,s)G^{\rm Lor}(w,\bar w,s;z,\bar z,t) \end{array} \label{loop-3d} \ee
Here we assumed that a particular $\mb p,\mb x$ in the top vertex were contracted with their conjugates in the bottom vertex; but it could just as well have been two $\mb p$'s in the top vertex with two $\mb x$'s in the bottom.
The key point is that the 3d propagator is symmetric $G^{\rm Lor}(z,\bar z,t;w,\bar w,s)=G^{\rm Lor}(w,\bar w,s;z,\bar z,t)$, and is a 1-form (in the odd variables $d\bar z,dt, d\bar w,ds$), so \eqref{loop-3d} vanishes.

Therefore, in perturbation theory, bulk local operators $\CA=\text{End}(\CB_0)$ are determined entirely at tree level, with quantum corrections from diagrams of the form \eqref{bdy-tree}.

We can't immediately use the same argument for $\CA^!=\text{End}(\CB_\infty)$, because the 3d lift of $\CB_\infty$ may be a nontrivial defect near infinity on $\cp^1$, as described in Section \eqref{sec:cp1}. If the defect is nontrivial, the bulk propagator would need to be slightly modified, and the modification could ruin the symmetry that led to a vanishing of \eqref{loop-3d}.  We'll use two alternative analyses to establish that there's no one-loop correction in $\CA^!$:
\begin{itemize}
\item Here, we'll do a calculation after reduction to 2d, which is slightly non-rigorous due to needing to sum infinite modes running in the loop. We'll explain that the sums look essentially identical for $\CA$ and $\CA^!$; and since the correction to $\CA$ vanishes by \eqref{loop-3d}, so must the correction to $\CA^!$.
\item In Section \ref{sec:proofs} we'll prove (by checking an MC equation) that our proposed tree-level computation of $\CA^!$ is actually the correct Koszul-dual of $\CA$.
\end{itemize}

The 2d analysis proceeds as follows. Consider diagram \eqref{bdy-loop}, and suppose (WLOG) we contract some $\mb p_i,\mb x^j$ in the top vertex with their $\mb x^i,\mb p_j$ conjugated in the bottom vertex. Expanding fields into modes, this looks like
\be \raisebox{-.4in}{\includegraphics[width=1.7in]{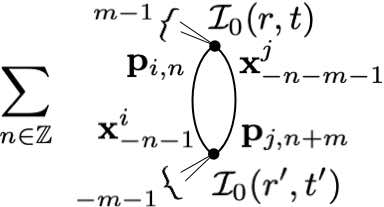}} =  \sum_{n\in \Z} \langle \mb p_{i,n}(r,t) \mb x^i_{-n-1}(r',t')\rangle  \langle \mb p_{j,n+m}(r',t') \mb x^j_{-n-m-1}(r,t) \rangle  \label{loop-2d}\ee
Namely, at each vertex $\CI_0=\oint\CI$ the mode-numbers of $\mb x,\mb p$ and the total `external' legs must sum to $-2$; and along each propagator the mode-numbers must sum to $-1$. We keep fixed the external mode number, here denoted $m$ (as these external edges will be further contracted with local operators whose $A_\infty$ product we're computing). We are left with a single of modes $n\in \Z$ running through the loop.

In the absence of boundary conditions, the 2d propagator in Lorentz gauge is
\be \langle \mb p_{i,n}(r,t) \mb x^j_m(r',t')\rangle  = \tfrac1{2\pi} \delta_i{}^j \delta_{n+m,-1} P(r,t;r',r')\,,\ee
where
\be \label{prop-2d-Lor} P(r,t;r',r') := \text{`$d(\theta-\theta')$'} = \frac{(t-t')d(r-r')-(r-r')d(t-t')}{(t-t')^2+(r-r')^2}\,. \ee
Notably, $P$ is a one-form (in the odd variables $dr,dt,dr',dt'$) and is symmetric $P(r,t;r',t')=P(r',t';r,t)$.

In the presence of the $\CB_0$ boundary, the propagator must be modified (essentially choosing a different gauge) so that it's compatible with the vanishing of all $n\geq 0$ modes of $\mb x$ and $\mb p$. An easy way to do this is the following. Let $\sigma$ (respectively $\sigma'$) a reflection across the boundary in the $r,t$ coordinates (resp. the $r',t'$ coordinates). It acts by pull-back on the one-form propagator. So if the boundary lies at $r=0$, we have $\sigma P(r,t;r',t'):=P(-r,t;r',t')$ and $\sigma'P(r,t;r',t'):= P(r,t;-r',t')$. Then we implement the correct boundary conditions on the propagator in the presence of $\CB_0$ by anti-symmetrizing with respect to reflections, for the fields that should vanish at the boundary:
\be  \langle \mb p_{i,n}(r,t) \mb x^j_m(r',t')\rangle  = \tfrac1{2\pi} \delta_i{}^j \delta_{n+m,-1}\times\begin{cases} 
  \tfrac12(\text{id}-\sigma) P(r,t;r',r') & n \geq 0 \\
  \tfrac12(\text{id}-\sigma') P(r,t;r',r') & m \geq 0\,. \end{cases} \ee
For $\CB_\infty$, the opposite $n<0$ modes must vanish, so instead we have
\be \langle \mb p_{i,n}(r,t) \mb x^j_m(r',t')\rangle  = \tfrac1{2\pi} \delta_i{}^j \delta_{n+m,-1}\times\begin{cases} 
  \tfrac12(\text{id}-\sigma) P(r,t;r',r') & n < 0 \\
  \tfrac12(\text{id}-\sigma') P(r,t;r',r') & m < 0\,. \end{cases} \ee
The result for \emph{both} $\CB_0$ and $\CB_\infty$ and for all modes can be combined in the single expression, if we notice that $\sigma P = -\sigma' P$ (this follows easily from \eqref{prop-2d-Lor})). Then 
\be  \langle \mb p_{i,n}(r,t) \mb x^j_m(r',t')\rangle =  \langle \mb x^j_{n}(r,t) \mb p_{i,m}(r',t')\rangle = \tfrac1{4\pi} \delta_i{}^j \delta_{n+m,-1} (\text{id}-\epsilon_n\epsilon_\CB\sigma )P(r,t;r't')\,. \label{prop-2d-bdy} \ee
where $\epsilon_n=1$ if $n\geq 0$ and $-1$ if $n<0$; and $\epsilon_B=1$ for $\CB_0$ and $-1$ for $\CB_\infty$.

Using \eqref{prop-2d-bdy} we find that the loop contribution \eqref{loop-2d} becomes
\begin{align} &  \sum_{n\in \Z} \langle \mb p_{i,n}(r,t) \mb x^i_{-n-1}(r',t')\rangle  \langle \mb x^j_{-n-m-1}(r,t)  \mb p_{j,n+m}(r',t')\rangle  \notag \\
& \qquad =\frac{1}{16\pi^2}  \sum_{n\in \Z} (P-\epsilon_n\epsilon_\CB \sigma P)(P-\epsilon_{-n-m-1}\epsilon_\CB \sigma P)  \notag \\
&\qquad  = \frac{1}{16\pi^2} \epsilon_\CB\, P\sigma P\sum_{n\in \Z} (\epsilon_n - \epsilon_{-n-m-1})  \notag \\
&\qquad = \frac{1}{8\pi^2} \epsilon_\CB \,P\sigma P\times \Big( \sum_{n\geq \max(0,-m)} 1 -  \sum_{n< \min(0,m)} 1\Big)\,. \label{eB-1loop}
\end{align}
Here $P$ is shorthand for $P(r,t;r',t')$; and we have used $P^2=(\sigma P)^2 = 0$ to simplify.

The virtue of \eqref{eB-1loop} is that is demonstrates that the 1-loop correction to $\CB_0$ and $\CB_\infty$ boundary algebras must be identical, up to the sign contained in $\epsilon_{\CB}$. However, we already know (by the 3d analysis in \eqref{loop-3d}) that the correction for $\CB_0$ must vanish. Thus the correction for $\CB_\infty$ vanishes as well. In terms of the formula \eqref{eB-1loop}, the vanishing should come about through a regularization of the sums in the right-most factored, so that the two divergent pieces with opposite signs exactly cancel each other. 

\subsection{Tree-level $A_\infty$ algebras}
\label{sec:Ainf-gen}

We can now describe the general form of the perturbative boundary algebras $\CA$ and $\CA^!$. Let's start with $\CA$. It is generated by $\{x_n^i,p_{i,n}\}_{n<0}$ as in a free theory \eqref{AA!-free}, with (say) $x$'s bosonic and $p$'s fermionic and charges
\be \label{A-charges}  \begin{array}{c|ccc} & R\;\text{(coh.)} & J\;\text{(spin)} & F\;\text{(parity)} \\\hline
 x^i_n & R_i & J_i -n-1 &  \text{even} \\
 p_{i,n} & 1-R_i & -J_i -n  & \text{odd} \end{array}
 \ee
where $J_i=\tfrac12 R_i$ if this comes from the HT twist of a 3d $\CN=2$ QFT.

For a free theory, the algebra is just the graded-commutative algebra $\C[x_n^i,p_{i,n}]$. Interactions correct this by adding $A_\infty$ operations, and from Sections \ref{sec:bdy-nonren} and \ref{sec:bdy-loop} we now know that the only diagrams to consider are the tree-level diagrams \eqref{bdy-tree} with a single bulk vertex $\CI_0$ and any number $d$ of  boundary vertices. Diagrams with $d$ boundary vertices add an $m_d$ operation to the $A_\infty$ algebra. An explicit computation of these diagrams, in a 2d B-model in twisted BV formalism, can be found in the lecture notes \cite{PIMS}.

Let's use $\{\mb Z^a\}$ to denote all 2d superfields (both $\mb x^i_n$ and $\mb p_{j,m}$ for all $i,j,n,m$) and $\hat{\mb Z}_a$ to denote their conjugates under the symplectic pairing in the kinetic term; \eg\
\be \mb Z^a = \mb x^i_{n} \quad \Rightarrow\quad \hat{\mb Z}_a = \mb p_{i,-n-1}\,;\qquad\text{or}\quad
 \mb Z^a = \mb p_{j,m} \quad\Rightarrow\quad \hat{\mb Z}_a = \mb x^j_{-m-1}\,. \ee
Then 2d propagators pair $\langle \mb Z^a  \hat{\mb Z}_b\rangle \sim \delta^a{}_b P$ (with additional projections as in \eqref{prop-2d-bdy} in the presence of boundary conditions). In the Feynman diagram \eqref{bdy-tree}, an insertion of $d$ boundary vertices $\mb Z^{a_1},...,\mb Z^{a_d}$ contracts with conjugates $\hat{\mb Z}_{a_1},...,\hat{\mb Z}_{a_d}$ in the bulk interaction vertex $\CI_0$ to produce a correction to $m_d(Z^{a_1},...,Z^{a_d})$ given by
\begin{align}
&m_d(Z^{a_1},...,Z^{a_d})  = \raisebox{-.5in}{\includegraphics[width=2in]{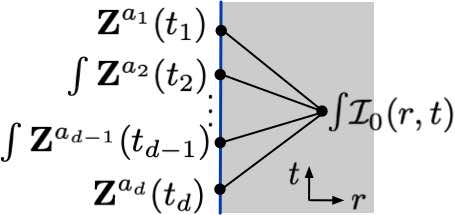}} \notag \\
&\hspace{.2in} = \frac{1}{d!} \sum_{\text{contractions}} \int_{\Delta^{d-2}\times (\R\times \R_+)}  \mb Z^{a_1}(t_1) \cdots \mb Z^{a_d}(t_d) \; :\CI_0(r,t): \notag \\
&\hspace{.2in} = \frac{(-1)^s}{d!} \int_{\Delta^{d-2}\times (\R\times \R_+)} P(0,t_1;r,t)\cdots P(0,t_d;r,t) \frac{\pd}{\pd \hat{\mb Z}_{a_1}}\cdots\frac{\pd}{\pd \hat{\mb Z}_{a_d}} :\CI_0(r,t): \label{bdy-treeZ}
\end{align}
where the sign $s=F(\hat Z_{a_{d-1}})+F(\hat Z_{a_{d-3}})+F(\hat Z_{a_{d-5}})+\ldots$ is a sum of fermion numbers of every second operator, and is due to commuting operators in $\CI_0$ past pairs of $\mb Z^a\mb Z_a$'s in order to get the propagators/contractions to appear in the order shown. The double-dots $::$ on $\CI_0$ denote normal-ordering; in particular, there are no self-contractions within $\CI_0$. The integral (over a half-space for the bulk insertion, and over a simplex $\Delta^{d-2}$ for all order-preserving configurations of intermediate boundary vertices) is convergent.

Up to a factor that can be absorbed in the normalization of $\CI_0$, one now finds corrections to the $m_d$ operations give by
\be \label{md-def} \CA:\qquad 
\begin{array}{r@{\;}c@{\;}l}
Q(Z^a):=m_1(Z^a) &=&\ds \frac{\pd}{\pd \hat Z_a} \CI_0 \bigg|_{\CB_0} \\[.3cm]
\delta m_2(Z^a,Z^b) &=&\ds (-1)^{F(\hat Z_a)} \frac12 \frac{\pd}{\pd \hat Z_a}\frac{\pd}{\pd \hat Z_b} \CI_0 \bigg|_{\CB_0}  \\[.3cm]
m_3(Z^a, Z^b, Z^c) &=&\ds (-1)^{F(\hat Z_b)} \frac16 \frac{\pd}{\pd \hat Z_a}\frac{\pd}{\pd \hat Z_b}\frac{\pd}{\pd \hat Z_c} \CI_0  \bigg|_{\CB_0} \\
&\vdots& \\
m_d(Z^{a_1},...,Z^{a_d}) &=&\ds (-1)^{F(\hat Z_{a_{d-1}})+F(\hat Z_{a_{d-3}})+...} \frac{1}{d!} \frac{\pd}{\pd \hat Z_{a_1}}\cdots \frac{\pd}{\pd \hat Z_{a_d}} \CI_0 \bigg|_{\CB_0} \,.
\end{array}
\ee
Note that the interaction vertex $\CI_0$ contains both positive and negative modes. The boundary variables $Z^{a_i}$ only contain negative modes (as appropriate on $\CB_0$), so the conjugate variables $\hat Z_{a_i}$ only contain positive modes; and the notation `$...\big|_{\CB_0}$' means that we should set all positive modes to zero (evaluating on the boundary condition $\CB_0$) after taking derivatives.

All operations except $m_2$ (the product) were zero in the free-field algebra $\C[x^i_n,p_{i,n}]_{n<0}$. Therefore, \eqref{md-def} simply defines what the $m_{d\neq 2}$ operations are. The product is graded-commutative in $\C[x^i_n,p_{i,n}]_{n<0}$, and \eqref{md-def} deforms it by introducing a nonzero graded commutator. Using $F(Z^a)+F(\hat Z_0)=1$, we find
\be [Z^a,Z^b] := m_2(Z^a, Z^b) - (-1)^{F(Z^a)F(Z^b)} m_2(Z^b,Z^a) = (-1)^{F(Z^a)+1}  \frac{\pd}{\pd \hat Z_a}  \frac{\pd}{\pd \hat Z_b}\CI_0\,. \label{md-comm} \ee

It is not at all obvious that operations \eqref{md-def} actually define the structure of an $A_\infty$ algebra. Indeed, for arbitrary interaction $\CI_0=\oint \CI$, they \emph{won't}! The condition that \eqref{md-def} consistently define an $A_\infty$ algebra is equivalent to the Maurer-Cartan equation (or the quantum BV master equation) for the 3d bulk interaction $\CI$; it generalizes the semi-classical condition $\{\int \CI,\int\CI\}_{\text{BV}}=0$ from Section \ref{sec:3dHT} and is required for a full perturbative quantization. The condition is guaranteed to hold for HT twists of 3d $\CN=2$ gauge/matter theories \cite{GRW-quantization,WilliamsWang}.

Let's also note that the 3d interaction has charges  $R(\CI)=2$, $J(\CI)=1$, $F(\CI)=0$, which implies that $\CI_0 = \oint \CI$ has $R(\CI_0)=2$, $J(\CI_0)=0$, $F(\CI_0)=0$. Moreover, 2d conjugate pairs satisfy $R(Z^a)+R(\hat Z_a)=1$, $J(Z^a)+J(\hat Z_a)=0$, $F(Z^a)+F(\hat Z_a)=1$ (mod 2). This correctly implies that the $A_\infty$ operations obtained from Feynman diagrams have $R(m_d)=2-d$, $J(m_d)=0$, $F(m_d)=d$ (mod 2).

\subsubsection{Modification for $\CA^!$}
\label{sec:A!-signs}

In a free theory, the algebra of local operators on $\CB_\infty$ starts out as the graded commutative algebra $\C[x^i_n,p_{i,n}]_{n\geq 0}$, involving positive rather than negative modes, with the same charges as in \eqref{A-charges}. In an interacting theory, it's again corrected by tree-level diagrams.
If we rotate $\CB_\infty$ by $180^\circ$, and reorient it, to make it an upward-oriented left boundary condition at $r=0$ just like $\CB_0$, then its algebra of local operators would be given by \emph{exactly the same} calculation as in \eqref{bdy-treeZ}, leading to the formulas in \eqref{md-def}. The only difference would be that the boundary operators $Z^a$ contain positive modes, the conjugates $\hat Z_a$ contain negative modes, and we evaluate $(...)\big|_{\CB_\infty}$ after taking derivatives, setting negative modes to zero.

We do, however, need to compensate for relative positions/orientations, since we want to view $\CB_\infty$ as the right boundary condition shown in \eqref{AABB} in order for Koszul duality to work out. This amounts to 1) taking the ``opposite'' of the naive boundary algebra due to the flipped orientation; and 2) choosing an oriented volume measure for the integral in \eqref{bdy-treeZ} (and thus a corresponding order for the propagators) that's appropriate for the right boundary.

Altogether, this will just introduce some extra overall signs in the $m_d$ operations, which depend in an intricate way on the fermion numbers of the fields involved. Rather than giving a universal formula for the signs here, we will state what they are for examples we care about: HT twists of 3d $\CN=2$ gauge theories with CS terms, matter, and superpotentials. Then we'll justify in Section \ref{sec:proofs} that the signs are correct, by verifying the MC equation for the universal MC element in $\CA^!\otimes \CA$.

There is one other potential difference between $\CA^!$ and $\CA$. Both boundary algebras could acquire an anomaly, turning them mathematically into curved $A_\infty$ algebras, from a diagram with zero boundary vertices:
\be \CA:\quad m_0 = \CI_0\big|_{\CB_0}\,,\qquad\text{or}\qquad \CA^!:\quad m_0=\pm \CI_0\big|_{\CB_\infty}\,. \ee
Setting non-negative modes to zero, the residue $\CI_0\big|_{\CB_0}=0$ will always vanish, so $\CA$ has no curvature term. However, if $\CI$ has a term linear in the 3d fields, \eg\ $\CI = \mb x^i+...$ for some $i$, then $\CI_0$ will not vanish on $\CB_\infty$, \eg\  $\CI_0\big|_{\CB_\infty} = \oint \CI\big|_{\CB_\infty}  = \mb x^i_0$, making $\CA^!$ a curved $A_\infty$ algebra. This is the same anomaly discussed in Footnote \ref{foot:anomaly}. We'll assume in the following that $\CI$ has no linear terms, though the abstract representation theory of $\CA^!$ would work equally well with a curvature term.

\subsection{Matter with superpotential: chiral critical locus}
\label{sec:Ainf-matter}

We'll now specialize the general computation of perturbative $\CA,\CA^!$ algebras from Section \ref{sec:Ainf-gen} to 3d $\CN=2$ theories. We begin with 3d Landau-Ginzburg models, with chiral multiplets $(\BX^i,\BPsi_i)_{i=1}^N$ and a polynomial superpotential of degree $d$, depending only on the $\BX$'s,
\be W(\mb X) = \tfrac12 a^{(2)}_{ij} \mb X^i\mb X^j + \tfrac 16 a^{(3)}_{ijk} \mb X^i\mb X^j\mb X^k + \ldots + \tfrac1{d!} a^{(d)}_{i_1...i_d} \mb X^{i_1}...\mb X^{i_d}\,,
\ee
with some fully symmetric tensors of coefficients $a^{(k)}$. Reduction to 2d gives $\CI_0=\oint W$,
\be \CI_0 = \tfrac12 a^{(2)}_{ij} \hspace{-.1in}\sum_{n+m=-1} \hspace{-.1in} \mb X^i_n\mb X^j_m + \tfrac 16 a^{(3)}_{ijk}  \hspace{-.2in}\sum_{n+m+p=-2} \hspace{-.1in} \mb X^i_n\mb X^j_m\mb X_p^k + \ldots + \tfrac1{d!} a^{(d)}_{i_1...i_d} \hspace{-.3in}\sum_{n_1+...+n_d=1-d}  \hspace{-.3in}\mb X^{i_1}_{n_1}...\mb X^{i_d}_{n_d}\,. \ee

The algebra of bulk local operators $\CA$ now becomes a chiralized version of the derived critical locus of $W$. Applying \eqref{md-def}--\eqref{md-comm}, we find that its underlying vector space is the same as the graded-commutative algebra
\be \CA = \C[X^i_n,\psi_{i,n}]_{n<0}\,, \ee
but there are now nonzero operations involving the $\psi$'s, which take the general form
\begin{align} m_k(\psi_{i_1,n_1},...,\psi_{i_k,n_k}) &= \frac{1}{k!} \frac{\pd}{\pd X^{i_1}_{-n_1-1}}...\frac{\pd}{\pd X^{i_k}_{-n_k-1}} \CI_0(X)\bigg|_{\CB_0} \\
&= \frac{1}{k!}  \oint z^{n_1+...+n_k}  \pd_{X^{i_1}}...\pd_{X^{i_k}} W(X(z)) \Big|_{\CB_0}\ \notag \end{align}
for all $1\leq k < d$ (with $m_2$ correcting the graded-commutative product). Explicitly:
\begin{align}
Q(\psi_{i,n}) 
&= a^{(2)}_{ij} X^j_n + \frac12 a^{(3)}_{ijk} \hspace{-.1in} \sum_{m+p=n-1}\hspace{-.1in}  X^j_m X^k_p + \ldots + \frac{1}{(d-1)!} a^{(d)}_{ii_2...i_d}\hspace{-.3in} \sum_{n_2+...+n_d=n+2-d}\hspace{-.3in} X^{i_2}_{n_2}...X^{i_d}_{n_d} \notag \\
\tfrac12 [\psi_{i,n},\psi_{j,m}] 
 &= \frac12 a^{(3)}_{ijk} X^k_{n+m} +\ldots + \frac{1}{2!(d-2)!} a^{(d)}_{iji_3...i_d} \hspace{-.4in}\sum_{n_3+...+n_d=n+m+3-d}\hspace{-.4in} X^{i_3}_{n_3}...X^{i_d}_{n_d}  \notag \\
m_3(\psi_{i,n},\psi_{j,m},\psi_{k,p}) &= \frac{1}{3!}a^{(4)}_{ijkl} X^l_{n+m+p} +\ldots + \frac{1}{3!(d-3)!} a^{(d)}_{ijki_4...i_d} \hspace{-.4in}\sum_{n_4+...+n_d=n+m+p+4-d}\hspace{-.4in} X^{i_4}_{n_4}...X^{i_d}_{n_d}  \notag \\
&\hspace{.4in} \vdots \notag \\
m_{d-1}(\psi_{i_1,n_1},...,\psi_{i_d,n_d}) &= \frac{1}{(d-1)!} a^{(d)}_{i_1...i_d} X^{i_d}_{n_1+...+n_{d-1}} \label{A-W-modes}
\end{align}
Note that all the sums on the RHS are \emph{finite}, since the indices are restricted to be strictly negative. They could be interpreted as sums over lattice simplices. Higher operations $m_k$ for $k\geq d$ all vanish. Any operations involving $X$'s (or both $X$'s and $\psi$'s) vanish as well; the $X$'s remain standard commuting variables.

The relation between $\CA$ and the derived critical locus of $W$ is through a quotient operation. Let
\be \CA\to \CA^{(-1)} = \C[X^i_{-1},\psi_{i,-1}] \ee
be the map that sends $X^i_n,\psi_{i,n}\mapsto 0$ for $n\leq -2$. Inspecting the relations \eqref{A-W-modes}, one finds that this is a morphism of $A_\infty$ algebras, where the RHS is simply the graded-commutative dg algebra with differential
\be  Q(\psi_{i,-1}) = \frac{\pd}{\pd X^i_{-1}} W(X_{-1})\,, \ee
and vanishing higher operations. This makes $\CA^{(-1)}\simeq \text{dCrit}(W)$ isomorphic to the usual derived critical locus.

The Koszul-dual algebra $\CA^!$ is nearly identical in structure. It's generated by the non-negative modes
\be \CA^! = \C[X^i_n,\psi_{i,n}]_{n\geq 0}\,,\ee 
with additional relations among the odd variables $\psi_{i,n}$. After a twist by $(-1)^F$ as in Section \ref{sec:KD-summary}, we find (and will verify in Section \ref{sec:proofs}) that all signs in the relations can be neatly repackaged by simply flipping the sign of the superpotential $W\mapsto -W$. Thus the relations are
\begin{align} m_k(\psi_{i_1,n_1},...,\psi_{i_k,n_k}) &= - \frac{1}{k!} \frac{\pd}{\pd X^{i_1}_{-n_1-1}}...\frac{\pd}{\pd X^{i_k}_{-n_k-1}} \CI_0(X)\bigg|_{\CB_\infty} \\
&=  - \frac{1}{k!}  \oint z^{n_1+...+n_k}  \pd_{X^{i_1}}...\pd_{X^{i_k}} W(X(z)) \Big|_{\CB_\infty}\ \notag \end{align}
for all $1\leq k< d$; or explicitly
\begin{align}
-Q(\psi_{i,n})  &= a^{(2)}_{ij} X^j_n + \frac12 a^{(3)}_{ijk} \hspace{-.1in} \sum_{m+p=n-1}\hspace{-.1in}  X^j_m X^k_p + \ldots + \frac{1}{(d-1)!} a^{(d)}_{ii_2...i_d}\hspace{-.3in} \sum_{n_2+...+n_d=n+2-d}\hspace{-.3in} X^{i_2}_{n_2}...X^{i_d}_{n_d} \notag \\
-\tfrac12 [\psi_{i,n},\psi_{j,m}] 
 &= \frac12 a^{(3)}_{ijk} X^k_{n+m} +\ldots + \frac{1}{2!(d-2)!} a^{(d)}_{iji_3...i_d} \hspace{-.4in}\sum_{n_3+...+n_d=n+m+3-d}\hspace{-.4in} X^{i_3}_{n_3}...X^{i_d}_{n_d}  \notag \\
-m_3(\psi_{i,n},\psi_{j,m},\psi_{k,p}) &= \frac{1}{3!}a^{(4)}_{ijkl} X^l_{n+m+p} +\ldots + \frac{1}{3!(d-3)!} a^{(d)}_{ijki_4...i_d} \hspace{-.4in}\sum_{n_4+...+n_d=n+m+p+4-d}\hspace{-.4in} X^{i_4}_{n_4}...X^{i_d}_{n_d}  \notag \\
&\hspace{.4in} \vdots \notag \\
-m_{d-1}(\psi_{i_1,n_1},...,\psi_{i_d,n_d}) &= \frac{1}{(d-1)!} a^{(d)}_{i_1...i_d} X^{i_d}_{n_1+...+n_{d-1}} \label{A!-W-modes}
\end{align}
All the sums on the RHS are still finite, as all the modes are now restricted to be non-negative.

\subsubsection{Reformulation via fields}
\label{sec:W-fields}

It's convenient, especially for checking the relations of a dg-shifted Yangian in the next section, to repackage the above $A_\infty$ relations in terms of generating functions
\be \begin{array}{ll} \CA:\quad  &X^i(z)_+ := \sum_{n< 0} X^i_n z^{-n-1}\,,\qquad \psi_i(z)_+ := \sum_{n< 0} \psi_{i,n} z^{-n-1} \\[.2cm]
 \CA^!:\quad & X^i(z)_- := \sum_{n\geq 0} X^i_n z^{-n-1}\,,\qquad \psi_i(z)_- := \sum_{n\geq 0} \psi_{i,n} z^{-n-1}\,, \end{array}  \label{XPsi-gen} \ee
as introduced in Section \ref{sec:KD-Yangian}. Physically, $X^i(z)_+$ and $\psi_i(z)_+$ represent bulk local operators inserted at a point $z\in \C_z$, and a time coordinate that's been suppressed. The $A_\infty$ relations govern products and higher products in the locally constant time direction.

Let us also denote
\be X^i(z) := X^i(z)_+ + X^i(z)_-\,,\qquad \psi_i(z) := \psi_i(z)_+ + \psi_i(z)_-\,. \ee
Then the 2d interaction is
\be \CI_0 = \oint_y W(X(y)) := \text{Res}_{y=0} W(X(y)) = - \text{Res}_{y=\infty} W(X(y))\,.  \ee
Note that the only singularities in $W(y)$ are at $y=0$ and $y=\infty$, so the contour can be thought of as placed on any counterclockwise circle around the origin. Moreover, the symplectic pairing of conjugate variables $X^i_n$ and $\psi_{i,-n-1}$ can be written
\be  \sum_{n\in \Z} \delta X^i_n \wedge \delta \psi_{i,-n-1} = \oint_y \delta X(y) \wedge \delta \psi(y) = \oint_y\Big[ \delta X(y)_+ \wedge \delta \psi(y)_- + \delta X(y)_-\wedge \delta \psi(y)_+ \Big]\,. \ee

Now, for $\CA$, we have
\be Q( \psi_i(z)_+) = \frac{\delta}{\delta X^i(z)_-} \CI_0 \bigg|_{\CB_0} = \oint_y \frac{\pd}{\pd X^i(y)_-}W(X(y))\delta(y-z)\bigg|_{\CB_0} =  \oint_{y>z} \frac{\pd_{X^i_+}W(X(y)_+)}{y-z}\,.  \ee
In the second step, we've rewritten a functional derivative as a ordinary derivative times a delta function; and in the final step we note that after restricting to $\CB_0$, so the function $\pd_{X^i_+}W(X(y)_+)$ only has a singularity at $y=\infty$, the delta function can be represented by
\be \delta(y-z) = \frac{1}{y-z} =  \sum_{n=0}^\infty \frac{z^n}{y^{n+1}} \ee
in the expansion regime $|y|>|z|$. Algebraically, the final integral just takes the residue at $y=z$; it can also be thought of as a contour integral on a circle with $|y|>|z|$. The result of the integral is of course $Q( \psi_i(z)_+)= \pd_{X^i_+}W(X(z)_+)$, in agreement with \eqref{A-W-modes}. 

The remaining operations can now be recast the same way, and one can verify algebraically that the answer is equivalent to \eqref{A-W-modes}:
\begin{align}   Q( \psi_i(z)_+)& =  \oint_{y>z} \frac{\pd_{X^i_+}W(X(y)_+)}{y-z} = \pd_{X^i_+}W(X(z)_+)\,, \notag \\
[\psi_i(z)_+,\psi_j(w)_+] &= \oint_{y>z,w}  \frac{\pd_{X^i_+}\pd_{X^j_+}W(X(y)_+)}{(y-z)(y-w)} = \frac{\pd_{X^i_+}\pd_{X^j_+}W(X(z)_+)-\pd_{X^i_+}\pd_{X^j_+}W(X(w)_+)}{z-w}\,, \notag \\
\hspace{-.3in}  m_n(\psi_{i_1}(z_1)_+,...,\psi_{i_n}(z_n)_+) &= \frac{1}{n!}\oint_{y>z_1,...,z_n} \frac{\pd_{X^{i_1}_+}...\pd_{X^{i_n}_+}W(X(y)_+)}{(y-z_1)...(y-z_n)} = \frac{1}{n!} \sum_{j=1}^n  \frac{\pd_{X^{i_1}_+}...\pd_{X^{i_n}_+}W(X(z_j)_+)}{\prod_{k\neq j}(z_j-z_k)}\,. 
\label{A-W-fields}
\end{align}
A special case of this is the equal-space commutator
\be [\psi_i(z)_+,\psi_j(z)_+] = \pd_z \pd_{X^i_+}\pd_{X^j_+}W(X(z)_+)\,. \ee
This relation was found in \cite[Eqn. 5.18]{CDGbdry} via 3d bulk Feynman diagrams.

For the Koszul-dual $\CA^!$, the basic functional manipulations look like 
\be -Q( \psi_i(z)_-) = \frac{\delta}{\delta X^i(z)_+} \CI_0 \bigg|_{\CB_\infty} = \oint_y \frac{\pd}{\pd X^i(y)_+}W(X(y))\delta(y-z)\bigg|_{\CB_\infty} =  \oint_{y<z} \frac{\pd_{X^i_-}W(X(y)_-)}{z-y}\,.  \ee
In particular, once we restrict to $\CB_\infty$ so that $W(X(y)_-)$ is only singular at the origin, the delta function can be represented by
\be \delta(y-z) = \frac{1}{z-y} = \sum_{n=0}^\infty \frac{y^n}{z^{n+1}}\,, \ee
expanded in the regime $|y|<|z|$. The final integral picks up the negative of the residue at $y=z$. Taking into account the sign change $W\mapsto -W$, the operations now look like
\begin{align}   Q( \psi_i(z)_-)& = - \oint_{y<z} \frac{\pd_{X^i_-}W(X(y)_-)}{z-y} = -\pd_{X^i_-}W(X(z)_-)\,, \notag \\
[\psi_i(z)_-,\psi_j(w)_-] &= - \hspace{-.1in} \oint_{y<z,w}  \hspace{-.1in} \frac{\pd_{X^i_-}\pd_{X^j_-}W(X(y)_-)}{(z-y)(w-y)} = -\frac{\pd_{X^i_-}\pd_{X^j_-}W(X(z)_-)-\pd_{X^i_-}\pd_{X^j_-}W(X(w)_-)}{w-z}\,, \notag \\
\hspace{-.3in}  m_n(\psi_{i_1}(z_1)_-,...,\psi_{i_n}(z_n)_-) &= -\frac{1}{n!}\oint_{y<z_1,...,z_n} \frac{\pd_{X^{i_1}_-}...\pd_{X^{i_n}_-}W(X(y)_-)}{(z_1-y)...(z_n-y)} = -\frac{1}{n!} \sum_{j=1}^n  \frac{\pd_{X^{i_1}_-}...\pd_{X^{i_n}_-}W(X(z_j)_-)}{\prod_{k\neq j}(z_k-z_j)}\,. 
\label{A!-W-fields}
\end{align}

\subsection{Gauge theory and loop algebras}
\label{sec:Ainf-gauge}

We'll now consider HT-twisted perturbative gauge theory, with Lie algebra $\fg$ and matter representation $V$. We allow a Chern-Simons level, but assume for the moment that the matter superpotential vanishes, so the action takes the form (\cf\ Section \ref{sec:3dN2})
\be S = \int_{\C\times \R} \BA d' \BB +\BPsi d' \BX + \CI\,,\quad \CI = \tfrac12 f^{ab}{}_c \BA_b\BA_a \BB^c +\tfrac12 k^{ab} \BA_a \pd\BA_b + (\varphi^a_V)^i{}_j \BPsi_i \BA_a \BX^j\,. \ee
It will be easy to add a matter superpotential at the end.

To simplify the structure a bit, it's useful to observe that this perturbative gauge theory can be rewritten as gauge theory for a semi-direct product algebra
\be \fh:= \fg\ltimes V\,,\qquad \text{basis}\;\; \{t^I\}=\{t^a\}\cup \{v^i\}\,, \label{hgV} \ee
with basis $\{t^I\}$ combining the bases of $\fg$ and $V$. The fields of the new gauge theory are
\be \BA_{I=a}=\BA_a\,,\quad \BA_{I=i} = \BPsi_i\,,\qquad \BB^{I=a}=\BB^a\,,\qquad \BB^{I=i} = \BX^i\,, \ee
and the structure constants $f^{IJ}{}_K=-f^{JI}{}_K$ and level $k^{IJ}=k^{JI}$ of $\fh$ are such that
\be f^{ab}{}_c = f^{ab}{}_c\,,\quad f^{ai}{}_j = (\varphi_V^a)^i{}_j \,,\quad f^{ij}{}_a=f^{ij}{}_k=0\,;\qquad k^{ab}=k^{ab}\,,\quad k^{ai}=k^{ij}=0\,. \label{k-hgV} \ee
In this formulation, the action is simply
\be S = \int_{\C\times \R} \BA_I d'\mb \BB^I +\CI\,,\qquad \CI = \tfrac12 f^{IJ}{}_K \BA_J\BA_I \BB^K +\tfrac12 k^{IJ}\BA_I\pd \BA_J\,, \ee
with reduced 2d interaction term
\be \CI_0 = \tfrac12 f^{IJ}{}_K \hspace{-.2in} \sum_{m+n+p=-2} \hspace{-.2in}\BA_{J,m}\BA_{I,n}\BB^K_p + \tfrac12 k^{IJ} \hspace{-.1in}\sum_{n+m=-2}\hspace{-.1in} (n+1)\, \BA_{I,n}\BA_{J,m}\,.\ee

Since the interaction is cubic, both $\CA$ and $\CA^!$ will be dg algebras, with $m_3$ and higher $A_\infty$ operations vanishing. The algebras are generated by fermions $c_{I,n}$ (the leading components of the superfields $\BA_{I,n}$, so denoted because they are BRST ghosts) and bosons $B^I_n$ (the leading components of $\BB^I_n$),
\be \CA = \C[B^I_n,c_{I,n}]_{n<0}\,,\qquad \CA^! =  \C[B^I_n,c_{I,n}]_{n\geq 0}\,. \ee
From \eqref{md-def}--\eqref{md-comm} we find that the relations in $\CA$ are
\begin{subequations} \label{gauge-A-modes}
\be
 \begin{array}{c} [B^I_n,B^J_m] = -f^{IJ}{}_K B^K_{n+m}\,, \qquad [B^I_n,c_{J,m}] = f^{IK}{}_J c_{K,m+n}\,, \\[.2cm]
\ds Q(c_{I,n}) = -\tfrac12 f^{IJ}{}_K \hspace{-.1in} \sum_{m+p=n-1}\hspace{-.1in} c_{I,m}c_{J,p}\,,\qquad Q(B^I_n) = -n\, k^{IJ} c_{J,n-1} - f^{IJ}{}_K \hspace{-.1in}\sum_{m+p=n-1}\hspace{-.1in} :c_{J,m}B^K_p:\,.
 \end{array}
\ee 
The normal ordering in the final equation, coming from normal ordering of the interaction vertex $\CI_0$, is important, since $c$ and $B$ don't commute. Indeed, for each $m+p=n-1$ we have $:c_{J,m}B_p^K\!:\,=c_{J,m}B_p^K+\tfrac12 f^{KL}{}_J c_{L,n-1} = B_p^Kc_{J,m} - \tfrac12 f^{KL}{}_J c_{L,n-1}$. There are $|n|=-n$ terms in the sum, whence
\begin{align}  Q(B^I_n) & \textstyle = -n(k^{IJ}+h^{IJ}) c_{J,n-1} - f^{IJ}{}_K \hspace{0in}\sum_{m+p=n-1}\hspace{0in} c_{J,m}B^K_p
 \notag \\
 & \textstyle = -n(k^{IJ}-h^{IJ}) c_{J,n-1} - f^{IJ}{}_K \hspace{0in}\sum_{m+p=n-1}\hspace{0in}B^K_p c_{J,m}\,, \end{align}
\end{subequations}
where
\be h^{IJ} := \tfrac12 f^{IK}{}_L f^{JL}{}_K\,. \ee
is a generalization of the dual Coxter ``number'' for $\fh$, expressed as an invariant bilinear form. When $\fg$ is a simple Lie algebra, we could more simply write
\be h^{ab} = \big(h^\vee+\tfrac12 I_2(V)\big)\delta^{ab}\,,\qquad h^{ai}=h^{ij} = 0\,, \ee
where $h^\vee$ is the usual dual Coxeter number of $\fg$, $I_2(V)$ is the quadratic index of the matter representation, and $\delta^{ab}$ is the Killing form on $\fg$.

The Koszul-dual $\CA^!$ should take the same form, up to replacing negative modes with positive modes, and swapping some signs. We claim that the correct choice of signs --- after a twist by $(-1)^F$ as in Sec \ref{sec:KD-summary} --- gives
\begin{subequations} \label{gauge-A!-modes}
\be
 \begin{array}{c} [B^I_n,B^J_m] = f^{IJ}{}_K B^K_{n+m}\,, \qquad [B^I_n,c_{J,m}] = -f^{IK}{}_J c_{K,m+n}\,, \\[.2cm]
\ds Q(c_{I,n}) = \tfrac12 f^{IJ}{}_K \hspace{-.1in} \sum_{m+p=n-1}\hspace{-.1in} c_{I,m}c_{J,p}\,,\qquad Q(B^I_n) = n\, k^{IJ} c_{J,n-1} + f^{IJ}{}_K \hspace{-.1in}\sum_{m+p=n-1}\hspace{-.1in} :c_{J,m}B^K_p:\,.
 \end{array}
\ee 
Namely, just as for matter with a superpotential, the sign of every operation is simply reversed. To undo the normal ordering, we now have $:c_{J,m}B_p^K\!:\,=c_{J,m}B_p^K-\tfrac12 f^{KL}{}_J c_{L,n-1} = B_p^Kc_{J,m} + \tfrac12 f^{KL}{}_J c_{L,n-1}$ for each of the $|n|=n$ terms with $m+p=n-1$, so
\begin{align}  Q(B^I_n) & \textstyle = n(k^{IJ}+h^{IJ}) c_{J,n-1} + f^{IJ}{}_K \hspace{0in}\sum_{m+p=n-1}\hspace{0in} c_{J,m}B^K_p
 \notag \\
 & \textstyle = n(k^{IJ}-h^{IJ}) c_{J,n-1} + f^{IJ}{}_K \hspace{0in}\sum_{m+p=n-1}\hspace{0in}B^K_p c_{J,m}\,. \end{align}
\end{subequations}

\subsubsection{Loop algebra, Lie algebra, and Wilson lines}
\label{sec:gauge-loop}

The Koszul-dual algebra \eqref{gauge-A!-modes} in gauge theory (with vanishing superpotential) has a nice geometric interpretation. Observe that there are successive quotients of dg algebras
\be   \CA^! \to \CA^!/(c_{I,n\geq 0}) \to \CA^!/(c_{I,n\geq 0},B^I_{n> 0}) \ee
The right-most algebra, containing only the zero-modes $\{B^I_0\}$ with zero differential and relations $[B^I_0,B^J_0]=f^{IJ}{}_K B^K_0$, is isomorphic to the enveloping algebra of the finite Lie algebra $\fh$. The middle algebra, containing all $\{B^I_{n\geq 0}\}$ with zero differential and relations $[B^I_n,B^J_m]=f^{IJ}{}_K B^K_{n+m}$, is isomorphic to the enveloping algebra of the positive loop algebra $\fh[\lambda]$, where `$\lambda$' is a formal loop parameter and we've identified $B^I_n\leftrightarrow B^I\lambda^n$.  $\CA^!$ itself is a deformation of (the enveloping algebra of) the loop algebra of the shifted \emph{cotangent} Lie algebra $(T^*[1]\fh)[\lambda]$. Thus, in summary
\be \begin{array}{ccccc}
   \CA^! &\to& \CA^!/(c_{I,n\geq 0}) &\to& \CA^!/(c_{I,n\geq 0},B^I_{n> 0}) \\
   \rotatebox{90}{$\approx$} & &  \rotatebox{90}{$\simeq$}  &&   \rotatebox{90}{$\simeq$}    \\
  U\big((T^*[1]\fh)[\lambda]\big) & \to  & U\big(\fh[\lambda]\big) &\to & U(\fh)
\end{array} \ee

Mathematically, the standard convention for the subalgebra of $\CA^!$ generated by $\{c_{I,n}\}_{n\geq 0}$, which is a copy of Chevalley-Eilenberg cochains for the negative loop algebra $\lambda^{-1}\fh[\lambda^{-1}]$, would swap the signs of all $c_I$'s. In particular, $ Q(c_{I,n}) =  \sum_{m+p=n-1}(- c_{I,m} \tfrac12 f^{IJ}{}_K)c_{J,p}$ as well as $Q(B^I_n) = ...+  \sum_{m+p=n-1} :c_{J,m}f^{JI}{}_KB^K_p:$, so that $Q$ acts as a gauge transformation with coefficient $c$. The standard math convention is related to our current one by undoing our $(-1)^F$ twist, which precisely negates all the $c_I$'s.

Physically, an important and immediate consequence of the map $\CA^!\to U(\fh)$ is the existence of Wilson lines. The map can be composed with a further quotient
\be \CA^! \to U(\fh) \to U(\fg) \ee
Wilson lines are the finite-dimensional representations of $\CA^!$ that factor through this quotient, thus determined by finite-dimensional representations of $\fg$. Given such a representation $\rho:\fg\to \text{End}(W)$ on a vector space $W$, extending to
\be \rho:\CA^!\to \text{End}(W)\,,\qquad \rho(c_{I,n\geq 0})=\rho(B^I_{n>0}) = \rho(B^i_0)=0\,,\quad \rho(B^a_0)=\rho(t^a) \ee
the Wilson line itself is constructed by the MC element
\be   \rho(\mu) = \rho\Big( \sum_{n\geq 0} B^I_n c_{I,-n-1}+c^I_n B_{I,-n-1}\Big)  = \rho(t^a) c_{a,-1}\,, \ee
in agreement with \eqref{Wilsonline}.

When there is matter, we also see that any representation of $\fh$ will define a line operator as well, through the map $\CA^!\to U(\fh)$. Such lines combine Wilson lines with matter vortices, induced by couplings of the form $\rho(t^i)c_{i,-1}=\rho(v^i)\psi_{i,-1}$, which set
\be X^i(z) \sim \frac{\rho(v^i)}{z} +\text{regular} \ee
near the support of a line.

\subsubsection{Reformulation via fields}
\label{sec:gauge-fields}

We can recast the relations in $\CA$ and $\CA^!$ in terms of generating functions
\be \begin{array}{ll} \CA:\quad  &B^I(z)_+ := \sum_{n< 0} B^I_n z^{-n-1}\,,\qquad c_I(z)_+ := \sum_{n< 0} c_{I,n} z^{-n-1}\,, \\[.2cm]
 \CA^!:\quad & B^I(z)_- := \sum_{n\geq 0} B^I_n z^{-n-1}\,,\qquad c_I(z)_- := \sum_{n\geq 0} c_{I,n} z^{-n-1}\,, \end{array}  \label{Bc-gen} \ee
just like we did in Section \ref{sec:W-fields} for matter with a superpotential. Relations \eqref{gauge-A-modes} in $\CA$ translate to
\begin{subequations} \label{gauge-A-fields}
\be \begin{array}{c}
\ds [B^I(z)_+,B^J(w)_+] = -f^{IJ}{}_K \frac{B^K(z)_+-B^K(w)_+}{z-w}\,,\qquad [B^I(z)_+,c_J(w)_+] = f^{IK}{}_J \frac{c_K(z)_+-c_K(w)_+}{z-w} \\[.4cm]
Q\, c_I(z)_+ =-\tfrac12 f^{IJ}{}_K c_I(z)_+c_J(z)_+\,,\qquad Q\, B^I(z)_+ = k^{IJ}\pd c_J(z)_+ - f^{IJ}{}_K :c_J(z)_+B^K(z)_+:
\end{array}\ee
We resolve the normal ordering using the equal-space commutator $[B^I(z)_+,c_J(z)_+]=f^{IK}{}_J\pd c_K(z)$, which gives $:\!c_J(z)_+B^K(z)_+\!:\,= c_J(z)_+B^K(z)_+ +\tfrac12 f^{KL}{}_J \pd c_L(z) = B^K(z)_+c_J(z)_+ -\tfrac12 f^{KL}{}_J \pd c_L(z)$, whence
\begin{align} Q\, B^I(z)_+ &= (k^{IJ}+h^{IJ})\pd c_J(z)_+ - f^{IJ}{}_K c_J(z)_+B^K(z)_+ \notag \\
&= (k^{IJ}-h^{IJ})\pd c_J(z)_+ - f^{IJ}{}_K B^K(z)_+c_J(z)_+\,.
\end{align}
\end{subequations}

Relations \eqref{gauge-A!-modes} in $\CA^!$ translate to
\begin{subequations} \label{gauge-A!-fields}
\be \begin{array}{c}
\ds [B^I(z)_-,B^J(w)_-] = f^{IJ}{}_K \frac{B^K(z)_--B^K(w)_-}{w-z}\,,\qquad [B^I(z)_-,c_J(w)_-] = -f^{IK}{}_J \frac{c_K(z)_--c_K(w)_-}{w-z} \\[.4cm]
Q\, c_I(z)_- =\tfrac12 f^{IJ}{}_K c_I(z)_-c_J(z)_-\,,\qquad Q\, B^I(z)_- = -k^{IJ}\pd c_J(z)_- + f^{IJ}{}_K :c_J(z)_-B^K(z)_-:
\end{array}\ee
The equal-space commutator still evaluates to $[B^I(z)_-,c_J(z)_-]=f^{IK}{}_J\pd c_K(z)$, giving $:\!c_J(z)_+B^K(z)_+\!:\,= c_J(z)_-B^K(z)_- +\tfrac12 f^{KL}{}_J \pd c_L(z) = B^K(z)_-c_J(z)_- -\tfrac12 f^{KL}{}_J \pd c_L(z)$,
\begin{align} Q\, B^I(z)_- &= -(k^{IJ}+h^{IJ})\pd c_J(z)_- + f^{IJ}{}_K c_J(z)_-B^K(z)_- \notag \\
&= -(k^{IJ}-h^{IJ})\pd c_J(z)_- + f^{IJ}{}_K B^K(z)_-c_J(z)_-\,.
\end{align}
\end{subequations}

\subsubsection{Adding a superpotential}
\label{sec:Ainf-full}

To finish, we re-introduce a superpotential $W(X)$ in the matter sector. The $A_\infty$ operations from Section \ref{sec:Ainf-matter} simply add to the gauge-theory operations above. The condition that $W$ is gauge-invariant is essential to obtain a consistent $A_\infty$ algebra.

We write down the explicit result for easy reference, using the generating-function formulation. We've now got to treat gauge and matter fields separately, so the generators of the algebras are $B^a(z)_\pm,c_a(z)_\pm,X^i(z)_\pm,\psi_i(z)_\pm$, with $+$ for $\CA$ and $-$ for $\CA^!$. In $\CA$ the relations are
\be \label{A-full}
\begin{array}{l}
\begin{array}{l}
  Q\,c_a(z)_+ = -\tfrac12 f^{ab}{}_c c_a(z)_+ c_b(z)_+   \\[.2cm]
  Q\,X^i(z)_+ = (\varphi_V^a)^i{}_j c_a(z)_+ X^j(z)_+ \\[.2cm]
  Q\,\psi_i(z)_+ = - (\varphi_V^a)^j{}_i c_a(z)_+ \psi_j(z)_+  + \pd_{X^i}W(X(z)_+)  \\[.2cm]
\end{array}
\begin{array}{l}
 [B^a(z)_+,B^b(w)_+] = -f^{ab}{}_c \frac{B^a(z)_+-B^b(w)_+}{z-w} \\[.1cm]
  [B^a(z)_+,c_b(w)_+] = f^{ac}{}_b \frac{c_c(z)_+ - c_c(w)_+}{z-w} \\[.1cm]
  [B^a(z)_+, X^i(w)_+] = -(\varphi^a_V)^i{}_j \frac{X^j(z)_+-X^j(w)_+}{z-w} \\[.1cm]
   [B^a(z)_+,\psi_i(w)_+] = (\varphi^a_V)^j{}_i \frac{\psi_j(z)_+-\psi_j(w)_+}{z-w} \\[.1cm]
   [X^i(z)_+,\psi^j(w)_+] = -(\varphi^a_V)^i{}_j \frac{c_a(z)_+-c_a(w)_+}{z-w}
 \end{array} \\[1.2cm]
 \;\; Q\,B^a(z)_+ = (k^{ab}+h^{ab})\pd c_b(z)_+ - f^{ab}{}_c c_b(z)_+B^c(z)_+ - (\varphi_V^a)^i{}_j \psi_i(z)_+X^i(z)_+ \\[.2cm]
 [\psi_i(z)_+,\psi_j(w)_+] = \frac{\pd_{X^i}\pd_{X^j} W(X(z)_+) - \pd_{X^i}\pd_{X^j} W(X(w)_+)}{z-w} \\[.2cm]
 m_n(\psi_{i_1}(z_1),...,\psi_{i_n}(z_n)) = \oint_{y>z_1,...,z_n} \frac{\pd_{X^{i_1}}...\pd_{X^{i_n}} W(X(y)_+)}{(y-z_1)...(y-z_n)} = \sum_{j=1}^n \frac{\pd_{X^{i_1}}...\pd_{X^{i_n}} W(X(z_j)_+)}{\prod_{k\neq j} (z_j-z_k)}
 \end{array}
\ee
with higher operations for $n\leq \text{deg}(W)-1$. 

In $\CA^!$, after a twist by $(-1)^F$, the relations have all signs reversed. In addition, the denominators representing delta functions in contour integrals are reversed and expanded in the opposite regime. Thus,
\be \label{A!-full}
\begin{array}{l}
\begin{array}{l}
  Q\,c_a(z)_- = \tfrac12 f^{ab}{}_c c_a(z)_- c_b(z)_-   \\[.2cm]
  Q\,X^i(z)_- = -(\varphi_V^a)^i{}_j c_a(z)_- X^j(z)_- \\[.2cm]
  Q\,\psi_i(z)_- =  (\varphi_V^a)^j{}_i c_a(z)_- \psi_j(z)_-  - \pd_{X^i}W(X(z)_-)  \\[.2cm]
\end{array}
\begin{array}{l}
 [B^a(z)_-,B^b(w)_-] = f^{ab}{}_c \frac{B^a(z)_--B^b(w)_-}{w-z} \\[.1cm]
  [B^a(z)_-,c_b(w)_-] = -f^{ac}{}_b \frac{c_c(z)_- - c_c(w)_-}{w-z} \\[.1cm]
  [B^a(z)_-, X^i(w)_-] = (\varphi^a_V)^i{}_j \frac{X^j(z)_--X^j(w)_-}{w-z} \\[.1cm]
   [B^a(z)_-,\psi_i(w)_-] = -(\varphi^a_V)^j{}_i \frac{\psi_j(z)_--\psi_j(w)_-}{w-z} \\[.1cm]
   [X^i(z)_-,\psi^j(w)_-] = (\varphi^a_V)^i{}_j \frac{c_a(z)_--c_a(w)_-}{w-z}
 \end{array} \\[1.2cm]
 \;\; Q\,B^a(z)_- = -(k^{ab}+h^{ab})\pd c_b(z)_- + f^{ab}{}_c c_b(z)_-B^c(z)_- + (\varphi_V^a)^i{}_j \psi_i(z)_-X^i(z)_- \\[.2cm]
 [\psi_i(z)_-,\psi_j(w)_-] = -\frac{\pd_{X^i}\pd_{X^j} W(X(z)_-) - \pd_{X^i}\pd_{X^j} W(X(w)_-)}{w-z} \\[.2cm]
 m_n(\psi_{i_1}(z_1),...,\psi_{i_n}(z_n)) = -\oint_{y<z_1,...,z_n} \frac{\pd_{X^{i_1}}...\pd_{X^{i_n}} W(X(y)_-)}{(z_1-y)...(z_n-y)} = -\sum_{j=1}^n \frac{\pd_{X^{i_1}}...\pd_{X^{i_n}} W(X(z_j)_-)}{\prod_{k\neq j} (z_k-z_j)}\,.
 \end{array}
\ee

\subsection{Trivializing translations in Chern-Simons theory}
\label{sec:TQFT-gauge}

With explicit formulas on hand, we finish by giving an example of how translations are trivialized in an HT QFT that's topologically invariant --- manifesting the structure discussed in Section \ref{sec:TQFT}.

We consider twisted $\CN=2$ gauge theory with any Lie algebra~$\fg$ and a \emph{nondegenerate} Chern-Simons level
\be  k \in (\text{Sym}^2\mathfrak g^*)^{\mathfrak g}\,,\qquad \det k \neq 0\,. \label{detk} \ee
We assume there's no matter. (Note that if matter is incorporated into an extended Lie algebra $\mathfrak g\ltimes V$ as in \eqref{hgV}, the extended Chern-Simons level \eqref{k-hgV} will always be degenerate, violating \eqref{detk}.) Perturbatively, this theory is topological, and equivalent to pure bosonic Chern-Simons at level $k'=k-h^\vee$ \cite{ACMV}; it admits Kac-Moody $\hat{\mathfrak g}_{k'}$  as a perturbative boundary vertex algebra \cite{Gwilliam:2019cbp, CDGbdry}.

Recall from Section \ref{sec:trans} that $\CA^!$ has translation isomorphisms $\tau_z=\text{exp}(z\,T)$, generated by a derivation $T$ that acts as
\be T(B^a(z)_-)= - \pd_z B^a(z)_-\,,\qquad T(c_a(z)_-) = -\pd_z c_a(z)_-\,. \ee
The derivation $T$ has degrees $R(T)=F(T)=0$ and $J(T)=1$.

Consider a second algebra derivation $\CL$ of $\CA^!$, given on generators by
\be
\CL (B^a(z)-)=0\,, \qquad \CL (c_{a}(z)_-)=(k^{-1})_{ab} B^b(z)_-\,.
\ee
Observe that $\CL$ has $R(\CL)=-1$, $F(\CL)=\text{odd}$, and $J(\CL)=1$. We claim that $[Q, \CL]=T$.

The proof is straightforward. When applied to $c_a(z)_-$, we have
\begin{align}
    [Q, \CL]c_a(z)_-&=(k^{-1})_{ab} QB^b(z)_-+\CL Qc_a(z)_-\\
    &=(k^{-1})_{ab}\lp -k^{bc}\pd_z c_c(z)_- + f^{bc}{}_d:B^d(z)_-c_c(z)_-:\rp
    +\CL \lp \tfrac{1}{2}f^{bc}{}_a c_b(z)_-c_c(z)_-\rp \notag
\end{align}
The second term above is equal to
\be
\tfrac{1}{2} f^{bc}{}_a (k^{-1})_{bd} B^d(z)_- c_c(z)_--\tfrac{1}{2} f^{bc}{}_a c_b(z)_- (k^{-1})_{cd} B^d(z)_-\,. \label{fBc}
\ee
Using the fact that $k^{-1}$ is $\mathfrak g$-invariant, we can write $f^{bc}{}_a (k^{-1})_{bd} =-f^{bc}{}_d (k^{-1})_{ab}$ and $-f^{bc}{}_a(k^{-1})_{cd} = f^{bc}{}_d (k^{-1})_{ac} = - f^{cb}{}_d (k^{-1})_{ac}$, letting us transform \eqref{fBc} into
\begin{align} & -\tfrac{1}{2} f^{bc}{}_d (k^{-1})_{ab} B^d(z)_- c_c(z)_--\tfrac{1}{2}  f^{cb}{}_d (k^{-1})_{ac} c_b(z)_-  B^d(z)_- \notag  \\
&=  -\tfrac{1}{2} (k^{-1})_{ab}  f^{bc}{}_d\big[B^d(z)_- c_c(z)_- + c_c(z)_-B^d(z)_-\big]  \notag \\ &= -(k^{-1})_{ab}  f^{bc}{}_d :B^d(z)_- c_c(z)_-:
\end{align}
This exactly cancels out the contribution $+ f^{bc}{}_d:B^d(z)_-c_c(z)_-:$ in the first term, and we find
\be
 [Q, \CL]c_a(z)_-=-\pd_z c_a(z)_-=Tc_a(z)_-.
\ee
A very similar calculation shows that $[Q, \CL]B^a(z)_-=-\pd_z B^a(z)_-=T B^a(z)_-$ as well.

\section{Dg-shifted Yangians in 3d $\CN=2$ theories}
\label{sec:proofs}

In this final section, we will give algebraic proofs that the perturbative Koszul-dual $A_\infty$ algebras $\CA^!$ in HT-twisted 3d $\CN=2$ QFT's do have the structure of dg-shifted Yangians, as defined in Section \ref{sec:KD-Yangian} and summarized in Section \ref{sec:Yangian-summary}.

The first half of the section will focus on pure matter with a polynomial superpotential, and the second half on gauge theory. We'll consider gauge theory with arbitrary matter and Chern-Simons level but with zero superpotential for simplicity; the results of the two halves can be easily combined to establish the general case.

In each type of theory, we will first show that the algebras $\CA^!$ defined in Sections \ref{sec:Ainf-matter}--\ref{sec:Ainf-gauge} really are the Koszul duals of local operators $\CA$, by proving that the universal Maurer-Cartan element
\be   \mu = \sum_{n\geq 0} \Big(x^i_n p_{i,-n-1}+p_{i,n}x^i_{-n-1}\Big) = \oint_u \Big[x^i(u)_- p_i(u)_+ + p_i(u)_-x^i(u)_+\Big] \in \CA^!\otimes \CA \label{proof-mu-gen} \ee
satisfies the Maurer-Cartan equation.  Note that we work in conventions where $\CA^!$ has been twisted by the automorphism $(-1)^F$; otherwise there would be minus sign in the second term of $\mu$.

We then consider the universal r-matrix, which takes the form
\be r(z) = - \oint_{s<z} \Big[ x^i(s)_- p_i'(s+z) - p_i(s)_- x^i{}'(s+z)_-\Big]  \in \CA^!\otimes \CA^! [\![z^{-1}]\!]\,, \label{proof-r}\ee
where $x^i,p_i$ are the bosonic and fermionic fields of the theory, respectively, and we use primes to distinguish the two copies of $\CA^!$. There's also an overall minus sign due to the $(-1)^F$ twist, which we were not careful about in Section \ref{sec:KD-Yangian}, but which will be important here. We will prove that the r-matrix satisfies a Maurer-Cartan equation in $ \CA^!\otimes \CA^! [\![z^{-1}]\!]$.

Finally, we will consider the sum of MC elements that controls OPE's of line operators,
\be \mu(z) + \mu'(0) + r(z)  \in \CA^!\otimes \CA^!\otimes \CA [\![z^{-1},z]\!]\,, \label{proof-sum}  \ee
where
\be \begin{array}{l} \mu(z) = \oint_u \big[x^i(u)_- p_i(u+z)_+ + p_i(u)_-x^i(u+z)_+\big] \\[.3cm]
 \mu'(0) =  \oint_u \big[x^i{}'(u)_- p_i(u)_+ + p_i'(u)_-x^i(u)_+\big]\,, \end{array} \ee
and $r(z)$ is as above. We'll prove that the sum \eqref{proof-sum} satisfies an MC equation. By the fundamental correspondece of Koszul duality \eqref{corresp-MC}, this is algebraically equivalent to the coproduct
\be \Delta_z: \CA^!\to \CA^!\otimes_{r(z)} \CA^!  [\![z^{-1},z] \ee
being a map of $A_\infty$ algebras, where $\Delta_z$ is the free-field coproduct, acting on generators as in \eqref{D-formula-gen}.

We summarize the results of this section in
\begin{Thm} \label{thm:N2}
The Koszul-dual algebras $\CA^!$ in perturbative HT-twisted 3d $\CN=2$ gauge theory (with CS terms and matter), and in $\CN=2$ matter with polynomial superpotential, have the full structure of dg-shifted Yangians.
\end{Thm}

\noindent\textbf{Caveat:}\label{caveat:MC}  We comment that one must interpret carefully the MC equation satisfied by $\mu(z) + \mu'(0) + r(z)$ in equation \eqref{proof-sum}, since the RHS of this equation is not really an algebra. However, this kind of issue is standard in the study of vertex algebras and spectral R-matrices. The standard solution to this problem is to consider the MC equation not as an equation in these algebras but on a tensor product of smooth modules. In that case, smoothness truncates either the negative powers of $z$ in $r(z)$, or the infinite positive powers of $z$ in $\mu(z)$, making the resulting expression into a Laurent series (in $z$ or $z^{-1}$, depending on whether we use the topology of $\CA$ or $\CA^!$). In the following proofs, any equality should be understood in this sense.

\subsection{Matter with superpotential}
\label{sec:Yangian-W}

Suppose we have $N$ chiral multiplets $(\BX^i,\BPsi_i)_{i=1}^N$ with a polynomial superpotential $W(\BX)$ of degree $d+1$. Recall from Section \ref{sec:W-fields} that our algebras are generated by the modes of the generating functions $X^i(z)_\pm,\psi_i(z)_\pm$, with $+$ for $\CA$ and $-$ for $\CA^!$, and the $A_\infty$ operation take the form
\be \label{A-W-forMC} \begin{array}{l}
\CA:\quad  \ds m_k(\psi_{i_1}(z_1)_+,...,\psi_{i_k}(z_k)_+) = \frac{1}{k!}\oint_{y>z_1,...,z_k} \frac{\pd_{X^{i_1}_+}...\pd_{X^{i_k}_+}W(X(y)_+)}{(y-z_1)...(y-z_k)} \\
\CA^!:\quad \ds m_k(\psi_{i_1}(z_1)_+,...,\psi_{i_k}(z_k)_+) = -\frac{1}{k!}\oint_{y<z_1,...,z_k} \frac{\pd_{X^{i_1}_+}...\pd_{X^{i_k}_+}W(X(y)_+)}{(z_1-y)...(z_k-y)} \end{array}
 \ee
for all $1\leq k\leq d$, with $m_n=0$ for $n>d$. As usual, $m_1=Q$ is the differential and $m_2$ represents a deformation of the graded-commutative multiplication already present in the free-field algebra, rather than the entire product. 

\subsubsection{The MC equation for $\mu$}
\label{sec:W-MC-mu}

We'd like to verify that the universal MC element $\mu = \oint_u \big[ \psi_i(u)_- X^i(u)_+ + X^i(u)_-\psi_i(u)_+\big]$ does indeed satisfy the $A_\infty$ MC equation in $\CA^!\otimes \CA$, thus proving that $\CA^!$ is in fact the Koszul dual of $\CA$. 
The first term in the MC equation is $Q(\mu)=m_1(\mu)$, which we can write as
\begin{subequations} \label{W-MC-d}
\begin{align}
Q(\mu) &= - \oint_{y<u} \frac{\pd_{X^i} W(X(y)_-) X^i(u)_+}{u-y}   +  \oint_{y>u}   \frac{X^i(u)_-\pd_{X^i} W(X(y)_+)}{y-u} \notag \\
&= - \oint_y \pd_{X^i} W(X(y)_-) X^i(y)_+ + \oint_y X^i(y)_-\pd_{X^i}W(X(y)_+)\,,
\end{align}
In the first line, the integration is over both $u$ and $y$, with the constraint $|y|<|u|$ (resp. $|y|>|u|$) initially imposed on $y$; we swap orders of integration and do the integral over $u$ in the domain $|u|>|y|$ (resp. $|u|<|y|$) to get to the second line.

Similarly, $\mu^2=m_2(\mu,\mu) = \frac12[\mu,\mu]$ is given by
\begin{align}
\mu^2 &= \frac{-1}{2} \hspace{-.15in}\oint_{y<u_1,u_2} \hspace{-.15in}  \frac{\pd_{X^{i_1}}\pd_{X^{i_2}} W(X(y)_-)  X^{i_1}(u_1)_+ X^{i_2}(u_2)_+}{(u_1-y)(u_2-y)}  + \frac12 \hspace{-.15in} \oint_{y>u_1,u_2} \hspace{-.15in} \frac{X^{i_1}(u_1)_-X^{i_2}(u_2)_-\pd_{X^{i_1}}\pd_{X^{i_2}} W(X(y)_+)}{(y-u_1)(y-u_2)}  \notag \\
&=  -\frac12 \oint_y  \pd_{X^{i_1}}\pd_{X^{i_2}} W(X(y)_-)  X^{i_1}(y)_+ X^{i_2}(y)_+ +\frac12 \oint_y X^{i_1}(y)_-X^{i_2}(y)_-\pd_{X^{i_1}}\pd_{X^{i_2}} W(X(y)_+)\,.
\end{align}
Again we swap orders of integration, doing all the integrals over $u$'s in the prescribed domains $|u_i|>|y|$ (resp. $|u_i|<|y|$) to get the second line. The pattern continues in a straightforward way, up to
\begin{align} 
& \hspace{-.3in}m_d(\mu^{\otimes d}) = \frac{-1}{d!} \hspace{-.2in} \oint_{y<u_1,...,u_d} \hspace{-.2in} \frac{\pd_{X^{i_1}}...\pd_{X^{i_d}} W(X(y)_-)  X^{i_1}(u_1)_+\cdots X^{i_d}(u_d)_+}{(u_1-y)\cdots (u_d-y)}
 + \frac{1}{d!} \hspace{-.2in} \oint_{y>u_1,...,u_d} \hspace{-.2in}  (\ldots)  \\
& \hspace{-.2in} = \frac{-1}{d!}\oint_y \pd_{X^{i_1}}...\pd_{X^{i_d}} W(X(y)_-)  X^{i_1}(y)_+\cdots X^{i_d}(y)_+ + \frac{1}{d!} \oint_y X^{i_1}(y)_-\cdots X^{i_d}(y)_- \pd_{X^{i_1}}...\pd_{X^{i_d}} W(X(y)_+)  \notag
\end{align}
\end{subequations}

Now we sum up the contributions in \eqref{W-MC-d}. We find that each ``column'' just comprises a Taylor expansion of the finite polynomial $W\big(X(y)_-+X(y)_+\big)$, either around $X(y)_-$ (left column) or around $X(y)_+$ (right column).  Therefore,
\begin{align} \sum_{k=1}^d m_k(\mu^{\otimes k}) = - \oint_yW\big(X(y)_-+X(y)_+\big) + \oint_y W\big(X(y)_++X(y)_-\big) = 0\,, \end{align}
which vanishes, proving that the $A_\infty$ MC equation is satisfied.

\subsubsection{The MC equation for $r$}
\label{sec:W-MC-r}

Next, consider the universal r-matrix. It's useful to shift by $z$ and flip the integration contour in \emph{one} of the two terms in $r(z)$, so that $z$ only appears in $X$'s rather than in $\psi$'s --- because only the $\psi$'s will participate in $A_\infty$ operations. Thus:
\begin{align} r(z)
  &= \oint_{s<z} \Big[ \psi_i(s)_-X'{}^i(s+z)_- - X{}^i(s)_-\psi_i'(s+z)_- \Big]  \notag \\
  &=  \oint_{s<z} \Big[ \psi_i(s)_-X'{}^i(s+z)_- + \psi_i'(s)_-X{}^i(s-z)_- \Big]\,. \label{r-W-forMC} \end{align}
Using the second expression, we find that the $A_\infty$ operations are
\begin{align} & m_k(r(z)^{\otimes k}) = - \frac{1}{d!} \hspace{-.2in} \oint_{y<s_1,...,s_k<z}  \hspace{-.2in}  \frac{\pd_{X^{i_1}}...\pd_{X^{i_k}} W(X(y)_-)X'{}^{i_1}(s_1+z)_-\cdots X'{}^{i_k}(s_k+z)_-}{(s_1-y)\cdots (s_k-y)}  \notag \\
& \hspace{1in} -\frac{1}{d!}  \hspace{-.2in} \oint_{y<s_1,...,s_k<z}  \hspace{-.2in}  \frac{\pd_{X'{}^{i_1}}...\pd_{X'{}^{i_k}} W(X'(y)_-)X{}^{i_1}(s_1-z)_-\cdots X{}^{i_k}(s_k-z)_-}{(s_1-y)\cdots (s_k-y)}  \\
&\quad =  -\frac{1}{d!}  \hspace{-.05in} \oint_{y<z} \!\! \Big[ \pd_{X^{i_1}}...\pd_{X^{i_k}} W(X(y)_-)X'{}^{i_1}(y+z)_-\cdots X'{}^{i_k}(y+z)_-  \hspace{-1.8in}\raisebox{-.3in}{$+\, \pd_{X'{}^{i_1}}...\pd_{X'{}^{i_k}} W(X'(y)_-)X{}^{i_1}(y-z)_-\cdots X{}^{i_k}(y-z)_- \Big]\,.$} \notag 
\end{align}
In the first line, the integrals are over $y$ and $s_1,...,s_k$ within the indicated domains. We swap orders of integration, doing all the $s_i$ integrals, to obtain the second line.

Summing up all the products, we again find two Taylor expansions of $W$:
\begin{align} \hspace{-.2in} \sum_{k=1}^d m_k (r(z)^{\otimes k}) &= - \oint_{y<z} \Big[W\big( X(y)_-+X'(y+z)_-\big)+ W\big( X'(y)_-+X(y-z)_-\big)\Big] \\
&= - \oint_{y<z} \Big[W\big( X(y)_-+X'(y+z)_-\big)- W\big( X'(y+z)_-+X(y)_-\big)\Big]  = 0  \notag \end{align}
After flipping the contour in the second term to shift $y\mapsto y+z$, we see that the two terms cancel, so the MC equation for $r(z)$ in $\CA^!\otimes\CA^![\![z^{-1}]\!]$ is indeed satisfied.

\subsubsection{The twisted coproduct as algebra morphism}
\label{W-MC-sum}

Finally, we'd like to show that the sum of terms $\mu(z)+\mu'(0)+r(z)$ satisfies the MC equation in $\CA^!\otimes \CA^!\otimes \CA[\![z^{-1},z]\!]$, in order to establish that the twisted coproduct $\Delta_z:\CA^!\to \CA^!\otimes_{r(z)} \CA^![\![z^{-1},z]$ is an $A_\infty$-algebra morphism.

We use the same form of $r(z)$ as in \eqref{r-W-forMC}. We also shift $z$ into the $X$ fields in $\mu(z)$,
\be \mu(z) = \oint_{u>z}\Big[ \psi_i(u)_- X^i(u+z)_+ +  X^i(u)_-\psi_i(u+z)_+ \Big]  = \oint_{u>z}\Big[ \psi_i(u)_- X^i(u+z)_+ + \psi_i(u)_+ X^i(u-z)_- \Big]\,,\ee
where, in contrast to $r(z)$, no flip of contour is required during the shift. And we have
\be \mu'(0) =  \oint_{v>z}\Big[ \psi_i'(v)_- X^i(v)_+ + \psi_i(v)_+ X'{}^i(v)_- \Big]\,. \ee
Each $A_\infty$ product $m_k(\mu(z)+\mu'(0)+r(z))^{\otimes k}$ contains just three types of terms: either $m_k$ is applied to $k$ $\psi_-$'s, or $k$ $\psi'_-$'s, or $k$ $\psi_+$'s (there are no other cross-terms). Evaluating each term using \eqref{A-W-forMC}, then swapping orders of integration to do all the respective $s,u,v$ integrals, and summing up over all $k$, we find:
\begin{align}  \hspace{-.3in} & \sum_{k=1}^d m_k(\mu(z)+\mu'(0)+r(z))^{\otimes k} = - \oint_{y<z} W\Big( X(y)_- +X'(y+z)_- + X(y+z)_+\Big) \quad \text{(from $\psi_-^{\otimes k}$)} \notag \\
 &\hspace{1in} -  \oint_{y<z} W\Big( X(y-z)_-+X'(y)_- + X(y)_+\Big) \quad \text{(from $\psi'_-{}^{\otimes k}$)} \label{mu-sum-W}  \\
 &\hspace{1in} + \oint_{y>z} W\Big( X(y-z)_- +X'(y)_- + X(y)_+\Big) \quad \text{(from $\psi_+{}^{\otimes k}$)} \notag
\end{align}
A graphical depiction of the three terms appearing in \eqref{mu-sum-W} looks like this:
\be \raisebox{-.6in}{ \includegraphics[width=4.2in]{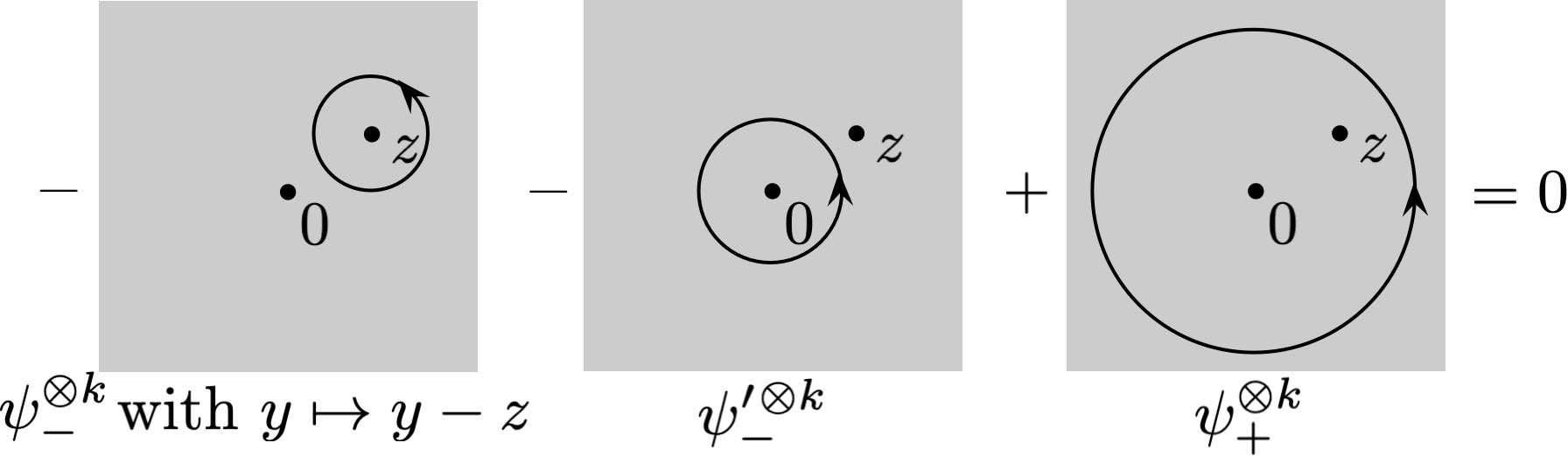}}  \label{residues} \ee

The different domains for the integration contours are (as usual) an algebraic prescription for how to expand the integrands and which residues to compute. For example, in the first term (from $\psi_-^{\otimes k}$) the integrand has singularities at $y=0$ (from $X(y)_-$), $y=-z$ (from $X'(y+z)_-$) and $y=\infty$ (from $X(y+z)_+$), and we are meant to take the residue \emph{only} at $y=0$. Thus the first term is
\be -\text{Res}_{y=0} W\Big( X(y)_- +X'(y+z)_-+ X(y+z)_+\Big)\,. \label{Res-0-W} \ee
The second and third terms in \eqref{mu-sum-W} have singularities at $y=0$, $y=z$, and $y=\infty$. They nearly cancel against each other; the third term includes the residue at $y=z$ while the second term does not, so their sum is
\be \text{Res}_{y=z} W\Big( X(y-z)_- +X'(y)_- + X(y)_+\Big)\,. \label{Res-z-W} \ee
After shifting $y\mapsto y+z$, \eqref{Res-z-W} perfectly cancels \eqref{Res-0-W}.  Therefore, $ \sum_{k=1}^d m_k(\mu(z)+\mu'(0)+r(z))^{\otimes k} =0$ and the MC equation is satisfied.

\subsection{Gauge theory}
\label{sec:Yangian-gauge}

Now consider perturbative gauge theory with Lie algebra $\fg$, matter representation $V$, and Chern-Simons level $k$. We assume for simplicity that the matter superpotential vanishes; the following arguments combine with those of Section \ref{sec:Yangian-W} in a straightforward way to handle the case of $W\neq 0$ as well.

When the superpotential vanishes, we can treat gauge theory with matter as a pure gauge theory for superalgebra $\fh = \fg\ltimes \Pi V$, as discussed in Section \ref{sec:Ainf-gauge}. The generators of $\CA$ and $\CA^!$ are the modes of $B^I(z)_\pm,c_I(z)_\pm$, where `$I$' indexes a combined basis of $\fg$ and $V$. Since the interaction is cubic, all $A_\infty$ operations $m_n$ with $n\geq 3$ vanish, and we just get dg algebras with relations summarized in Section \ref{sec:gauge-fields}.  The differentials are 
\be \label{gauge-Q-forMC} \begin{array}{ll}
Q\, c_I(z)_+ =-\tfrac12 f^{IJ}{}_K c_I(z)_+c_J(z)_+\,,&  Q\, B^I(z)_+ = (k^{IJ}+h^{IJ})\pd c_J(z)_+ - f^{IJ}{}_K c_J(z)_+B^K(z)_+\,, \\[.2cm]
Q\, c_I(z)_- =\tfrac12 f^{IJ}{}_K c_I(z)_-c_J(z)_-\,,& Q\, B^I(z)_- = -(k^{IJ}+h^{IJ})\pd c_J(z)_- + f^{IJ}{}_K c_J(z)_-B^K(z)_-\,.
\end{array}
\ee
It's convenient to express the commutators as contour integrals (to enable a swap of orders of integration); they are
\be \label{gauge-comm-forMC} \begin{array}{ll}
\ds  [B^I(z)_+,B^J(w)_+] = \hspace{-.05in} \oint_{y>z,w} \hspace{-.05in} \frac{  -f^{IJ}{}_KB^K(y)_+}{(y-z)(y-w)}\,, & \ds [B^I(z)_+,c_J(w)_+] =\hspace{-.05in} \oint_{y>z,w}\hspace{-.05in} \frac{f^{IK}{}_J c_K(y)_+}{(y-z)(y-w)}\,, \\ {}
  \ds [B^I(z)_-,B^J(w)_-] =  \hspace{-.05in}\oint_{y<z,w}\hspace{-.05in} \frac{f^{IJ}{}_K B^K(y)_-}{(z-y)(w-y)}\,, & \ds [B^I(z)_-,c_J(w)_-] =  \hspace{-.05in}\oint_{y<z,w} \hspace{-.05in}\frac{-f^{IK}{}_J c_K(y)_-}{(z-y)(w-y)}\,.
\end{array}
\ee

The MC equations for $\mu$, $r(z)$, and the sum $\mu(z)+\mu'(0)+r(z)$ all follow from manipulations very similar to matter with a superpotential in Section \ref{sec:Yangian-W}. In particular, terms in the MC equations re-sum into Taylor expansions of the interaction terms. However, since the interaction terms in gauge theory contain both bosons and fermions (in particular, both parts of the conjugate $\BB$ and $\BA$ fields), signs are quite tricky to keep track of in these expansions. As the interaction is just cubic, and the MC equations have just two terms to sum, we'll analyze them directly rather than trying to organize them into Taylor expansions.

\subsubsection{The MC equation for $\mu$}

For the universal MC element $\mu =  \oint_u \big[ c_I(u)_- B^I(u)_+ +  B^I(u)_- c_I(u)_+ \big] $, we want to show that the MC equation $Q(\mu)+\tfrac12[\mu,\mu]=0$ is satisfied.

From \eqref{gauge-Q-forMC}, relabeling the integration variables $u\mapsto y$, we've got
\begin{align} \hspace{-.3in}
Q(\mu) &=\oint_y\Big[ \tfrac12 f^{JK}{}_I c_J(y)_-c_K(y)_- B^I(y)_+ - c_I(y)_-(k+h)^{IJ} \pd c_J(y)_++ c_I(y)_-f^{IJ}{}_K c_J(y)_+B^K(y)_+ \notag \\
 &\qquad -(k+ h)^{IJ}\pd c_J(y)_-c_I(y)_+ + f^{IJ}{}_K c_J(y)_-B^K(y)_-c_I(y)_+  -\tfrac12 B^I(y)_-f^{JK}{}_I c_J(y)_+c_K(y)_+ \Big]
\end{align}
The two terms proportional to $(k+ h)$ cancel against each other after an algebraic integration by parts: $\oint_y \pd c_J(y)_-c_I(y)_+ = - \oint_y c_J(y)_-\pd c_I(y)_+$. For the remaining cubic terms, we look at $\frac12 [\mu,\mu]$. We compute this by using \eqref{gauge-comm-forMC} and swapping orders of integration, just like in \eqref{W-MC-d}:
\begin{align} \hspace{-.3in}
&\tfrac12[\mu,\mu] = \notag \\
& \hspace{-.3in}  \oint_{u,v} \tfrac12 \Big[ c_I(u)_- c_J(v)_- [B^I(u)_+,B^J(v)_+] + \underbrace{[c_I(u)_-B^I(u)_+,B^J(v)_-c_J(v)_+]}_{c_{I-}B^J_-[B^I_+,c_{J+}]+[c_{I-},B^J_-]c_{J+}B^I_+} + \tfrac12[B^I(u)_-,B^J(v)_-]c_I(u)_+c_J(v)_+\Big] \notag \\
& =  \oint_y\Big[ -\tfrac12 c_I(y)_-c_J(y)_- f^{IJ}{}_K B^K(y)_+ + c_I(y)_-B^J(y)_-f^{IK}{}_Jc_K(y)_+  \\
 & \hspace{1.5in} + f^{JK}{}_I c_K(y)_- c_J(y)_+ B^I(y)_+  + \tfrac12 f^{IJ}{}_K B^K(y)_-c_I(y)_+c_J(y)_+ \Big]\,. \notag
 \end{align}
This exactly cancels the cubic part of $Q(\mu)$ after using antisymmetry $f^{IJ}{}_K = - f^{JI}{}_K$.

\subsubsection{The MC equation for $r$}

Next, we'd like to show that the universal r-matrix 
\be r(z) = \oint_{s<z} \Big[ c_I(s)_- B'{}^I(s+z)_- - B^I(s)_- c'_I(s+z)_-\Big]  \label{rz-G} \ee
satisfies the MC equation $Q(r(z))+\tfrac12[r(z),r(z)]=0$ in $\CA^!\otimes \CA^![\![z,z]$.

We compute $Q(r(z))$ directly using \eqref{gauge-Q-forMC},
\begin{align} \label{r-gauge-quadratic}
 Q\,r(z) &= \oint_{s<z}\Big[ \tfrac12 f^{JK}{}_I c_J(s)_-c_K(s)_- B'{}^I(s+z)_- + c_I(s)_-(k+ h)^{IJ}\pd c'_J(s+z)_-   \\[-.3cm] & \hspace{.5in}  -c_I(s)_-f^{IJ}{}_K c'_J(s+z)_-B'{}^K(s+z)_- +(k+ h)^{IJ}\pd c_J(s)_-c'_I(s+z)_- \notag \\[.2cm]
& \hspace{.5in}   -f^{IJ}{}_K c_J(s)_-B^K(s)_-c'_I(s+z)_- - B^I(s)_- \tfrac12 f^{JK}{}_I c'_J(s+z)_-c'_K(s+z)_- \Big]   \notag
\end{align}
The quadratic terms, proportional to $(k+h)$, vanish after an algebraic integration by parts $\oint \pd c_J(s) c'_I(s+z) = -\oint c_J(s)\pd c'_I(s+z)$.

For the cubic terms, we compare add the $[r(z),r(z)]$ commutator. To compute $[r(z),r(z)]$, we shift integration variables (and do contour flips) in $r(z)$ to ensure that the individual fields involved in each term of the commutator are independent of $z$. This assures that after applying \eqref{gauge-comm-forMC} we can swap orders of integration and perform all the $s,t$ integrals below cleanly to simplify the result. We get:
\begin{align}
&\tfrac12[r(z),r(z)] = \notag \\
&\oint_{s,t<z}\Big[ \tfrac12 c_I(s-z)_-c_J(t-z)_- [B'{}^I(s)_-, B'{}^J(t)_-] -c_I(s-z)_-B^J(t-z)_-[B'{}^I(s)_-,c'_J(t)_-] \notag \\[-.3cm]
&\hspace{.5in} - [c_I(s)_-,B^J(t)_-]c'_J(t+z)_-B'{}^I(s+z)_- +\tfrac12 [B^I(s)_-,B^J(t)_-]c'_I(s+z)_-c'_J(t+z)_-\Big] \notag \\[.2cm]
&= \oint_{y<z} \Big[ \tfrac12  c_I(y-z)_-c_J(y-z)_- f^{IJ}{}_K B'{}^K(y)_- + c_I(y-z)_-B^J(y-z)_- f^{IK}{}_J c'_K(y)_- \notag \\[-.3cm]
&\hspace{.5in} - f^{JK}{}_I c_K(y)_- c'_J(y+z)_-B'{}^I(y+z)_- + \tfrac12 f^{IJ}{}_K B^K(y)_- c'_I(y+z)_-c'_J(y+z)_-\Big] \notag \\[.2cm]
&=\oint_{y<z} \Big[-f^{IJ}{}_K \tfrac12  c_I(y)_-c_J(y)_-  B'{}^K(y+z)_- - f^{IK}{}_Jc_I(y)_-B^J(y)_-  c'_K(y+z)_- \notag \\[-.3cm]
&\hspace{.5in} - f^{JK}{}_I c_K(y)_- c'_J(y+z)_-B'{}^I(y+z)_- + \tfrac12 f^{IJ}{}_K B^K(y)_- c'_I(y+z)_-c'_J(y+z)_-\Big]  \label{r-gauge-cubic}
\end{align}
In the final line, we've shifted the first two terms back to $(y,y+z)$ form, flipping contours and adding minus signs. The four terms in \eqref{r-gauge-cubic} exactly cancel the remaining cubic terms in \eqref{r-gauge-quadratic}.

\subsubsection{The twisted coproduct as algebra morphism}

Finally, we consider the sum $\mu(z)+\mu'(0)+r(z)$, with
\begin{align} 
&\hspace{-.2in} r(z) = \oint_{s<z} \Big[ c_I(s)_- B'{}^I(s+z)_- - B^I(s)_- c'_I(s+z)_-\Big] = \oint_{s<z} \Big[  -c_I(s-z)_- B'{}^I(s)_- + B^I(s-z)_- c'_I(s)_-\Big]\,,  \notag \\
& \hspace{-.2in} \mu(z) = \oint_u \Big[ c_I(u)_-B^I(u+z)_++B^I(u)_-c_I(u+z)_+\Big] = \oint_{u>z} \Big[ c_I(u-z)_-B^I(u)_++B^I(u-z)_-c_I(u)_+\Big]\,, \notag \\
& \hspace{-.2in} \mu'(0) = \oint_v \Big[ c_I'(u)_-B^I(u)_++B'{}^I(u)_-c_I(u)_+\Big]\,. \label{Ms-G}
\end{align}
We want to show that $Q(\mu(z)+\mu'(0)+r(z))+(\mu(z)+\mu'(0)+r(z))^2 = 0$. We already know that $\mu,\mu',$ and $r$ separately satisfy the MC equation, so it remains to show that
\be [\mu(z),\mu'(0)]+[r(z),\mu'(0)]+[r(z),\mu(z)]=0\,. \ee

$[\mu(z),\mu'(0)]$ gets contributions from commutators of $B_+,c_+$ fields. We shift $u\mapsto u-z$ in $\mu(z)$, using the second expression in \eqref{Ms-G}, so that the fields in the commutators are independent of $z$. Applying \eqref{gauge-comm-forMC} and doing the $u,v$ integrals then gives
\be \hspace{-.0in} \raisebox{.3in}{$\ds [\mu(z),\mu'(0)] = \!\! \oint_{y>z} \!\! \Big[ - f^{IJ}{}_K c_I(y-z)_-c'_J(y)_- B_K(y)_+ +f^{IK}{}_J c_I(y-z)_-B'{}^J(y)_-c_K(y)_+ 
   \hspace{-3in}\raisebox{-.35in}{$ +  f^{JK}{}_I B^I(y-z)_-c'_J(y)_- c_K(y)_+ \Big]$}  $}
\ee
Similarly, $[r(z),\mu'(0)]$ gets contributions from commutators of $B_-',c_-'$ fields. We first shift $s\to s-z$  in $r(z)$ so that these fields are independent of $z$. Then
\be \hspace{-.0in} \raisebox{.3in}{$\ds [r(z),\mu'(0)] = \!\! \oint_{y<z} \!\! \Big[  f^{IK}{}_J c_I(y-z) c'_K(y)_- B^J(y)_+ - f^{IJ}{}_K c_I(y-z) B'{}^K(y)_- c_J(y)_+      \hspace{-3in}\raisebox{-.35in}{$    + f^{JK}{}_I B^I(y-z)_- c'_K(y)_- c_I(y)_+    \Big]$}  $}
\ee
Observe that the integrands in $[\mu(z),\mu'(0)]$ and $[r(z),\mu'(0)]$ are identical up to a sign, but integration domains differ. Taking the difference leaves behind the algebraic residue at $y=z$, schematically $\oint_{y>z} f(y) - \oint_{y<z}f(y) = \text{Res}_{y=z} f(y)$.  After shifting $y\mapsto y+z$ to make this the residue at zero $\text{Res}_{y=0} f(y+z)$, we get
\begin{align} & [\mu(z),\mu'(0)] + [r(z),\mu'(0)] =    \oint_{y<z} \!\! \Big[ -f^{IJ}{}_K c_I(y)_-c'_J(y+z)_- B_K(y+z)_+ \notag \\
& +f^{IK}{}_J c_I(y)_-B'{}^J(y+z)_-c_K(y+z)_+ 
 +  f^{JK}{}_I B^I(y)_-c'_J(y+z)_- c_K(y+z)_+\Big] 
\end{align}
This precisely cancels $[r(z),\mu(z)]$, which gets its contributions from $c_-,B_-$ commutators. We use the form of $r(z)$ on the LHS so that $c_-,B_-$ fields are independent of~$z$,
\begin{align} & [r(z),\mu(z)] = \oint_{y<z} \Big[ f^{JK}{}_I c_K(y)_- B'{}^I(y+z)_- c_J(y+z)_+ \\
 &\hspace{.3in} - f^{IK}{}_J c_K(y)_- c'_I(y+z)_- B^J(y+z)_+  - f^{IJ}{}_K B^K(y) c'_I(y+z)_- c_J(y+z)_+ \Big] \notag \end{align}
obtaining $ [\mu(z),\mu'(0)] + [r(z),\mu'(0)] +  [r(z),\mu(z)]  = 0$.

Geometrically, the cancellations among these three terms can be depicted as
\be \raisebox{-.8in}{\includegraphics[width=3.9in]{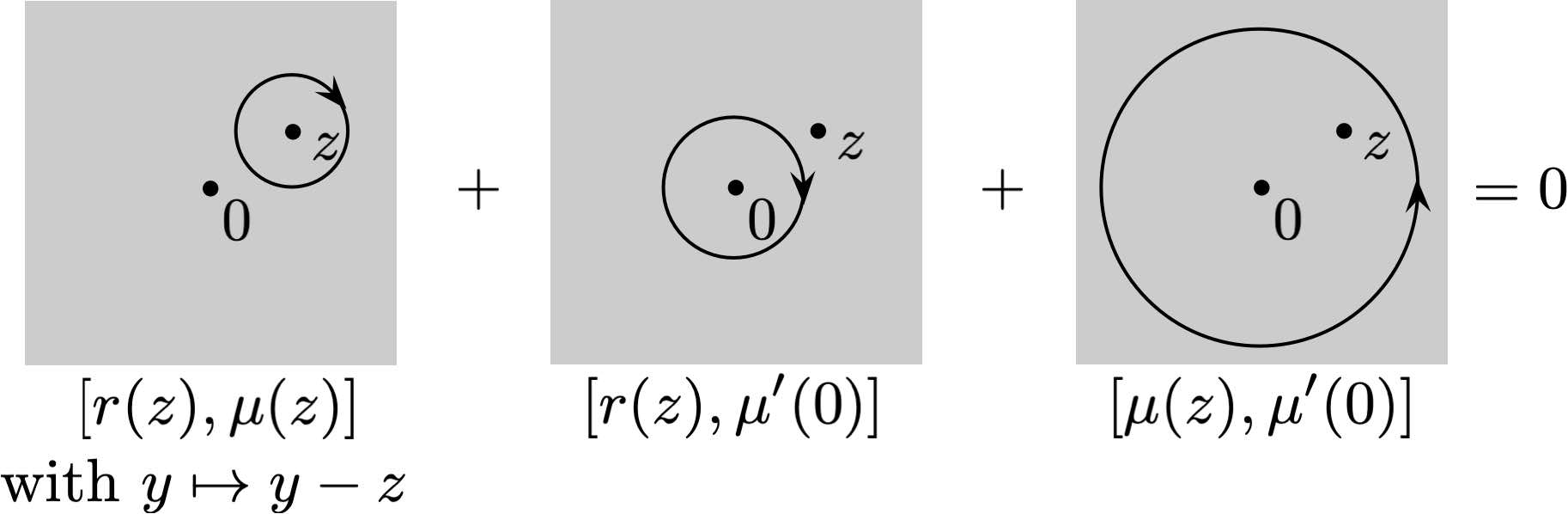}} \ee

\appendix

\section{The r-matrix and the $\lambda$-bracket}
\label{app:r-lambda}

We'd like to explain a particular relation between the universal r-matrix $r(z)\in \CA^!\otimes \CA^![\![z^{-1}]\!]$ and the $\lambda$-bracket $\{-{}_\lambda-\}:\CA\otimes \CA\to \CA[\![\lambda]\!]$ in a perturbative 3d HT QFT --- roughly that the two are inverses of each other. We'll do this via explicit formulas. 
We note, however, that the result parallels other known (or expected) transfers of structure across Koszul duality; in particular:
\begin{itemize}
\item In a 3d perturbative, topological QFT, local operators $\CA$ form an $E_3$ algebra, with a non-degenerate Poisson bracket of cohomological degree $-2$, \cf\ \cite{Lurie-DAGVI,AyalaFrancis,descent} and references therein for a modern and physical perspective. Their Koszul dual $\CA^!$ then naturally forms a quasi-triangular bialgebra, with R-matrix given by the inverse of the Poisson bracket \cite{Tamarkin-formality, tamarkin2007quantization, Lurie-DAGVI, AyalaFrancis-Koszul,CostelloFrancisGwilliam}. See \cite[Appendix~C]{sparks} for a geometric derivation of the correspondence, using integrated descendants.

\item In 4d perturbative HT QFT, the structure is similar, except one of the topological directions becomes chiral. Local operators $\CA$ form an ($E_2$,\,chiral) algebra with a $\lambda$-bracket (the chiral extension of a Poisson bracket) of  degree $-2$ \cite{OhYagi-Poisson}. The Koszul dual $\CA^!$ forms a generalized Yangian (a chiralization of a quasi-triangular bialgebra), and one expects that its R-matrix $R(z)$ is the inverse of the $\lambda$-bracket.

\item In our 3d perturbative HT QFT's, local operators $\CA$ form an ($E_1$,\,chiral) algebra, with a $\lambda$-bracket of degree $-1$ \cite{OhYagi-Poisson,GarnerWilliams,AlfonsiKimYoung}. Following the pattern, one would \emph{expect} that the Koszul dual $\CA^!$ has an R-matrix given by the inverse of the $\lambda$-bracket --- except that due to the degree shift, the R-matrix lands in cohomological degree $+1$. Being an odd element, it's necessarily infinitesimal (an ``r-matrix''), and we claim that it's precisely our $r(z)$.
\end{itemize}

Notice that we always have $\text{deg($R/r$-matrix)}-\text{deg}(\text{bracket}) = 2$\,; 
the $R/r$-matrix and the brackets are chiral precisely when there's a $\C$ direction; and there's a quantum $R$ (vs. infinitesimal $r$) matrix when there are three (vs. two) cohomologically trivial spacetime directions (either $\R$ or $\C$).

This suggests another basic case one might consider: in a perturbative 2d topological QFT, local operators $\CA$ form an $E_2$ algebras, with nondegenerate Poisson bracket of degree -1, better known as the Gerstenhaber bracket \cite{WittenZwiebach, LianZuckerman, PenkavaSchwarz, Getzler-2d}. Extrapolating from the current paper, one might expect the Koszul dual $\CA^!$ to be a ``dg-shifted quantum group,'' \ie\ to have an MC element $r\in \CA^!\otimes\CA^!$ of degree $1$, that behaves like an infinitesimal r-matrix and inverts the bracket on $\CA$. 
%
%
We hope to explore this in future work.

Now let's come back to perturbative 3d HT QFT's. Choose a basis $\{x^i_{-n-1},p_{i,-n-1}\}_{n\geq 0}=\big\{\tfrac1{n!}\pd^n x^i, \tfrac1{n!}\pd^n p_i\big\}$ for the linear subspace $\CA_1\subset \CA$ that generates $\CA$, which is element-wise dual to the basis  $\{p_{i,n},x^i_n\}_{n\geq 0}$ for the linear subspace $\CA_1^!$ that generates $\CA^!$. Specifically, these bases are dual under the shifted-symplectic pairing in the BV action \eqref{S2d-gen}, which identifies
\be   \CA^!_1 \simeq (\Pi \CA_1[1])^*\,, \label{AA*} \ee
where $\Pi$ denotes parity shift, $[1]$ denotes cohomological shift, and $(\,)^*$ is linear dual.

Both the $\lambda$-bracket and the r-matrix are ultimately controlled by the free-field propagator.
We can compute the $\lambda$-bracket of basis elements using \eqref{def-lambda} to get
\begin{align}  \{x^i_{-n-1} {}_{\,\lambda\,} p_{j,-m-1}\}(0) &:=  \oint_{S^2} e^{\lambda z} \frac1{n!m!} \big\langle\pd^n  x^i{}^{(1)}(z,t) \pd^m p_j(0)\big\rangle \notag \\
&= \oint_{S^2} e^{\lambda z} \frac{(n+m)!}{n!m!} (-1)^n \delta^i{}_j\frac{1}{z^{n+m+1}}\delta^{(1)}(t)dz \notag \\
&= \frac{(n+m)!}{n!m!} (-1)^n \delta^i{}_j \oint_{S^1} e^{\lambda z} z^{-n-m-1} dz = \frac{(-1)^n}{n!m!} \delta^i{}_j\,\lambda^{n+m}
\end{align}
Similarly, $\{p_{i,-n-1} {}_{\,\lambda\,} x^j_{-m-1}\} = - \frac{(-1)^n}{n!m!} \delta^j{}_i\,\lambda^{n+m}$\,. Therefore, we may re-write $r(z)$ (using the form \eqref{KD-univ-r}, after a $(-1)^F$ twist in $\CA^!$) as
\begin{align}
r(z) &= \sum_{m,n\geq 0} \bigg( \begin{array}{@{}c@{}} n+m \\ n \end{array} \bigg) \frac{ x^i_n\otimes  p_{i,m} -p_{i,n} \otimes x^i_m   }{(-1)^n z^{n+m+1}}  \notag \\
 &= \sum_{n,m=0}^\infty x^i_n\otimes p_{j,m} \int_0^\infty d\lambda \, e^{-\lambda z}\, \{ p_{i,-n-1} {}_{\,\lambda\,} x^j_{-m-1}\} + \sum_{n,m=0}^\infty p_{i,n}\otimes x^j_m \int_0^\infty d\lambda \, e^{-\lambda z}\, \{ x^i_{-n-1} {}_{\,\lambda\,} p_{j,-m-1}\}
\end{align}

In greater generality, let $\{\Phi_\alpha\}$ denote a basis for the linear generators of $\CA$ and $\{\Phi^\alpha\}$ the dual linear generators for $\CA^!$, canonically paired under the identification \eqref{AA*}. Then
\be r(z) = \sum_{\alpha,\beta} \Phi^\alpha \otimes \Phi^\beta \int_0^\infty d\lambda \, e^{-\lambda z}\, \{ \Phi_\alpha {}_{\,\lambda\,} \Phi_\beta \}\,. \label{r-lambda} \ee

\newpage

\bibliographystyle{ytamsalpha}

\bibliography{OPE}

\end{document}